\documentclass[10pt,a4paper,twoside]{report}

\usepackage[utf8]{inputenc}   %
\usepackage{acro}

\usepackage[english]{babel} %

\newcommand{\acknowledgments}{@undefined} %
\addto\captionsenglish{\renewcommand{\acknowledgments}{Acknowledgments}}

\addto\captionsportuguese{\renewcommand{\acknowledgments}{Agradecimentos}}
\addto\captionsportuguese{} %

\newcommand{\coverThesis}{@undefined} %
\newcommand{\coverSupervisors}{@undefined} %
\newcommand{\coverExaminationCommittee}{@undefined} %
\newcommand{\coverChairperson}{@undefined} %
\newcommand{\coverSupervisor}{@undefined} %
\newcommand{\coverMemberCommittee}{@undefined} %
\addto\captionsenglish{\renewcommand{\coverThesis}{Thesis to obtain the Master of Science Degree in}}
\addto\captionsenglish{\renewcommand{\coverSupervisors}{Supervisors}}
\addto\captionsenglish{\renewcommand{\coverExaminationCommittee}{Examination Committee}}
\addto\captionsenglish{\renewcommand{\coverChairperson}{Chairperson}}
\addto\captionsenglish{\renewcommand{\coverSupervisor}{Supervisor}}
\addto\captionsenglish{\renewcommand{\coverMemberCommittee}{Member of the Committee}}
\addto\captionsportuguese{\renewcommand{\coverThesis}{Disserta\c{c}\~{a}o para obten\c{c}\~{a}o do Grau de Mestre em}}
\addto\captionsportuguese{\renewcommand{\coverSupervisors}{Orientador(es)}}
\addto\captionsportuguese{\renewcommand{\coverExaminationCommittee}{J\'{u}ri}}
\addto\captionsportuguese{\renewcommand{\coverChairperson}{Presidente}}
\addto\captionsportuguese{\renewcommand{\coverSupervisor}{Orientador}}
\addto\captionsportuguese{\renewcommand{\coverMemberCommittee}{Vogal}}

\def\FontLb{%
  \usefont{T1}{ptm}{b}{n}\fontsize{16pt}{16pt}\selectfont}

\def\FontMb{%
  \usefont{T1}{ptm}{b}{n}\fontsize{14pt}{14pt}\selectfont}
\def\FontSn{%
  \usefont{T1}{ptm}{m}{n}\fontsize{12pt}{12pt}\selectfont}

\usepackage{geometry}	
\usepackage{setspace}

\usepackage{graphicx}
\usepackage{amsmath}  
\usepackage{amsthm}   
\usepackage{amsfonts}

\usepackage{dcolumn}
\newcolumntype{d}{D{.}{.}{-1}} %
\newcolumntype{e}{D{E}{E}{-1}} %

\usepackage{nomencl}
\makenomenclature

\RequirePackage{ifthen} 
\ifthenelse{\equal{\languagename}{english}}%
    { %
    \renewcommand{\nomgroup}[1]{%
      \ifthenelse{\equal{#1}{R}}{%
        \item[\textbf{Roman symbols}]}{%
        \ifthenelse{\equal{#1}{G}}{%
          \item[\textbf{Greek symbols}]}{%
          \ifthenelse{\equal{#1}{S}}{%
            \item[\textbf{Subscripts}]}{%
            \ifthenelse{\equal{#1}{T}}{%
              \item[\textbf{Superscripts}]}{}}}}}%
    }{%
    \renewcommand{\nomgroup}[1]{%
      \ifthenelse{\equal{#1}{R}}{%
        \item[\textbf{Simbolos romanos}]}{%
        \ifthenelse{\equal{#1}{G}}{%
          \item[\textbf{Simbolos gregos}]}{%
          \ifthenelse{\equal{#1}{S}}{%
            \item[\textbf{Subscritos}]}{%
            \ifthenelse{\equal{#1}{T}}{%
              \item[\textbf{Sobrescritos}]}{}}}}}%
    }%

\usepackage{rotating}

\usepackage{hyperref}    
\hypersetup{colorlinks=true,       %
            linkcolor=red,  %
            anchorcolor=black,%
            citecolor=blue,  %
            filecolor=black,  %
            menucolor=black,  %
            urlcolor=black,   %
	          bookmarksopen=false,    %
	          bookmarksnumbered=true, %
	          pdftitle={Thesis},
            pdfauthor={Andre C. Marta},
            pdfsubject={Thesis Title},
            pdfkeywords={Thesis Keywords},
            pdfstartview=FitV,
            pdfdisplaydoctitle=true}

\usepackage[figure,table]{hypcap}

\usepackage[numbers,sort&compress]{natbib} %

\usepackage{notoccite}

\usepackage{multirow}

\usepackage{booktabs}

\usepackage{pdfpages}

\usepackage{pifont}
\usepackage{tabulary}
\usepackage{amssymb}

\newcommand{\degree}{\ensuremath{^\circ\,}} %

\def\bm#1{\mathchoice                             %
  {\mbox{\boldmath$\displaystyle#1$}}%
  {\mbox{\boldmath$#1$}}%
  {\mbox{\boldmath$\scriptstyle#1$}}%
  {\mbox{\boldmath$\scriptscriptstyle#1$}}}

\usepackage[utf8]{inputenc}
\usepackage[T1]{fontenc}

\let\iff\Leftrightarrow

\newenvironment{nalign}{
    \begin{equation}
    \begin{aligned}
}{
    \end{aligned}
    \end{equation}
    \ignorespacesafterend
}

\newenvironment{nalignat}[1][2]{
    \begin{equation}
    \begin{alignedat}{#1}
 }{
    \end{alignedat}
    \end{equation}
    \ignorespacesafterend
}

\usepackage{physics}
\usepackage{tikz}
\usepackage{float}
\usepackage{comment}

\usepackage{subcaption}

\usepackage{cancel}

\usepackage{algpseudocode,algorithm,algorithmicx}
\usepackage{setspace}
\algtext*{EndWhile}%
\algtext*{EndIf}%
\algtext*{EndFor}
\algnewcommand{\algorithmicand}{\textbf{ and }}
\algnewcommand{\algorithmicor}{\textbf{ or }}
\algnewcommand{\OR}{\algorithmicor}
\algnewcommand{\AND}{\algorithmicand}
\algnewcommand{\Var}{\texttt}

\usepackage{siunitx}

\newcommand{\CG}[1]{\text{#1}\textsubscript{CG}}

\usepackage[section]{placeins}

\usepackage[export]{adjustbox}

\newcommand{\E}{\mathcal{P}}

\newcommand{\Lag}{\mathcal{L}}
\newcommand{\G}{\mathcal{G}}
\newcommand{\V}{\mathcal{V}}
\newcommand{\nodesnum}{N}
\newcommand{\sumnodes}[1]{\sum_{#1 \in V}}
\newcommand{\edgesnum}{M}

\newcommand{\Neigh}{\mathcal{N}}
\newcommand{\R}{\mathbb{R}}

\newcommand{\dt}{\Delta \tau}

\newcommand{\lat}{\Phi}
\newcommand{\lon}{\lambda}

\newcommand{\Phy}{\textit{Physarum}}
\newcommand{\Pp}{\textit{Physarum polycephalum}}

\newcommand{\HP}{Hagen-Poiseuille}

\begin{document}

\pagestyle{plain}
\pagenumbering{roman}

\thispagestyle {empty}

\begin{flushleft}
\includegraphics[width=5cm]{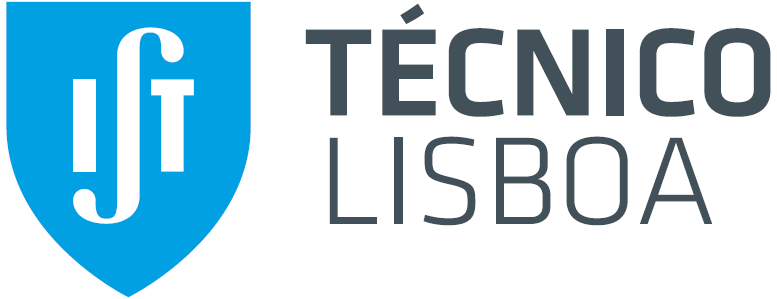}
\end{flushleft}

\begin{center}

\includegraphics[width=7cm]{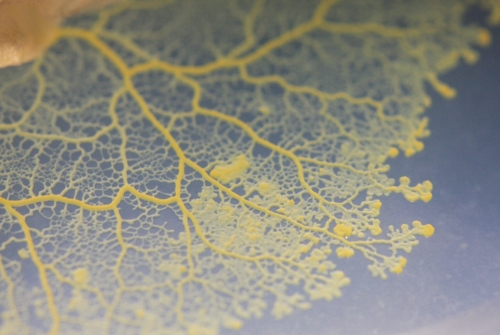}

\vspace{1.cm}

{\FontLb Formation and Optimisation of Vein Networks in \textit{Physarum} } \\
\vspace{2.5cm}
{\FontMb Rodrigo Fernandes Ferreira de Almeida} \\
\vspace{2.75cm}
{\FontSn \coverThesis} \\
\vspace{0.3cm}
{\FontLb Engineering Physics} \\
\vspace{1.5cm}
{\FontSn 
\begin{tabular}{ll}
 \coverSupervisors: & Prof. Dr. Rui Manuel Agostinho Dilão
\end{tabular}} \\
\vspace{1.0cm}
{\FontMb \coverExaminationCommittee} \\
\vspace{0.3cm}
{\FontSn
\begin{tabular}{rl}
\coverChairperson:     & Prof. Dr. Ilídio Lopes \\
\coverSupervisor:      & Prof. Dr. Rui Manuel Agostinho Dilão \\
\coverMemberCommittee: & Profa. Dra. Ana Maria Ribeiro Ferreira Nunes 
\end{tabular} } \\
\vspace{2.cm}
{\FontMb July 2021} \\
\end{center}

\cleardoublepage

\null\vskip5cm%
\begin{flushright}
    To my mother and my girlfriend. 
\end{flushright}
\vfill\newpage

\cleardoublepage

\section*{\acknowledgments}

\addcontentsline{toc}{section}{\acknowledgments}

I would like to express my profound appreciation and gratitude to my supervisor Professor Rui Dilão for the opportunity,  his immense dedication and continued guidance, and for all the valuable insights provided. Most importantly, for never giving up on me. 

I am extremely grateful to my amazing and truly supportive girlfriend which has always been there for me, helped me with my struggles and celebrated my victories. Without your love, constant support and motivation I definitely couldn't have done it.

I'm also eternally grateful to my parents and family for their sacrifices and the  unconditional support they provided me with in every moment of my life. In particular, to my mother, for always believing in me and for all the wise words and encouragement in this last year. 

Last but not least, I would like to thank all my dearest friends for helping me keep my sanity, and for all the pep talks and constant 
support throughout this long journey. 

I truly admire you all, and you inspired me to be who I am today. I couldn't have asked for better.

\cleardoublepage

\section*{Resumo}

\addcontentsline{toc}{section}{Resumo}

O \Pp{} é um bolor limoso acelular que se desenvolve como uma rede de veias altamente adaptativa onde circula o protoplasma. À medida que procura nutrientes, o \textit{Physarum} mostra uma contínua reestruturação da sua rede como resposta a estímulos locais, otimizando assim as ligações entre fontes de alimento. Este comportamento de alto nível já foi explorado para resolver vários problemas de otimização. Este trabalho foca-se na construção de um modelo para a formação da rede adaptativa do \textit{Physarum} que resolve algumas inconsistências de modelos anteriores. Começamos por derivar uma classe geral de equações que descrevem a adaptação e o fluxo de uma rede estática, composta por canais elásticos que transportam um fluido incompressível submetido a um escoamento de Hagen-Poiseuille. Uma parametrização específica do modelo é obtida através da minimização da potência dissipada pela rede. Considerando uma parametrização mais genérica, descobrimos uma transição de fase no sistema. O modelo é aplicado à resolução de labirintos, e na construção de redes eficientes e robustas numa geometria que representa Portugal continental. 
Comparando as redes resultantes com o sistema ferroviário português, verificou-se que as redes produzidas pelo modelo são em geral mais eficientes quando são consideradas flutuações nos fluxos dos canais. 
Finalmente, o modelo de adaptação é estendido para incorporar o crescimento da rede na presença de múltiplas fontes de alimento. O acoplamento de ambos os processos produz redes com características semelhantes a vários sistemas de rede encontrados na natureza. Os resultados revelam que, quando as fontes de alimento operam alternadamente, o modelo consegue replicar as conexões diretas entre fontes de alimento observadas no \textit{Physarum}.

\vfill

\textbf{\Large Palavras-chave:} \Pp, Rede adaptativa, Optimisação de redes, Crescimento de redes, \HP{}

\cleardoublepage

\section*{Abstract}

\addcontentsline{toc}{section}{Abstract}

\Pp{} is an acellular slime mould that grows as a highly adaptive network of veins filled with protoplasm. As it forages, \textit{Physarum} dynamically rearranges its network structure as a response to local stimuli information, optimising the connection between food sources. This high-level behaviour was already exploited to solve numerous optimisation problems. 

We develop a flow-based model for the adaptive network formation of \textit{Physarum}, which solves some inconsistencies of previous models. We first derive a general class of equations describing the adaptation and flow dynamics of a static network comprised of elastic channels filled with an incompressible fluid undergoing a Hagen-Poiseuille flow.

An explicit form of the model is obtained by minimising the total power dissipated by the network. Considering a more general 
functional form 
of the adaptive equations, a phase transition in the system is also found. The model is used for maze-solving and 
to build efficient and resilient  networks in an arena mimicking mainland Portugal. By comparing the resulting networks with the real Portuguese railway system, we found that the model produced networks with a better overall performance when considering fluctuations in the network flows.

Finally, the adaption model is extended to incorporate the network growth in the presence of multiple food sources. 
The coupling of both processes produces networks with similar traits 
to several network systems found in nature. 

We found that when the food sources operate alternately, 
the model can replicate the direct connections between the food sources observed in \textit{Physarum}. 

\vfill

\textbf{\Large Keywords:} \Pp, Adaptive network, Network optimisation, Network growth, \HP{} flow

\cleardoublepage

\tableofcontents
\cleardoublepage 

\phantomsection
\addcontentsline{toc}{section}{\listtablename}
\listoftables
\cleardoublepage 

\phantomsection
\addcontentsline{toc}{section}{\listfigurename}
\listoffigures
\cleardoublepage 

\phantomsection
\addcontentsline{toc}{section}{List of Abbreviations}
\chapter*{Nomenclature}

\subsection*{Acronyms}

\begin{tabular}{@{}l@{\hskip 0.5in}l@{}}
    \textbf{MST}  & Minimum Spanning Tree  \\
    \textbf{CG}   & Complete Graph  \\
    \textbf{TL}   & Total Length  \\
    \textbf{TE}   & Transport Efficiency \\
    \textbf{FT}   & Fault Tolerance
\end{tabular}

\subsection*{Variables}

\begin{tabular}{@{}l@{\hskip 0.5in}l@{}}
    $L_{ij}$ & Length of the edge $(i,j)$ \\
    $D_{ij}$ & Conductivity of the edge $(i,j)$ \\
    $Q_{ij}$ & Volumetric flux flowing through the edge $(i,j)$ \\
    $\V_{ij}$ & Volume of the edge $(i,j)$ \\
    $\V$      & Network's total volume \\
    $\E_{ij}$ & Power dissipated by the channel $(i,j)$ \\
    $\E$      & Total power dissipated by the network \\
    $p_{i}$  & Pressure of the node $i$ \\
    $q_{i}$  & Net volumetric flux of the node $i$ \\
    $I_0$    & Total inlet flux given by the source nodes \\
    $E$ & Set of edges of a graph \\
    $V$ & Set of vertices (nodes) of a graph \\
    $\eta$ & Fluid's dynamic viscosity \\
    $\beta$ & $\sqrt{8\pi\eta}$ \\
    $\dt$ & Time step of the network optimisation algorithm 
\end{tabular}

\cleardoublepage

\setcounter{page}{1}
\pagenumbering{arabic}

\chapter{Introduction}
\label{chapter:introduction}

\section{Motivation}

Transport networks appear in a variety of forms in nature, from river networks to the organ systems of multicellular organisms.
Examples of the latter include the leaf venation in plants, composed by the xylem and phloem; the vascular system in animals made up of arteries, veins and capillaries; the mycelial cords of  fungi, and the plasmodial veins of slime moulds. 
In all these cases, the networks play a key role in distributing resources and information throughout the entire organism in a rapid and efficient manner, overcoming  the size limitations of purely diffusive transport. These flow systems are thus indispensable for the organisms' development, fitness and survival.

These flow networks are composed of tubular vessels with different lengths and thicknesses which are typically organised in a hierarchical tree fashion. In general, they contain redundant connections, forming loopy structures that make them more robust and tolerant to damage and also improve transport efficiency under fluctuating loads \cite{Katifori2010,Corson2010}. The architecture of the networks depends on its functionality and generally has a decentralised nature, in the sense that it emerges from local responses to environmental stimuli. 
The vessel dimensions and hierarchical organisation have a profound impact on the transport efficiency of the nutrients and other resources by affecting the local fluid flow.

Transport networks are also present in different aspects of human life, from road, railway and communication networks, to irrigation systems and power grids, which  are crucial for industrial development and to satisfy human needs. 

Recently, attention has focused on the acellular slime mould,
\Pp, as an ideal model organism to study the interplay  between structure and function in biological transport networks, and to understand the mechanism underlying 
several complex behaviours displayed by simple organisms, such as the adaptive network formation. \textit{Physarum} is a single-celled amoeboid organism that grows as an extensive and highly adaptive network of protoplasmic veins. As it forages and progressively accommodates new food sources, \textit{Physarum}  dynamically optimises the connections between them, by adapting the thickness of the network veins.
The adaptation is believed to be related to local changes in the flux flow, driven by rhythmical contractions of the veins walls whose frequency and amplitude are regulated by food sources and other external stimuli \cite{Alim2013}. However, the  main biochemical and physical oscillator underlying the rhythmic behaviour and mobility of \textit{Physarum} has not been yet identified \cite{Teplov_2017}. 

Despite lacking any kind of neural circuit, \textit{Physarum} displays high-level behaviours, arising from the network adaptation. For instance, it can solve mazes \cite{Nakagaki2000}, and in the presence of multiple food sources, it can build networks with a trade-off between total length, transport efficiency and fault tolerance, comparable with real human-made networks \cite{Tero1}. In this regard, different toy models have been proposed to describe the network optimisation, based on the current-reinforcement principle \cite{TeroPS,Tero1}, where the flow modifies the network architecture, which in turn affects the flows. But in general, they reveal to be inconsistent with their own assumptions and don't incorporate the network formation.

Understanding the basic rules underlying its complex behaviours is of interdisciplinary interest. Not only as a guide to design efficient decentralised networks in different domains, but the slime mould computational abilities have  been already proved useful to solve more complex network optimisation problems \cite{PpReview,Gao2019}, design bioelectronic circuits and unconventional computing devices\cite{Adamatzky2016}. Furthermore, network formation and amoeboid locomotion are intrinsic features to processes of wound healing and metastasis formation. The study of these mechanisms in \textit{Physarum}may thus reveal possible insights into cancer research.  Until now, there isn't a single model which can capture all the features of\textit{Physarum}'s network self-organisation, as it's a complex task and the mechanism underlying its behaviour is not clear yet.  
 
\section{Objectives}

The goal of this thesis is to build a generic model for the network formation and optimisation  as observed during the growth of \Pp. 
The current state-of-the-art models lack many of the \textit{Physarum}’s biological features, such as growth, or  violate some basic physical principles \cite{Tero1}. Some modelling approaches are even purely phenomenological \cite{Jones2010}, lacking, therefore, any biological insights. We aim to construct a more realistic flow-based model which addresses these issues. 

\textit{Physarum} will be modelled as a flow network composed of adaptive channels filled with an incompressible and viscous fluid (protoplasm) and whose diameters change in response to the flux flowing through them. 
The flow is assumed to have a \HP{} profile and is driven by a set of flux sources and sinks which mimic stimulated regions of the organism. 
The model is inspired by a previous flow-based model, but 
takes the conservation of volume of the circulating fluid into account, as required by a \HP{} flow. Previous models \cite{Tero1} describe the channels' adaptation through \textit{ad hoc} local evolution laws which violate the above physical assumptions.  

We will begin by modelling the adaptation of static organisms, where the network growth is at first neglected. Afterwards, we explore some of the model's main features and applications. Finally, we attempt to mimic the \textit{Physarum}'s dynamic adaptive network formation, by incorporating a growth mechanism into the adaptation  model explicitly dependent on nutrients supplied by food sources.  

\section{Thesis Outline}

This thesis is organised into six chapters.
In chapter \ref{chapter:Physarum} we begin with a biological description of \textit{Physarum} and discuss its main high-level behaviours. The two seminal models proposed to describe the adaptive network formation are also explained in detail.

In chapter \ref{chapter:New Model}, we formulate the new  model for the network optimisation, founded on physical principles of a \HP{} flow. 
As a first insight, we start by performing simple tests on the model, considering a fixed set of terminals and present some of its main features.
In particular, we study the adaptation dynamics for a particular 
functional form of the model on a configuration of terminals mimicking the case of \textit{Physarum}.

In chapter \ref{chapter: Applications}, we explore some  applications of the model to path-finding and network design. 
We first test its ability to solve mazes and to find the shortest path between two nodes in an arbitrary graph. 
Then, on an arena with the shape of mainland Portugal, we study the importance of flux fluctuations to build efficient and resilient networks similar to those of \textit{Physarum} by considering time-dependent distributions of sources and sinks. The terminals represent the geographic location of major Portuguese cities. 
The performance of the resultant networks is compared with that of the real railway system connecting those cities.

In chapter \ref{chapter:growth}, we extend the adaptation model to include the network formation, giving a better representation of \textit{Physarum}'s foraging behaviour. In particular, we simulate the growth and optimisation in the presence of one or more food sources.

Finally, in chapter \ref{chapter:conclusions}, we give a general overview of the results and discuss some of the main conclusions made and possible improvements to the model.

\cleardoublepage

\chapter{Phenomenology of \Pp}
\label{chapter:Physarum}

\section{Life-cycle}

\begin{figure}[htb]
\centering
\includegraphics[width=0.6\textwidth]{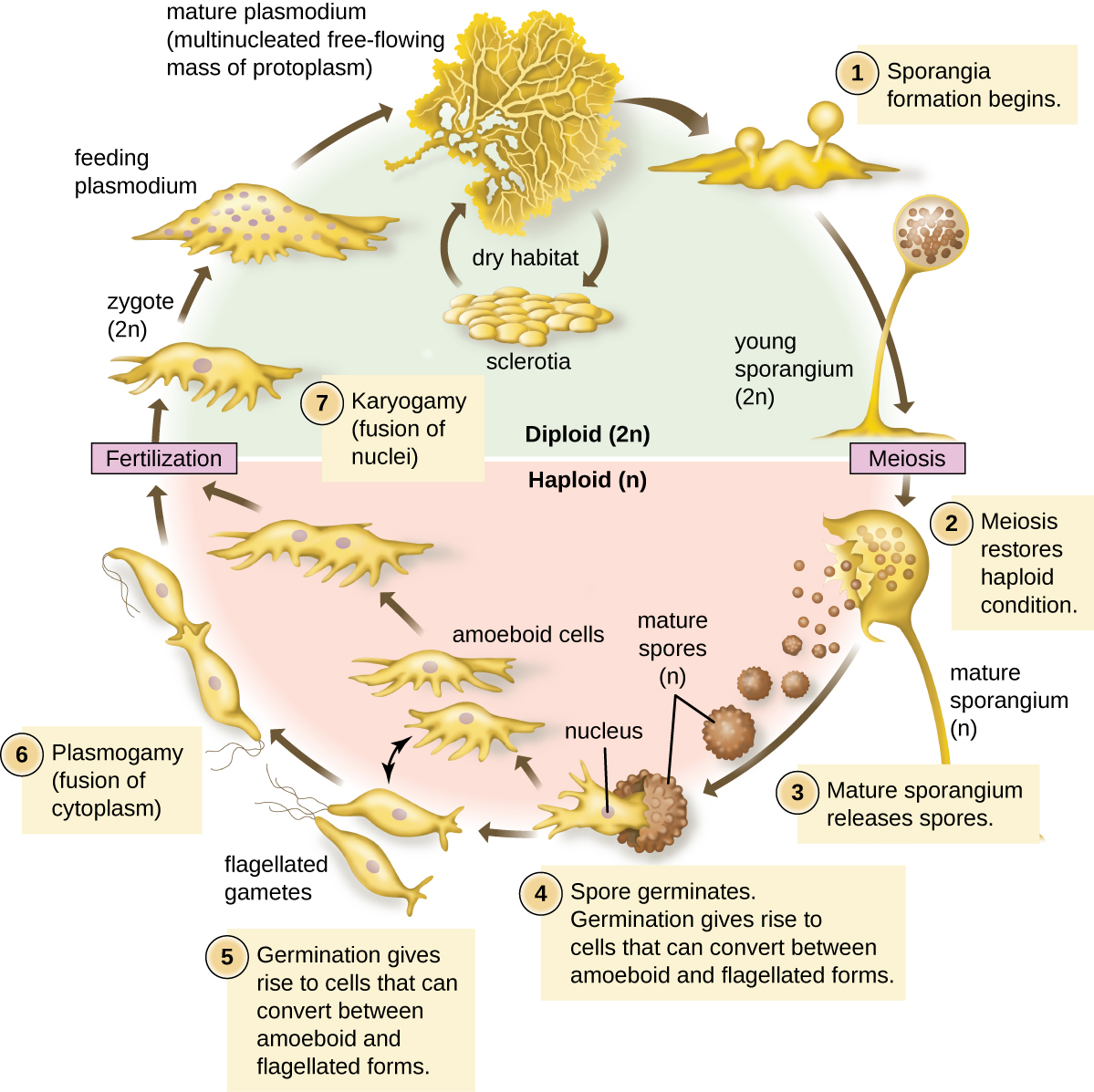}
\caption{The life-cycle of \Pp{}. The life-cycle begins with a haploid phase, followed by a diploid phase where \textit{Physarum} reaches its main stage, the plasmodium. Source: \cite{lifecycle}.}
\label{fig:lifecycle}
\end{figure}

\textit{Physarum polycephalum}  is a macroscopic slime mould of the family \textit{Physaraceae}, order \textit{Physarales}, class \textit{Myxomycetes}.
Despite their name, slime moulds are not fungi, but rather amoeboid protists (phylum \textit{Amoebozoa}, infraphylum \textit{Mycetozoa}), sharing, therefore, common traits with plants, fungi and animals. These organisms exhibit a complex life cycle (Figure \ref{fig:lifecycle}), which provide them with great adaptability to environmental changes \cite{Adamatzky2016,Oettmeier_2017}.

In the main vegetative phase of their life cycle (plasmodium), the slime moulds of class \textit{Myxomycetes}, commonly known as the true or plasmodial slime moulds, exist as a syncytium, i.e., a giant cell enclosed by a single membrane which contains typically millions of diploid nuclei. The \textit{Physarum} plasmodium consists of an amorphous yellow mass endowed with an amoeba-like behaviour. While foraging, it spreads as a network of vein-like protoplasmic tubes in the direction of the food source, being able to move at speeds higher than 1 cm/h \cite{Adamatzky2016}. The food source is covered by extensions of its protoplasm, and it's digested with the help of enzymes. Typical foodstuffs include bacteria, fungal spores and decaying matter. The spreading fronts adopt a fan-like shape, and the number of fronts increases with the nutrient level of the environment, providing it with a larger and more efficient area of absorption.   

\begin{figure}[htb]
\centering
\subfloat[\label{phy_img_a}]{\includegraphics[height=0.43\textwidth]{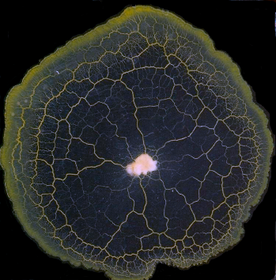}} %
\hspace{0.05\textwidth}%
\subfloat[\label{phy_img_b}]{\includegraphics[height=0.43\textwidth, trim={0 0 2.5cm 0}, clip]{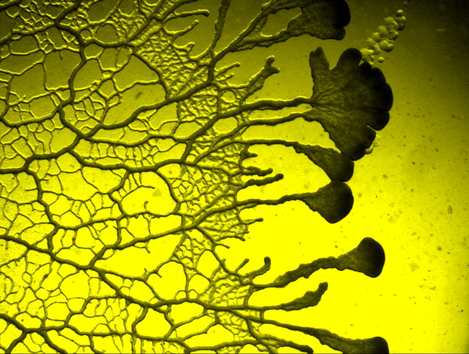}} 
\caption{
\Pp's  plasmodium phase.
\textbf{(a)} \textit{Physarum} growing radially from a central food source. As it grows, a highly ramified network is formed, which is subject to continuous optimisation. Some veins are reinforced while others shrink, resulting in a hierarchical and reticulated structure of veins. \textbf{(b)} Close look at \textit{Physarum}'s growth fronts with a fan-like shape. Images adapted from \cite{Physarum_images}.}
\label{fig:phy_network}
\end{figure}

Under extreme adverse environmental conditions, the plasmodium changes into an inactive dormant state, forming a hard compact mass know  as sclerotium. This hibernation state can be sustained for several years, and when moist, the plasmodium state is gradually recovered \cite{Adamatzky2016}. 

Like all true slime moulds, \textit{Physarum} reproduces by sporulation. Certain stimuli like starvation and light irradiation trigger the plasmodium to grow sporangia: clusters of black globulose enclosures that produce and protect the spores (haploid cells), that are then released. When the conditions become favourable, a spore hatches into an amoeba-like single-cell (myxamoeba), or, if in the presence of water, into a flagellated version of the former (swarm cell), being these two forms exchangeable \cite{Adamatzky2016}. Each of these cells can mate with another through the fusion of their protoplasm, forming a diploid zygote, which after successive nuclear divisions develops into a new plasmodium.

\section{Network Flows Dynamics}

The most commonly observed form of \textit{Physarum} is the plasmodium. The network tubes are made of a gel-like outer layer (ectoplasm) that contains the cytoskeleton and encloses the endoplasmic fluid. The cytoskeleton consists of a system of proteins, essentially composed of actin and myosin. The filaments of actin provide the structural support of the tube walls, and, together with the myosin, the motor protein, are responsible for the unique mobility and growth of \textit{Physarum} \cite{Adamatzky2016,Oettmeier_2017}.

\subsection{Shuttle streaming}

The interactions between actin and myosin generate relaxation-contraction cycles of the walls, which result in a rhythmic back-and-forth propulsion of the endoplasm over the entire network within a period of around two minutes \cite{Alim2013,Teplov_2017}. This type of periodic back-and-forth streaming, known as shuttle streaming, enables the transport and distribution of food supplies, organelles and other substances throughout the organism. Furthermore, these oscillations are locally well-coordinated in such a manner that allows the net movement of the slime mould as it forages. Collectively, they establish a gradient of pressure so that the flow is driven towards the leading edges (anterior margin), where the growth of structural proteins occurs simultaneously. The amplitude and frequency of contractions are regulated according to external stimuli: attractants (e.g., food source) increase them, swelling the stimulated edges which thrive the network, while repellents (e.g., light exposure) decrease both, resulting in the shrinkage of the affected tubes to avoid those negative stimuli \cite{attract_repell}. In this way, \textit{Physarum} can dynamically rearrange the structure of its network and optimise it, in response to local stimuli information. The timescale of morphological rearrangements ($\sim$ 1\,h) is much larger than the timescale of flow generation ($\sim$ 2\,min) \cite{Alim2013}.

\subsection{\textit{Physarum} Oscillator}

It's clear that the rhythmic contractions and force generation are produced by the interactions between the actin and myosin. However, the underlying mechanism which regulates these interactions isn't well understood. Synchronous oscillations of the membrane potential, intracellular Ca$^{2+}$ and other chemicals, like ATP, NADH, H$^{+}$, are observed in real organisms along with the contraction-relaxation cycles. But the set of essential and independent variables responsible for generating the rhythm and the collective movement of \textit{Physarum} hasn't been identified \cite{Teplov_2017}. What is certain is that Ca$^{2+}$ plays a prominent role in the contraction-relaxation cycle, like in smooth muscles contraction. The plasmodium contains vesicles capable of sequestering and releasing calcium through stretch-activated channels, and data confirms that a rise in Ca$^{2+}$ concentration can trigger the contractions \cite{Teplov_2017,Oster1984,Oettmeier_2017}.

\section{\textit{Physarum}'s Intelligent Behaviours}

Despite lacking any kind of neural circuit, the morphological adaptation displayed by \textit{Physarum} provides it with high-level behaviours. For instance, the slime mould can find the shortest path between two food sources in a maze \cite{Nakagaki2000}, and to replicate optimised, man-made transport networks \cite{Tero1}. In contrast to animals, where complex behaviours can be assigned to their evolved nervous system, this type of intelligence displayed by many brainless organisms, like other species of slime moulds and even fungi, is still poorly understood. In particular, in the case of \textit{Physarum}, the means of communication through its entire body, and how it processes that information to coordinate its movement and growth remain unknown.
Recent studies  \cite{Alim2017} suggest that the transport of signalling molecules is involved in the coordination of the fluxes.
The control of the internal fluid flow is likely crucial to the coordination of its behaviour, including the continuous network self-organisation \cite{Alim2013}.

\subsection{Maze-Solving }

\begin{figure}[htb]
\centering
\subfloat[\label{maze_a}]{\includegraphics[width=0.3\textwidth]{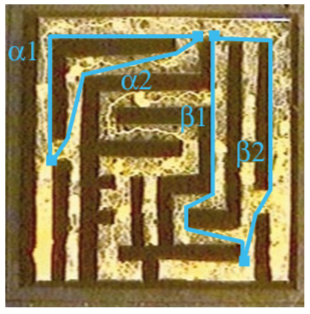}} \hfill
\subfloat[\label{maze_b}]{\includegraphics[width=0.3\textwidth]{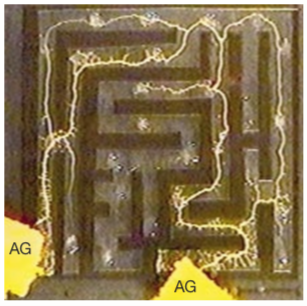}} \hfill
\subfloat[\label{maze_c}]{\includegraphics[width=0.3\textwidth]{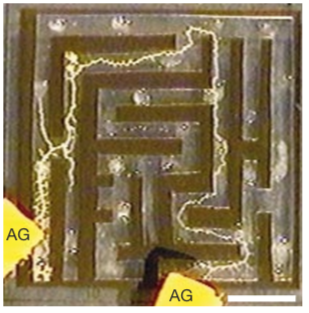}}
\caption{\Pp{} maze solving experiment. \textbf{(a)} The slime mould (yellow) was let cover all the maze. Brown blocks correspond to the maze walls and blue lines indicate the  segments of the possible solutions of the maze, $\alpha_1$ ($41\pm1\;mm$), $\alpha_2$ ($33\pm1\;mm$), $\beta_1$ ($44\pm1\;mm$), $\beta_2$ ($45\pm1\;mm$). \textbf{(b)} After placing the agar blocks (AG), it explored all possible routes and shrank the vessels which led to dead ends. \textbf{(c)} Four hours later, only the shortest path remained. Source: \cite{Nakagaki2000}.} 
\label{fig:maze} 
\end{figure}

Nakagaki et al. \cite{Nakagaki2000} were the first to report the maze-solving skills of \textit{Physarum}, showing its ability to find the shortest path between two food sources. The  maze (Figure \ref{fig:maze}) consisted of an agar substrate with plastic films as walls, which are dry surfaces that the slime mould tends to avoid.

Initially, a sample of plasmodium was allowed to fill the entire maze (Figure \ref{maze_a}), and only then, two nutrient-rich agar blocks  were placed on the endpoints (Figure \ref{maze_b}). There were four possible routes connecting the two, $\alpha_1-\beta_1$, $\alpha_1-\beta_2$, $\alpha_2-\beta_1$, $\alpha_2-\beta_2$, being $\alpha_2$ ($\beta_1$) about $22\%$ ($2\%$) shorter than $\alpha_1$ ($\beta_2$). The slime mould quickly restructured its network, removing the redundant vessels and the ones branching along the dead ends, and reinforcing the optimal ones, until only one path connecting the food sources remained. The path that survived was different between experiments, but the shortest segment  $\alpha_2$ was always selected. The segments $\beta_1$ and $\beta_2$ were chosen about the same number of times, due to their negligible difference in length, which is lost  by the  natural undulations of the tubular trajectory. This means that in all the experiments, the path selected was always approximately the shortest one.

\subsection{Network Optimisation}
\label{section:Tokyo_experiment}

In later experiments, Tero et al. \cite{Tero1} studied the robustness of the slime mould network adaptability when it distributes itself over several food sources. From an evolutionary perspective, this appears as a competitive advantage for organisms that forage as large contiguous  networks, since, to optimise their strategy, they must be able to balance the network efficiency with the cost of producing it.

\begin{figure}[hbt]
 \subfloat[\label{tokyo_a}]{\includegraphics[width=0.3\textwidth]{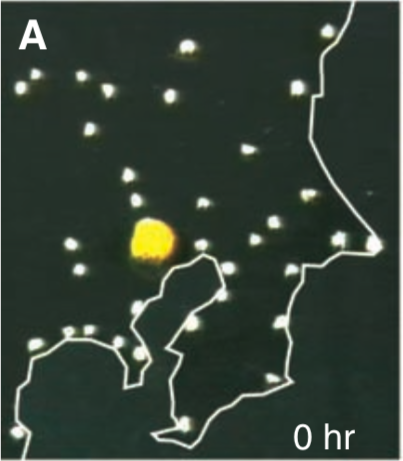}}\hfill
 \subfloat[\label{tokyo_b}]{\includegraphics[width=0.3\textwidth]{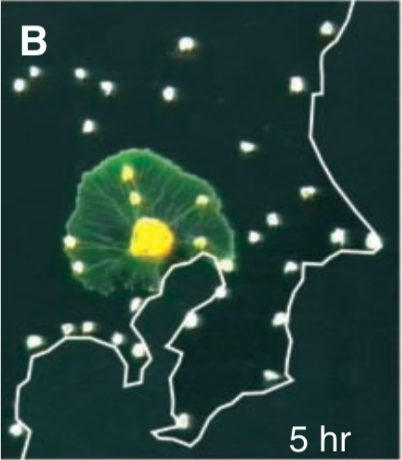}}\hfill
 \subfloat[\label{tokyo_c}]{\includegraphics[width=0.3\textwidth]{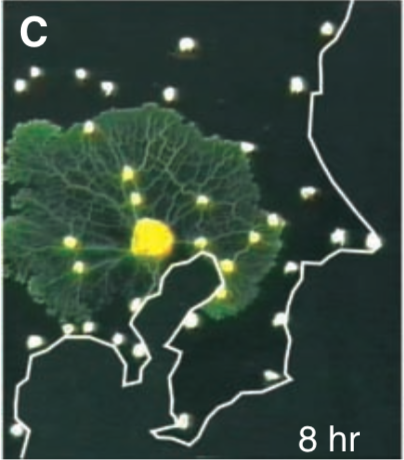}}\hfill
 \subfloat[\label{tokyo_d}]{\includegraphics[width=0.3\textwidth]{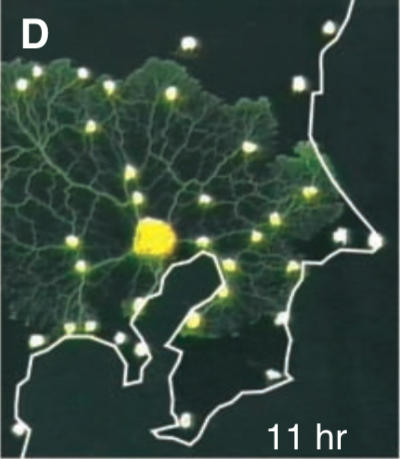}}\hfill
 \subfloat[\label{tokyo_e}]{\includegraphics[width=0.3\textwidth]{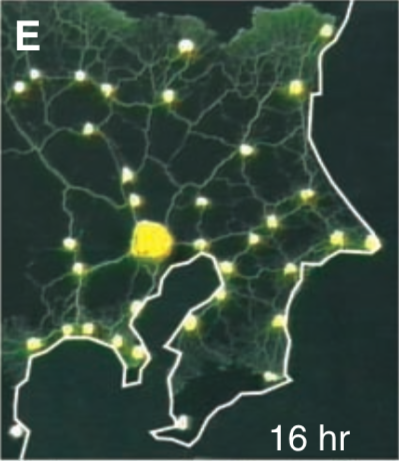}}\hfill
 \subfloat[\label{tokyo_f}]{\includegraphics[width=0.3\textwidth]{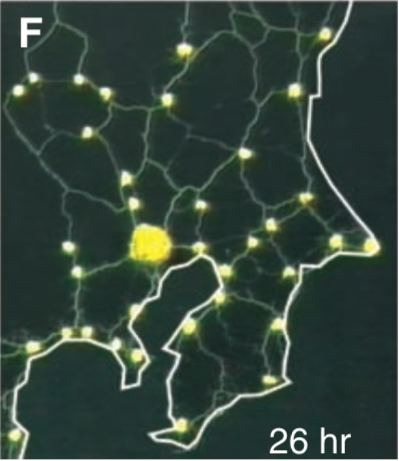}}\hfill
\caption{\Pp{} Tokyo Experiment. \textbf{(a)} Initially a plasmodium sample (yellow) was placed at Tokyo location in an area surrounded by Pacific coastline (white border) and filled with other 35 food sources  representing the neighbour major cities (white dots). \textbf{(b to f)} The plasmodium progressively colonised each food source, and simultaneously optimised the connections between them. Adapted from \cite{Tero1}.}
\label{fig:tokyo} 
 \end{figure}

Thirty-six food sources were arranged in a substrate, portraying the geographic locations of the surrounding cities of Tokyo (Figure \ref{fig:tokyo}). The plasmodium grew out from Tokyo's food source and explored gradually the surroundings with a contiguous front until it accommodated all food sources, and simultaneously optimised its network in a way that only the more efficient food source connections survived. 

In a more realistic setting, the geographical constraints, namely, high-altitude areas, lakes and the Pacific Ocean, were replicated by increasing the luminosity of those regions, restricting the growth of the plasmodium to shaded areas that it tends to prefer. The resultant minimal networks achieved by \textit{Physarum} were compared to the real rail network of Tokyo in terms of cost-efficiency and fault tolerance. Here an optimal cost-efficiency means a low total length of vessels with a short average minimum distance between the food sources, whereas fault tolerance is defined as the probability of disconnecting part of the network when a single link is removed, evaluating, therefore, its robustness.

\begin{figure}[htb]
\centering
\begin{tabular}[t]{ccc}
\begin{subfigure}{0.33\textwidth}
\centering
\includegraphics[width=0.97\linewidth,height=2.4in]{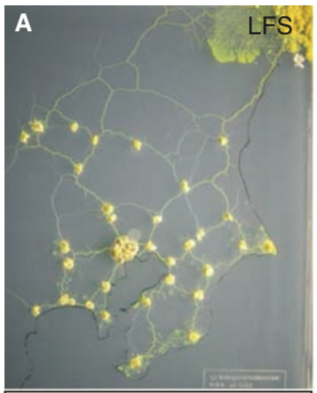}
\caption{} 
\end{subfigure}
 &
\begin{subfigure}{0.33\textwidth}
\centering
\includegraphics[width=0.97\linewidth,height=2.4in]{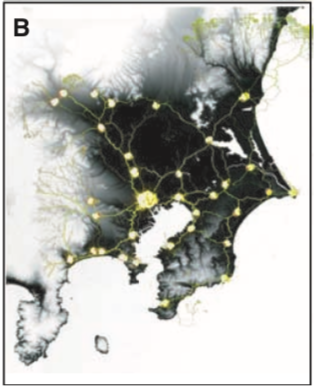}
\caption{} 
\end{subfigure}
&
\begin{tabular}{c}
\smallskip
    \begin{subfigure}[t]{0.25\textwidth}
        \centering
        \includegraphics[width=0.84\textwidth]{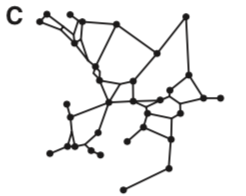}
        \caption{}
    \end{subfigure}\\
    \begin{subfigure}[t]{0.25\textwidth}
        \centering
        \includegraphics[width=0.84\textwidth]{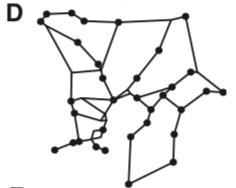}
        \caption{}
    \end{subfigure}
\end{tabular}\\
\end{tabular}
\caption{Comparison of the \textit{Physarum} networks with the Tokyo rail network. \textbf{(a)} Resultant network without geographical constraints. \textbf{(b)} In another experiment, an illumination mask was applied to simulate the geographical constraints of the Japan rail network. \textbf{(c and d)} The resultant network (c) was compared with the real rail network (d). Adapted from \cite{Tero1}.} 
\label{fig:rail} 
\end{figure}

As shown in Figure \ref{fig:rail}, the shape of the optimised network displayed by the \textit{Physarum} is quite similar to the actual man-made railway system. Surprisingly, the authors concluded that \textit{Physarum} networks showed a slightly better  cost-efficiency, yet marginally lower robustness. 
This is an astonishing achievement given that \textit{Physarum} builds the networks without centralised control, in contrast to human-made infrastructure networks. Since then, other real-world transportation networks have also been approximated by \textit{Physarum}, such as the highways of the UK, Germany and USA \cite{Adamatzky_UK,Adamatzky_USA}.

\section{Modelling \textit{Physarum}'s Network Adaptation}

Until now, there is no single model which can describe the whole behaviour of \textit{Physarum}, even considering only the plasmodium stage. Due to the complexity of the task, current approaches focus on modelling a specific issue at a time, namely the growth mechanism, the contraction patterns of the actin-myosin cortex, and the network formation and adaptation. Naturally, these mechanisms are all inter-connected, and together are responsible for the complex behaviours exhibited by \textit{Physarum}. In particular, the coordination of the flows arising from the synchronisation of the contractions, as a response to external stimuli, seems to be a key feature underlying the network optimisation.  

This work focuses on modelling the network formation and optimisation. In this respect, different modelling techniques were proposed, including cellular automaton models \cite{CELL}, agent-based models \cite{Jones2010}, and mathematical flow-based models \cite{Tero1}. Each one assumes that the optimisation is achieved through a different set of simple rules and, in general, don't take the coordination of contractions into account. Some are even purely phenomenological models or bio-inspired algorithms designed to solve complex graph problems \cite{Gao2019,PpReview}, lacking therefore meaningful biological insight. In the following, we describe the two main models found in the literature.

\subsection{The Multi-Agent System Model}

Jones \cite{Jones2010} proposed a multi-agent bottom-up approach to model the formation and the optimisation of \textit{Physarum} transport networks. According to this method, the macroscopic network adaptation arises as an emergent phenomenon from simple microscopic interactions between small portions of the plasmodium. 
The plasmodium is represented by a hypothetical population of  mobile particle-like agents, each occupying a single cell of a two-dimensional lattice. The lattice stores the current agent positions and local concentrations of stimuli which diffuse through the environment and may evaporate. Positive stimuli, referred to as chemoattractants, are released by food sources and by moving agents. Hazardous stimuli that agents tend to avoid may be also considered. Each agent uses three forward sensors (\texttt{FL}, \texttt{F}, \texttt{FR}) to sense the local concentration of these stimuli in three regions in front of its position, and responds to this information by moving towards the strongest local source of chemoattractant. The sensed regions are parametrized by the width of the sensors, \texttt{SW} (usually one cell), the distance from the agent, \texttt{SO} (sensor offset), and the angle relative to the direction of the agent, \texttt{SA} (Figure \ref{fig:JonesSensor}). By depositing stimuli while they move, the agents not only adapt to the environment but also influence each other's behaviours. A minimum \texttt{SO} of 3 cells  is required for strong local coupling to occur and for complex patterns to emerge. Increasing the \texttt{SO} value results in thicker networks, faster network adaptation, and coarser-grained networks.

\begin{minipage}[b]{0.45\textwidth}
\begin{figure}[H]
\centering
\includegraphics[width=0.8\textwidth]{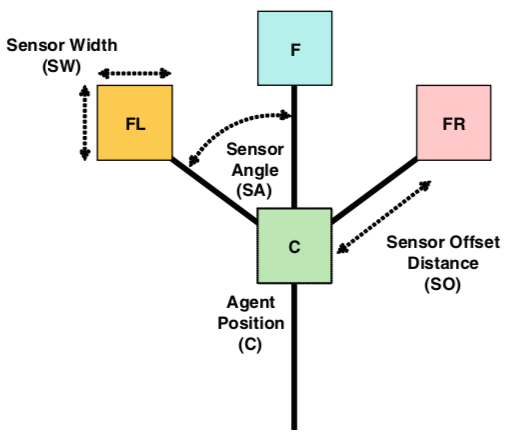}
\caption{Agent morphology. An agent consists of three sensors and a body. Source: \cite{Jones2010}.}
\label{fig:JonesSensor}
\end{figure}
\end{minipage}
\hfill
\scalebox{0.9}{%
\begin{minipage}[b]{0.5\textwidth}
\begin{algorithm}[H]
\footnotesize
\setstretch{1.}
\begin{algorithmic}[1]
\Procedure{MotorStage}{}
\State Attempt move forward in the current direction 
    \If{(moved successfully)}
        \State Deposit trail in new location
    \Else 
        \State Choose random new orientation 
    \EndIf
\EndProcedure    
\Statex
\Procedure{SensoryStage}{}
\State Sample trail map values
    \If{(\Var{F} $>$ \Var{FL}) \AND (\Var{F} $>$ \Var{FR})}
        \State Stay Facing Same Direction 
    \ElsIf{(\Var{F} $<$ \Var{FL}) \AND (\Var{F} $<$ \Var{FR})}
        \State Rotate randomly left or right by \Var{RA}
    \ElsIf{(\Var{FL} $<$ \Var{FR})}    
        \State Rotate right by \Var{RA}
    \ElsIf{(\Var{FR} $<$ \Var{FL})}    
        \State Rotate left by \Var{RA}    
    \Else 
        \State Continue facing the same direction
    \EndIf
\EndProcedure   
\end{algorithmic}
\caption{Multi-Agent Algorithm}
\label{alg:Jones}
\end{algorithm}
\end{minipage}%
}

\begin{figure}[htb]
\centering
\includegraphics[width=1\textwidth]{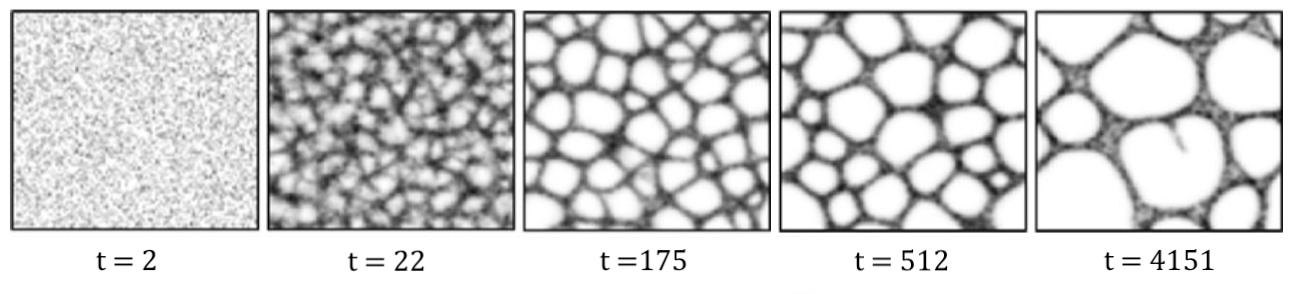}
\caption{Formation and self-organisation of the multi-agent networks in the absence of external positive stimuli. Simulation on a $200\times200$ lattice, with $15\%$ of the area filled with agents initially placed at random locations. Simulation parameters: \texttt{RA} $=45\degree$, \texttt{SA} $=22.5\degree$, \texttt{SO} $=9$. Adapted from \cite{Jones2010}.}
\label{fig:Jones_NoFood}
\end{figure}

\begin{figure}[htb]
\centering
\includegraphics[width=0.7\textwidth]{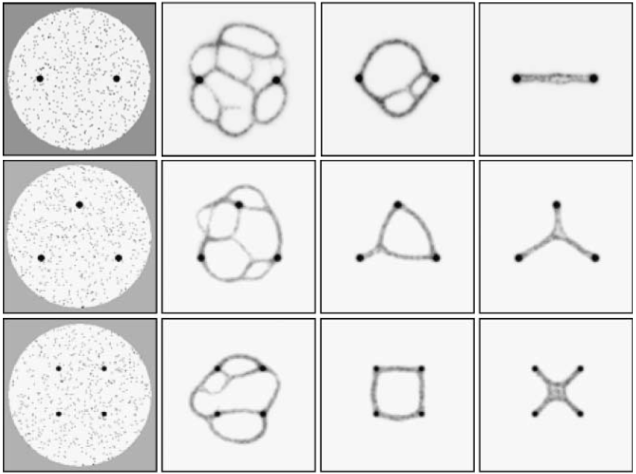}
\caption{Formation and self-organisation of the multi-agent networks under the influence of external positive stimuli (black dots). Simulation on a $200\times200$ lattice, with $2\%$ of the area filled with agents initially placed at random locations. Simulated with \texttt{RA} $=45\degree$, \texttt{SA} $=22.5\degree$, \texttt{SO} $=9$. In the case of two stimuli sources (top row), the optimisation process converges to the shortest path connecting them. In the presence of multiple stimuli, the final network closely approximates the Steiner minimum tree solution. Adapted from \cite{Jones2010}.}
\label{fig:Jones_steiner}
\end{figure}

The simulation starts with a  chosen number of agents placed at random unoccupied locations and with random orientation (from zero to 360 degrees). At each time step, based on perceived sensory information, an agent rotates itself towards the direction covered by the sensors with the highest chemoattractant concentration. The agent is allowed to move one step forward in that direction,  only if the new site is not already occupied, and deposits a constant concentration of chemoattractant, which attracts nearby agents. Otherwise, if the movement isn't allowed, the agent remains in its current position without leaving any trail, and in the next step, a new orientation is randomly selected. In this way, mobile agents can be interpreted as the endoplasmic flux, while immobile agents represent the actin-myosin cortex. 
 A time step is concluded when every agent is given a change move. The update of the agents is performed randomly at each step  to avoid any correlation from sequential ordering. 
 After a long run, a stable collective pattern of the positions of agents may arise, forming a continuous network that connects the food sources. Typical simulations of the model in the absence and the presence of external positive stimuli are shown in Figure \ref{fig:Jones_NoFood} and Figure \ref{fig:Jones_steiner}, respectively.

However, the model is biologically unrealistic, since the agents act autonomously and not as a continuous network, especially in the first stages when they randomly fill the lattice. Even ignoring this issue, the agents are always biased towards gradients of positive stimuli, which doesn't depict the true foraging behaviour of the slime mould. Nevertheless, the model has been successfully applied to solve several graph problems \cite{Adamatzky2016,Gao2019}.

\subsection{\textit{Physarum Solver}}
\label{section:Physarum Solver}

Tero et al. \cite{TeroPS} proposed a flow-based model known as \textit{Physarum Solver}, initially to describe the maze-solving ability of the slime mould \cite{Nakagaki2000} (Figure \ref{fig:maze}). According to the model, the path-finding is achieved through feedback loops between the thickness of the tube and the rate of internal flows: higher fluxes trigger an increase of the tube radius, while lower fluxes lead to their shrinkage. 
The plasmodium is modelled as a graph, where the edges represent the tubes in which endoplasm flows, and the nodes the connections between them. The nodes $i$ and $j$ are kept at pressures $p_i$ and $p_j$, and are linked by a cylindrical tube  of constant length $L_{ij}$ and a variable radius $r_{ij}$. The problem is reduced to finding the shortest path between two special nodes, $N_1$ and $N_2$, that represent the food sources (FS). It's assumed that the flow in each tube is laminar and follows the Hagen-Poiseuille Law, and so the flux through tube $(i,j)$ is given by 

\begin{equation}
    Q_{ij}=\frac{\pi r_{ij}^4(p_i-p_j)}{8\eta L_{ij}}=\frac{D_{ij}(p_i-p_j)}{L_{ij}} \;,
    \label{eq:PSflux}
\end{equation}

\noindent where $\eta$ is the viscosity of the fluid and $D_{ij}=\pi r_{ij}^4/8\eta$ is the conductivity of the tube. 

Experimental observations show that most of the plasmodium lies over the FS and that the fluid is rhythmically exchanged between those regions through the network. Hence, 
it's assumed that only the FS nodes can drive the flow and that the tubes are passive elements that regulate their thickness accordingly \cite{TeroPS}.
One of the FS always acts like the source ($N_1$) and the other as the sink ($N_2$).  

At each time step, a constant flux $I_0$ flows from the source and into the sink. Since the total flux must be conserved, the inflow and outflow at each node must be balanced, which implies

\begin{equation}
    \sum_jQ_{ij}=
    \left\{
    \begin{aligned}
     0 \quad &\text{for } i\neq 1,2 \\
    -I_0 \quad   &\text{for } i=1 \\
    I_0 \quad &\text{for } i=2    \;.
    \end{aligned}\right.
    \label{eq:PSfluxconservation}
\end{equation}

Letting the pressure at the sink node be 0 and knowing the tube lengths, $L_{ij}$, and conductivities, $D_{ij}$, the flux through each one can be computed in each step  by solving the linear system (\ref{eq:PSfluxconservation}) together with the equation (\ref{eq:PSflux}).

The plasmodium adapts its network through the change of the tubular thickness, $r_{ij}$, in response to the magnitude of the flux $|Q_{ij}|$ flowing through each vessel. In the model, this adaptation is captured by the conductivities $D_{ij}\propto r_{ij}^4$, which are updated at each step according to

\begin{equation}
    \frac{dD_{ij}}{dt}=f(|Q_{ij}|)-\mu D_{ij} \;,
    \label{eq:Tero_update_rule}
\end{equation}

\noindent where $f$ is a monotonously increasing function, satisfying $f(0) = 0$, that describes the tubes' expansion response to the flux, and $\mu$ is a positive constant. The second term represents the rate of tube shrinkage, implying that in absence of flux the tube disappears at an exponential rate $\mu$. Together this implies that conductivities tend to increase in edges with big flux.  New conductivities are fed back to (\ref{eq:PSflux}) to calculate new fluxes and pressures. The network's flux conversation induces competition between edges for more flux. Shorter tubes are favoured through a positive feedback loop:  shorter tubes carry more flow, and therefore grow, which in turn increases their flow in subsequent iterations and so on.
This iterative process continues until the network finally converges to a steady state. 

This algorithm was readily extended to include the network adaptation in the presence of multiple FS \cite{Tero1}, as an attempt to explain the Tokyo experiments (Figure \ref{fig:tokyo}). 
In this case, the plasmodium network is initialized as a random graph, and at each time the source and sink  are randomly chosen from FS nodes. 

Regarding the update rule (\ref{eq:Tero_update_rule}) two functional forms of $f$ are mainly used in literature:

\begin{equation}
\begin{aligned}
f(|Q|) &= |Q|^\gamma  \\
f(|Q|) &= \dfrac{Q^\gamma}{1 + Q^\gamma}
\end{aligned}
\qquad (\gamma > 0) \;.
\label{eq:Tero_adaptation_funcs}
\end{equation}

While the first performs better on maze solving (i.e., when there is only one source and one sink), the second one is more adequate to design efficient networks in the presence of multiple FS. Note that the latter describes a sigmoid response of the tube diameter to the flux flowing through. Biologically this is more realistic since the saturation mimics the maximum distensibility of the tubes. 

\begin{figure}[htb]
 \subfloat[\label{ps_a}]{\includegraphics[width=0.22\textwidth]{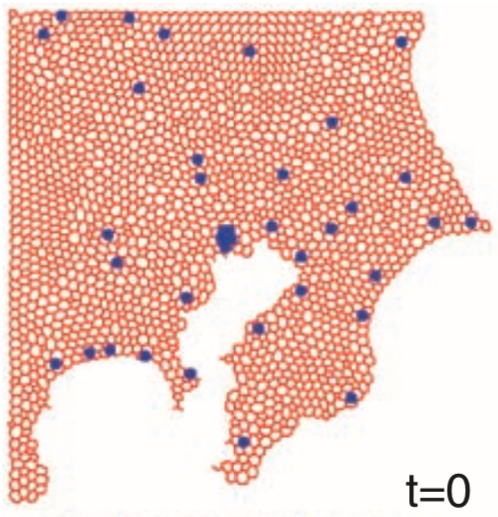}}\hfill
 \subfloat[\label{ps_b}]{\includegraphics[width=0.22\textwidth]{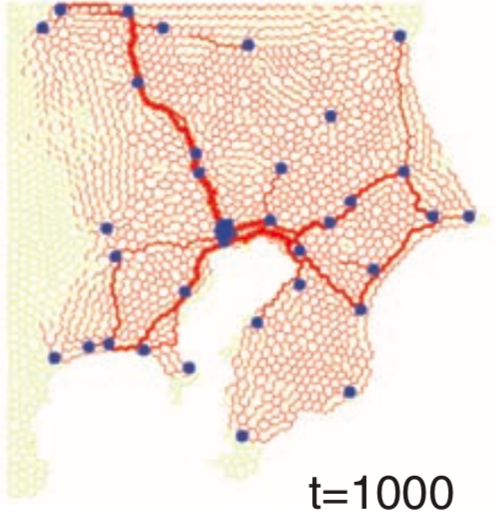}}\hfill
 \subfloat[\label{ps_c}]{\includegraphics[width=0.22\textwidth]{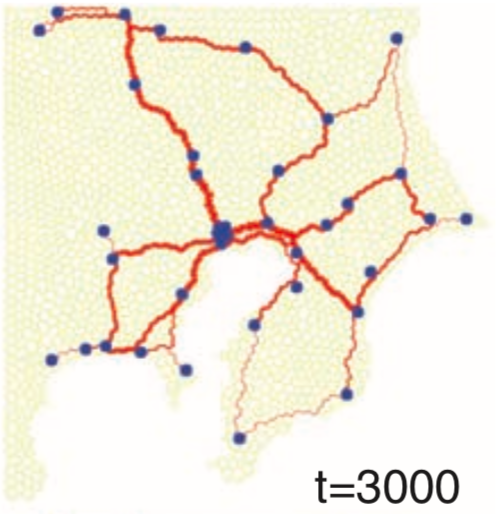}}\hfill
 \subfloat[\label{ps_d}]{\includegraphics[width=0.22\textwidth]{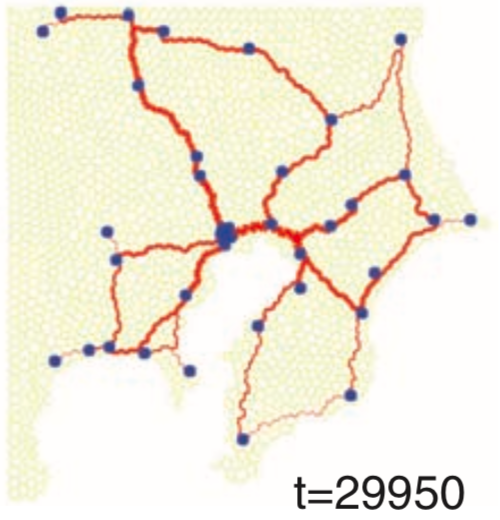}}\hfill
\caption{Simulation of the \textit{Physarum Solver}  ($\gamma = 1.8$, $I_0 = 2$.). Food sources are represented by blue dots. \textbf{(a)} Initially the space was populated with a finely meshed network of thin tubes. \textbf{(b,c)} Over time, many of these tubes died out, and the other few were selectively thickened resulting in a stable optimised network \textbf{(d)}. Adapted from \cite{Tero1}.}
\label{fig:PS} 
\end{figure}

A typical  simulation of the \textit{Physarum solver} is presented in Figure \ref{fig:PS}.
Despite the resulting networks resembling the ones produced by \textit{Physarum} (Figure \ref{fig:tokyo}), the model breaks some physical principles, namely the conservation of volume of the circulating
fluid, which will be discussed in more detail in the following chapter. Furthermore, the model only describes the network optimisation behaviour of full-grown plasmodium, represented by the initial random mesh, and thus, can't account for the formation of the network itself.

Nevertheless, the basic principle of the feedback mechanism between the flux and the veins thickness in which the model is based seems to agree with the experimental observations. Also, the modelling approach of \textit{Physarum} as a flow network is naturally more biologically realistic comparing to agent-based models which are in general purely phenomenological. 
For these reasons, in the following chapter, we propose a new flow-based model for the network optimisation, inspired by the \textit{Physarum Solver} principles, as an attempt to solve some of its problems.

\cleardoublepage

\chapter{\textit{Physarum}'s  Network Adaptation Model}
\label{chapter:New Model}

In this chapter, we formulate a new flow-based model to describe \textit{Physarum}'s network optimisation. As a starting point, we ignore the network growth and begin to model the network adaptation of static individuals. Note that we don't attempt to model the contractile activity of \textit{Physarum}, which is outside the scope of this work. This new model is inspired by the previous \textit{Physarum Solver} model, but we propose a generic class of evolution laws for the channels conductivities consistent with the assumptions of a \HP{} flow, namely the volume conservation of the circulating fluid. We start by showing how this is violated in the previous model, and from there we derive the new class of adaptation rules. Then, based on the principle of least dissipation of energy, we derive an explicit form of the adaptive equations. Finally, we explore some of the model features as a first overview.

\section{Hagen-Poiseuille Flow}

\textit{Physarum} can be naturally modelled as a flow transport network made of hydraulic-coupled channels with adapting radius inside which the protoplasm flows. The \textit{Physarum Solver} model presented in the last chapter, and other similar flow-based models, typically assume that the protoplasm follows a \HP{} flow.

The \HP{} equation describes the steady-state  
laminar flow of incompressible, Newtonian fluids through a channel with a constant, and much smaller than its length, cross-section. The flow through the channel is driven by a pressure drop between the two ends and the friction between the fluid layers due  to the viscosity of the fluid, $\eta$. The counterbalance between the two forces results in a parabolic velocity profile of the flow. For a cylindrical channel $(i, j)$ with radius $r_{ij}$ and length $L_{ij}$, the \HP{} equation states that the relation between the  volumetric flux flowing through it, $Q_{ij}$, and the pressures at both ends, $p_i$ and $p_j$, is given by

\begin{equation}
    Q_{ij}=\frac{\pi r_{ij}^4(p_i-p_j)}{8\eta L_{ij}}=\frac{D_{ij}(p_i-p_j)}{L_{ij}} \;,
    \label{eq:HP_flow}
\end{equation}

\noindent where $D_{ij}=\pi r_{ij}^4/8\eta$ is the conductivity of the channel.

A laminar flow is characterised by a low Reynolds number (Re), which is defined as the ratio of the inertial forces to the viscous forces acting on the fluid. The protoplasm is well described as a low-Reynolds-number incompressible fluid, i.e., with a constant density.  Recent measurements \cite{Haupt2020} estimate that the maximum Reynolds number of protoplasmic flow, obtained using the top speed of the shuttle streaming inside large veins of \textit{Physarum}, is Re $\sim 0.1$, which is four orders of magnitude lower than that required for the onset of turbulence in a cylindrical tube. Furthermore, the results of the last study and of former experiments \cite{Kamiya1950,Bykov2009} show that the flow velocity profile across the diameter of the vein is always parabolic regardless of the speed of shuttle streaming. These findings indicate that the protoplasm viscous forces dominate over the inertial forces, and the flow is indeed laminar, thus supporting the assumption of a \HP{} flow.

\subsection{Derivation}

In the following, the \HP{} equation is derived from first principles \cite{HP_flow_derivation, Katifori_BioFlows}.

Consider a cylindrical fluid element of radius $r$ and length $dz$. The surrounding fluid exerts pressure on the end faces of the cylinder, which is assumed to be constant over any chosen cross-section of the pipe (i.e., $p=p(z)$). At one end is acted a pressure $p$ and at the other, the same pressure lower by $dp>0$, $p-dp$. The resultant force on the volume element arising from the pressure, responsible for driving the flow, is given by

\begin{nalign}
    \vb*{dF_p}
       &=\left[ p\pi r^2 -(p-dp)\pi r^2 \right ]  \vu{z}   \\
       &=\pi r^2 dp \; \vu{z} \;.
\end{nalign}

The flow is assumed to be laminar, meaning that can be described as the relative motion of thin concentric cylindrical layers of fluid that slide over each other in the axial direction, without the occurrence of lateral mixing. The velocity of the fluid is then $\vb{v}=v\,\vu{z}$. Due to the viscosity of the fluid, $\eta$, there is friction between the fluid layers, which acts as flow resistance. The viscous drag force, $\vb*{dF_{vis}}$, is proportional to the area of contact between the layers, $2\pi r dz$, and to the shear stress, which for a Newtonian fluid varies linearly with velocity gradient perpendicular to the flow direction, 

\begin{nalign}
    \vb*{dF_{vis}}&=\left(-\eta\frac{dv}{dr}\right)2\pi r dz(-\vu{z}) = \eta \frac{dv}{dr}2\pi r dz \; \vu{z} \;.
\end{nalign}

The velocity gradient can be derived from the steady-flow condition, which requires the balance of the two counteracting forces, 

\begin{nalignat}[3]
\label{eq:dv/dr}
  && \vb*{dF_p} + \vb*{dF_{vis}}& = 0   &\quad \iff \\
\iff \quad &&
-\pi r^2 dp  & = \eta\frac{dv}{dr}2\pi r dz &\quad \iff \\
\iff \quad &&
\frac{dv}{dr} &= -\frac{1}{2\eta}\frac{dp}{dz}r \;.
\end{nalignat}

It's assumed that the flow is fully developed $\left(\dfrac{dv}{dz}=0\right)$ and axisymmetric $\left(\dfrac{dv}{d\theta}=0\right)$. The former implies that
right-hand side of the above equation can't depend on $z$, and thus pressure gradient, $\dfrac{dp}{dz}$, is a constant.
For a channel of length $L$, and denoting the pressure difference between the two ends by $\Delta p>0$ (high pressure minus low pressure), follows that $\dfrac{dp}{dz}=\dfrac{\Delta p}{L}$. By integrating the above equation, we find that the velocity profile, $v(r)$, is given by

\begin{nalign}
    \label{eq:v(R)}
    v(r)&=\int \frac{dv}{dr} dr = - \int \frac{1}{2\eta} 
    \dfrac{\Delta p}{L} r dr \\
       &= -\frac{1}{4\eta}\dfrac{\Delta p}{L}r^2 + C \;.
\end{nalign}

The constant of integration $C$ is found by imposing no-slip boundary conditions at the wall. In other words, it's assumed that the fluid particles adhere to the wall, and thus, there is no relative motion between the two, i.e.,  $v(R)=0$. This implies that

\begin{nalignat}[3]
\label{eq:C}
  && v(\,r=R\,) & = -\frac{1}{4\eta}\dfrac{\Delta p}{L}R^2 + C = 0 &\quad \iff \\
\iff \quad &&
C&= \frac{1}{4\eta} \dfrac{\Delta p}{L} R^2 \;. & 
\end{nalignat}

Substituting $C$ back in equation \ref{eq:v(R)} leads to 

\begin{equation}
    v(r)=\frac{1}{4\eta} \dfrac{\Delta p}{L} (R^2-r^2) \;, 
\end{equation}

\noindent which shows that the velocity of the fluid displays a parabolic profile, being zero at the wall, and attaining its maximum at the center of the pipe, $v_{max}=v(0)=\dfrac{1}{4\eta}\dfrac{\Delta p}{L}R^2$.

To compute the volumetric flow rate, consider a ring of fluid with thickness $dr$ at a distance $r$ from the centre of the pipe. Within a time $dt$, the fluid flowing through the ring covers a volume $dV= 2\pi rdr v(r)dt$. 
The flow rate $dQ$ of the ring fluid element is therefore 

\begin{equation}
    dQ=\frac{dV}{dt}
    =2\pi rv(r)dr \;.
    \label{eq:dQ}
\end{equation}

The total flow rate is finally obtained by integrating \ref{eq:dQ} over the entire cross-section of the pipe, yielding 

\begin{nalign}
    Q&=\int dQ 
     = \int^R_0 2\pi rv(r)dr \\
     &=\frac{2\pi}{4\eta} \dfrac{\Delta p}{L} \int^R_0 r (R^2-r^2) dr  \\
     &=\frac{\pi}{2\eta}\frac{dp}{dz} 
     \left.\left[ R^2\frac{r^2}{2}-\frac{r^4}{4}\right]
     \right|^{r=R}_{r=0} \\
     &=\frac{\pi R^4}{8\eta} \dfrac{\Delta p}{L} \;.
\end{nalign}

\section{Model Motivation}

As seen in section \ref{section:Physarum Solver}, the \textit{Physarum Solver} model assumes that the flow inside \Phy{} networks  can be described by the  \HP{} equation \ref{eq:HP_flow}. A fluid undergoing a \HP{} flow is assumed to be incompressible, which means that it has a constant density. Since the amount of fluid flowing through the network remains constant, this implies that no changes of the 
network's total volume may occur during the optimisation process. Therefore the adaptation dynamics proposed by Tero, 

\begin{equation}
    \frac{dD_{ij}}{dt}=f(|Q_{ij}|)-\mu D_{ij} \;, 
    \label{eq:Tero_update_rule_1} 
\end{equation}

\noindent should  ensure that the volume of the network is conserved over time.  

Each vessel is approximated by a cylindrical tube with radius $r_{ij}$ and length $L_{ij}$, corresponding to a volume $\V_{ij} =\pi r_{ij}^2 L_{ij}$. Given that $D_{ij}=\pi r_{ij}^4/8\eta$, the volume of each vessel can be rewritten as $\V_{ij}=\sqrt{8\pi\eta}L_{ij}\sqrt{D_{ij}}$. Hence, the conservation of the network's total volume reads

\begin{equation}
    \V = \sum_{(i,j)\in E} \V_{ij} = 
    \beta \sum_{(i,j)\in E} \sqrt{D_{ij}} L_{ij}  = \text{const} \;, 
    \label{eq:volume_conservation}
\end{equation}

\noindent where $\beta = \sqrt{8\pi\eta}$.

According to the model (\ref{eq:Tero_update_rule_1}), the change of the network volume over time is given by

\begin{nalign}
    \frac{d\V}{dt} =
    \beta \sum_{(i,j)\in E} \frac{1}{2\sqrt{D_{ij}}}\frac{dD_{ij}}{dt} L_{ij} 
    \propto \sum_{(i,j)\in E} \Big[ 
    f(|Q_{ij}|)-\mu D_{ij} \Big] \frac{L_{ij}}{\sqrt{D_{ij}}} \;.
    \label{eq:Tero_dVdt}
\end{nalign}

Since the choice of $f(|Q_{ij}|)$ is arbitrary, as long as $f(0)=0$ and $f'(|Q_{ij}|) > 0$, one can expect that not every choice results in (\ref{eq:Tero_dVdt}) being identically zero, and thus doesn't guarantee that the volume of the fluid is conserved. 

\begin{figure}[hbt!] 

\newcommand{\mysub}[2][]{%
    \subfloat[#1]{\includegraphics[trim={2.9cm 2.5cm 2.5cm 2.5cm}, clip, width=0.249\textwidth]{#2}}%
}
\newcommand{\myfig}[1]{%
    \includegraphics[trim={2.3cm 2.5cm 3cm 3.2cm}, clip, width=0.249\textwidth]{#1}%
}

\begin{subfigure}{\textwidth}   
    \centering
    \begin{tabular}{ccc}
    \myfig{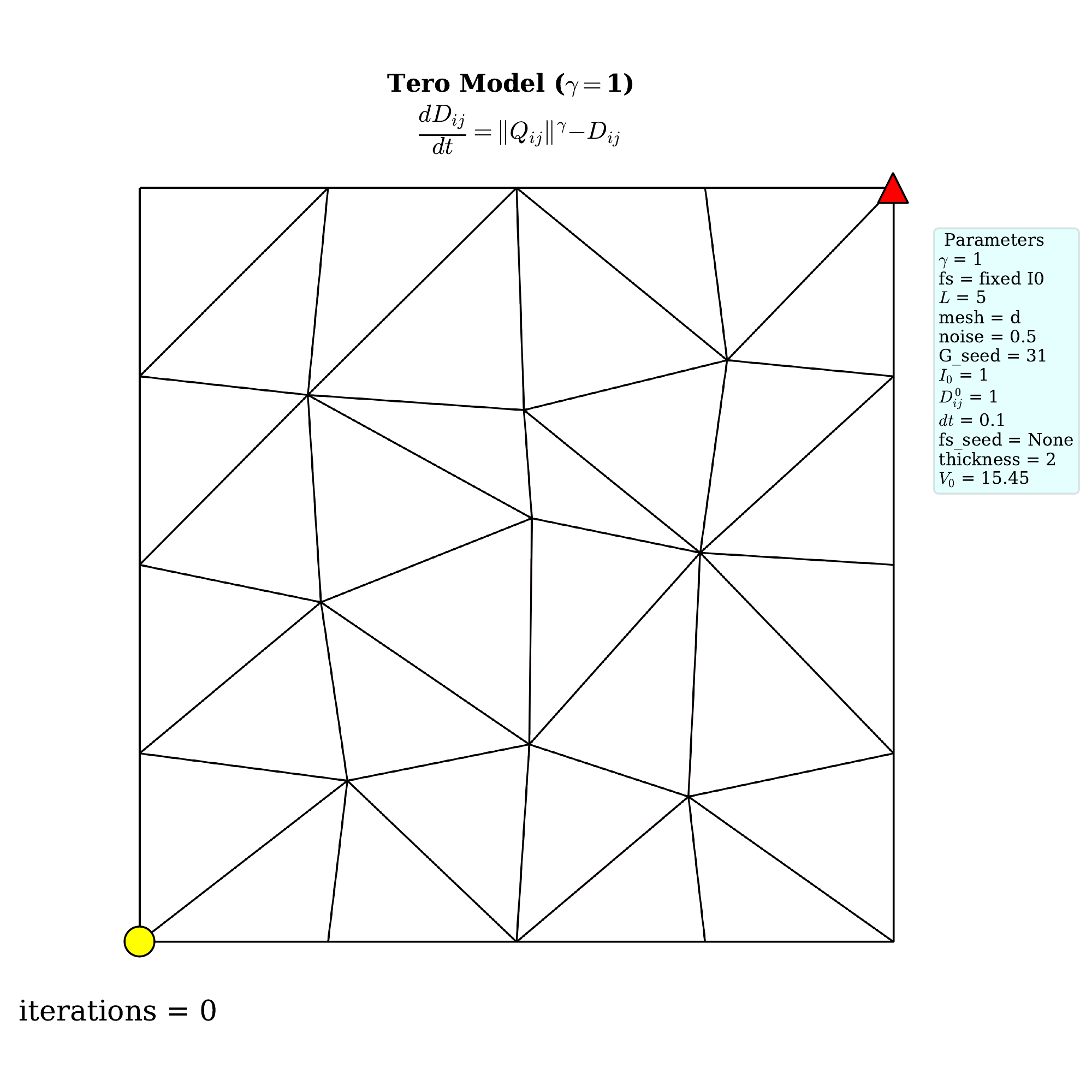} &
    \myfig{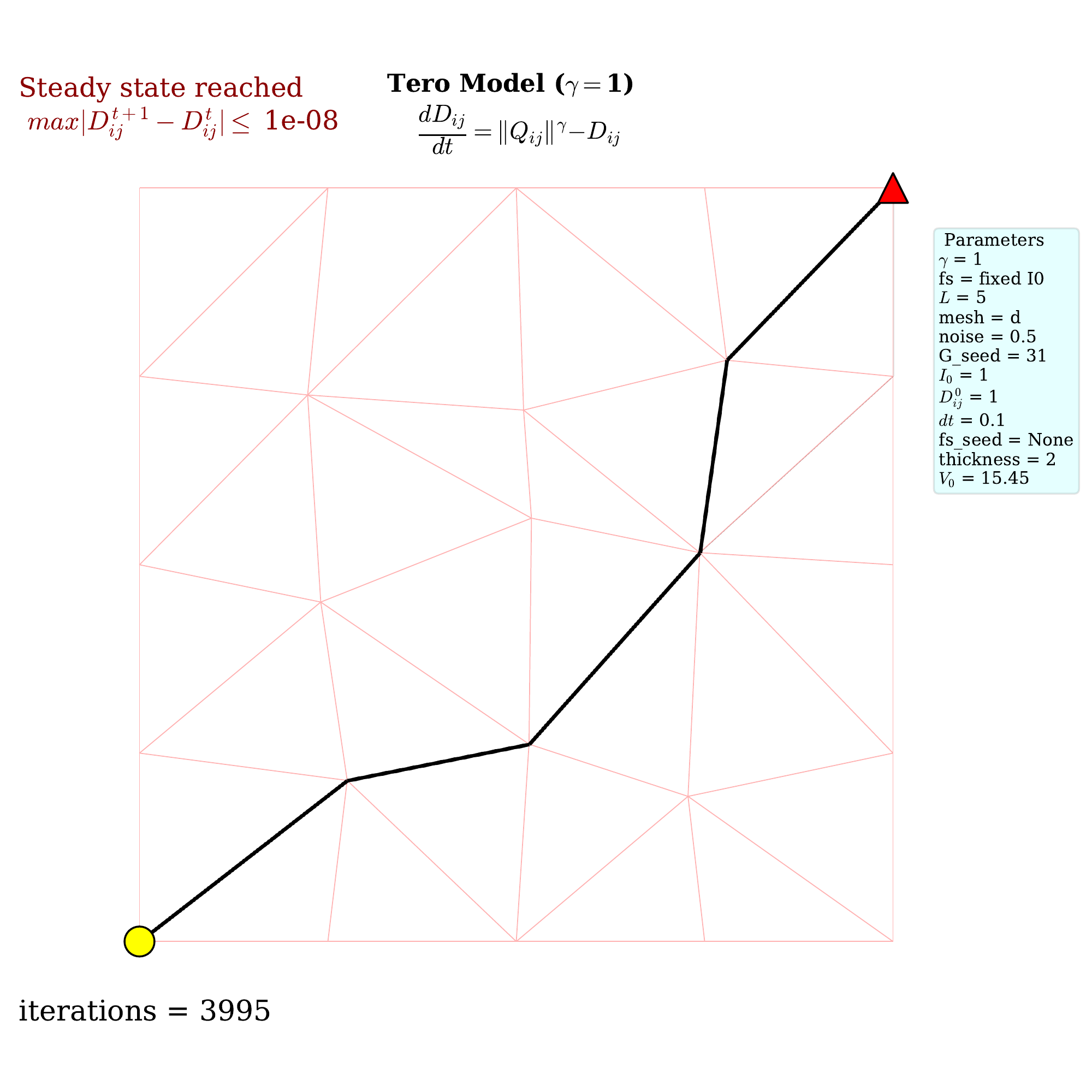} &
    \includegraphics[trim={0cm 0.2cm 0cm 0.1cm}, clip, width=0.34\textwidth]{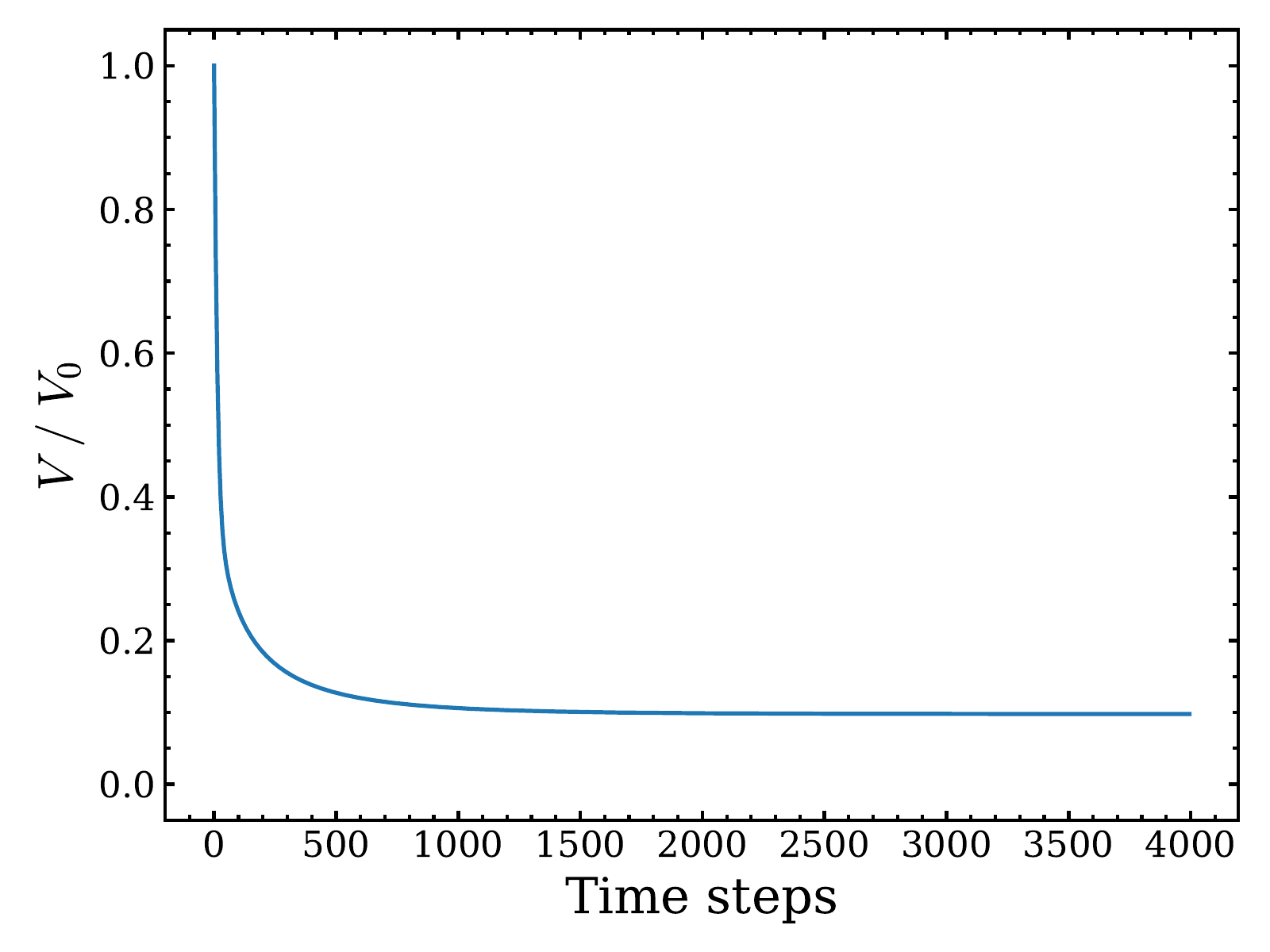} 
    \end{tabular}
    \caption{Simulation with fixed terminals: 1 source (yellow circle) and 1 sink (red triangle) with intensities $q_{source}= - q_{sink}=1$. Adaptation function: $f(|Q_{ij}|)=|Q_{ij}|$.
    }
    \label{fig:Tero_volume_a}
\end{subfigure}
\begin{subfigure}{\textwidth}   
    \centering
    \begin{tabular}{ccc}
    \myfig{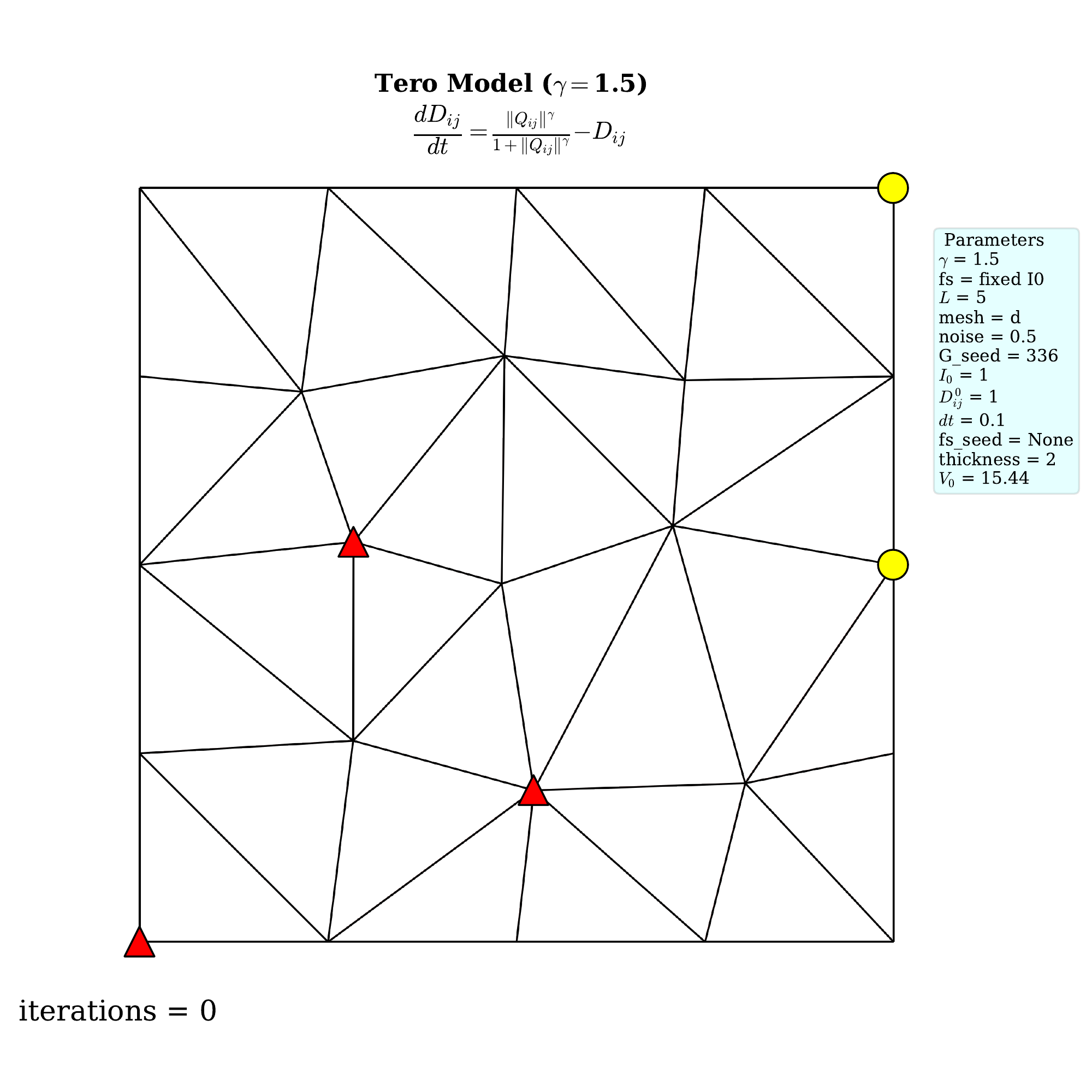} &
    \myfig{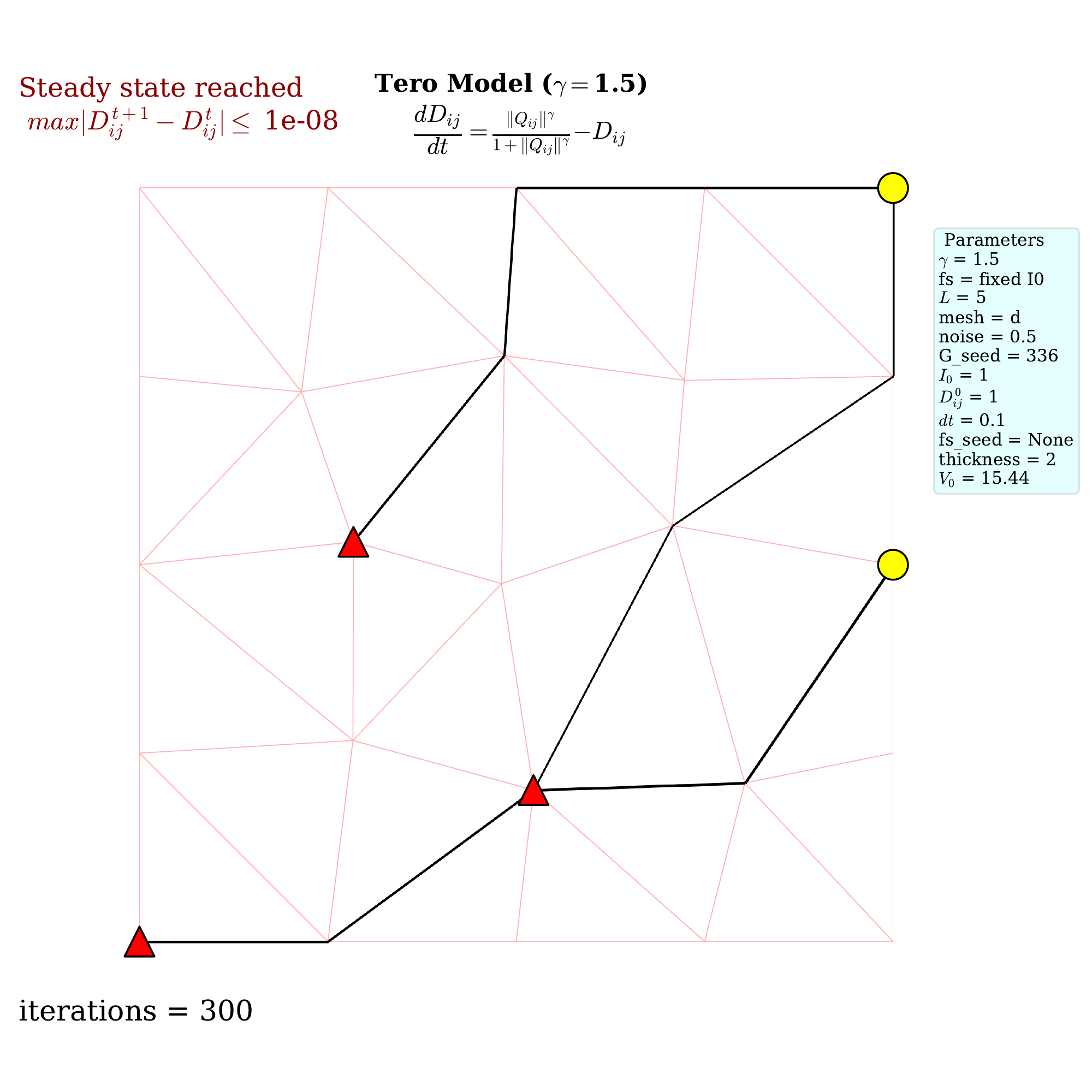} &
    \includegraphics[trim={0cm 0.2cm 0cm 0.1cm}, clip, width=0.34\textwidth]{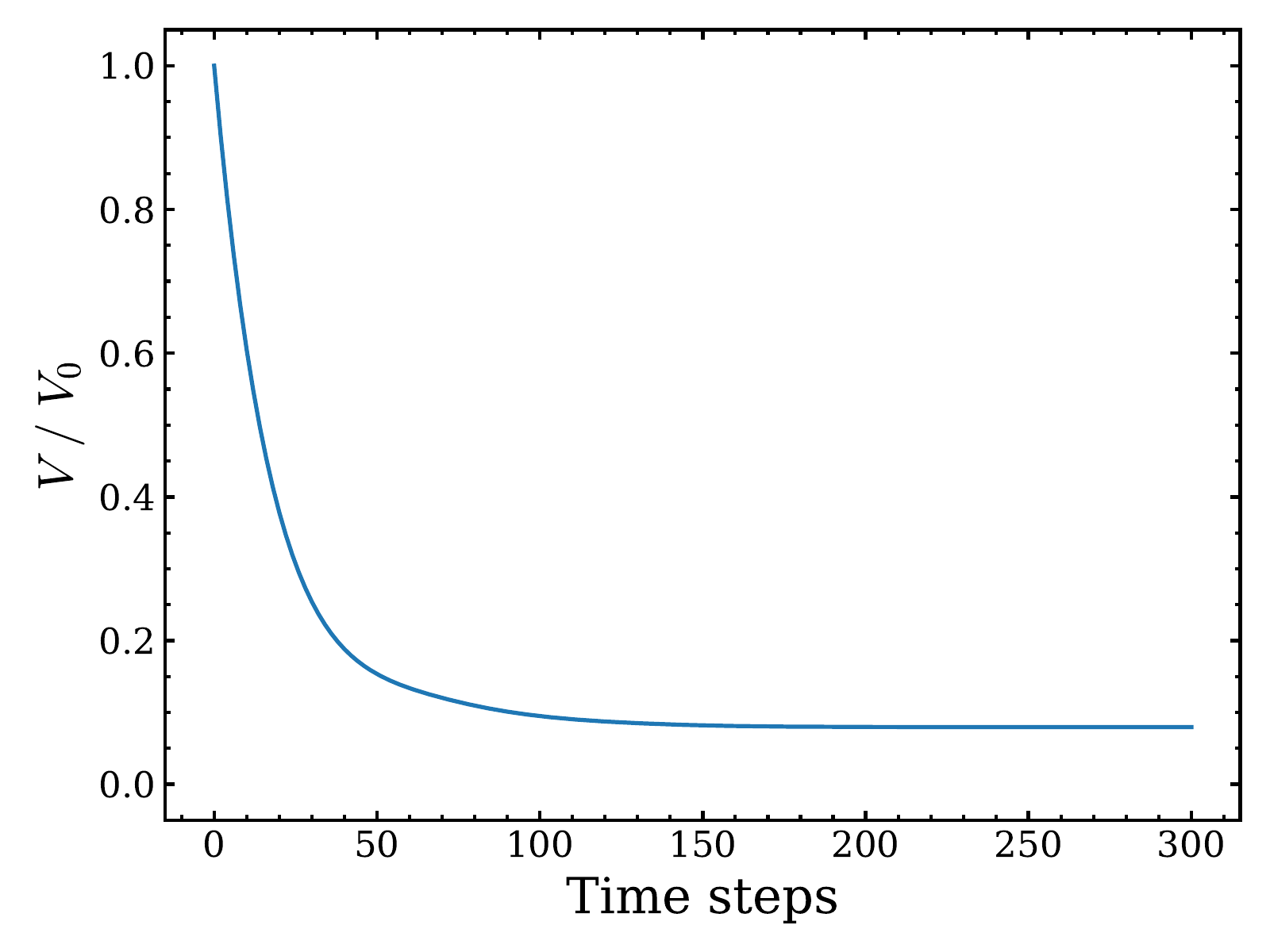} 
    \end{tabular}
    \caption{Simulation with 5 fixed terminals: 2 sources (yellow circles) and 3 sinks (red triangles), with intensities $q_{source}= - q_{sink}=1$.  Adaptation function: $f(|Q_{ij}|)=|Q_{ij}|^{1.5}/(1+|Q_{ij}|^{1.5})$.
    }
    \label{fig:Tero_volume_b}
\end{subfigure}
\begin{subfigure}{\textwidth}   
    \centering
    \begin{tabular}{ccc}
    \myfig{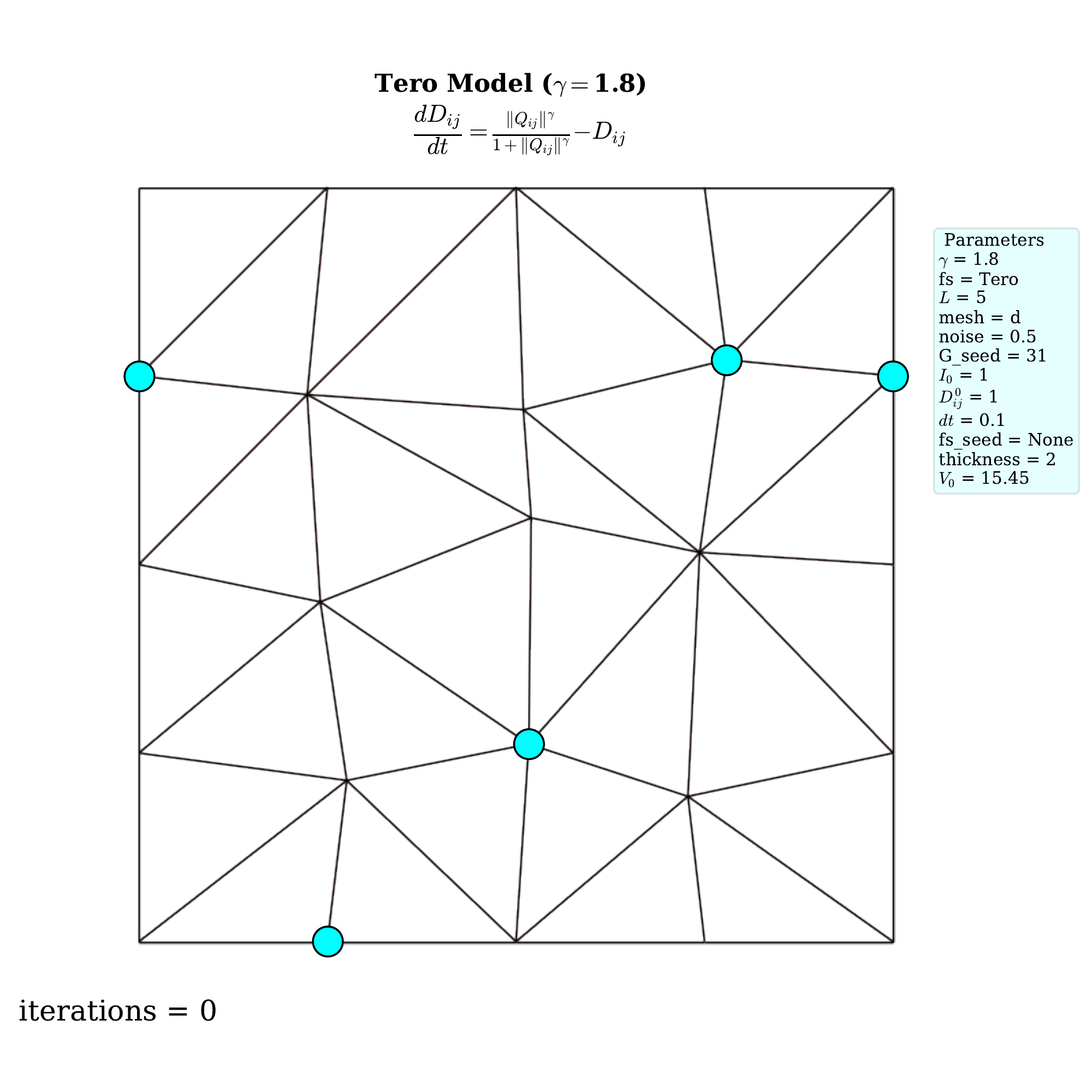} &
    \myfig{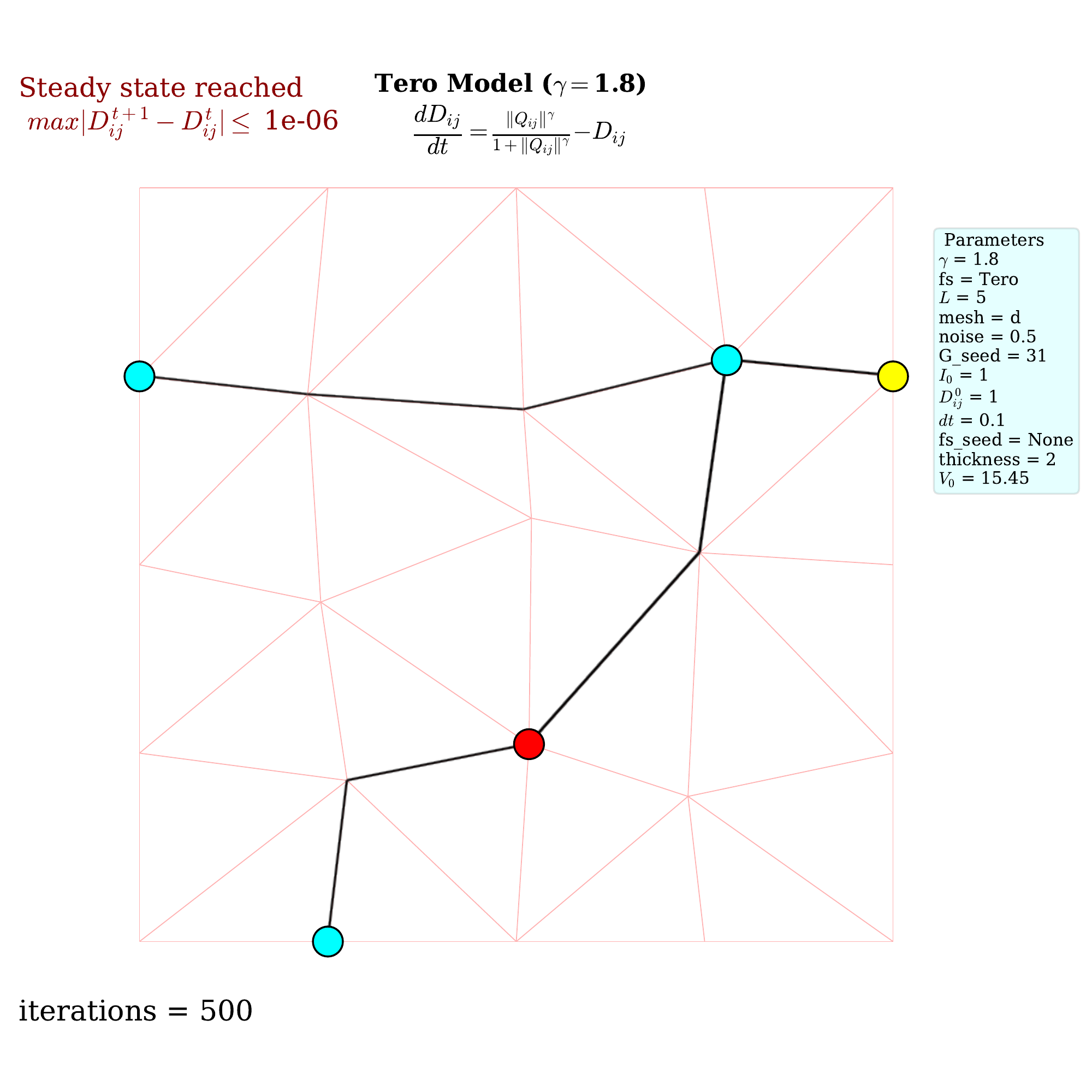} &
    \includegraphics[trim={0cm 0.2cm 0cm 0.1cm}, clip, width=0.34\textwidth]{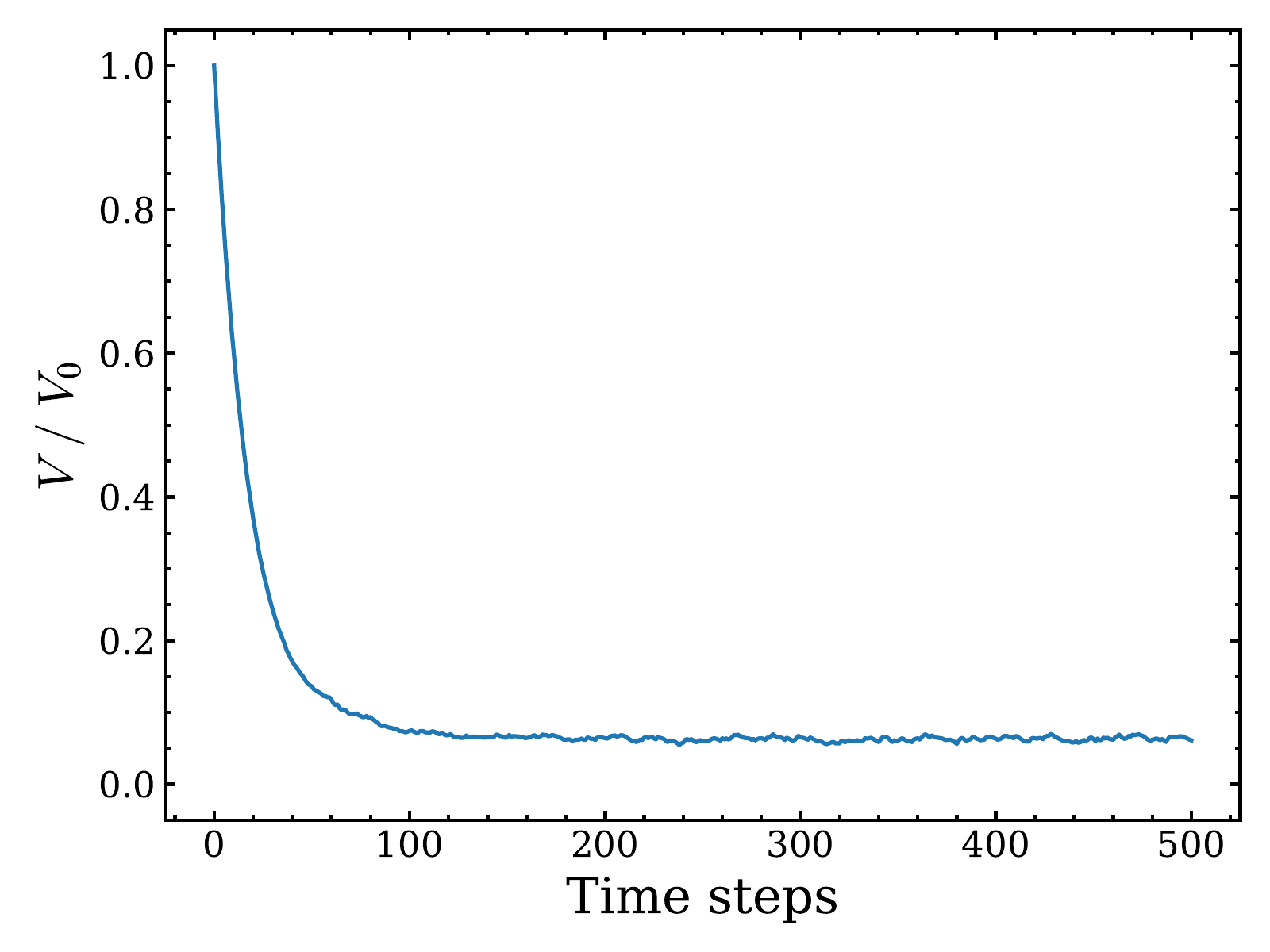} 
    \end{tabular}
    \caption{Simulation with 5 terminals (blue) and random choice of a source-sink pair in each step ($q_{source} = - q_{sink} = 1$) like the original algorithm.
    Adaptation function: $f(|Q_{ij}|)=|Q_{ij}|^{1.8}/(1+|Q_{ij}|^{1.8})$.
    }
    \label{fig:Tero_volume_c}
\end{subfigure}
\begin{subfigure}{\textwidth}   
    \centering
    \begin{tabular}{ccc}
    \myfig{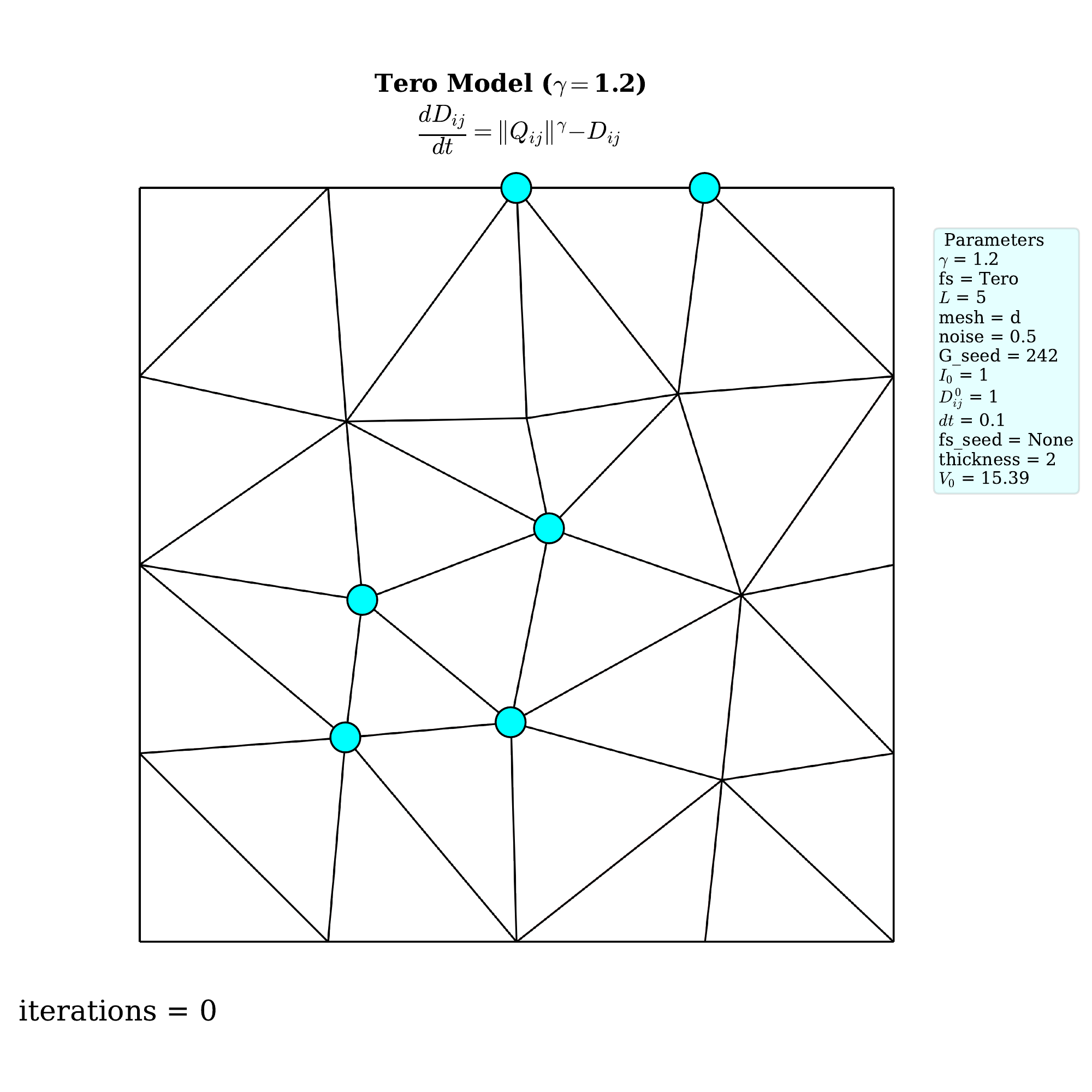} &
    \myfig{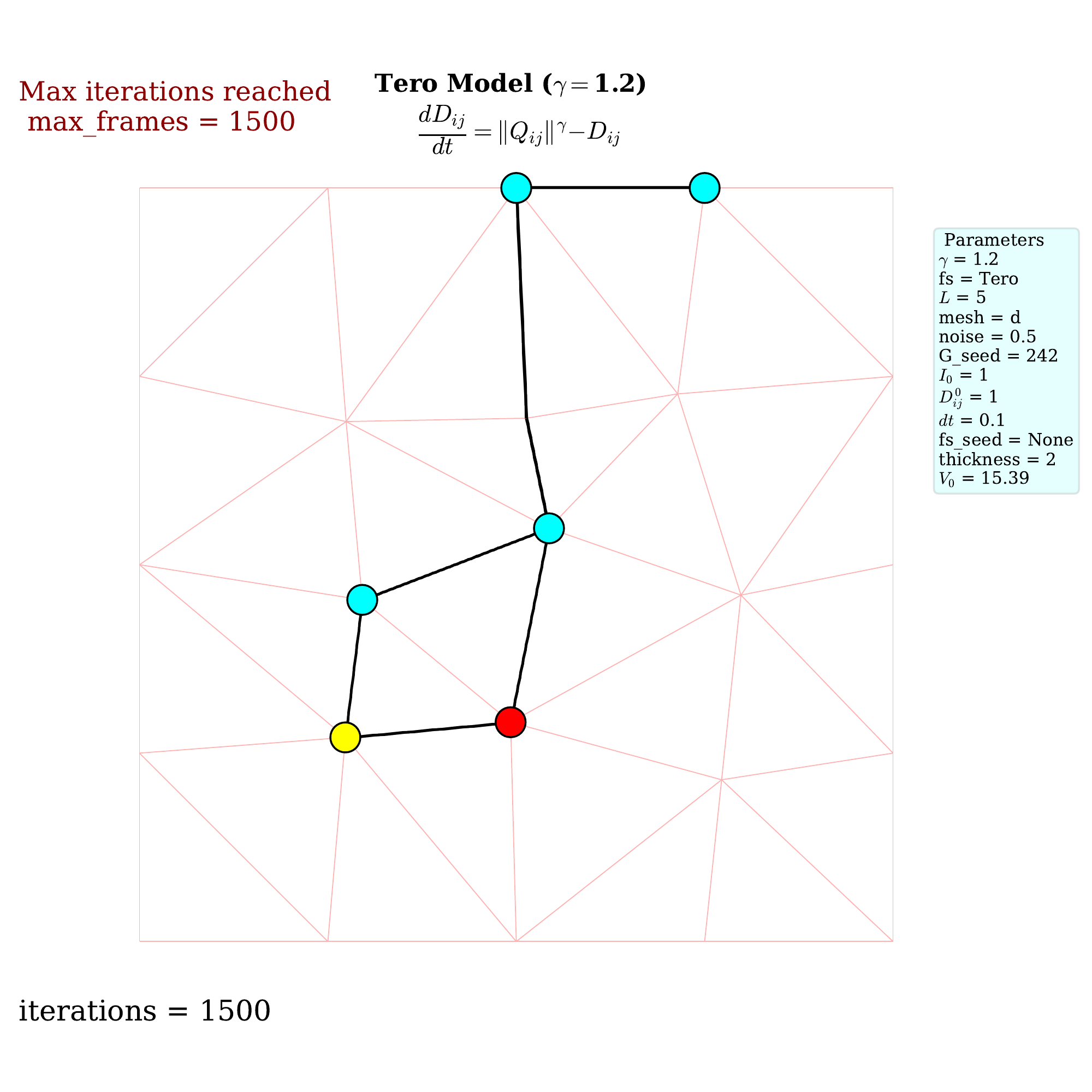} &
    \includegraphics[trim={0cm 0.2cm 0cm 0.1cm}, clip, width=0.34\textwidth]{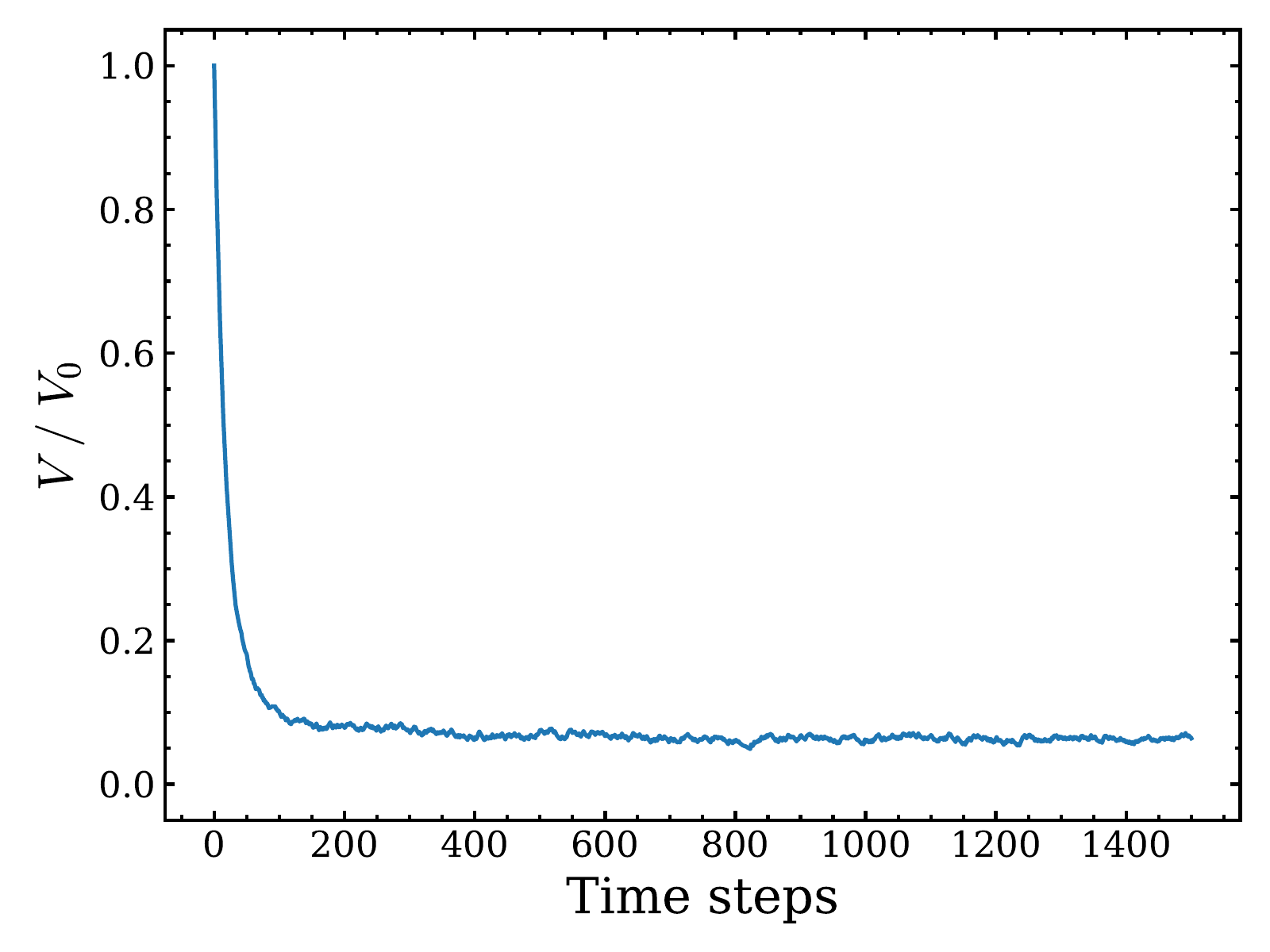} 
    \end{tabular}
    \caption{Simulation with 6 terminals (blue) and random choice of a source-sink pair in each step ($q_{source} = - q_{sink} = 1$) like the original algorithm.
    Adaptation function: $f(|Q_{ij}|)=|Q_{ij}|^{1.2}$.
    }
    \label{fig:Tero_volume_d}
\end{subfigure}

\caption{Simulations of the \textit{Physarum Solver} for different choices of the adaptation function $f$ in \eqref{eq:Tero_update_rule_1}, different distribution of sources and sinks,  considering $\mu=1$ and a total flux flowing through the network $I_0=1$. The simulations were carried out in planar graphs with $5\times5$ nodes, considering initial homogeneous conductivities, $D_{ij}(0)=1$. 
In each case, the left image depicts  the initial network geometry, the middle image corresponds to the steady state  of the adaptation mechanism, and the right plot represents the volume of the network over time, $\V$, normalised to the initial volume, $\V_0$. The thickness of the black lines is proportional to the radius of the edges $(\sim D_{ij}^{1/4})$.
The plots show that the volume is not conserved throughout any simulation, which implies that the \HP{} flow can no longer be applied as the fluid is compressible.}
\label{fig:Tero_volume} 
\end{figure}

This is confirmed by the results of Figure \ref{fig:Tero_volume}, showing four simulations of the \textit{Physarum Solver} in a small graph, 
for different distributions of sources and sinks, and different choices of adaptation functions, $f$, like the ones typically found in literature \eqref{eq:Tero_adaptation_funcs}. 
In the first two (Figure \ref{fig:Tero_volume_a} and Figure \ref{fig:Tero_volume_b}) we have considered  fixed configurations of sources and sinks, while in the other two (Figure \ref{fig:Tero_volume_c} and Figure \ref{fig:Tero_volume_d}), in each time step a source and a sink are randomly picked from the set of terminals, as described in section \ref{section:Physarum Solver}.
In each case, it's shown the plot of the network volume over the simulation time. As the plots show, the volume isn't conserved throughout the simulation, but rather tends to decrease over time, converging to a much lower value than the initial one, which corresponds to the volume of the steady-state solution. Note that in the last two simulations the volume exhibits some fluctuations due to the stochastic choice of the source-sink pair in each step, but the asymptotic behaviour is the same. 

We conclude that the adaptation dynamics proposed by Tero violates the volume conservation of the fluid, and thus the assumption that the fluid is incompressible. Therefore the model is physically inconsistent. This contradiction has been already reported in \cite{JFerreira_Tese}. In the following, we propose a more realistic evolution law for the channels conductivities that takes this assumption into account. 

\newpage
\clearpage

\section{Model Formulation}

\subsection{\textit{Physarum} as a Flow Network}

Similarly to the previous model, the geometry of \textit{Physarum}'s vein network is described as an undirected, planar and connected graph, $\G=(V, E)$, embedded in the Euclidean plane, where $V$ is the set of $\nodesnum$ nodes or vertices with coordinates $(x_i,y_i)$ for $i\in V$, and $E$ is the set of $\edgesnum$ straight edges $(i,j)$, connecting the node $i$ and $j$. The edges represent the network veins (channels), and the nodes the junctions between them.  

Each node $i$ is characterised by a pressure $p_i$. An edge $(i,j)$ is assumed to be a cylindrical elastic channel with a fixed length $L_{ij}$, and a radius $r_{ij}$ which can change in response to the magnitude of the flux flowing through it. The fluid in the network is viscous and incompressible, and undergoes a \HP{} flow, being the channel fluxes $Q_{ij}$ given by \ref{eq:HP_flow}.  If $Q_{ij}>0$, then the fluid flows from $i$ to $j$, while $Q_{ij}<0$ means that the flow is from $j$ to $i$. 

We assume that the network flows are driven by a set of sources and sinks (terminals), located at fixed nodes, which mimic stimulated regions of \textit{Physarum}.  Each node $i$ is  thus characterised by a net flux $q_{i}$. If a node $i$ is a source, it injects flow in the system, and $q_i>0$. If the node is a sink, it removes flow from the system,  and $q_i<0$;  otherwise $q_i=0$. The volume conservation of the fluid imposes that 

\begin{equation}
    \sum_{i\in V} q_i = \sum_{i \in \text{sources}} q_i +
    \sum_{i \in \text{sinks}} q_i = 0  \;.
    \label{eq:sum_qi}
\end{equation}

The channel fluxes can be determined by the conservation of the flux at each vertex $i$, which is also a direct consequence of the incompressibility of the fluid, 

\begin{equation}
    \sum_{j\in\Neigh(i)} Q_{ij} = 
    \sum_{j\in\Neigh(i)} \frac{D_{ij}(p_i-p_j)}{L_{ij}} = 
    q_i \quad, \quad i\in V
    \label{eq:sum_Qij}
\end{equation}

\noindent where $\Neigh(i)=\{j: (i,j)\in E\}$ is the set of the neighbour nodes of the node $i$, and $D_{ij}=D_{ji}$ is the conductivity of the channel ($i,j$).

\subsection{Adaptation Dynamics}

Experimental observations seem to support the current-reinforcement optimisation principle behind the \textit{Physarum Solver} model, i.e., the feedback between the flux and the vessel thickness. The conservation of the fluid's volume should play a crucial role in this regard, since it implies that vessels with higher flow rates expand at the expense of vessels with lower flow rates, which consequently shrink and eventually collapse. In this way, the volume conservation acts as a global constraint which enables the adaptation dynamics at one part of the network to affect the dynamics across the whole network. Therefore, the optimisation can't be described by a local mechanism, as in \textit{Physarum Solver}, and the role of volume constraint in the process must be acknowledged in the model.

In general, this problem is overlooked in the literature, so we aim to construct a more realistic model which tackles this issue by deriving a general adaptation rule which intrinsically conserves the volume \eqref{eq:volume_conservation}. According to our knowledge, the only solutions found in literature consist in projecting in each time step the solution of the adaptive equation \eqref{eq:Tero_update_rule_1} onto the surface defined by the volume constraint \cite{Takamatsu2017},  or through a simple global rescaling of the conductivities in each time step \cite{Jiang2019}. These constitute mathematical \textit{ad hoc} solutions which obfuscate the physical interpretation of the adaptation mechanism, and introduce unpredictable changes in the dynamics.  

The new adaptation rule should be able to capture the same current-reinforcement feedback dynamics as that of \textit{Physarum Solver}, but with the constraint that the total volume of the network must remain constant over time. This suggests basing our model on an adaptation rule very similar to the one proposed by Tero et al. \eqref{eq:Tero_update_rule_1}. However, since  we assume that the adaptation is described by variations of the vessels cross-section, $\pi r_{ij}^2 \propto \sqrt{D_{ij}}$, in response to the flux flowing through, it's more meaningful to derive an equation for the $\sqrt{D_{ij}}$ rather than for the conductivities $D_{ij}$. Therefore we make the \textit{ansatz} that

\begin{equation}
    \frac{d}{dt} \sqrt{D_{ij}} = f(\vb{Q}) - \mu\sqrt{D_{ij}} \;,
    \label{eq:new_ansatz}
\end{equation}

\noindent where $f(\vb{Q})$ is an unknown function of all the network fluxes $\vb{Q}$, with $f({\bf0})=0$, which generically describes the channel expansion due to the flux. The second term represents the tube shrinkage at a rate $\mu>0$ in the absence of flux. Below, following equation \eqref{eq:new_adapt_rule_tube}, we discuss the consistency of expression \eqref{eq:new_ansatz}.

Now we can ask under what conditions this adaption rule conserves the
volume. Naturally, this limits the choice of the function $f$, in contrast to Tero's model, where the choice is roughly arbitrary. By differentiating both sides of (\ref{eq:volume_conservation}), the conservation of the volume can be described by the following constraint 

\begin{equation}
    \frac{d\V}{dt} = \beta \sum_{(i,j)\in E}
    L_{ij} \frac{d}{dt} \sqrt{D_{ij}} = 0 \;.
    \label{eq:dV0/dt}
\end{equation}

Replacing (\ref{eq:new_ansatz}) in the last equation and using (\ref{eq:volume_conservation}) yields 

\begin{equation}
  \sum_{(i,j)\in E}L_{ij} f(\vb{Q}) = 
   \mu \underbrace{\sum_{(i,j)\in E}L_{ij} \sqrt{D_{ij}}}_{\V/\beta} = 
   \mu \frac{\V}{\beta} \;.
   \label{eq:Lijfij}
\end{equation}

To satisfy this condition, it's sufficient to 
define $f$ in terms of a new function $g$ by the relation

\begin{equation}
    f(\vb{Q}) :=  \frac{\mu}{\beta}\V 
    \frac{g(|Q_{ij}|)}{\sum\limits_{(k,m)\in E}L_{km} g(|Q_{km}|)}
    \label{eq:fij} \;.
\end{equation}

By introducing the last expression into the \textit{ansatz}  (\ref{eq:new_ansatz}), and redefining the time scale $\tau = \mu t$ we obtain 

\begin{equation}
\frac{d}{d\tau} \sqrt{D_{ij}} = 
\frac{\V}{\beta} \frac{g_{ij}}{\sum\limits_{(k,m)\in E}L_{km} g_{km}} - \sqrt{D_{ij}}
\quad, \quad (i,j)\in E
 \label{eq:new_adapt_rule}
\end{equation}

\noindent where $g_{ij}\equiv g(|Q_{ij}|)$. Therefore, we conclude that for any choice of the function $g$, the volume  of the fluid in a network with adaptive channel conductivities evolving according to (\ref{eq:new_adapt_rule}) is conserved over time. Thus, it correctly describes an optimisation process of  a network  filled with an incompressible fluid subjected to a \HP{} flow.  

Note that, the adaptation rule (\ref{eq:new_adapt_rule}) depends on the overall structure of the flows through the term  $Z\equiv \sum_{(k,m)}L_{km} g(|Q_{km}|)$, making the adaptation explicitly a non-local process, in contrast to \textit{the Physarum Solver} model  (\ref{eq:Tero_update_rule_1}), where the coupling of the system's dynamics stems only from conservation of the flux at the nodes \eqref{eq:sum_Qij}. This global coupling factor can be seen as a measure of an effective network length, where the contribution of each channel is weighted by a function of its local flux, $g$.  

For a better comparison with the previous model (\ref{eq:Tero_update_rule_1}), given that $d\sqrt{D_{ij}}=(2\sqrt{D_{ij}})^{-1}dD_{ij}$, the expression (\ref{eq:new_adapt_rule}) can be rewritten in terms of the conductivities $D_{ij}$ as

\begin{equation}
\frac{dD_{ij}}{d\tau}= 2\frac{\V}{\beta}\sqrt{D_{ij}} 
\frac{g_{ij}}{\sum\limits_{(k,m)\in E}L_{km} g_{km}} - 2D_{ij}
\quad, \quad (i,j)\in E \;.
\label{eq:continous_model_Dij}
\end{equation}

\section{Minimisation of Energy Dissipation}

In the previous section, we have derived a general class of adaption models describing the flow of incompressible fluids in networks of elastic channels with arbitrary geometry. However, to analyse the temporal evolution of a network following the adaptation dynamics (\ref{eq:new_adapt_rule}), the function $g$ must be chosen.
The choice of $g$ can't be completely arbitrary in order to preserve the physical meaning of $f$ in (\ref{eq:new_ansatz}). By (\ref{eq:fij}) it's required that $g(0)=0$ so that $f(0)=0$. Also, $g$ should be such that 
$\partial f/\partial|Q_{ij}|\geq 0$ for a given channel $(i,j)$, since a channel should expand if its local flux is increased until it eventually saturates. This last condition is harder to ensure, given that $f$ now depends on all the channel fluxes, which are inherently coupled through the conservation laws \eqref{eq:sum_Qij}. 

Here, the choice of $g$ is made by introducing the criterion of
minimisation of the total power dissipated during the flow (dissipation), subject to the constraint of the fixed volume of fluid, assuming a steady flow imposed by a fixed set of sources and sinks \cite{Bohn2007}. When a fluid flows  through a channel some energy is lost due to  friction. The energy dissipation rate depends on parameters such as the fluid’s speed and viscosity, and for a steady-state  flow through a channel $(i,j)$ is given by $\E_{ij} = \Delta p_{ij} Q_{ij}$. The total dissipation of a flow network, $\E$, is the sum of the dissipation at each channel, 

\begin{equation}
    \E= \sum_{(i,j)\in E} \Delta p_{ij} Q_{ij}
    =\sum_{(i,j)\in E} 
    \frac{Q_{ij}^2}{D_{ij}}L_{ij} \;.
    \label{eq:energy_dissipation}
\end{equation}

The principle of least energy dissipation is widely used in the context of the optimisation of biological transport networks, such as vascular systems, leaf venation in plants, and river networks. The notion of optimal network is typically defined as the one which minimises the energy dissipation under an optional set of constraints, e.g., a limited amount of resources \cite{Banavar2001,Bohn2007,Corson2010}, or other cost functionals involving the energy \cite{Hu2013}. The functional being minimised and the constraints to which it is subject have a profound impact on the structure and properties of the optimal networks \cite{Banavar2001}.

In the case of \textit{Physarum}, although the mechanism underlying the network adaptation is not well-understood, and is certainly more complex than that, it's reasonable to admit that is somehow related to the  optimisation of its network's energy consumption. However not in a straightforward way as we assume here since the organism is far from a state of equilibrium. \textit{Physarum} displays a continuous adaptation to the environmental stimuli mediated by the shuttle streaming, which introduces fluctuations in the flux. This means that the fluctuations should be accounted for in the minimisation process \cite{Corson2010}, but for simplicity, we ignore that and consider the minimisation under a steady flow regime generated by a constant set of sources and sinks. 

We seek to minimise the dissipation rate, $\E$, of a steady flow network with respect to $\bm{\sqrt{D}}\equiv \{\sqrt{D_{ij}}: (i,j) \in E \}$,  subject to the local constraints of flux conservation \eqref{eq:sum_Qij}, and the additional global constraint of a constant volume, $V$
(incompressible fluid). The problem consists of minimising the following Lagrangian

\begin{nalign}
    \Lag&=\E -\lambda (\V-
    \beta\sum_{(k,m)\in E}\sqrt{D_{km}}L_{km}) \\
    &= \sum_{(k,m)\in E} 
    \frac{Q_{km}^2}{D_{km}}L_{km} -
    \lambda (V-
    \beta\sum_{(k,m)\in E}\sqrt{D_{km}}L_{km}) \;,
\end{nalign}

\noindent where $\lambda$ is a Lagrangian multiplier. Note that the channel fluxes can't be regarded  as independent variables in the minimisation process as they are uniquely determined for a given distribution of the nodes' fluxes, $\vb{q}$, and channels conductivities through \eqref{eq:sum_Qij}. The set of conductivities that minimises  $\Lag$ is the solution of

\begin{equation}
    \frac{\partial \Lag}{\partial \sqrt{D_{ij}}} =0 
    \quad \text{ for } (i,j)\in E  \qquad, \qquad   
    \frac{\partial \Lag}{\partial \lambda} =0 \;.
\end{equation}

We start by considering the derivative with respect to $\sqrt{D_{ij}}$. Using the chain rule  we have that 

\begin{equation}
   \frac{\partial \Lag}{\partial \sqrt{D_{ij}}} = 2\sqrt{D_{ij}}\frac{\partial \Lag}{\partial D_{ij}}  \;,
\end{equation}

\noindent where 

\begin{nalign}
\frac{\partial \Lag}{\partial D_{ij}}&=
\frac{\partial \E}{\partial D_{ij}} - \lambda
\frac{\partial }{\partial D_{ij}}
(\V-\beta\sum_{(k,m)\in E}\sqrt{D_{km}}L_{km}) \\ 
&=\frac{\partial \E}{\partial D_{ij}}
 + \lambda\beta\frac{L_{ij}}{2\sqrt{D_{ij}}} \;.
\label{eq:dLdDij}
\end{nalign}

Expanding the first term of the last expression yields 

\begin{nalign}
\frac{\partial \E }{\partial D_{ij}}&= 
\frac{\partial}{\partial D_{ij}}
\sum_{(k,m)\in E} \frac{Q_{km}^2}{D_{km}}L_{km}  \\
&= -\frac{Q^2_{ij}}{D_{ij}^2}L_{ij}+2
\sum_{(k,m)\in E}\frac{Q_{km}}{D_{km}}
\frac{\partial Q_{km}}{\partial D_{ij}}L_{km} \;.
\label{eq:dPdDij}
\end{nalign}

Denoting the adjacency matrix of the network by $\bm{A}=[A_{ij}]$, i.e., the matrix with entries $A_{ij} =1$ if $(i,j)\in E$ and $A_{ij}=0$ otherwise, the second term of (\ref{eq:dPdDij}) can be written as  

\begin{nalign}
2\sum_{(k,m)\in E} \ 
\underbrace{\ \frac{Q_{km}}{D_{km}}L_{km} \ }_{p_m-p_k}
\frac{\partial Q_{km}}{\partial D_{ij}}
&=\sumnodes{k}\sumnodes{m} A_{km}
(p_m-p_k) \frac{\partial Q_{km}}{\partial D_{ij}} \;.
\end{nalign}

Using the symmetry of the adjacency matrix, $A_{ij}=A_{ji}$, (undirected graph), and the antisymmetry of the flux matrix, $Q_{ij} = - Q_{ji}$, this can be further simplified to (cf. Lemma 2.1 in \cite{Haskovec2019})

\begin{nalign}
\sumnodes{k}\sumnodes{m} A_{km} (p_m-p_k) \frac{\partial Q_{km}}{\partial D_{ij}} 
&= \underbrace{\sumnodes{k}\sumnodes{m} A_{mk} p_k 
\frac{\partial Q_{mk}}{\partial D_{ij}}}_{k\leftrightarrow m}
-\sumnodes{k}\sumnodes{m} A_{km} p_k
\frac{\partial Q_{km}}{\partial D_{ij}} \\
&=\sumnodes{k}\sumnodes{m} A_{km} p_k 
\frac{\partial (-Q_{km})}{\partial D_{ij}}
-\sumnodes{k}\sumnodes{m} A_{km} p_k
\frac{\partial Q_{km}}{\partial D_{ij}} \\
&=-2\sumnodes{k}p_k\sumnodes{m} A_{km}  
\frac{\partial Q_{km}}{\partial D_{ij}} \\
&=-2\sumnodes{k}p_k\frac{\partial}{\partial D_{ij}}
\sum_{m\in \Neigh(k)} Q_{km} \\
&=-2\sumnodes{k}p_k\frac{\partial q_k}{\partial D_{ij}} \;,
\end{nalign}

\noindent where the conservation of the flux \eqref{eq:sum_Qij} was used in the last step.  

For fixed sources and sinks, $\dfrac{\partial q_k}{\partial D_{ij}} = 0$, which implies that the second term in (\ref{eq:dPdDij}) is zero. Therefore, within the assumption of constant sources and sinks, the derivatives of $\Lag$ with respect to $\sqrt{D_{ij}}$ are given by 

\begin{equation}
\frac{\partial \Lag}{\partial \sqrt{D_{ij}}}
= \left(-\frac{Q^2_{ij}}{D_{ij}^2}L_{ij} 
+ \lambda'\frac{L_{ij}}{2\sqrt{D_{ij}}}\right) 2\sqrt{D_{ij}} \;.
\end{equation}

\noindent where we've redefined the Lagrangian multiplier $\lambda'=\beta\lambda$. The minima of $\Lag$ satisfy

\begin{equation}
\frac{\partial \Lag}{\partial \sqrt{D_{ij}}}=0 
\quad \iff \quad    
\begin{cases}
D_{ij}= \left(\dfrac{2}{\lambda'}\right)^{2/3} Q_{ij}^{4/3} \\
D_{ij} = 0 \;.
\end{cases} 
\end{equation}

The constant of proportionality of the non-trivial minima, $\alpha\equiv \left(\dfrac{2}{\lambda'}\right)^{2/3}$, can be determined by direct substitution in the constraint equation  \eqref{eq:volume_conservation}, resulting from $\dfrac{\partial \Lag}{\partial \lambda}=0$,

\begin{equation}
\frac{\partial \Lag}{\partial \lambda}=0 
\quad \iff \quad 
\alpha =\left(\frac{\V/\beta}{\sum\limits_{(k,m)\in E}Q_{km}^{2/3}L_{km}}\right)^2
\;.
\end{equation}

Therefore the non-trivial values of conductivities that minimise the total dissipation of the network are

\begin{equation}
D_{ij}=\left(\frac{\V}{\beta}
\frac{Q_{ij}^{2/3}}{\sum\limits_{(k,m)\in E}Q_{km}^{2/3}L_{km}}\right)^2 \;.
\label{eq:minEDij}
\end{equation}

This scaling relation between the conductivities and the fluxes in the minimal configuration of energy is very similar to the one obtained in \cite{Bohn2007}.

On the other hand, we find that the non-trivial steady states of the volume-preserving adaptation law (\ref{eq:new_adapt_rule}) satisfy 

\begin{equation}
\frac{d}{d\tau} \sqrt{D_{ij}^*} = 0
\quad \iff \quad
D_{ij}^* = \left(\frac{\V}{\beta}\frac{g_{ij}}{\sum\limits_{(k,m)\in E}L_{km} g_{km}} \right)^2 
\label{eq:new_steady_state} \;.
\end{equation}

Comparing the results (\ref{eq:minEDij}) and (\ref{eq:new_adapt_rule}), we conclude that for the choice of $g_{ij}=Q_{ij}^{2/3}$, the total dissipation of the network  at the steady state is minimal, assuming a constant distribution of nodes flux, $\vb*{q}$, during the adaptation process. With these choices, we obtain the following dynamics which preserves the volume and converges to a state of minimal dissipated energy,

\begin{equation}
\frac{d}{d\tau} \sqrt{D_{ij}} = 
\frac{\V}{\beta}\frac{Q_{ij}^{2/3}}{\sum\limits_{(k,m)\in E}L_{km} Q_{km}^{2/3}} - \sqrt{D_{ij}} \quad, \quad (i,j)\in E \;.
\label{eq:new_adapt_rule_minE}
\end{equation}

As a side note, for the simplest case of a single elastic channel with length $L_{12}$ and conductivity $D_{12}$, the equation \eqref{eq:new_adapt_rule} reduces to 

\begin{equation}
\frac{d}{d\tau} \sqrt{D_{12}} = \frac{\V}{\beta} \frac{1}{L_{12}} - \sqrt{D_{12}} \;,
\label{eq:new_adapt_rule_tube}
\end{equation}

\noindent which is independent of the choice of the function $g$. 
This differential equation has a unique stable fixed point for $\sqrt{D_{12}^*}=\V/(\beta L_{12})$ which coincides with \eqref{eq:minEDij}. Therefore, according to our model \eqref{eq:new_adapt_rule}, for a \HP{} flow on a single elastic tube, the radius of the tube at the steady state minimises the power dissipated by the flow, regardless of the choice of $g$. This property is consistent with the distribution of channel fluxes arising from the Kirchhoff Law \eqref{eq:sum_Qij}, which is by definition the one which minimises the total energy dissipated \cite{Feynman1975}, and thus justifies the \textit{ansatz} made in \eqref{eq:new_ansatz} by adding the term $\sqrt{D_{ij}}$.

\section{Methods}

\subsection{Algorithm}

The algorithm used to simulate our model is described in the following.
First, we generate a planar graph, $\G$, embedded in the two-dimensional Euclidean space, which represents the initial geometry of \Phy's network, or of any other transport network. The edges lengths $L_{ij}$ are obtained based on the node positions. These two remain fixed throughout the simulation, and only the conductivities $D_{ij}$ are the target of adaptation. Some of the nodes are assigned as sources or sinks and have a net flux different from zero, such that the constraint \ref{eq:sum_qi} is verified. The initial conditions of our dynamical system are the initial edges conductivities, $D_{ij}(0)$. Usually, we consider an initial homogeneous distribution, and if they aren't specified, it's assumed $D_{ij}(0)=1$ for all edges $(i,j)\in E$. The initial conductivities are used to compute the total volume of fluid through \eqref{eq:volume_conservation}, where we always consider the parameter $\beta=1$. 

Given the distribution of the nodes' net fluxes, $\vb*{q}$, and the initial set of edge conductivities, the temporal evolution of the system starts with the computation of the channel fluxes $Q_{ij}$ by solving the linear system \eqref{eq:sum_Qij}. Then, based on those fluxes, the conductivities of all the channels are updated according to \eqref{eq:new_adapt_rule_minE}, or more generally, according to  \eqref{eq:new_adapt_rule} given a predefined function $g$. The new channel conductivities are used to compute the new channel fluxes in the next time step of the algorithm. These steps are repeated over time until a steady state of channel conductivities is eventually reached.

From the numerical point of view, we consider that a steady state is reached when the change in the conductivity of all channels from one step to another is lower than $10^{-6}$. This can be translated into
the following stopping condition  

\begin{equation}
    \underset{(i,j)\in E}{max}|D_{ij}(n\dt) - D_{ij}((n-1)\dt)| \ \leq  \ 10^{-6} \;.
    \label{eq:stop_condition} 
\end{equation}

\noindent where $n$ is the first integer for which the inequality is verified, and $\dt$ is the time increment used to solve numerically the adaptation rule \eqref{eq:new_adapt_rule}. The time of convergence of the adaptation algorithm is thus $\tau*= n\dt <\infty$.

The simulations were all done in \textit{Python}, exploiting in particular the modules \textsf{NetworkX} for mesh representation and graph analysis; \textsf{SciPy} and \textsf{NumPy} for numerical computations; and \textsf{Matplotlib} and \textsf{seaborn} for the graphical interface and plotting. More exotic initial meshes were generated using \textit{Mathematica}.

\subsection{Computation of the network flows}
\label{section:computation_Qij}

The network can be seen as an edge-weighted graph, where the edge weights are given by $C_{ij}=\frac{D_{ij}}{L_{ij}}$,  measuring the degree of ease with which a channel can carry flow. The channel fluxes in each time step can be computed by first finding the pressures of the nodes. The linear system \eqref{eq:sum_Qij} can be rewritten as 

\begin{figure}[H]
\includegraphics[width=0.75\textwidth]{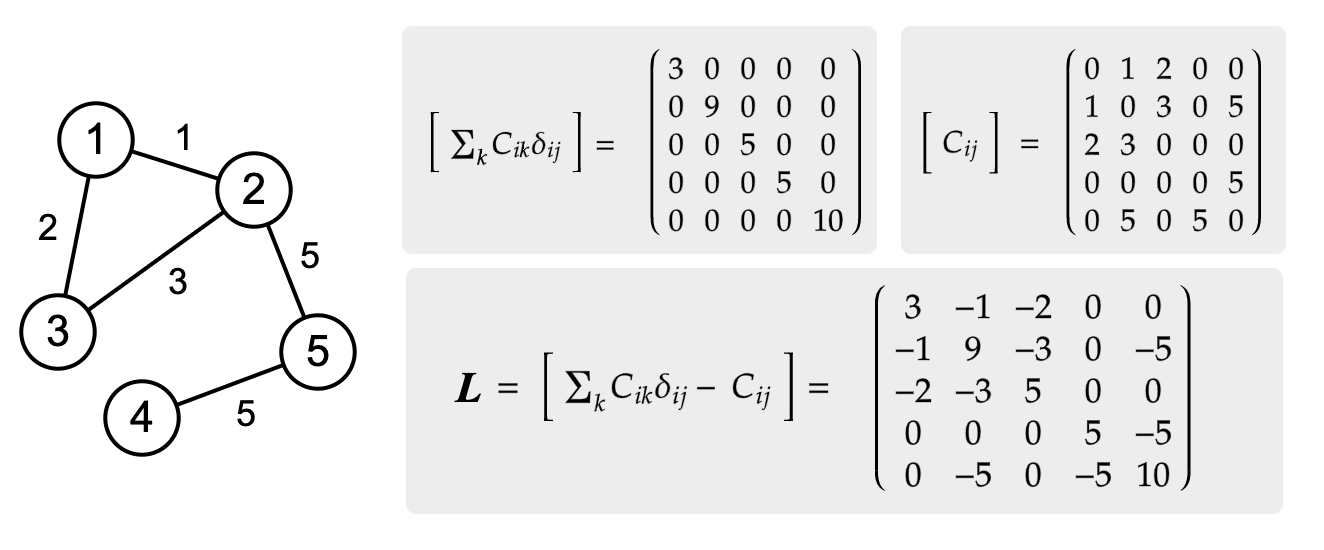}
\centering
\caption{Example of the calculation of the Laplacian matrix for a small weighted graph. On the left, the circles represent the nodes, labelled by their index, and lines represent the edges labelled by the corresponding weight.}
\label{fig:lap_matrix_diagram}
\end{figure}

\begin{nalign}
    \sum_j Q_{ij} &= \sum_j C_{ij}(p_i-p_j) \\
    &= \left(\sum_j C_{ij}\right)p_i- \sum_jC_{ij}p_j \\
    &= \left(\sum_k C_{ik} \right) \sum_j \delta_{ij} p_j -\sum_j C_{ij}p_j \\
    &= \sum_j \left[\left(\sum_k C_{ik}\right)\delta_{ij} - C_{ij} \right]p_j \\
    & = \sum_j \ell_{ij}p_j \;.
    \label{eq:Lijpj}
\end{nalign}

\noindent where $\delta_{ij}$ denotes the Kronecker delta function. Let  $\vb{p}$ be the $\nodesnum$-dimensional vector whose ith element is the pressure $p_i$ of the ith node, and $\vb{q}$ the  $\nodesnum$-dimensional vector whose ith element is the net current $q_i$ through the ith node.
If we define $\vb*{L}$ as the $N \times N$ symmetric matrix with entries 

\begin{equation}
    \ell_{ij}=\left(\sum_k C_{ik}\right)\delta_{ij} - C_{ij} \;,
    \label{eq:Lapij}
\end{equation}

\noindent the system of conservation laws (\ref{eq:Lijpj}) can be written in matrix form as

\begin{equation}
    \vb*{L}\vb{p}=\vb{q} \;.
    \label{eq:Lp=q}
\end{equation}
The  matrix $\vb*{L}$ is the generalisation of the Laplacian matrix for a weighted graph with edge weights $C_{ij}$. For a simple undirected graph, the Laplacian matrix is defined as $\vb*{L}=\vb*{D}-\vb*{A}$, where $\vb*{D}$ is the degree matrix \footnote{The degree matrix of a simple undirected graph $\G=(V,E)$ with $\nodesnum = |V|$ nodes is a $\nodesnum \times \nodesnum$ diagonal matrix $\vb*{D}=diag(deg_1,...,deg_\nodesnum)$ where $deg_i$ is the number of neighbour nodes (degree) of the node $i$.}
and $\vb*{A}$ the adjacency matrix of the graph. In Figure \ref{fig:lap_matrix_diagram} it's depicted the calculation of the Laplacian matrix for a small weighted graph.  

Note that the Laplacian matrix is singular since all the rows sum to zero, meaning  it has an eigenvector $\mathbf{1}=(1,1,...,1)$ with a zero eigenvalue ($\vb*{L}\mathbf{1}=\mathbf{0}$). In fact, it can be shown the dimension of the nullspace of the Laplacian and algebraic multiplicity of the zero eigenvalue is equal to the number of connected components of the graph \cite{Godsil2001}.  In the particular case of connected graphs that we are interested in, the former implies that $rank(\vb*{L}) = N - 1$. This is related to the fact that the linear system \eqref{eq:sum_Qij} is invariant to translations in the pressure, which implies that the pressures are defined up to an additive constant. The physical explanation is that one can only measure pressure differences, as with potential differences in electrical circuits. However, the fluxes are uniquely determined since the additive term is cancelled. 

This means that we can arbitrarily choose the pressure of one node as the reference pressure, $p=0$. This  can be done indirectly by adding a small arbitrary constant to one diagonal element of the Laplacian, which  makes the Laplacian invertible (cf. section 2.4 of \cite{Katifori_BioFlows}). In our case, in each time step, we randomly perturb the diagonal entry of \eqref{eq:Lapij} corresponding to one of the sink nodes. Finally, once the pressures of the nodes are determined, the channel fluxes can be computed by direct substitution of the former in \eqref{eq:sum_Qij}. 

\subsection{Numerical Scheme}

Some caution must be taken regarding the choice  of numerical scheme used to solve the differential equation (\ref{eq:new_adapt_rule_minE}), or more generally the equation (\ref{eq:new_adapt_rule}), since it has to guarantee that the volume is conserved from one step to another. In this work, we adopt a simple explicit Euler numerical scheme with time step $\dt$, which results in the following discretisation of (\ref{eq:new_adapt_rule}),

\begin{equation}
\sqrt{D_{ij}^{n+1}} = \sqrt{D_{ij}^{n}} + \dt 
\left(\frac{\V}{\beta} \frac{g^n_{ij}}{\sum\limits_{(k,m)\in E}L_{km} g^n_{km}} - \sqrt{D_{ij}^{n}} \right) \quad, \quad (i,j)\in E \;.
\label{eq:new_Euler}
\end{equation}

\noindent where $D_{ij}^n$ denotes the conductivity of the edge $(i,j)$ at the time $\tau = n \dt$, \textit{i.e} $D_{ij}^{n}\equiv D_{ij}(n\dt) $ and similarly $g_{ij}^n\equiv g(|Q_{ij}(n\dt)|)$. According to the discretisation (\ref{eq:new_Euler}), the volume of the network in the time step $n+1$ is 

\begin{nalign}
\V^{n+1} &= \beta\sum_{(i,j)\in E} L_{ij} \sqrt{D_{ij}^{n+1}} \\
&= \underbrace{\beta\sum_{(i,j)\in E} L_{ij}\sqrt{D_{ij}^{n}}}_{\V^n} + \dt\underbrace{\left(
\V - \beta\sum_{(i,j)\in E} L_{ij}\sqrt{D_{ij}^{n}}
\right)}_{0} \\
&= \V^{n} \;,
\end{nalign}

\noindent which implies that, for any choice of $g$, the Euler method  conserves the volume.

Most of the simulations were carried out with a relatively large time increment, $\dt = 0.1$, due to computational power limitations. It should be noted however that we've performed some tests with different time increments, while maintaining the remaining parameters fixed, and in some cases the final networks reached by the algorithm were different.

\subsection{Mesh Generation}

The geometry of the initial networks was generated through a Delaunay Triangulation of a set of points representing the nodes. Given a discrete set of points (nodes) $V$ arbitrarily placed in a plane, a two-dimensional Delaunay triangulation of $V$, $DT(V)$, is a triangulation such that no point in $V$ is inside the circumscribed circle of any triangle in $DT(V)$. This type of triangulation results in planar graphs where nodes can have an arbitrarily high degree depending on how they are initially distributed.

To generate the underlying initial networks we first considered an $L\times L$ square lattice of side length 1, with a total of $\nodesnum=L \times L$ nodes. Then, the position of the interior nodes of the lattice was randomly perturbed with Gaussian noise, with a given standard deviation which controls the distortion of the final mesh.
Finally, a Delaunay triangulation is performed on perturbed lattice points,  resulting in a graph network $\G$ with edge lengths, $L_{ij}$, randomly distributed in the interval $[1/(N-1) -\epsilon, 1/(N-1) + \epsilon]$, where $\epsilon$ is a small constant. We've performed tests on lattices of linear sizes ranging from $L=2$ up to $L=40$.

The method used to distribute the points over the plane resulted in networks with an average degree typically around 5 or 6. This high number of node neighbours increases the set of possible paths that the adaptation process can select from, which makes the results more robust. Furthermore, by using a Delaunay triangulation of a random set of points to represent the underlying \textit{Physarum} network instead of a structured mesh (rectangular, triangular, hexagonal), we also avoid that eventual lattice symmetries introduce some bias on the results. Some tests were also performed with these types of meshes as underlying networks, although they are not presented here. In these cases, some noise was added to the position of the vertices  to break the lattice symmetries. However, as the simulations showed, the optimised networks obtained weren't in general as natural as the ones obtained from an initial Delaunay mesh, which resembled more closely the networks displayed by \textit{Physarum}.

\section{Exploration of the Model}

We start by performing simple tests on the model. First, a brief insight into the typical network temporal evolution is given. Then it's investigated how the choice of sources and sinks and the initial distribution of edge conductivities affect the shape of the final network. Finally, it's studied the general adaptation dynamics \eqref{eq:new_adapt_rule} for the class of functions $g_\gamma(|Q_{ij}|) = |Q_{ij}|^\gamma$, where we analyse the final network dependency on the parameter $\gamma>0$.

\subsection{Network temporal evolution}

\begin{figure}[hbt!] 

\newcommand{\mysub}[2][]{%
    \subfloat[#1]{\includegraphics[trim={2.9cm 3cm 3.5cm 3.1cm}, clip, width=0.249\textwidth]{#2}}%
}

\newcommand{\myfig}[1]{%
    \includegraphics[trim={2.9cm 3cm 3.5cm 3.1cm}, clip, width=0.25\textwidth]{#1}%
}

\begin{subfigure}{\textwidth}    
    \centering
    \begin{tabular}{@{}c@{}c@{}c@{}c@{}}
    $t=0$ & $t=50$ & $t=75$ & $t=125$ \\[0.01cm]
    \myfig{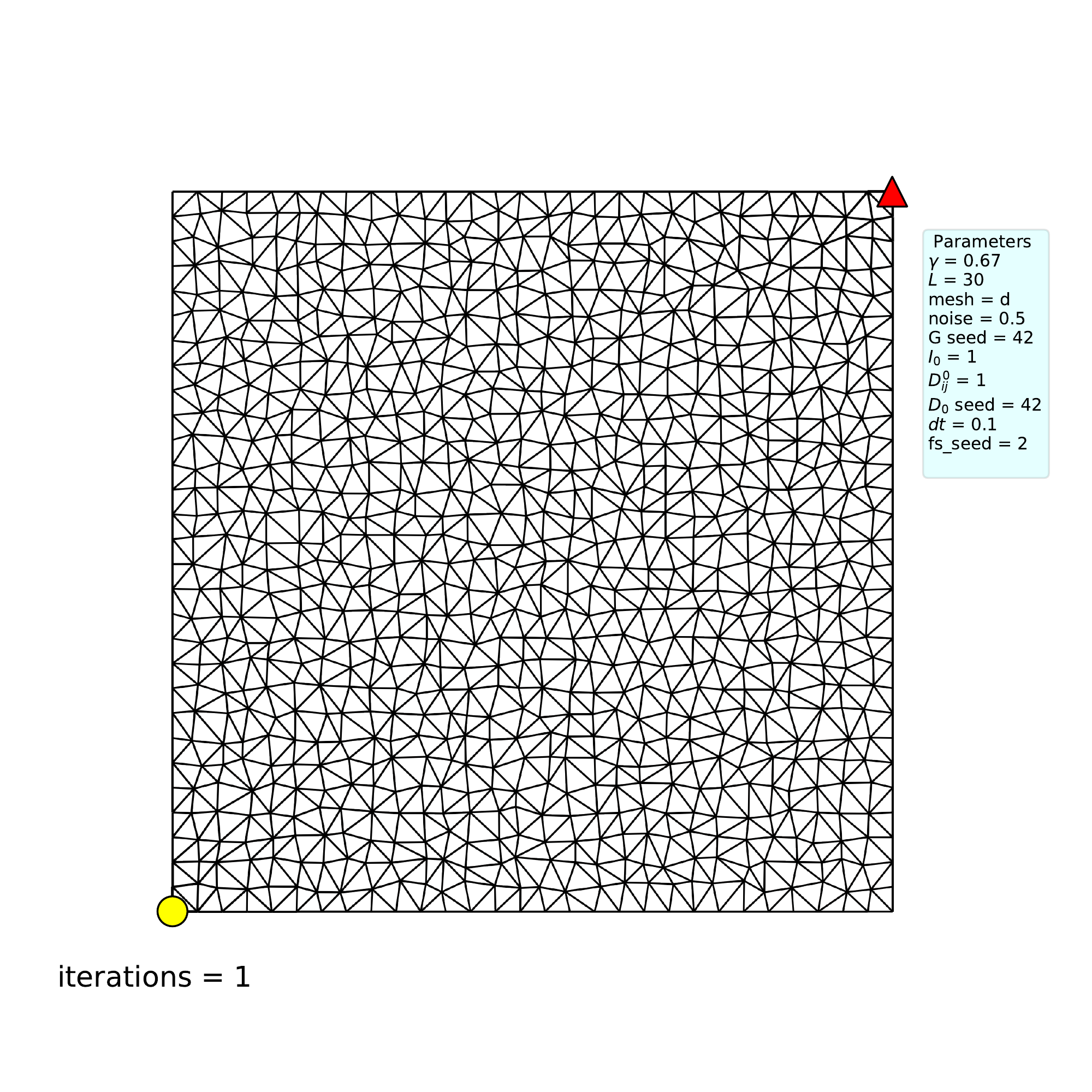} &
    \myfig{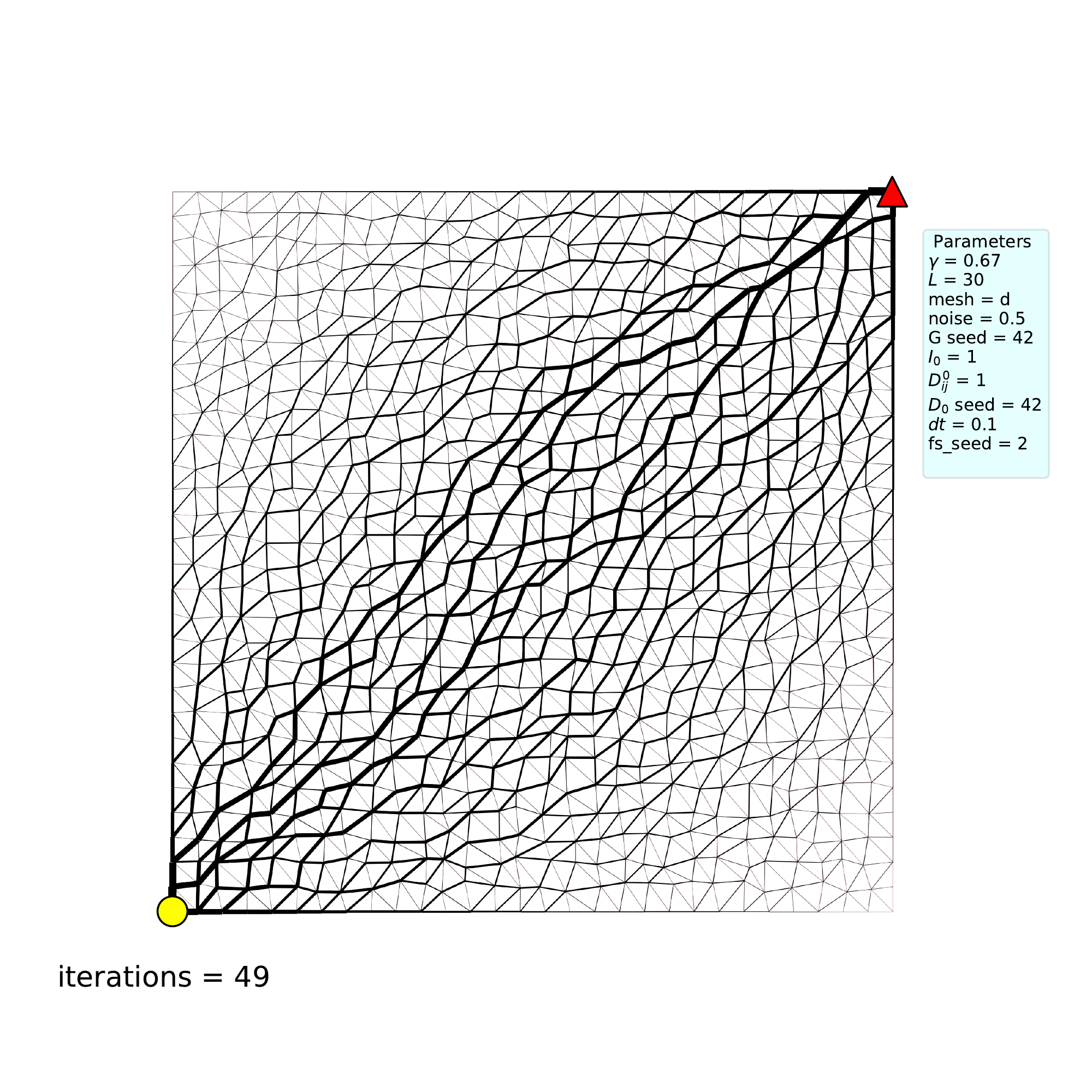} & 
    \myfig{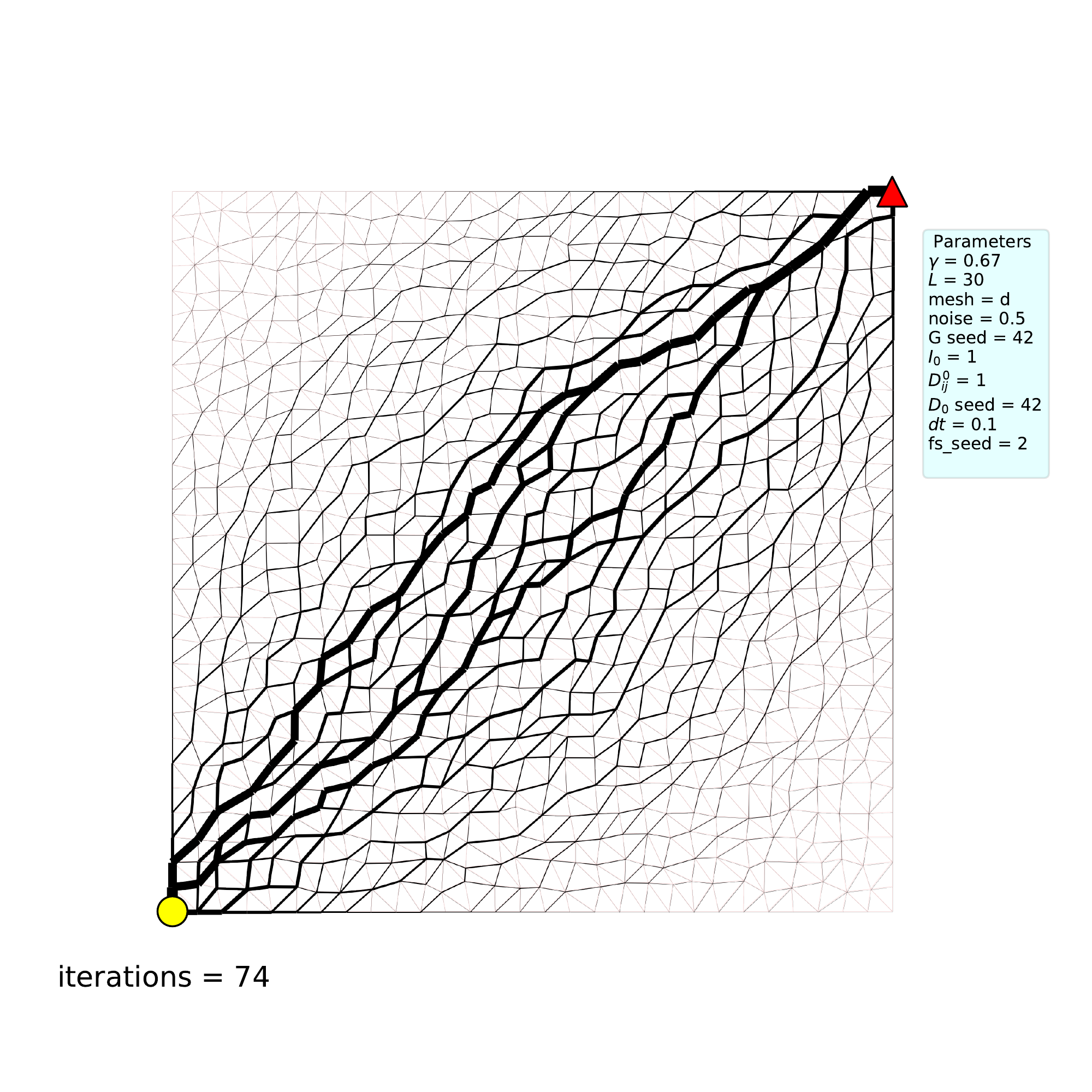} & 
    \myfig{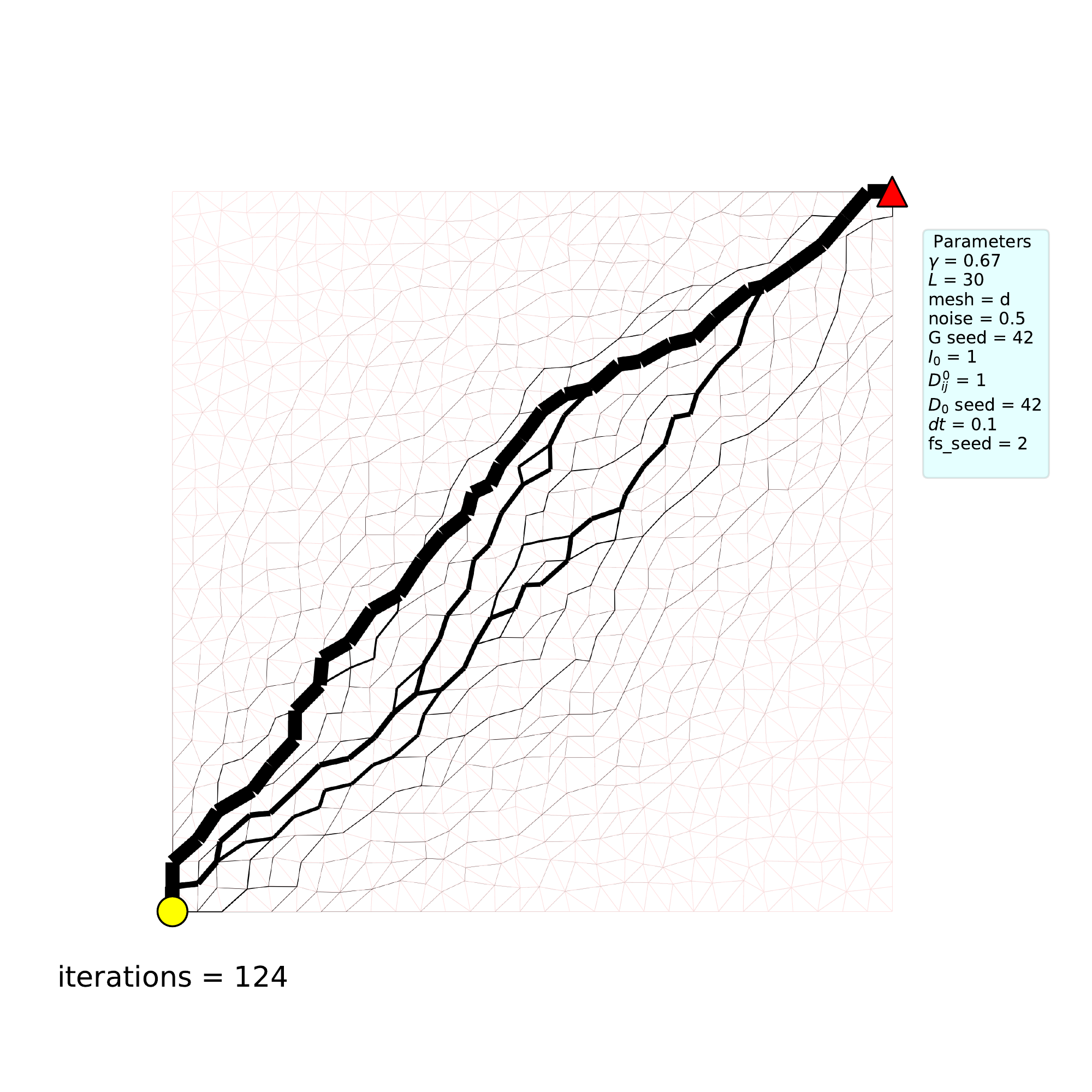} \\
    $t=150$ & $t=300$ & \multicolumn{2}{c}{Network's volume over time} \\[0.01cm]
    \myfig{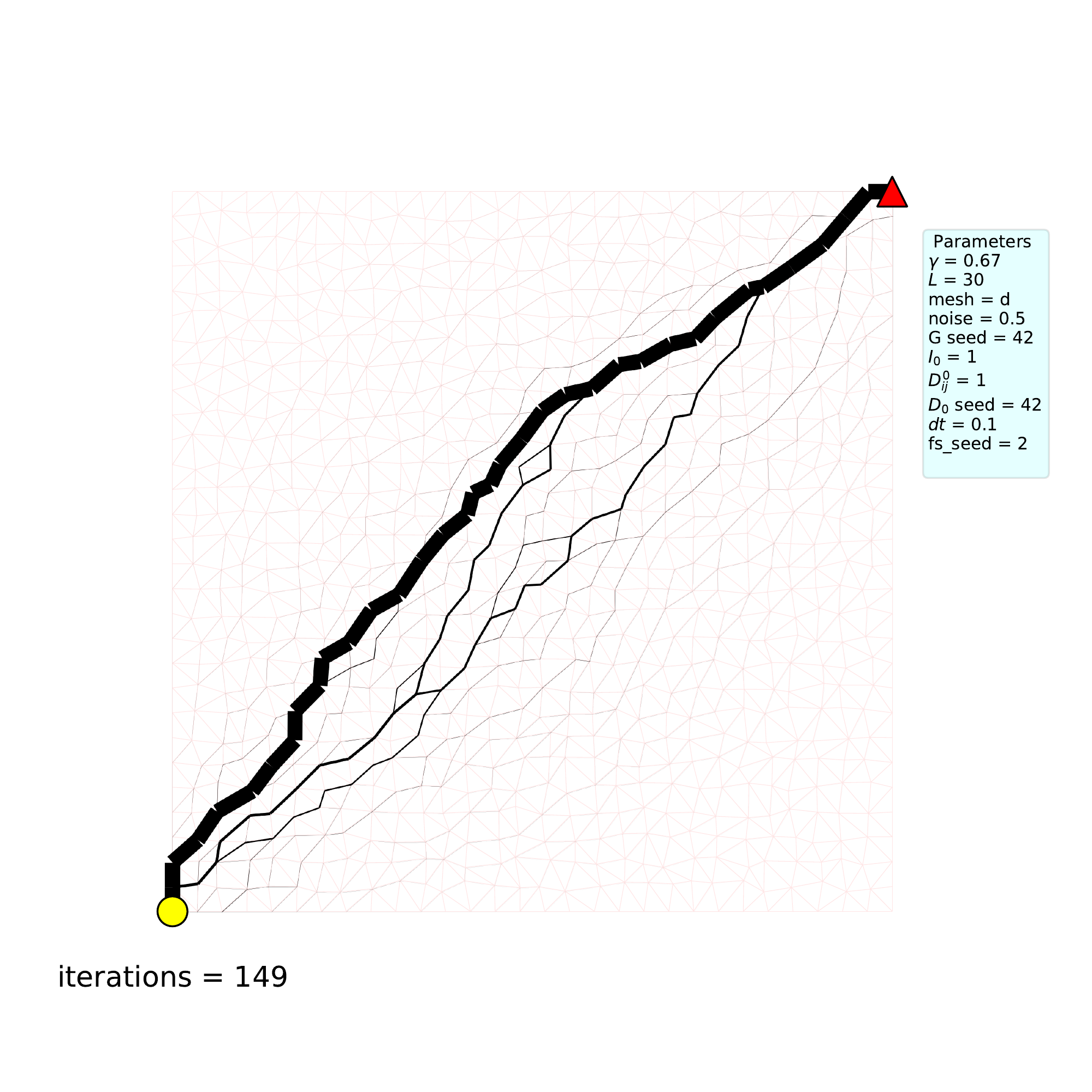} &
    \myfig{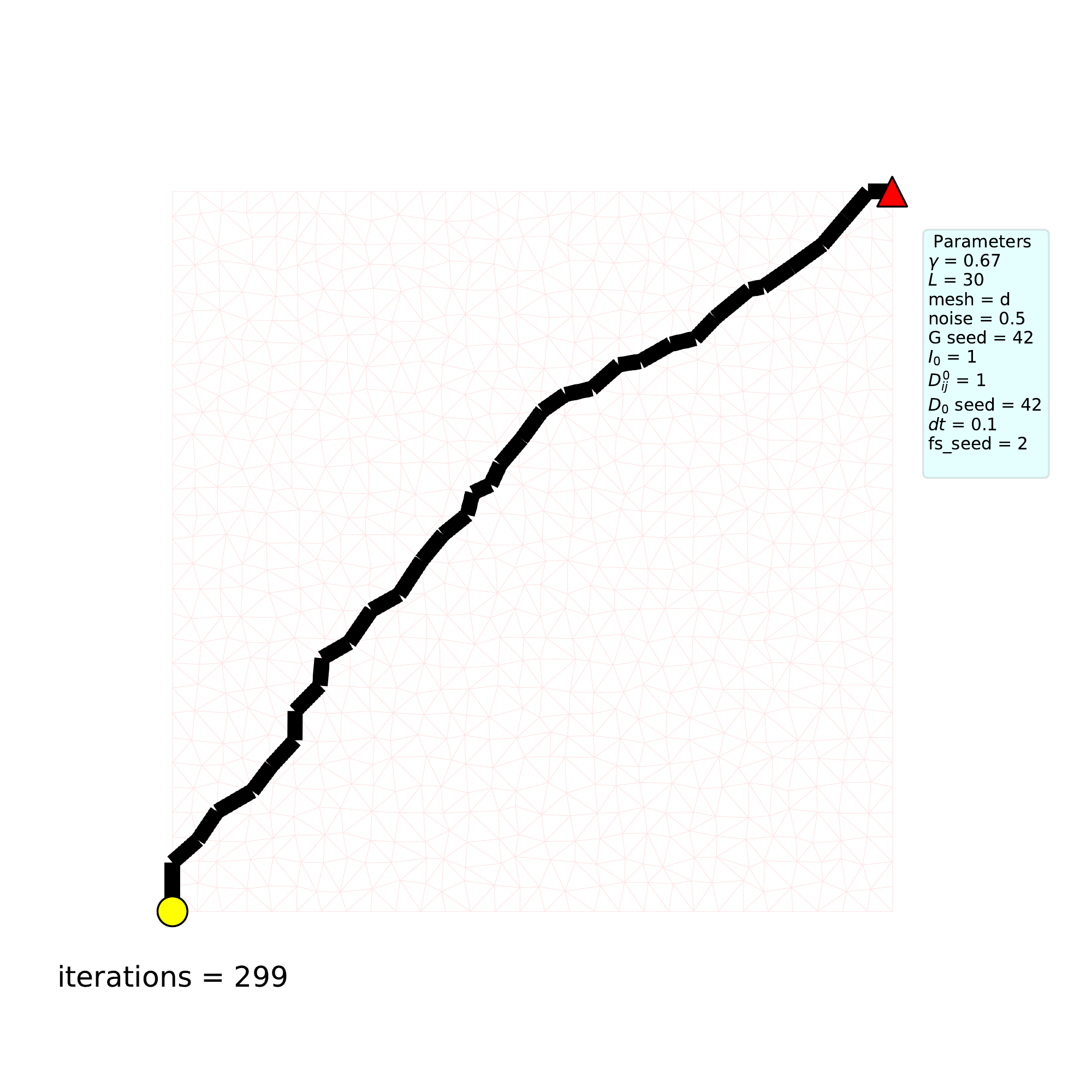} & 
    \multicolumn{2}{c}{
    \includegraphics[trim={0.3cm 0.2cm 0.3cm 0.2cm}, clip, height=0.25\textwidth]{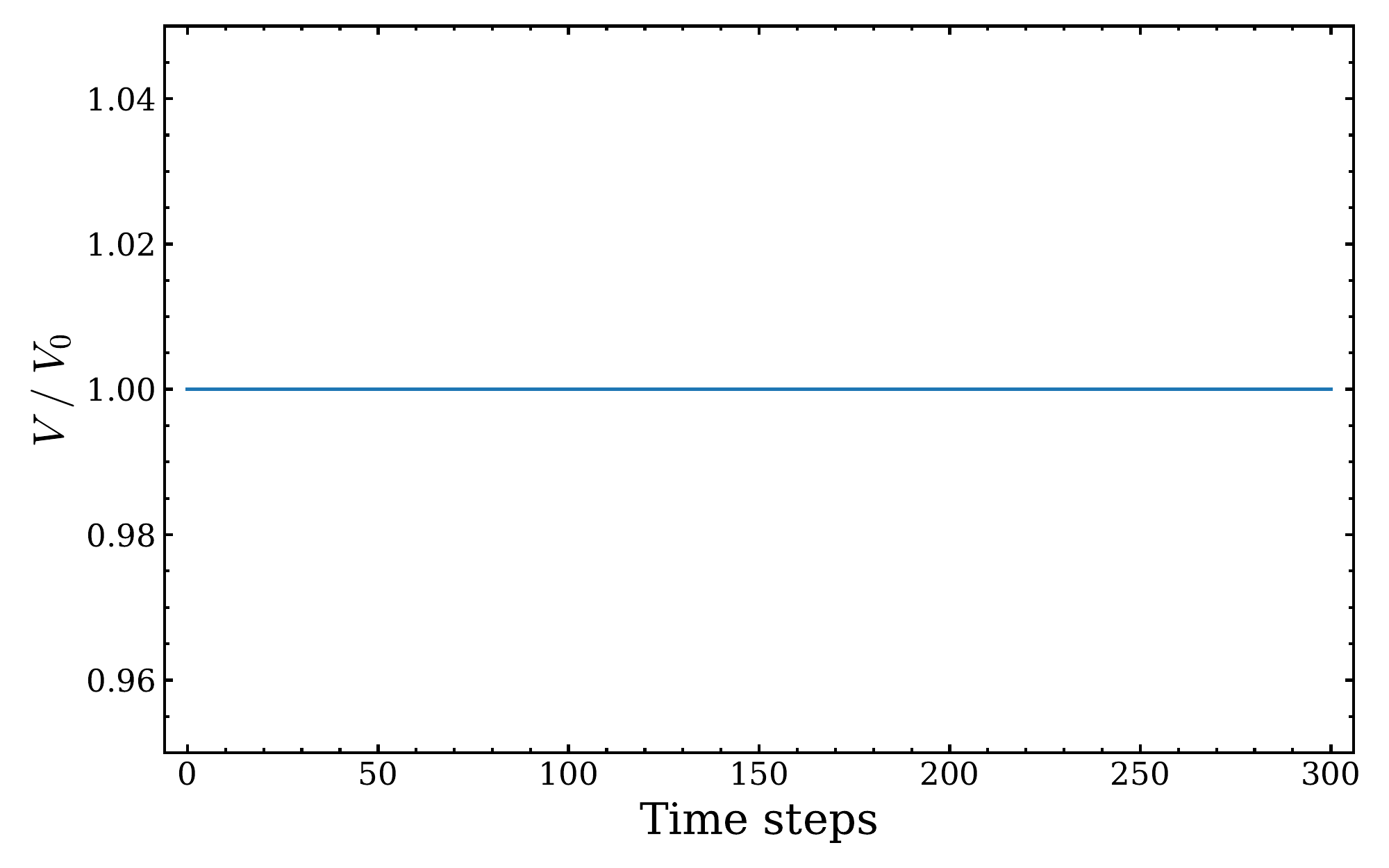} 
    }
    \end{tabular}
    \caption{Simulation for one source (yellow) and one sink (red) placed at diagonally opposite corners of a square. The intensity of the terminals are related by $q_{source}=-q_{sink}=1$.}
    \label{fig:net_evol_a}
\end{subfigure}  
\\[0.2cm]
\begin{subfigure}{\textwidth}    
    \centering
    \begin{tabular}{@{}c@{}c@{}c@{}c@{}}
    $t=0$ & $t=50$ & $t=75$ & $t=125$\\[0.01cm]
    \myfig{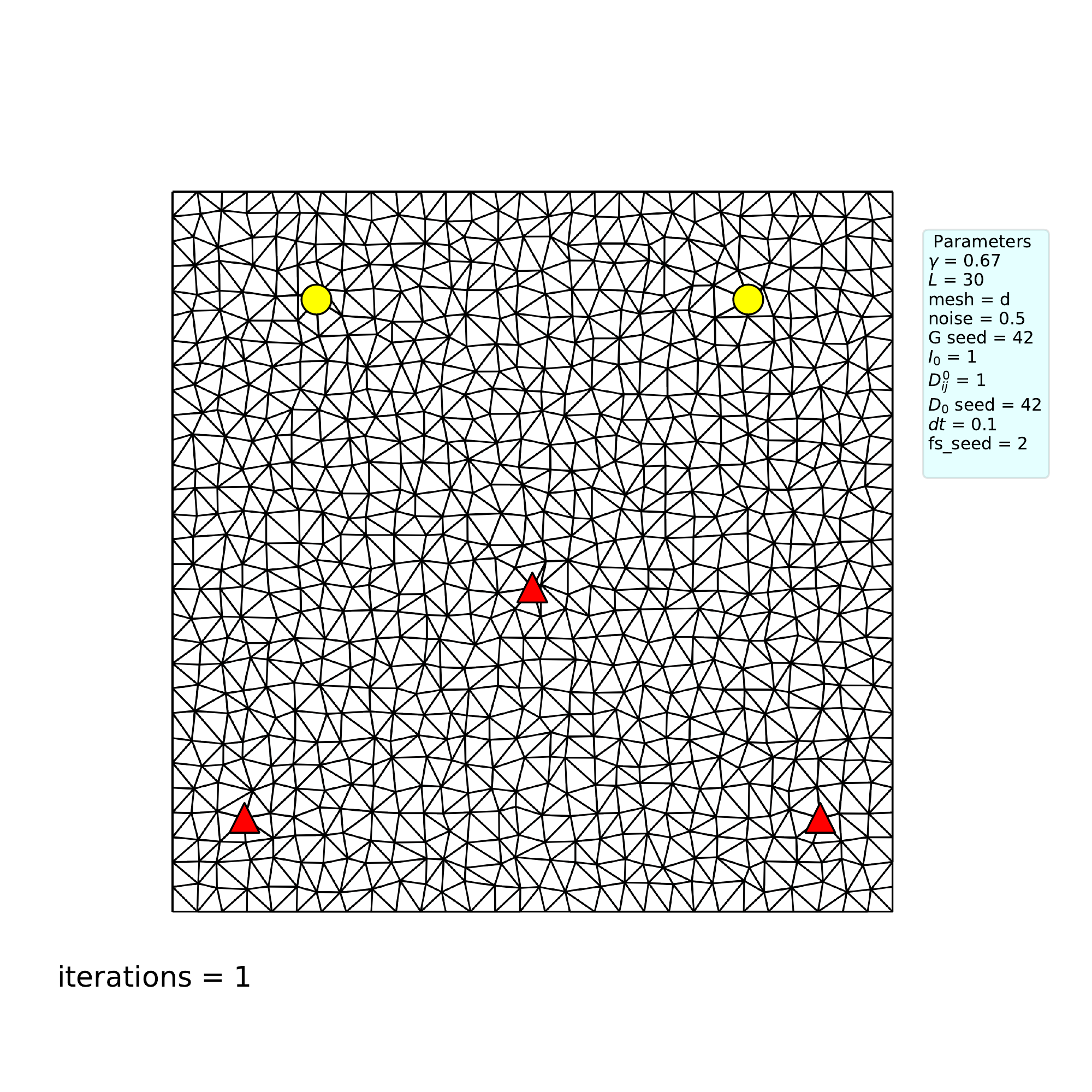} &
    \myfig{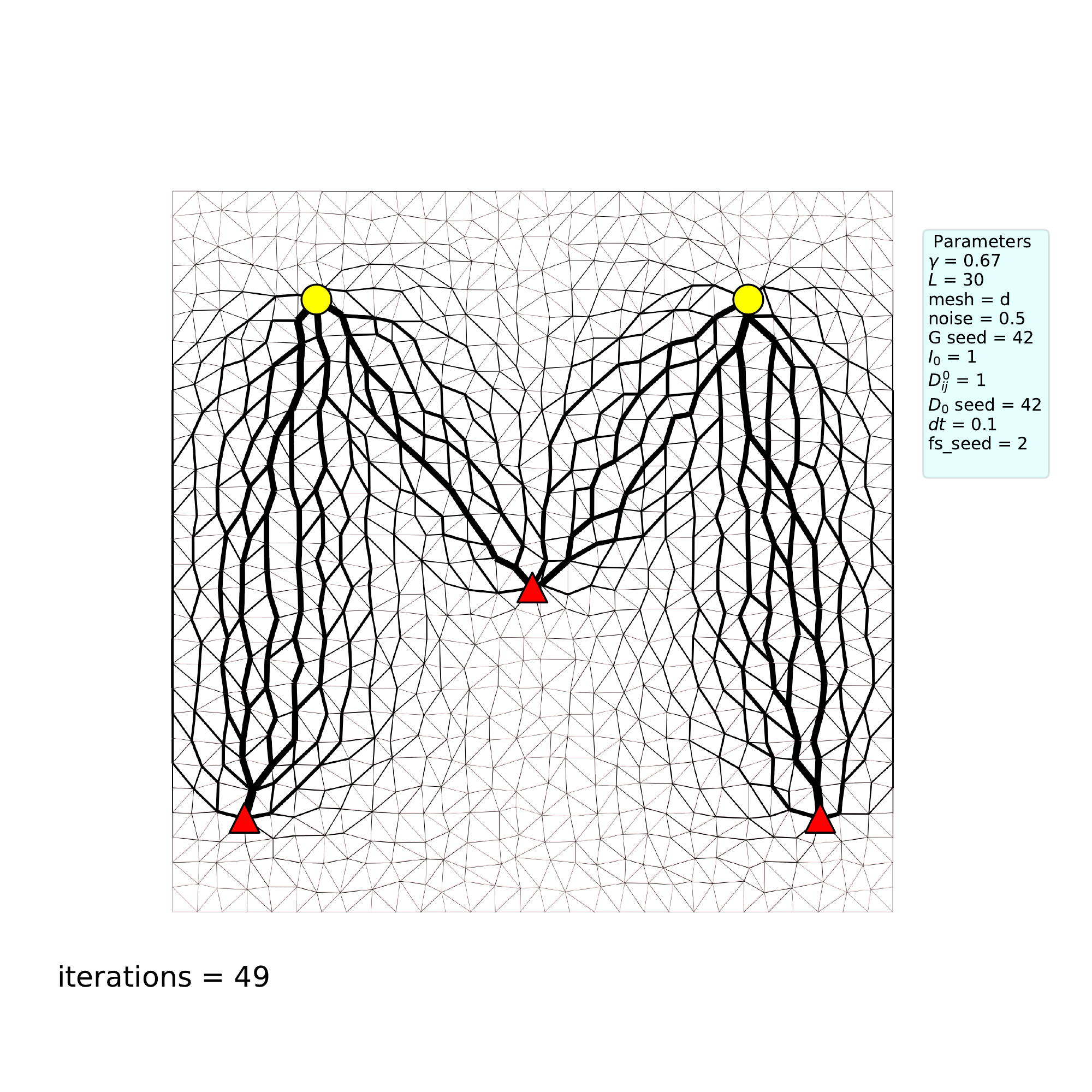} & 
    \myfig{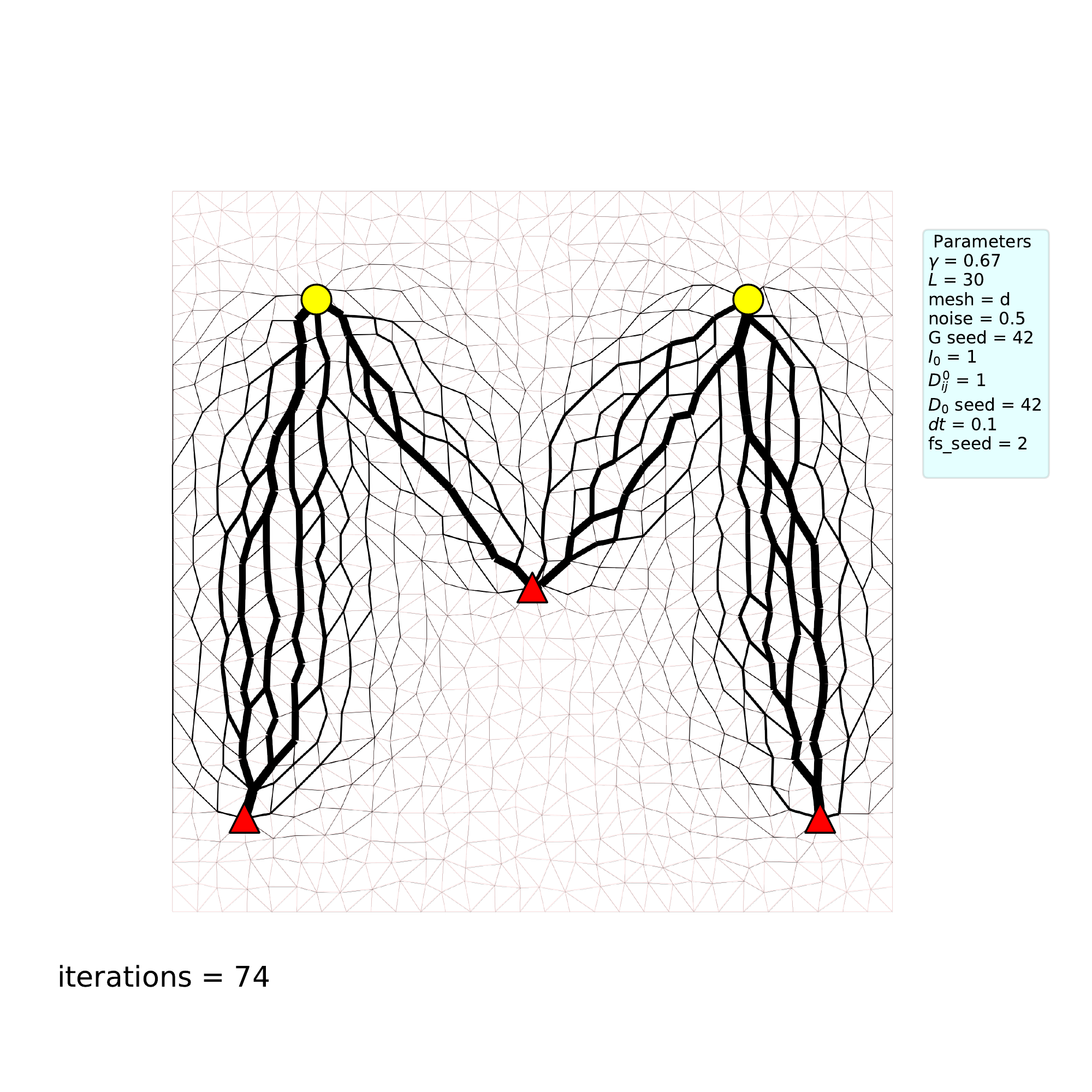} & 
    \myfig{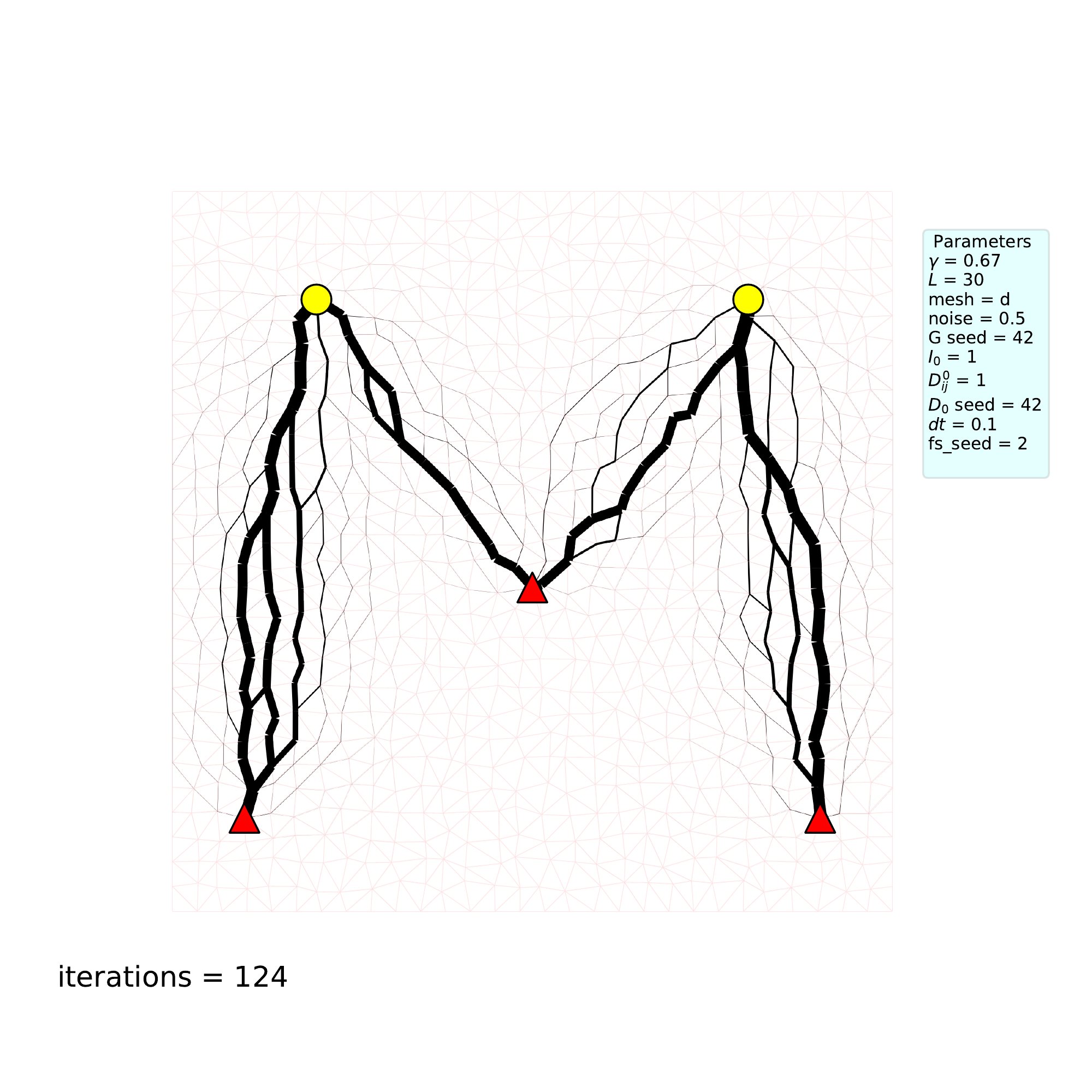}\\
    $t=200$ & $t=350$ & \multicolumn{2}{c}{Network's volume over time} \\[0.01cm]
    \myfig{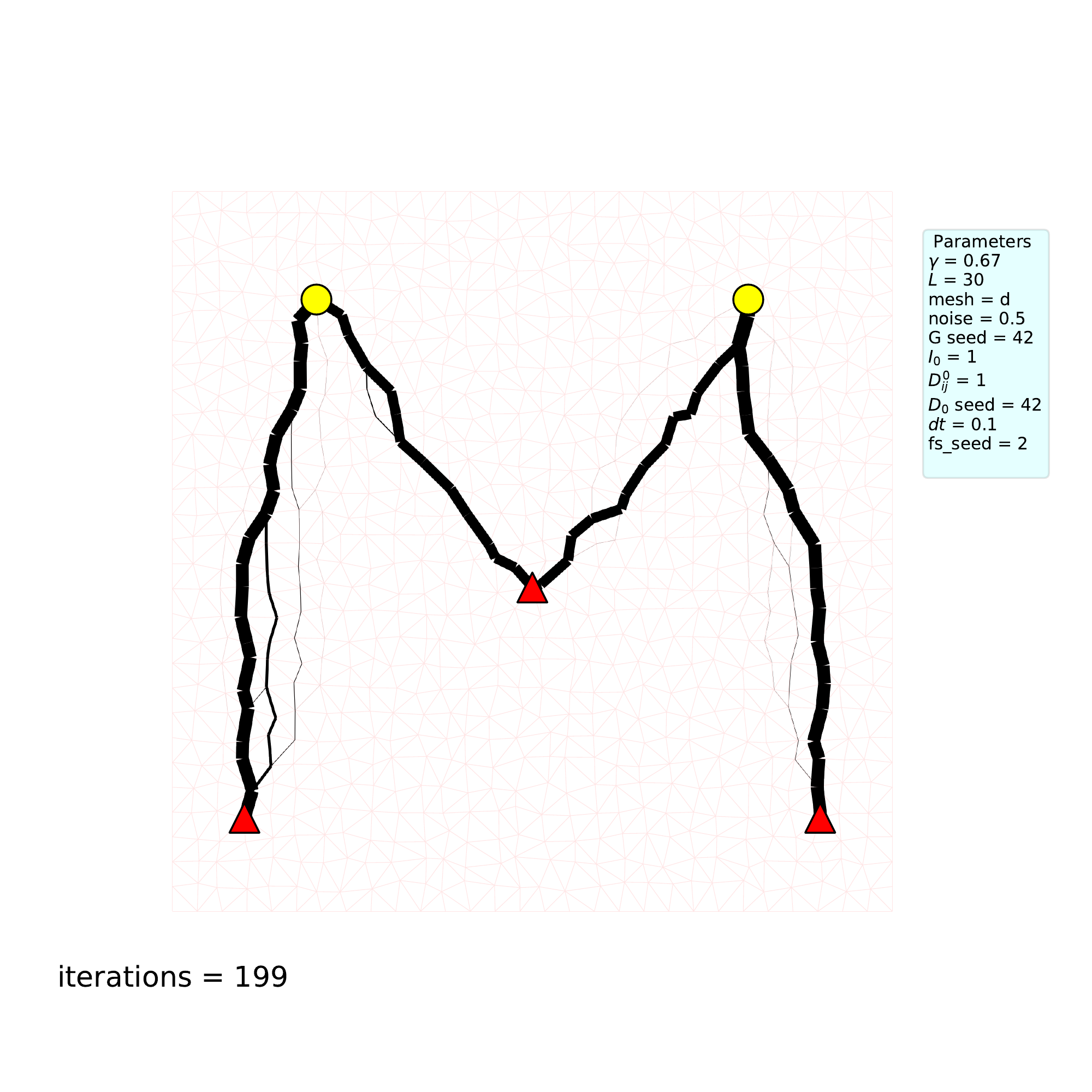} &
    \myfig{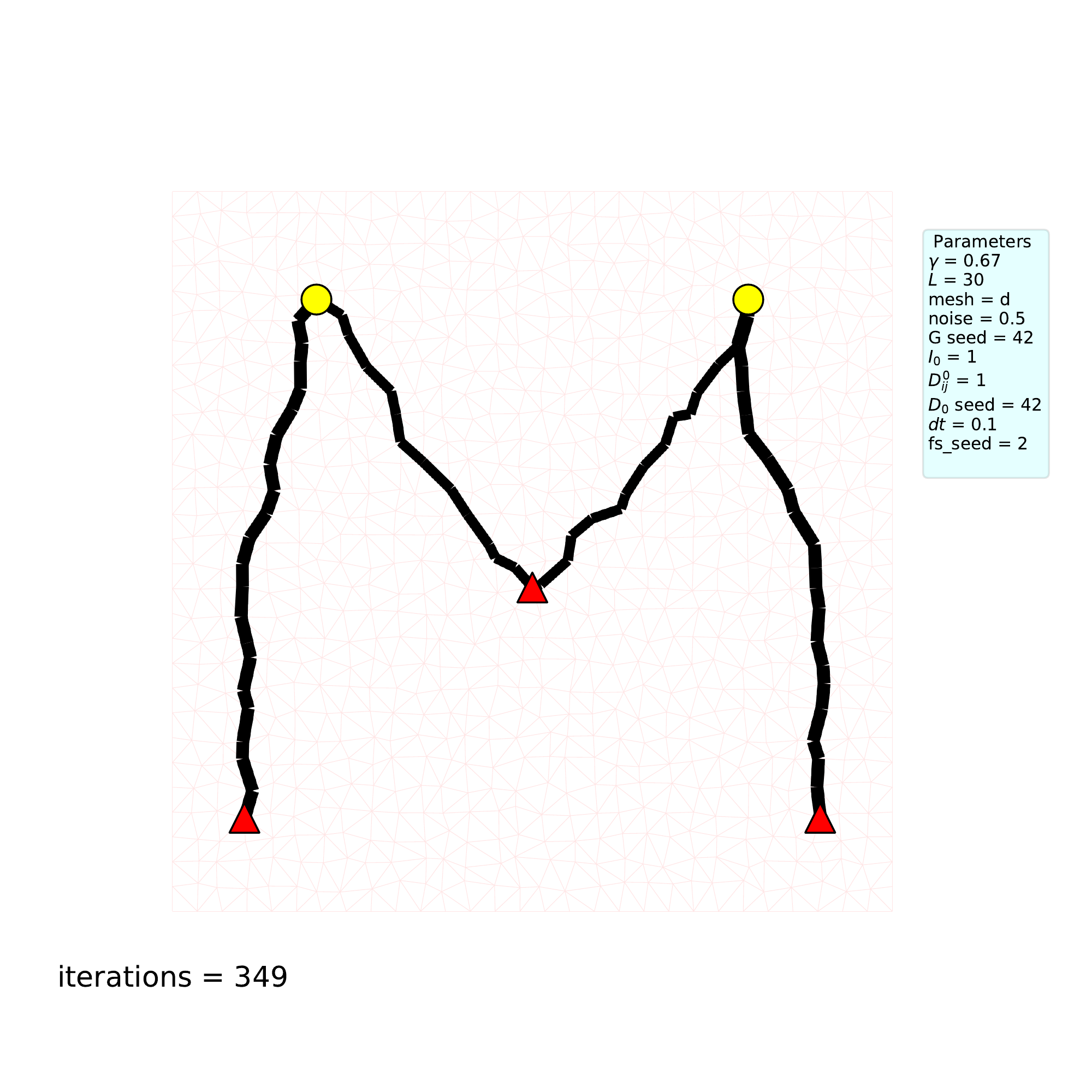} & 
    \multicolumn{2}{c}{
    \includegraphics[trim={0.3cm 0.2cm 0.3cm 0.2cm}, clip, height=0.25\textwidth]{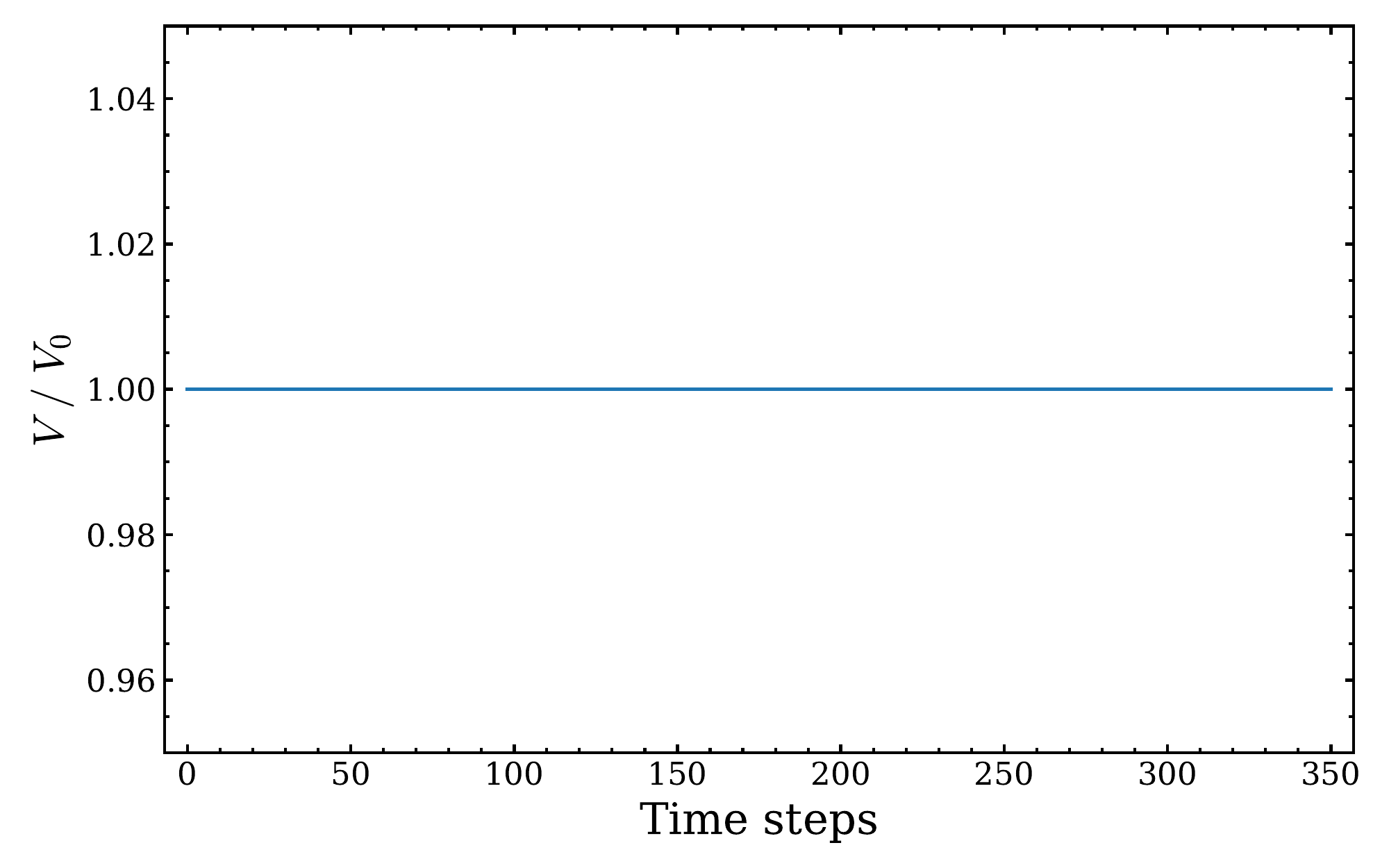} 
    }
    \end{tabular}
    \caption{Simulation for multiple terminals aligned in an M-shaped configuration: 2 sources (yellow) and 3 sinks (red). The intensity of the sources is $q_{source}=1/2$ and the intensity of the sinks is $q_{sink}=-1/3$.}
     \label{fig:net_evol_b}
\end{subfigure}  

\caption{Simulation examples of the model \eqref{eq:new_adapt_rule_minE}, depicting the network optimisation process at different time steps, for two different choices of terminals. The simulations start from an initial Delaunay triangulation of 900 nodes, shown in red, where the edges start with homogeneous conductivities, $D_{ij}(0)=1$. The thickness of the black lines is proportional to the radius of the channels ($\sim D_{ij}^{1/4}$). At the bottom right of each group, it's plotted the networks' volume throughout the simulation, which shows that the volume is conserved in both cases.}

\label{fig:network_evolution} 
\end{figure}

In Figure \ref{fig:network_evolution}, it's shown snapshots taken at different time steps of two simulations of the model \eqref{eq:new_adapt_rule_minE}, depicting the typical evolution of the networks over time, considering a fixed set of terminals. The simulations were carried out on a network resulting from a Delaunay triangulation of 900 randomly placed nodes, with uniform initial channel  conductivities, $D_{ij}(0)=1$.  Figure \ref{fig:net_evol_a} represents the case of only one source and one sink, while Figure \ref{fig:net_evol_b} represents the case of the adaptation in the presence of multiple terminals, where we've considered two sources and three sinks. In both cases, the sources give the same amount of flux  $q_{source}=1/N_{sources}$, which is evenly distributed between the sinks, $q_{sink}=-1/N_{sinks}$, where $N_{sources}$ and $N_{sinks}$ are the number of sources and sinks respectively. 

In this work we adopt the following conventions: the sources are represented by yellow circles, the sinks are represented by red triangles, and the thickness of the lines representing the edges is proportional to their radius ($~D_{ij}^{1/4}$). The initial network is also shown in light red.  

From the simulations, we observe that the channels farther away from   
the straight lines connecting each source-sink pair gradually shrink and eventually vanish, while the closest to those paths are thickened due to the volume conservation. The farthest they are the faster they disappear. On the bottom right of each figure, it's plotted  the volume of the networks over time, which confirms that the volume is conserved throughout the simulations.  

\begin{figure}[hbt!] 
\captionsetup[subfloat]{labelformat=empty,position=top,skip=0.5pt}

\newcommand{\mysub}[2][]{%
    \subfloat[#1]{\includegraphics[trim={2.8cm 2.8cm 3.2cm 2.8cm}, clip, width=0.25\textwidth]{#2}}%
}

\centering

\mysub[$t=0$]{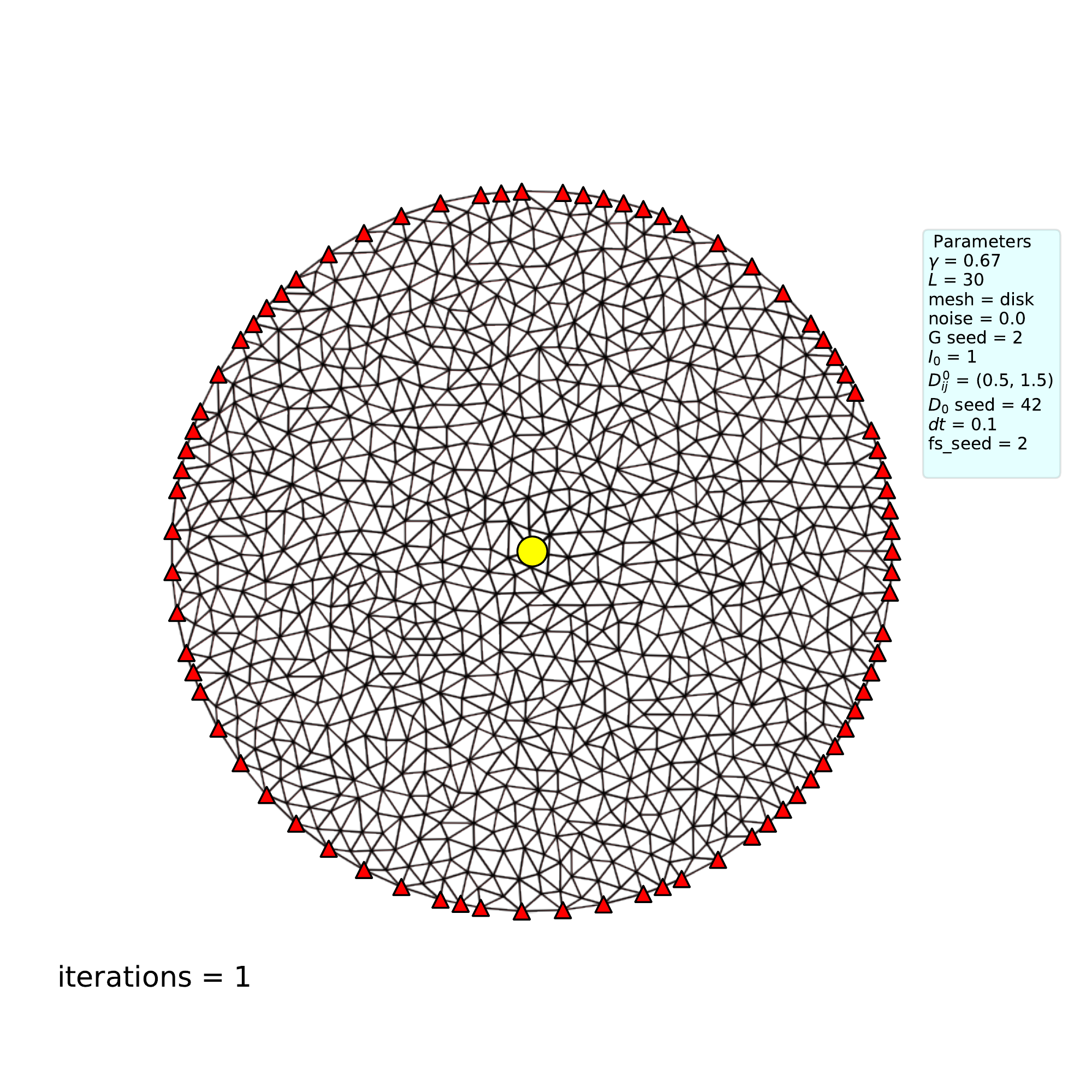} \hfill
\mysub[$t=60$]{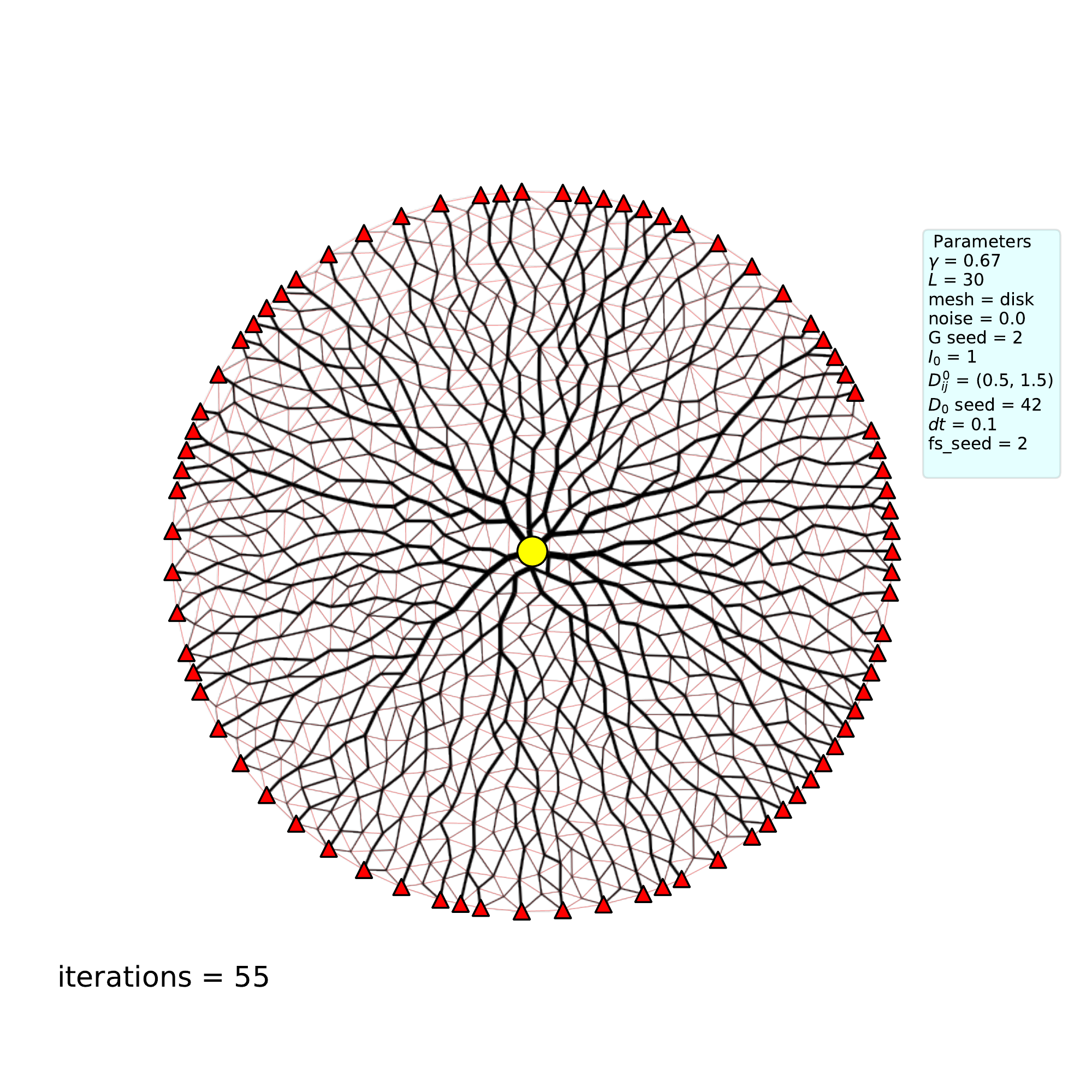} \hfill
\mysub[$t=120$]{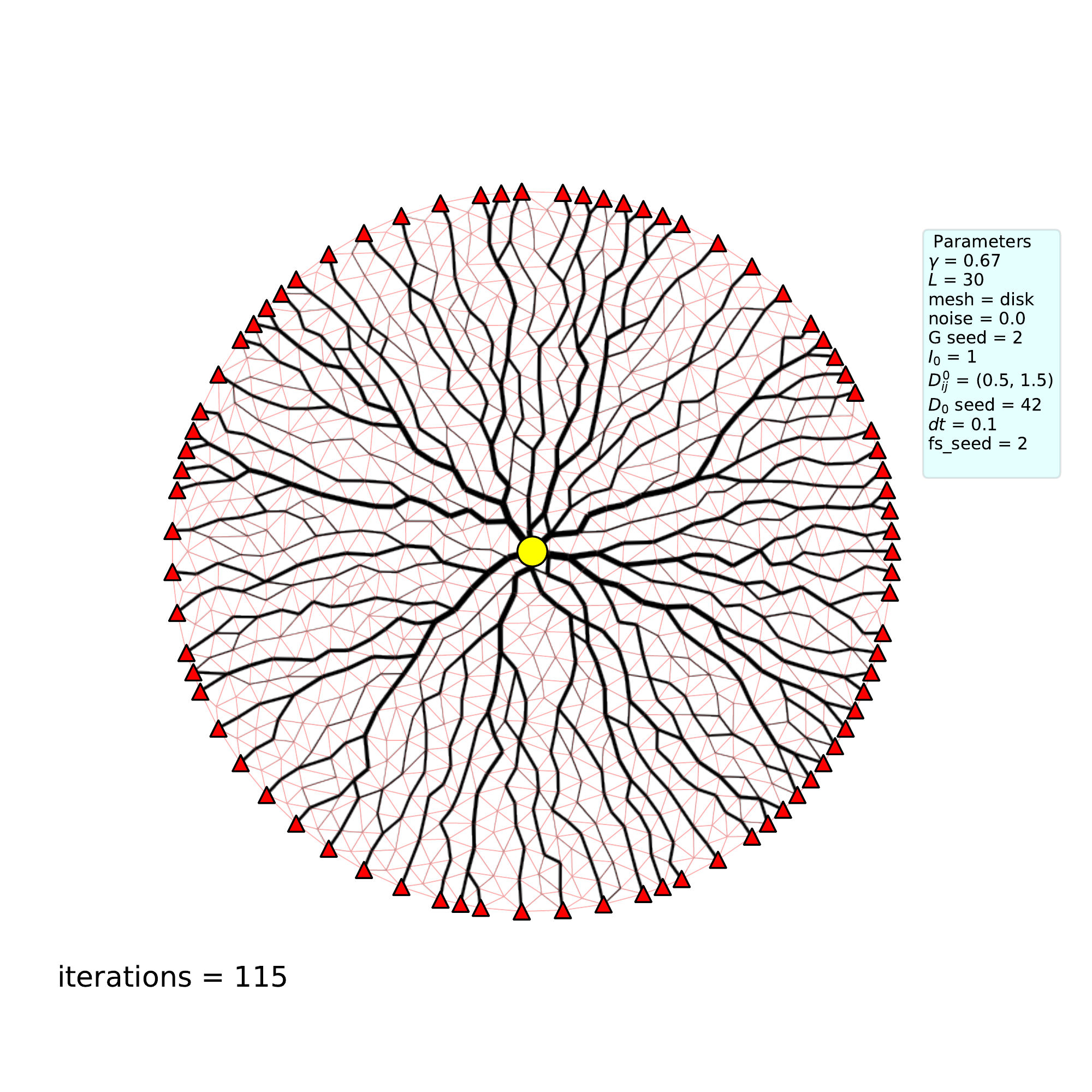} \hfill
\mysub[$t=300$]{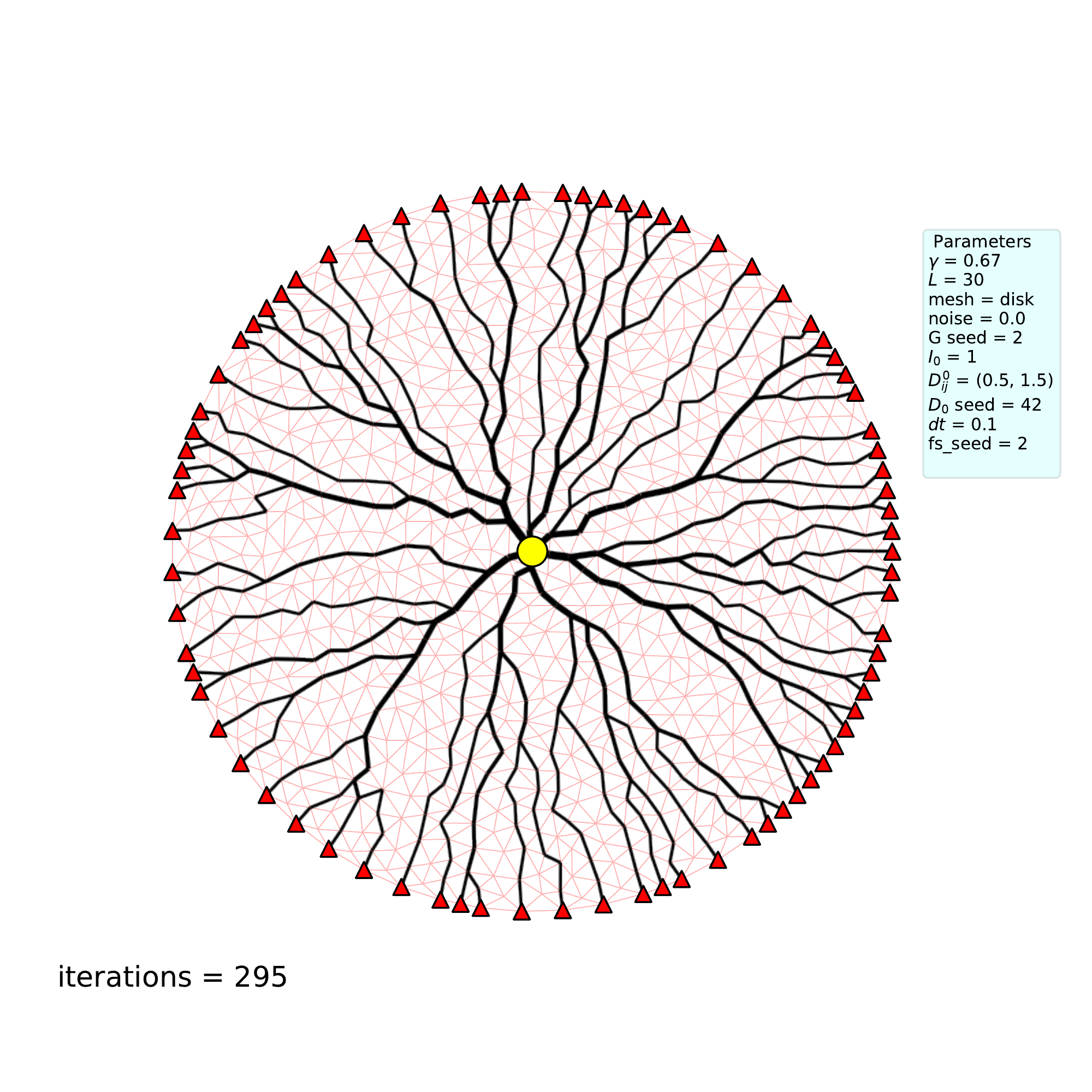} \hfill

\caption{Simulation of the model \eqref{eq:new_adapt_rule_minE} replicating the \Phy{}'s network adaptation, starting from initial homogeneous conductivities, $D_{ij}(0)=1$. The labels $t$ designate the iteration number in which the snapshots were taken. The flux is driven by a central source (yellow) and the sinks (red) placed at the boundary of the organism, which represent stimulated regions with high metabolic activity. The source gives $q_{source}=1$, which is evenly distributed between the sinks. The adaptation dynamics results in a tree-like steady-state  network (cf. Figure \ref{fig:phy_network}).}
\label{fig:phy_simulation_network} 
\end{figure}

In Figure \ref{fig:phy_simulation_network} we've simulated the adaptive mechanism on a more realistic setting mimicking the \textit{Physarum} scenario. We've considered an organism with a circular shape in the presence of a central food source which acts as a source of nutrient flux with intensity $q_{source}=1$. The nutrients are equally distributed between all the nodes at the boundary, which behave like sinks, simulating regions where there is a continuous uptake of nutrients so that the eventual growth of the organism  may occur (which is for now neglected). The final network shows a preferential radial orientation consistent with the flux direction and resembles to some degree the networks displayed by \textit{Physarum} depicted in Figure \ref{fig:phy_network}. However, the key difference is that \textit{Physarum}'s networks have some redundancy which is not observed in the simulations. Thinner secondary veins branch from thicker primary veins and connect on the other end to other main veins, and this pattern is repeated for even thinner veins, resulting in the formation of loopy structures. Those loops and redundant paths are important in the sense that they provide robustness and tolerance to damage to the network, and are characteristic of other biological systems as well like the leaf venation in plants. 

Based on extensive simulations considering different configurations of fixed terminals, we couldn't reproduce the formation of loops and redundant connections. In all the simulations, the adaptation dynamics \eqref{eq:new_adapt_rule_minE}  resulted in steady-state  networks which looked like \textit{trees}, i.e., acyclic connected graphs where any two vertices are connected by exactly one path. This is partly due to assuming fixed sources and sinks and thus  neglecting flux fluctuations that are believed to be important for the formation of those redundant paths  \cite{Corson2010, Katifori2010}. The impact of the flux fluctuations will be explored in the next chapter.

\subsection{Choice of sources and sinks}

\begin{figure}[hbt!] 

\newcommand{\mysub}[2][]{%
    \subfloat[#1]{\includegraphics[trim={3cm 2.9cm 2.6cm 3.1cm}, clip, width=0.249\textwidth]{#2}}%
}

\centering
\begin{tabular}{@{}c@{}c@{}c@{}c@{}}

\mysub[\label{fig:mult_term_init}]{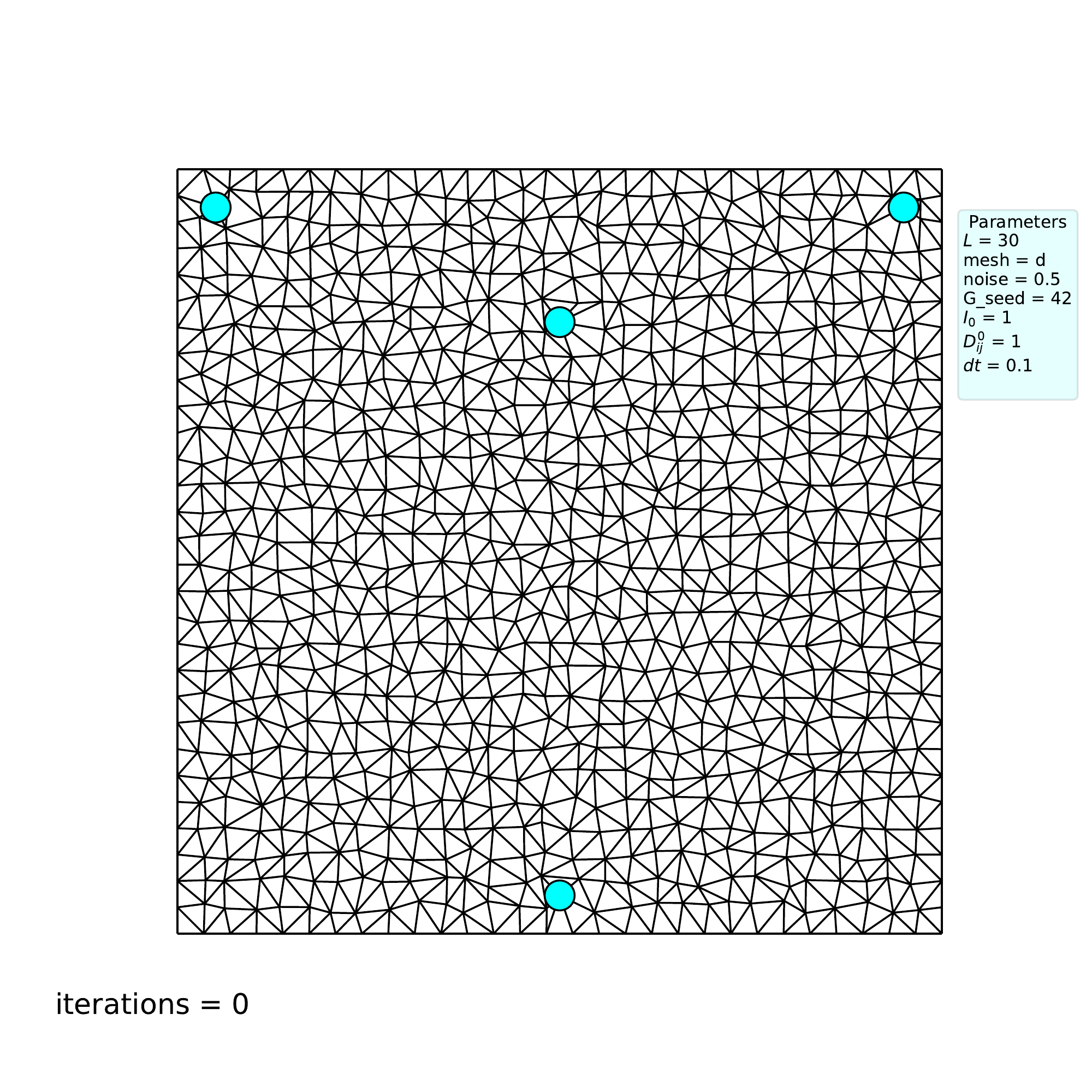} &
\mysub{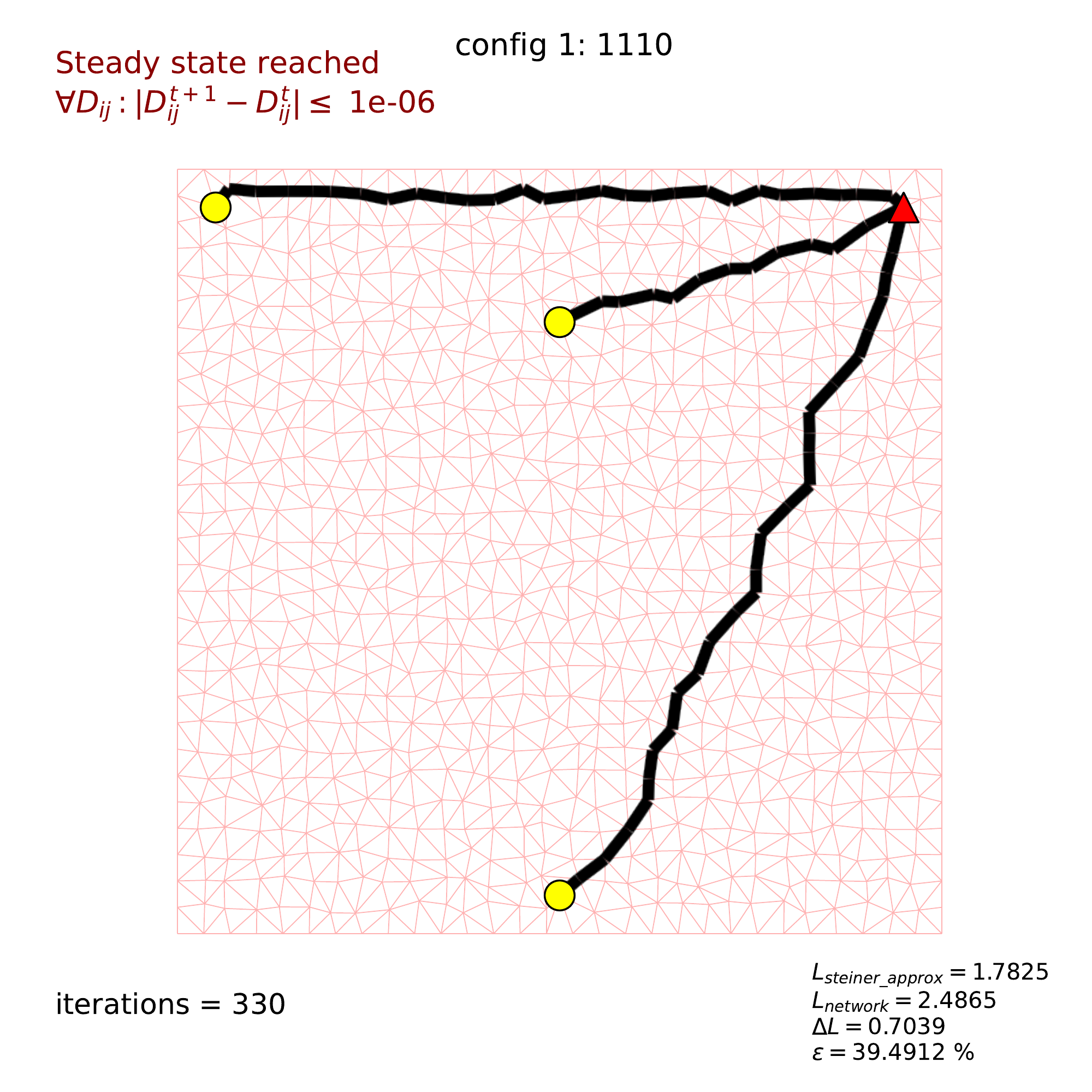} &
\mysub{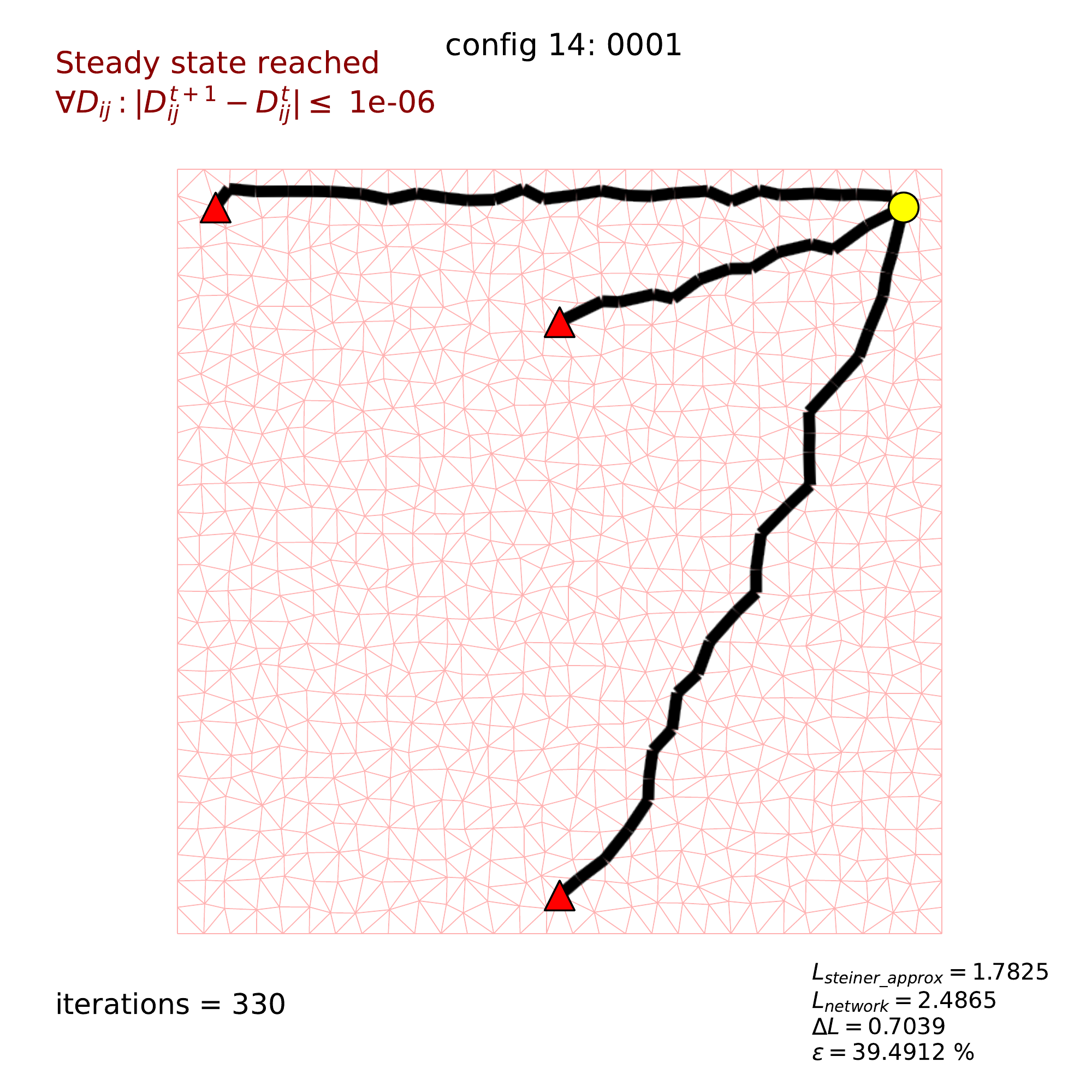} & \\

\mysub{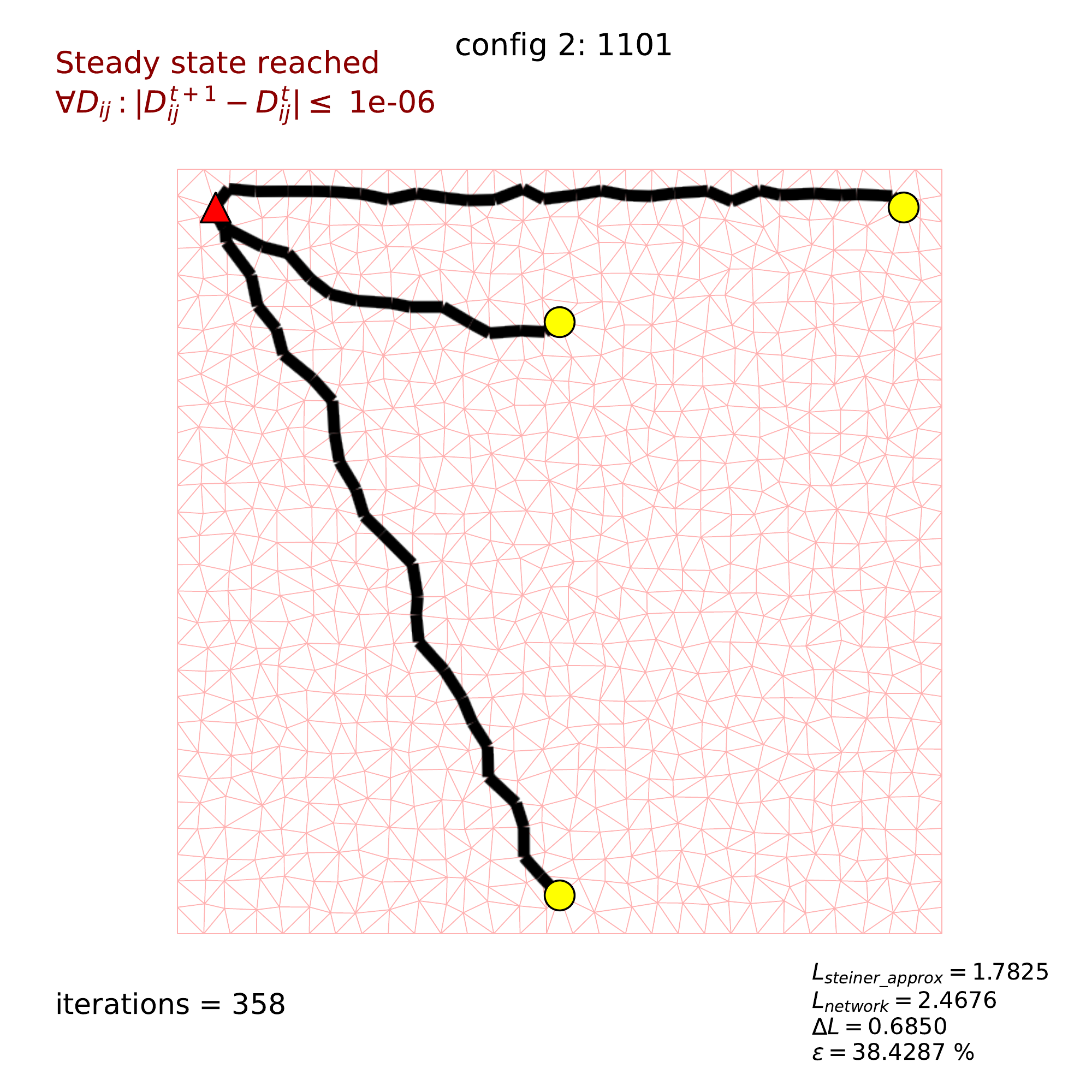} &
\mysub{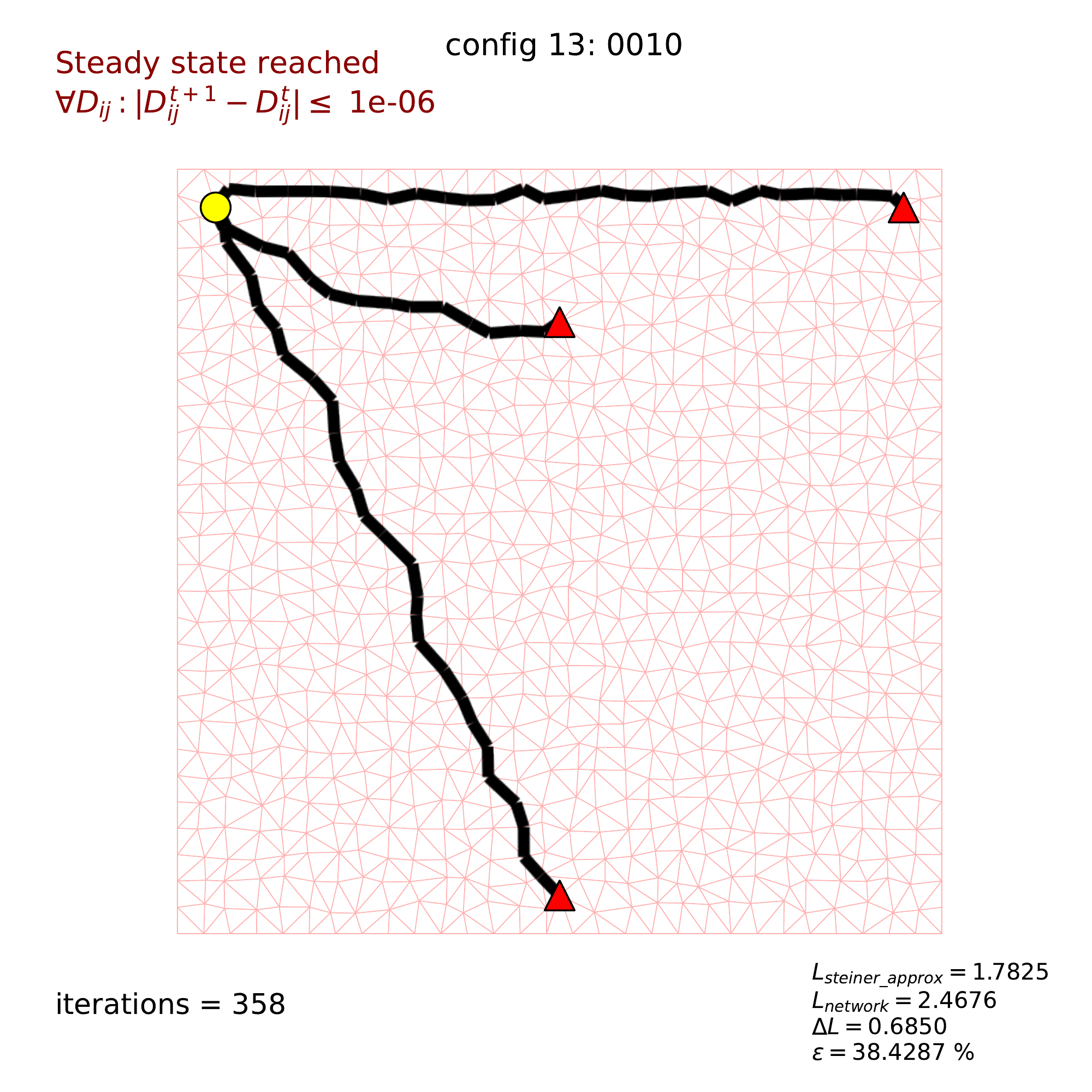} &
\mysub{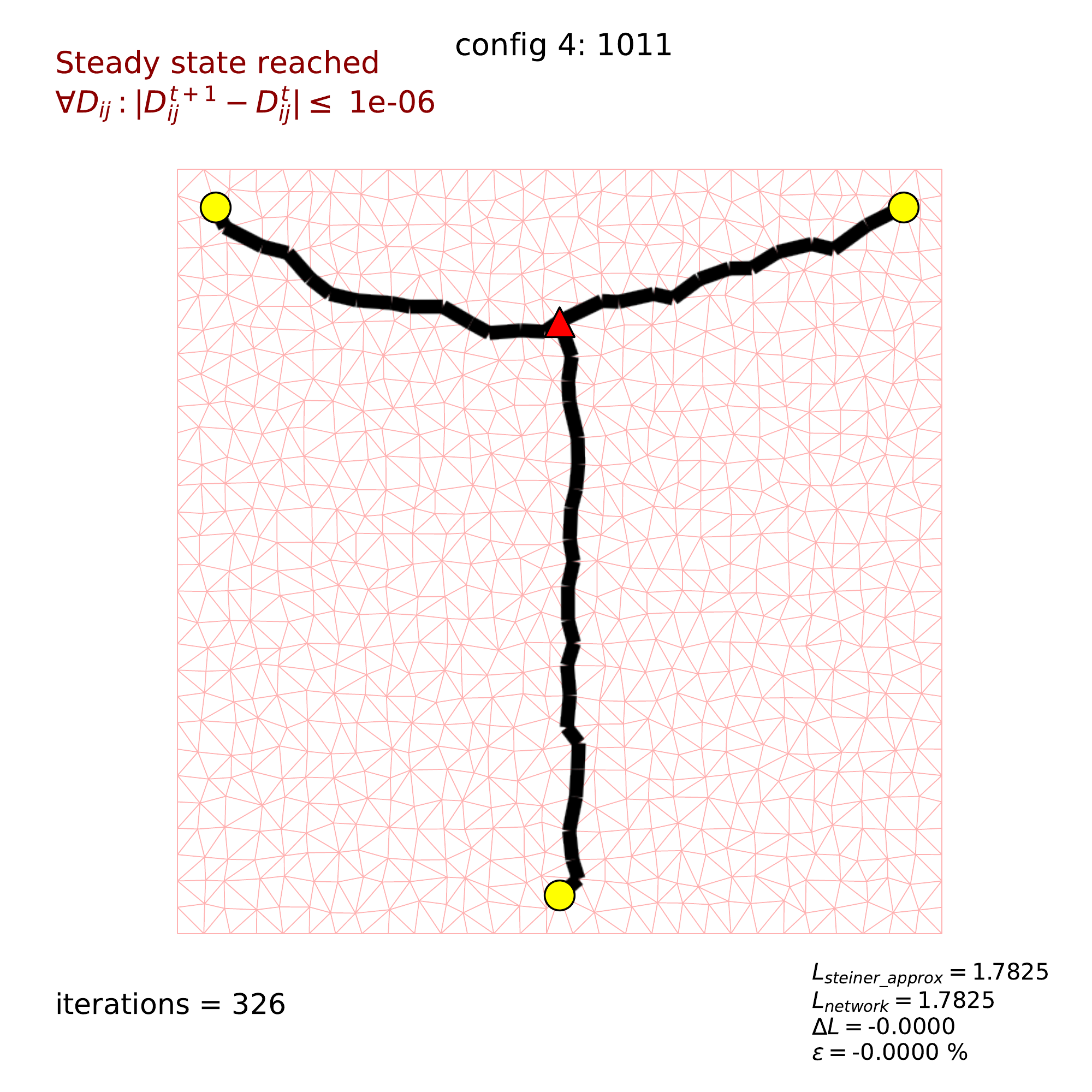} &
\mysub{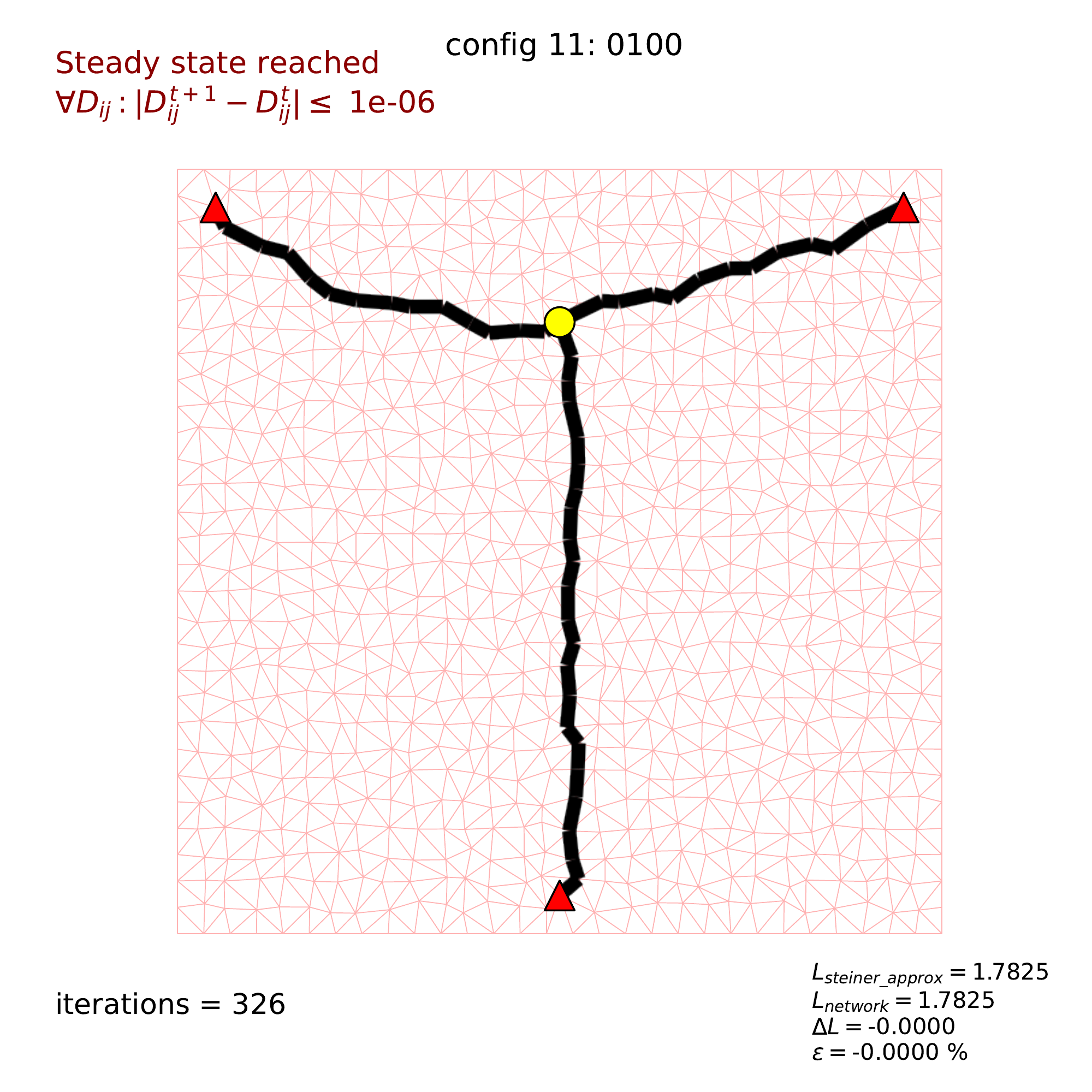} \\

\mysub{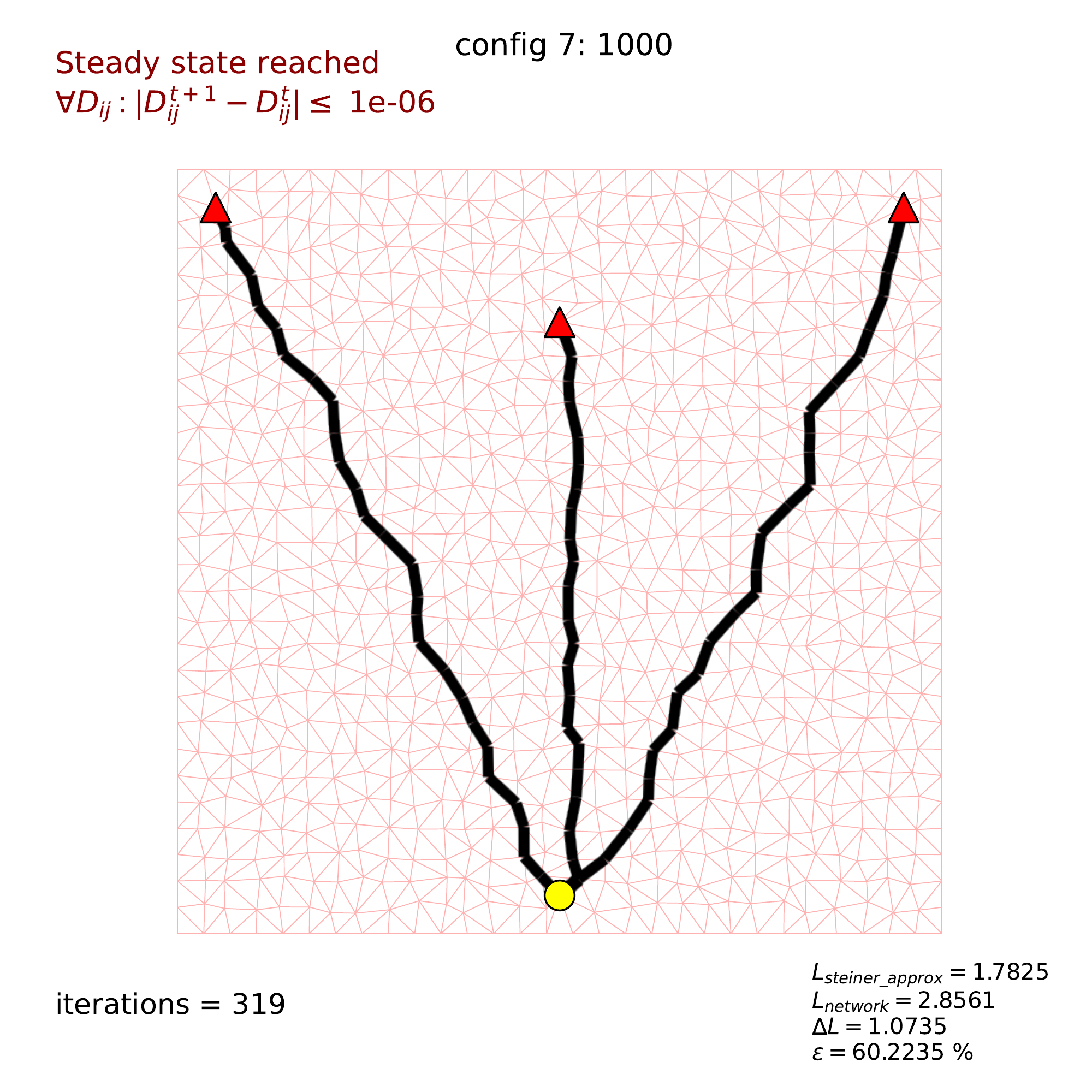} &
\mysub{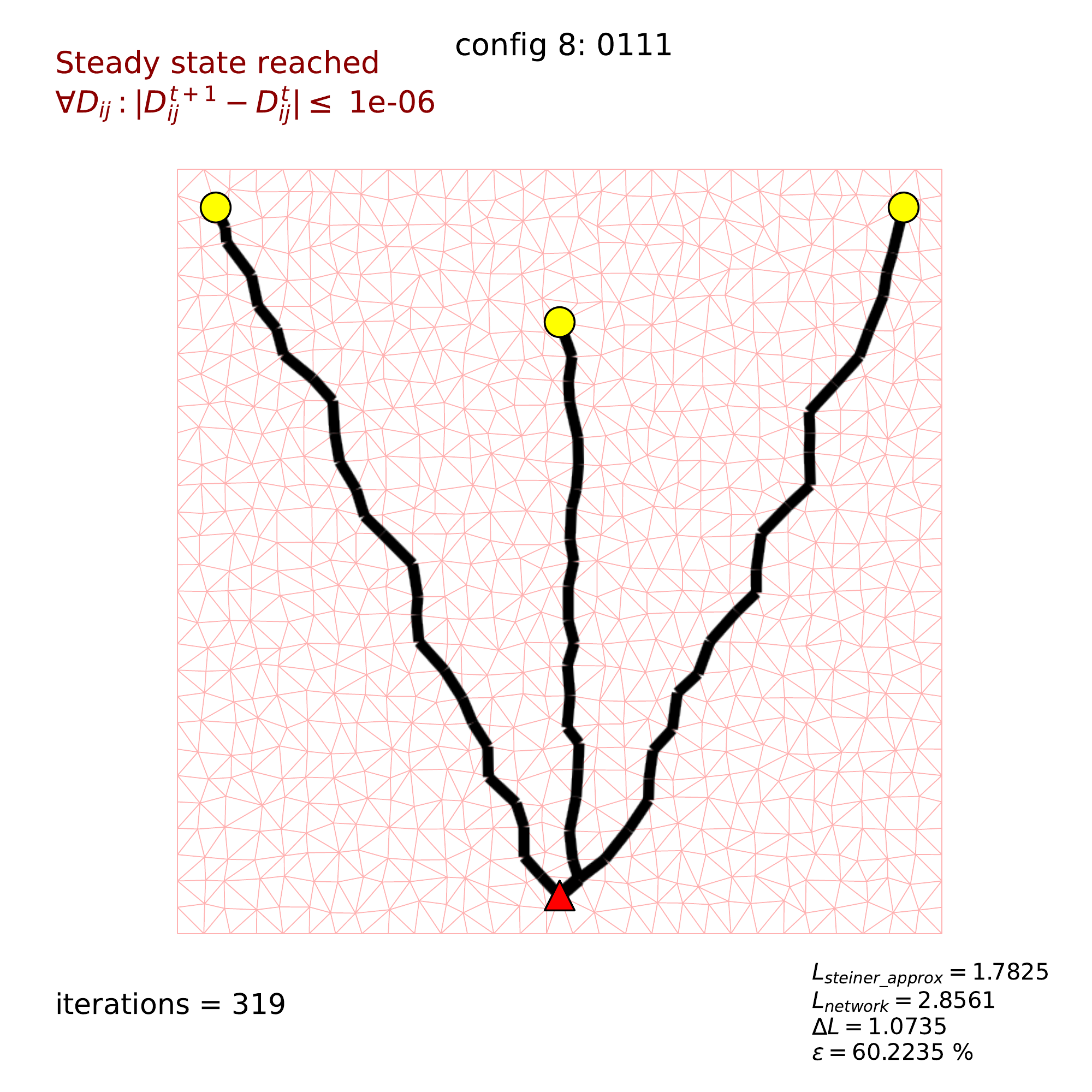} &
\mysub[\label{fig:mult_term_j}]{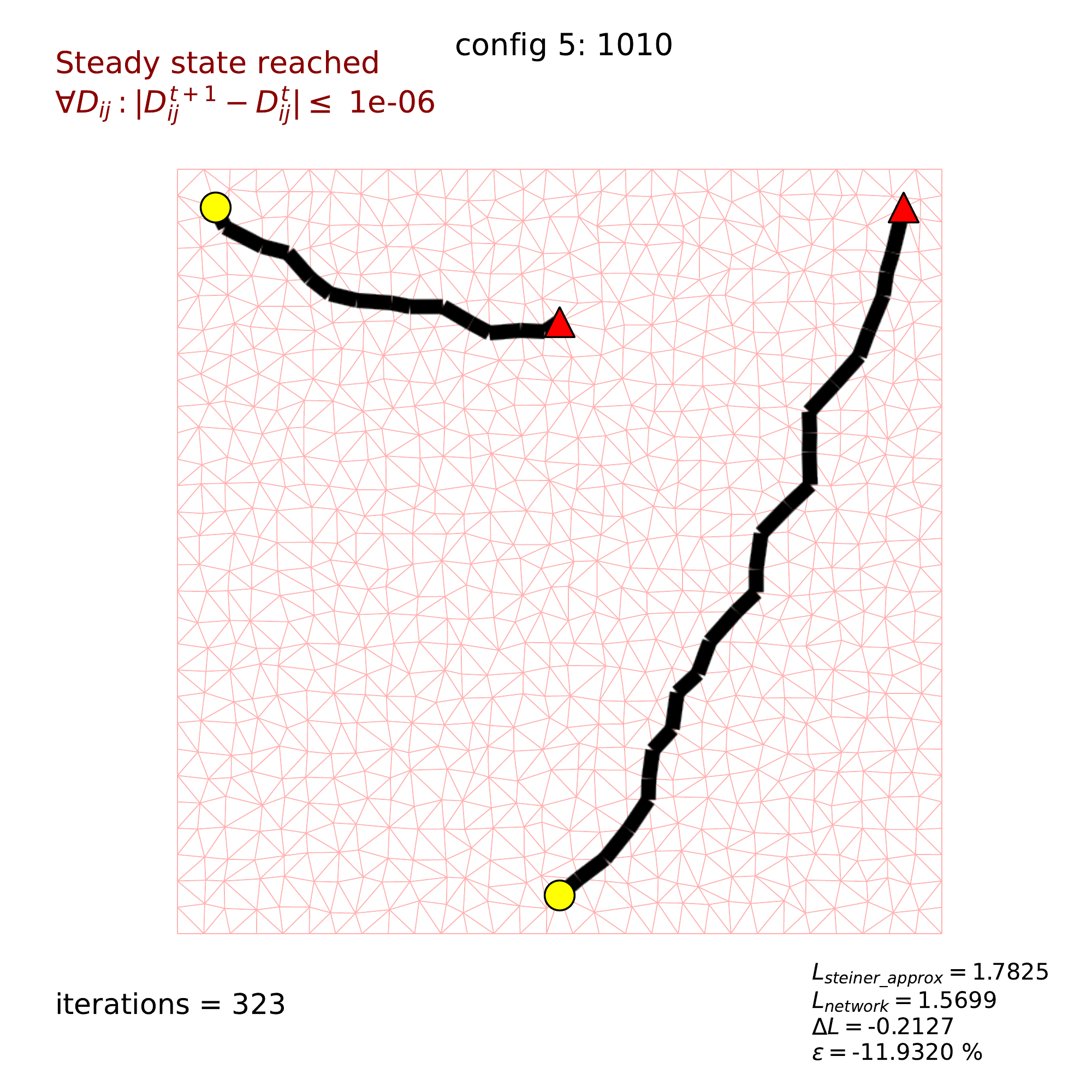} &
\mysub{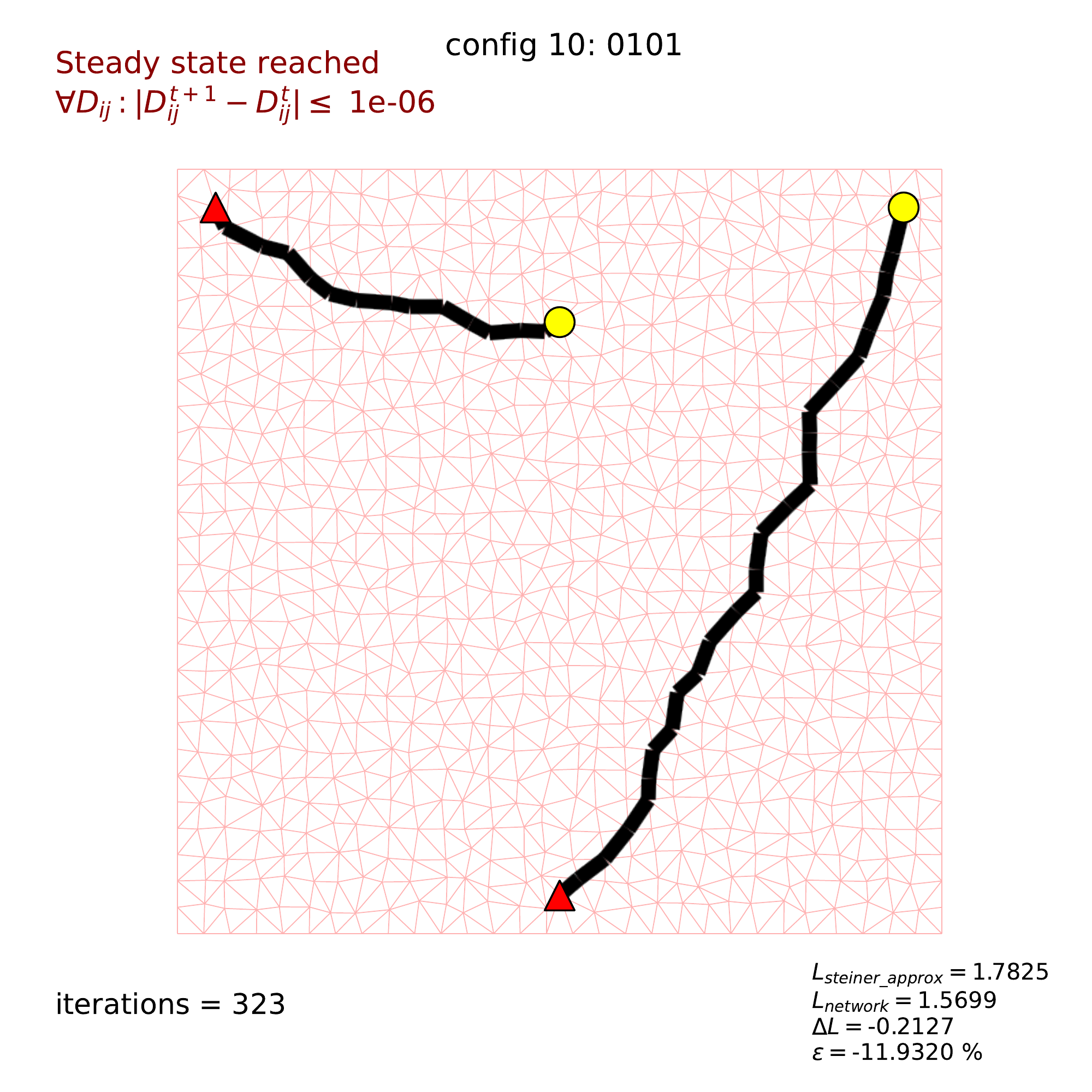} \\

\mysub{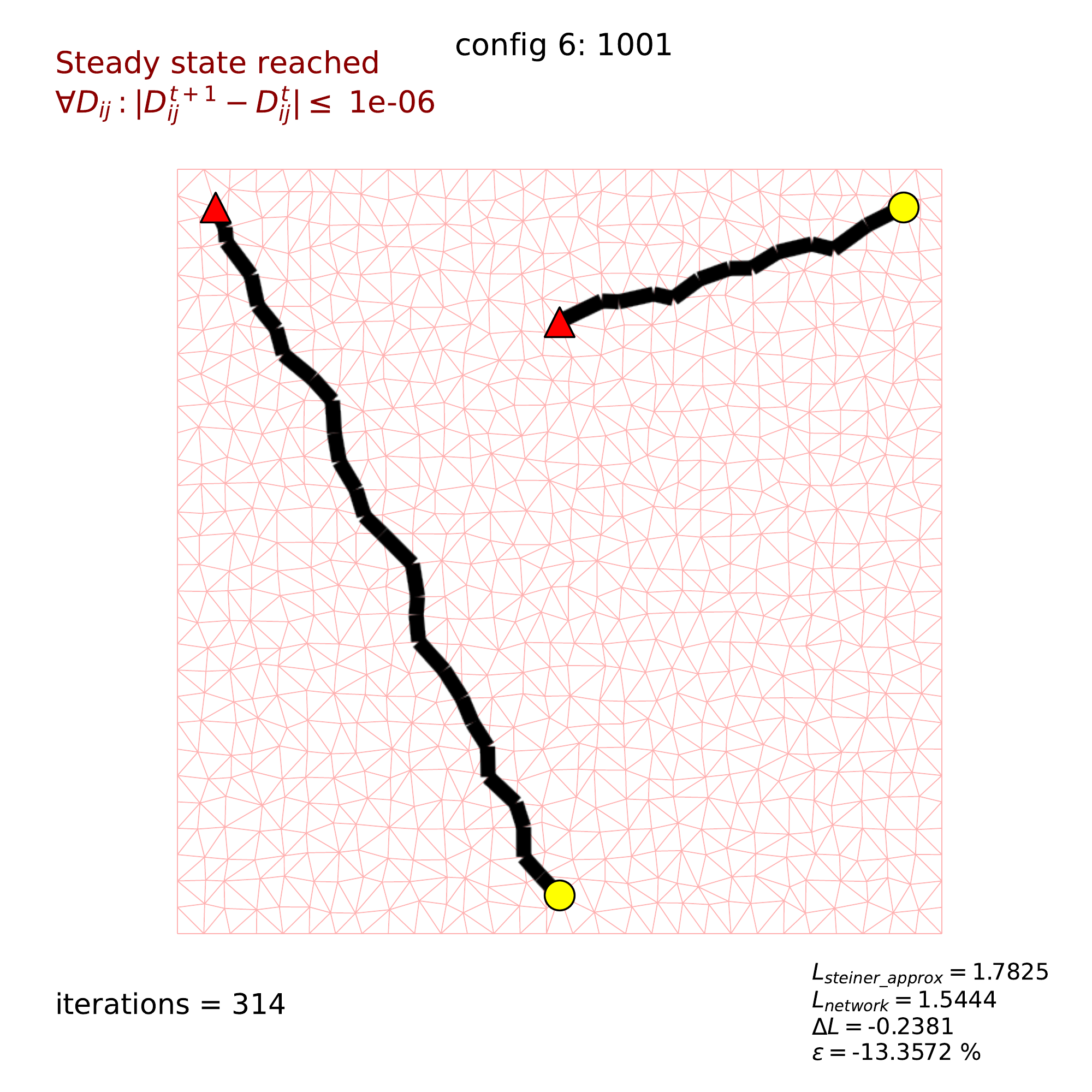} &
\mysub{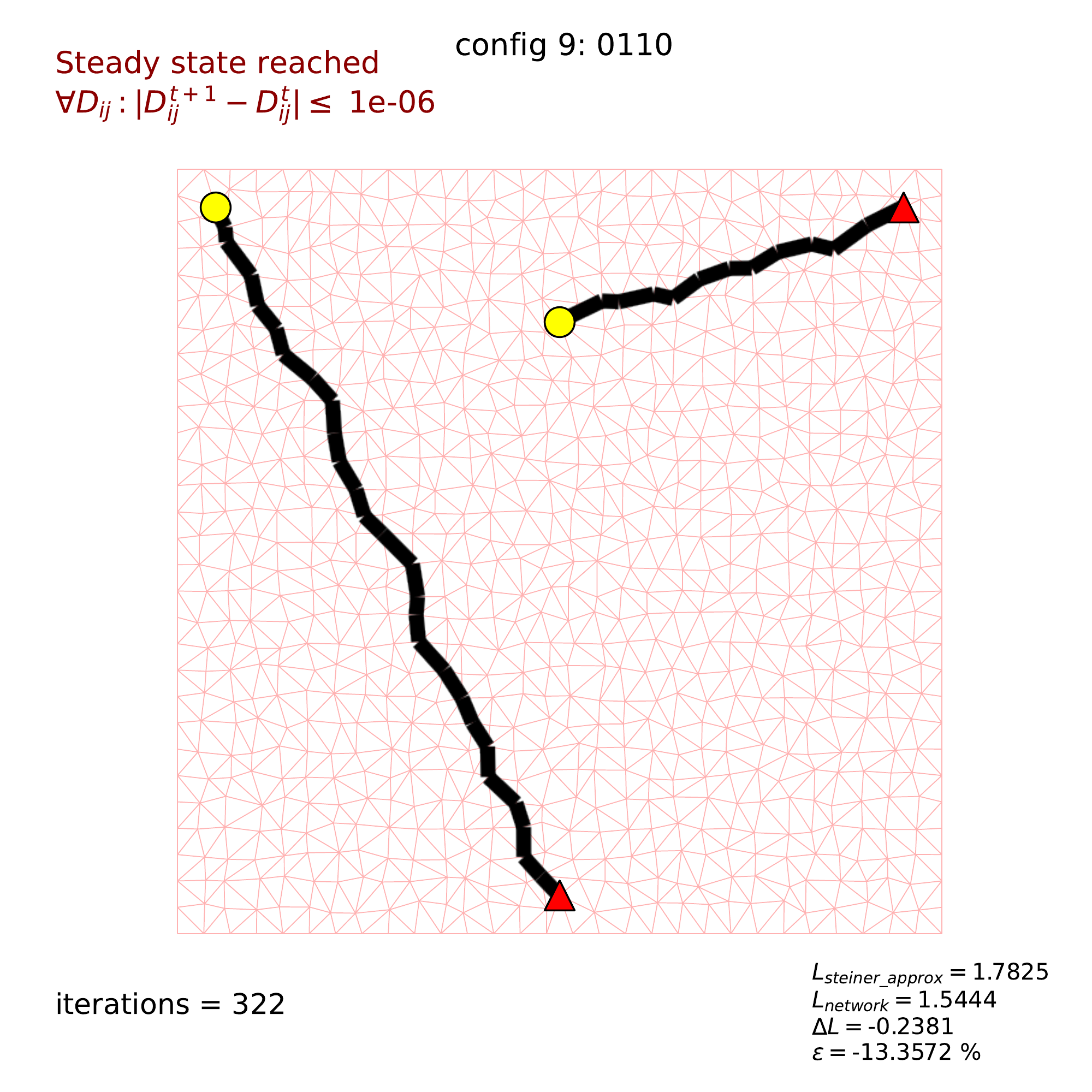} &
\mysub{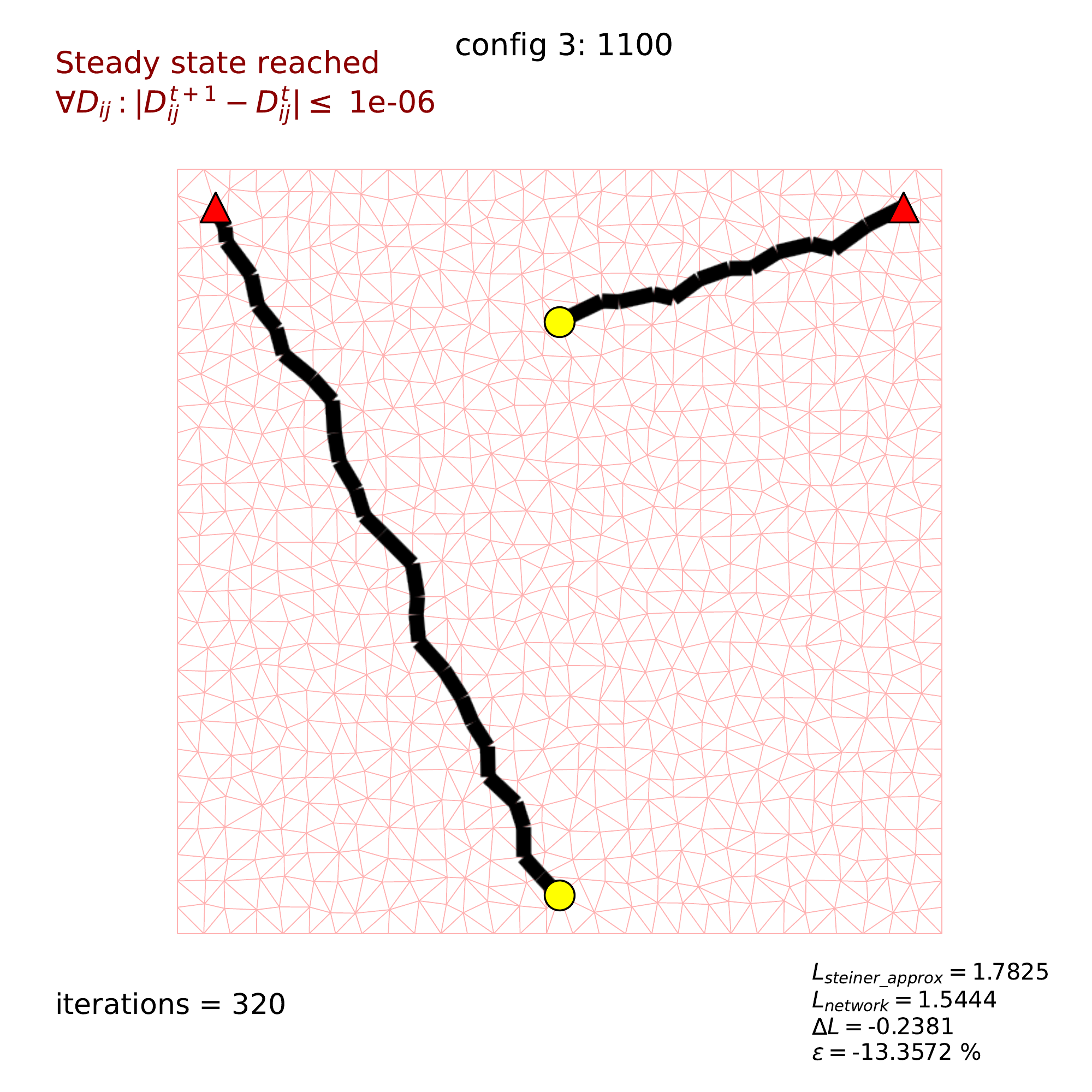} &
\mysub[\label{fig:mult_term_o}]{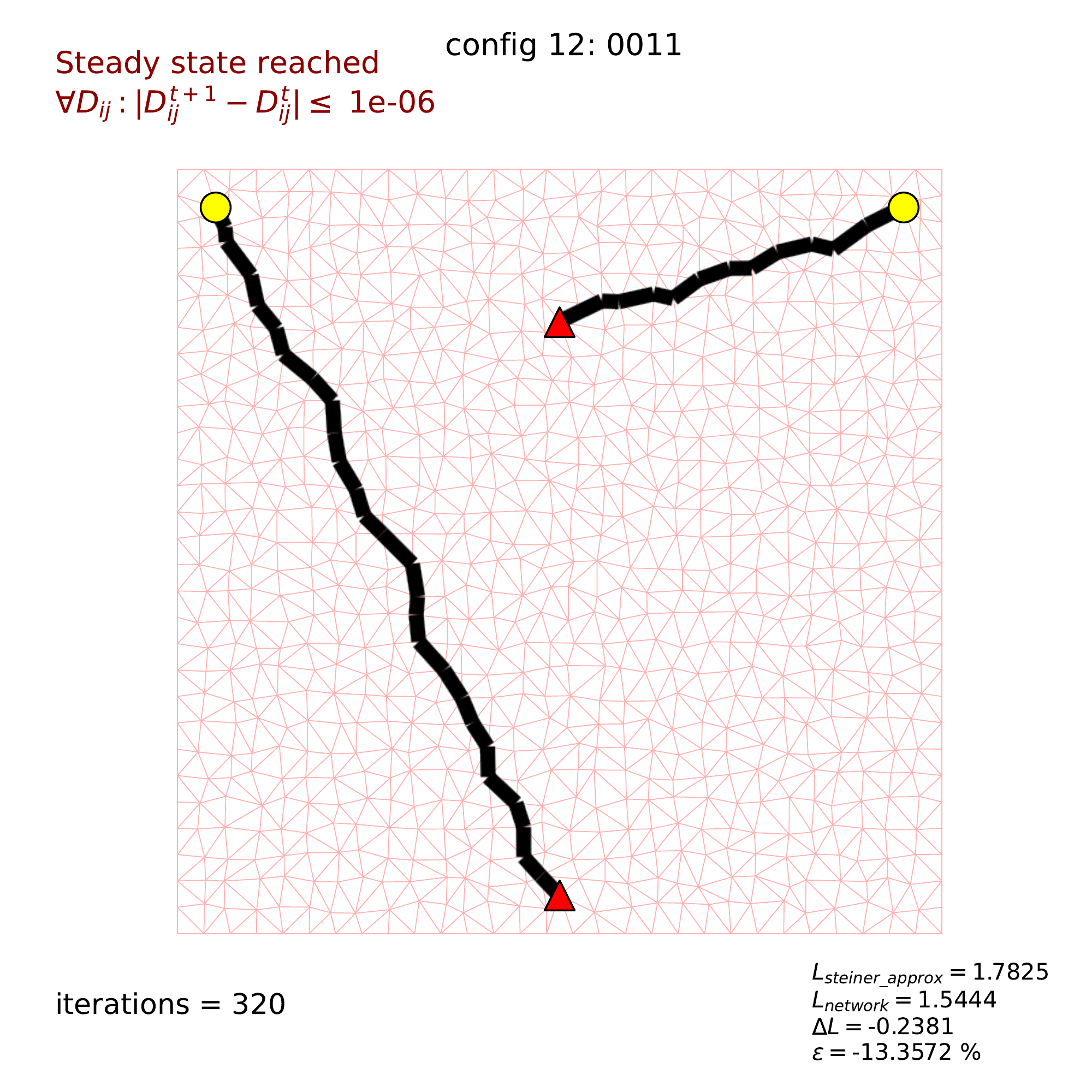} \\

\end{tabular}
\caption{Dependency of the optimal network geometry on the choice of sources and sinks. \textbf{(a)} Several simulations of the adaptation dynamics \eqref{eq:new_adapt_rule_minE} were carried on an initial network (red mesh) with $D_{ij}(0)=1$, considering a fixed Y-shaped arrangement of 4 terminals (blue circles).  \textbf{(b-o)} steady-state  networks reached considering all the different combinations of sources (yellow circles) and sinks (red triangles) states of the arrangement of terminals \textbf{(a)}. The thickness of the black lines is proportional to the radius of the channels. In each case, the distribution of node fluxes follows \eqref{eq:q_fixed_I0} ($I_0=1$). The geometry of the steady-state  network varies greatly with the choice of sources and sinks. By interchanging the sources with the sinks, the same steady state  is obtained. Some configurations can lead to apparently disconnected solutions \textbf{(j-o)}.  
}
\label{fig:Multiple Terminals} 
\end{figure}

We now investigate how the choice of terminals and their intensity affect the geometry of the optimal networks. For that, we've considered 
a fixed Y-shaped arrangement of 4 terminal nodes, and tested the adaptation mechanism \eqref{eq:new_adapt_rule_minE} for all the different combinations of source and sink states of the 4 nodes (Figure \ref{fig:Multiple Terminals}). Since the all-sinks and all-sources aren't valid states, as they don't satisfy \eqref{eq:sum_qi}, for 4 terminals there are $2^4-2=14$ possible states.  To make a fair comparison, the same initial mesh and same initial conductivities ($D_{ij}(0)=1$) were used in the simulations (Figure \ref{fig:mult_term_init}), and the nodes net fluxes were chosen according to 

\begin{equation}
    q_i = 
    \begin{cases}
    I_0 / N_{sources} \ \  &, \ \ i\in \text{sources} \\
    - I_0 / N_{sinks} \ \ &, \ \ i\in \text{sinks} \\
    0   \ \ &, \ \ \text{otherwise}
    \end{cases}
    \qquad , \  i\in V
    \label{eq:q_fixed_I0}
\end{equation}

\noindent where $I_0=1$, which implies that the same amount of flux flows through the network in each case. 

The resultant optimised networks obtained for each state are depicted in Figure \ref{fig:Multiple Terminals}. The results show that for a given set of terminals the geometry of the steady-state  network depends greatly on 
which of those are sources and which ones are sinks.

An interesting observation is that symmetrical configurations, i.e., configuration with the opposite choice of sources and sinks, lead to the same final network. This isn't due to the particular choice of the Y-shape arrangement, but it's a general result confirmed by numerous simulations, which is related with the choice of the nodes net fluxes distribution \eqref{eq:q_fixed_I0}. To demonstrate this, let a given configuration of sources and sinks be represented by a source vector $\vb*{q}$, and let the symmetrical configuration be described by the source vector $\vb*{q'}$, both following \eqref{eq:q_fixed_I0}. Since the number of sources in one is equal to the number of sinks in the other,  i.e., $N_{sources}=N'_{sinks}$ and $N_{sinks}=N'_{sources}$, we have that 

\begin{nalign}
q_{source} 
&\longrightarrow
&q'_{sink} = \frac{I_0}{N'_{sinks}} = -\frac{I_0}{N_{sources}} 
= -q_{source} \\
q_{sink}
&\longrightarrow
&q'_{source} = \frac{I_0}{N'_{sources}} = -\frac{I_0}{N_{sinks}}
= -q_{sink} \;.
\end{nalign}

Opposite configurations can thus be seen as a transformation $\vb*{q'}\rightarrow -\vb*{q}$ in the system. Given the linearity of the conservation laws \eqref{eq:sum_Qij}, this is translated into a global flow reversal, i.e., $Q_{ij}' \rightarrow - Q_{ij}$ for $(i,j)\in E$. However, since the adaptation dynamics \eqref{eq:new_adapt_rule_minE} depends only on the magnitude of the flux in each vessel, the flow reversal has no effect on the dynamics. Therefore, we conclude that opposite configurations yield the same final network. 

\begin{figure}[!hbt] 

\newcommand{\mysub}[2][]{%
    \subfloat[#1]{\includegraphics[trim={2.9cm 3cm 3.5cm 3.1cm}, clip, width=0.249\textwidth]{#2}}%
}

\newcommand{\myfig}[1]{%
    \includegraphics[trim={2.9cm 3cm 3.5cm 3.1cm}, clip, width=0.249\textwidth]{#1}%
}

    \centering
    \begin{tabular}{@{}c@{}c@{}c@{}c@{}}
    $I_0=0.1$ & $I_0=1$ & $I_0=10$ & $I_0=100$ \\[0.1cm]
    \mysub[]{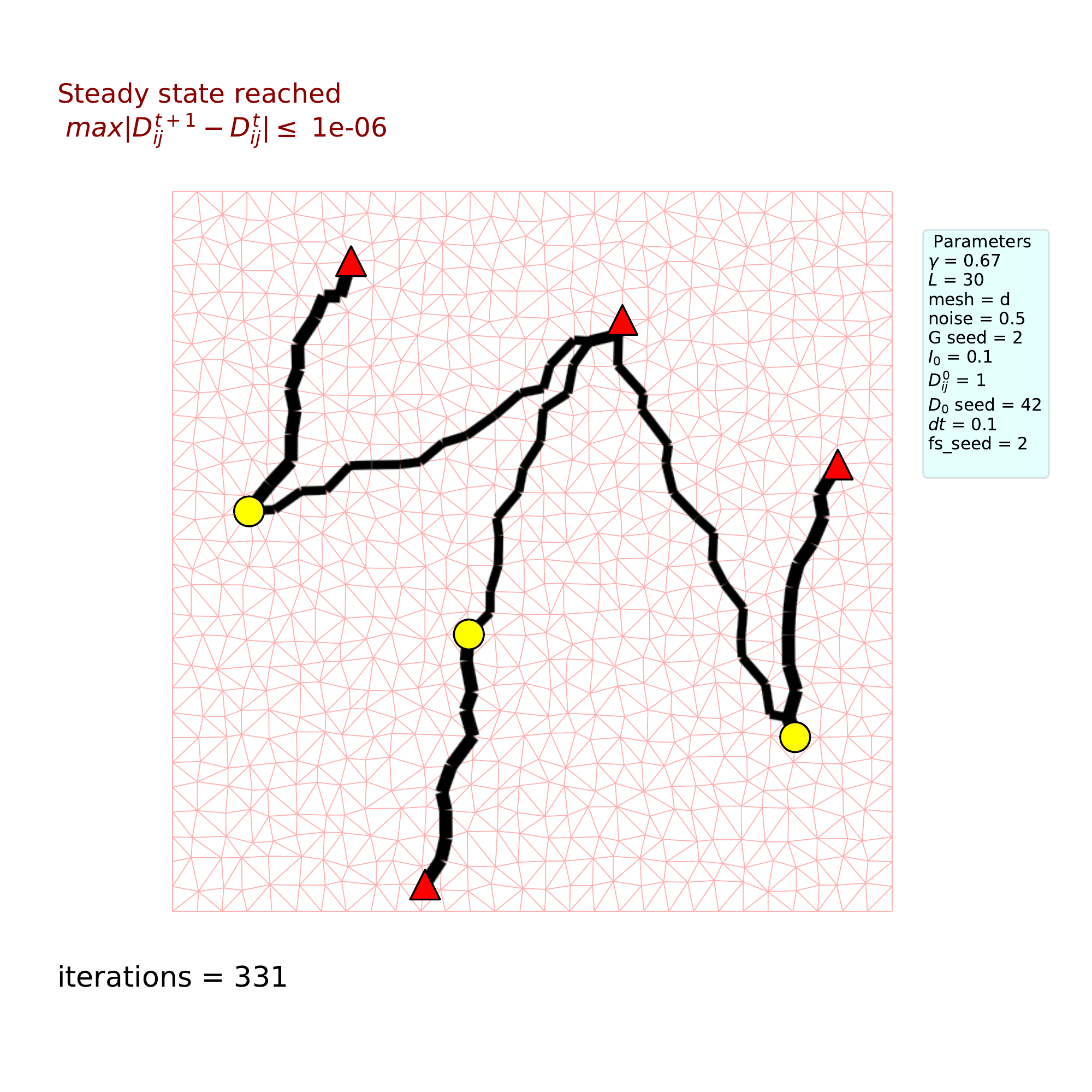} &
    \mysub[]{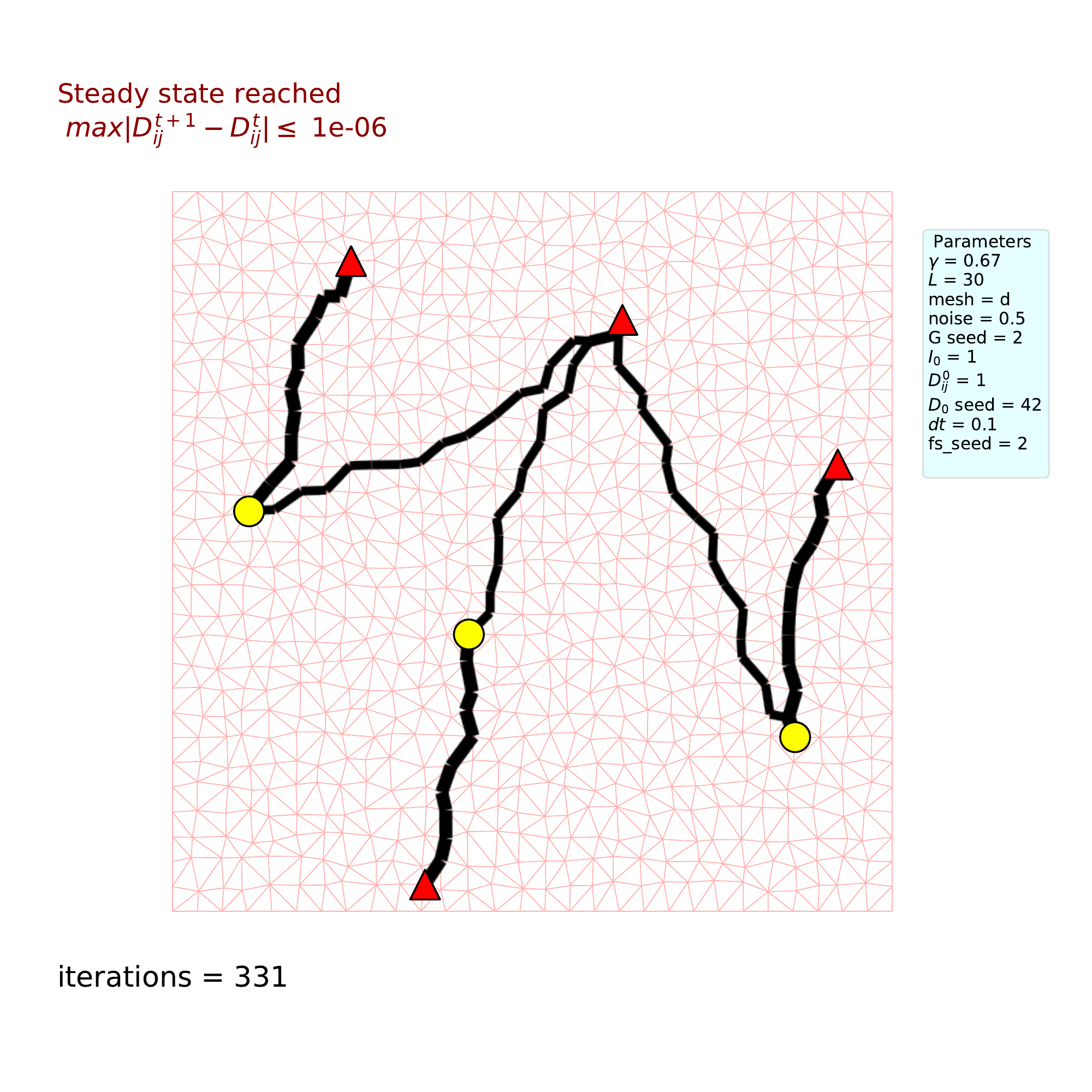} &
    \mysub[]{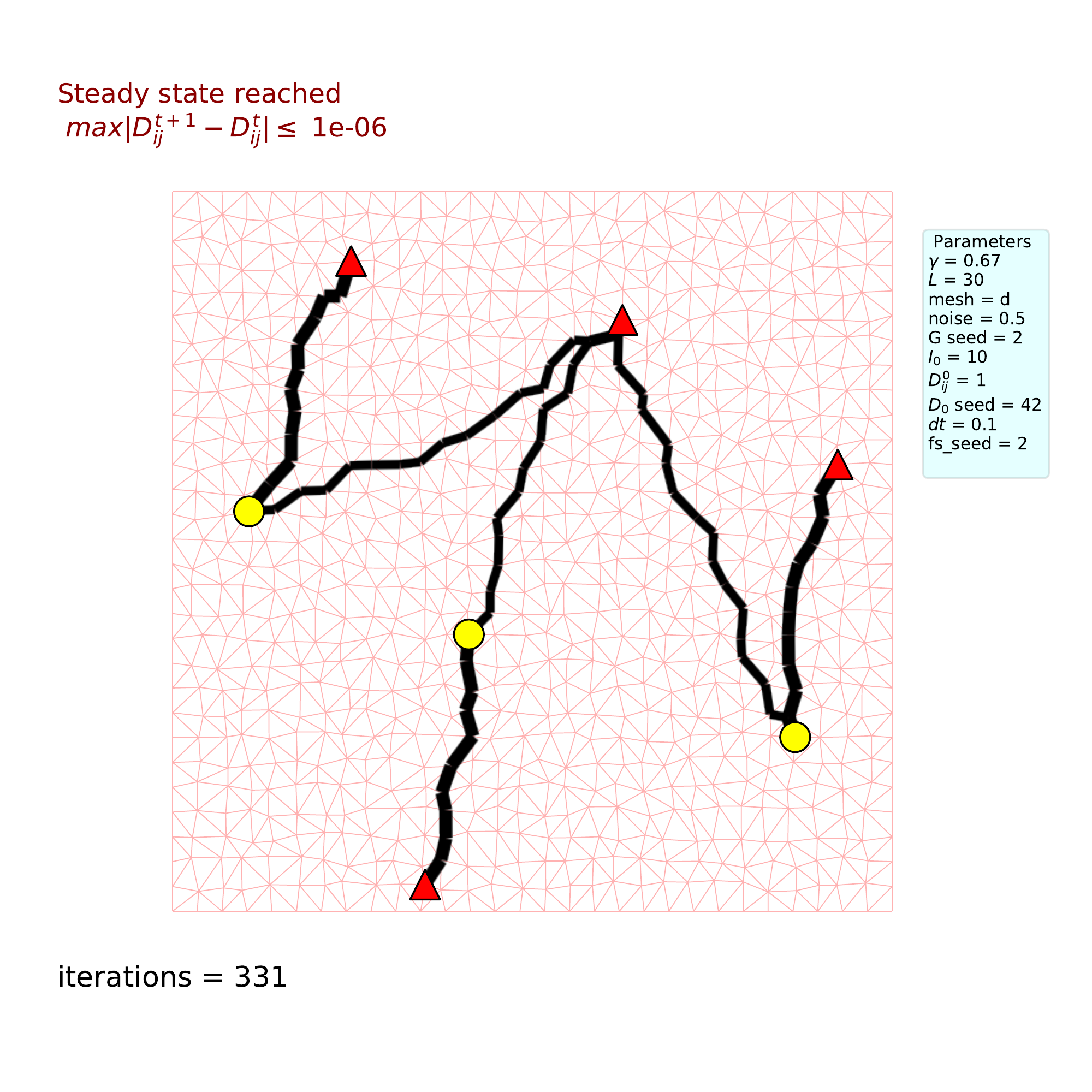} &
    \mysub[]{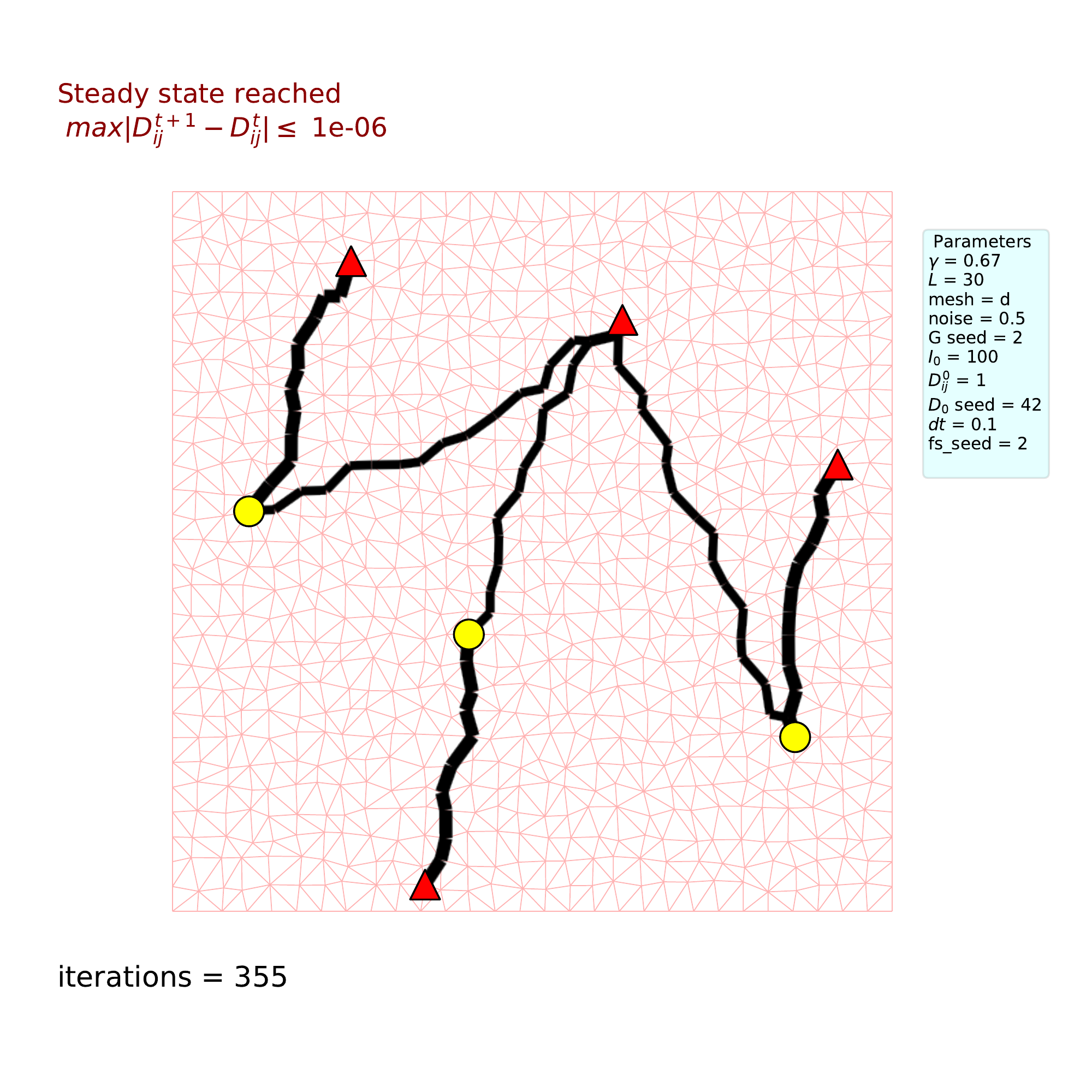} 
    \end{tabular}

\caption{ 
Dependency of the steady states of the dynamics \eqref{eq:new_adapt_rule_minE} on a global scaling of the nodes' net fluxes, considering the same configuration of terminals (3 sources and 4 sinks), initial conditions ($D_{ij}(0)=1$) and initial mesh. The images show the steady states obtained considering the nodes flux distribution \eqref{eq:q_fixed_I0} for different choices of the parameter $I_0$. The dynamics converged to the same steady-state  conductivities 
regardless of the value of $I_0$, which implies that it's invariant for scaling transformation of $\vb*{q}$.} 
\label{fig:q_initial_scale} 
\end{figure}

More generally, following a similar approach, we can demonstrate that the adaptation dynamics \ref{eq:new_adapt_rule} is invariant for  scale transformations of the nodes' net fluxes, $\vb*{q}\rightarrow \alpha \vb*{q}$, with $\alpha\in \R$, as long as the function $g$ satisfies $g(|\alpha Q_{ij}|)=|\delta|g(|Q_{ij}|)$ for some non-zero constants $\alpha$  and $\delta$ (e.g., $g$ polynomial). This means that the parameter $I_0>0$ in \eqref{eq:q_fixed_I0} has no effect on the conductivities of the steady-state  optimal networks. These observations are supported by the results of Figure \ref{fig:q_initial_scale}, where different choices of the parameter $I_0$ were tested for the same geometry and initial conditions.

\begin{figure}[!hbt] 

\newcommand{\mysub}[2][]{%
    \subfloat[#1]{\includegraphics[trim={2.9cm 7cm 3.5cm 8cm}, clip, width=0.249\textwidth]{#2}}%
}

\centering

\mysub[]{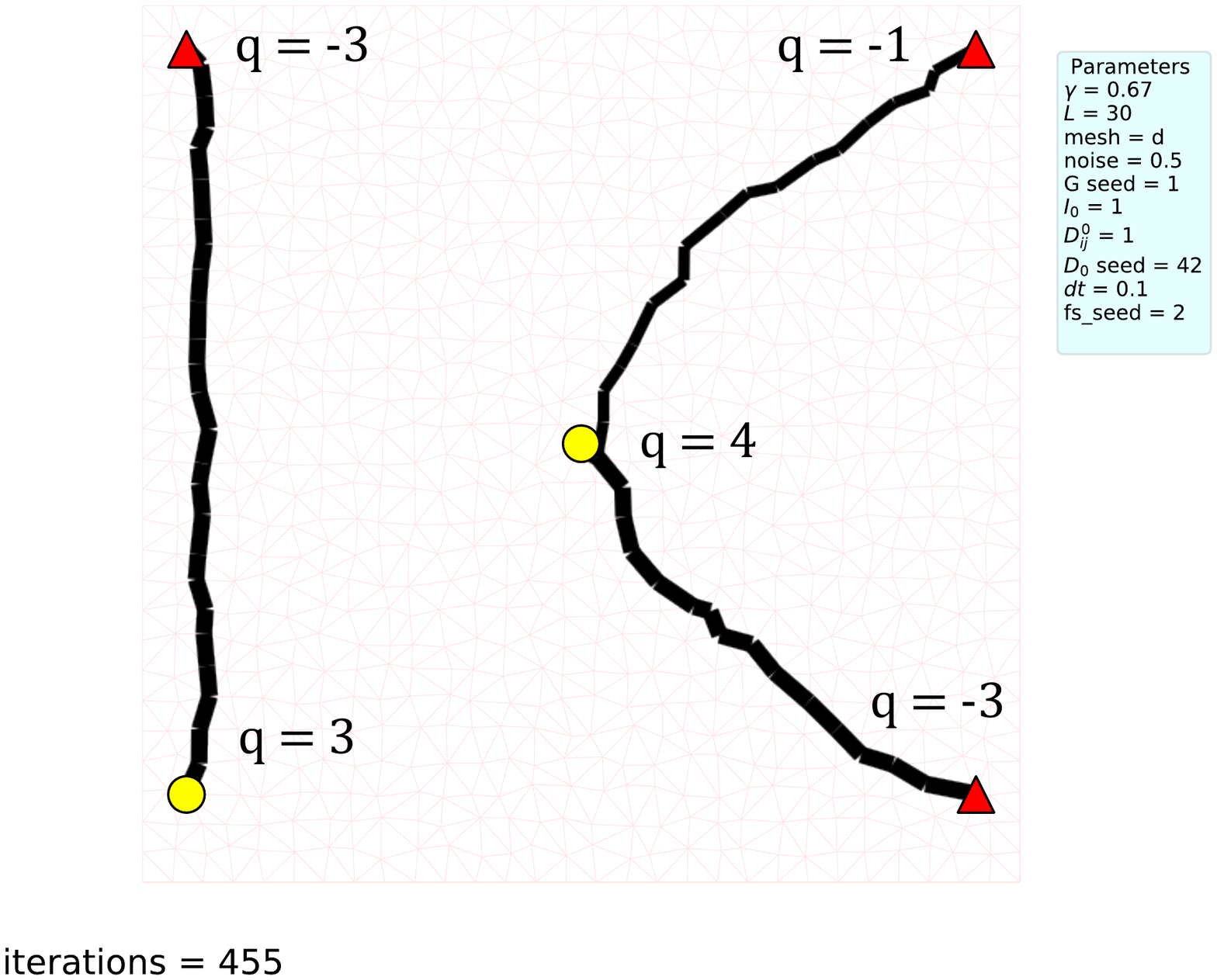} \hfill
\mysub[]{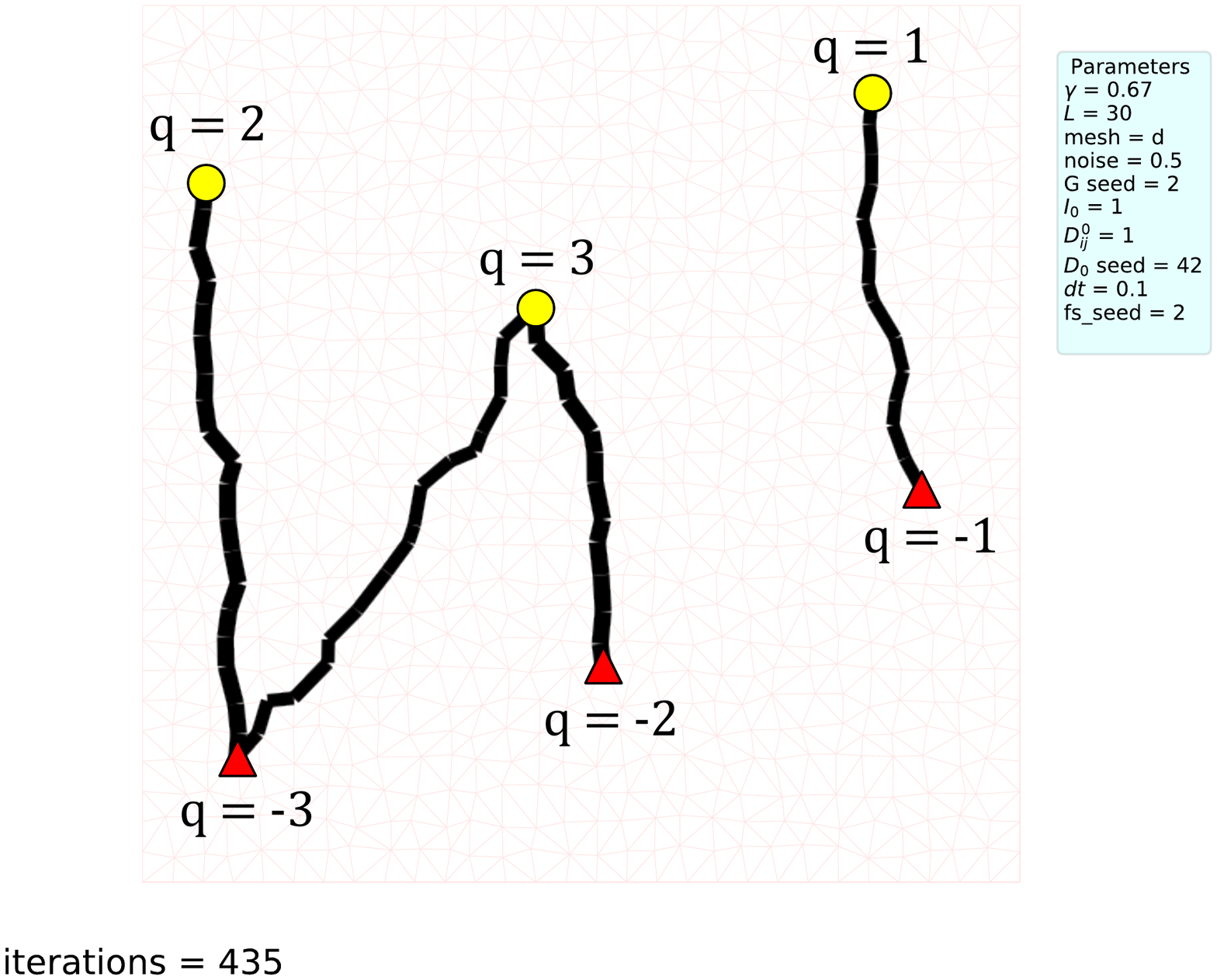} \hfill
\mysub[]{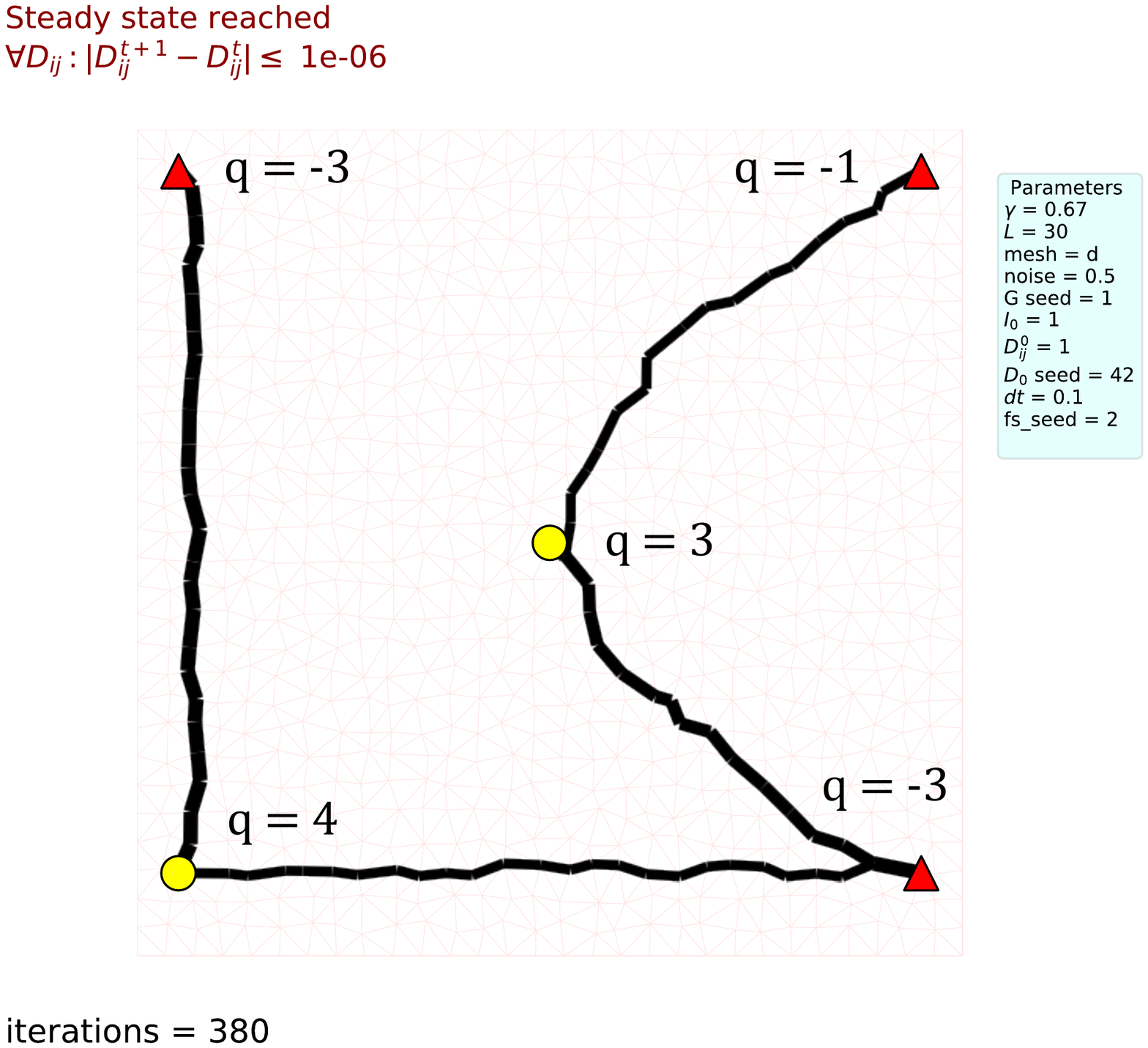} \hfill
\mysub[]{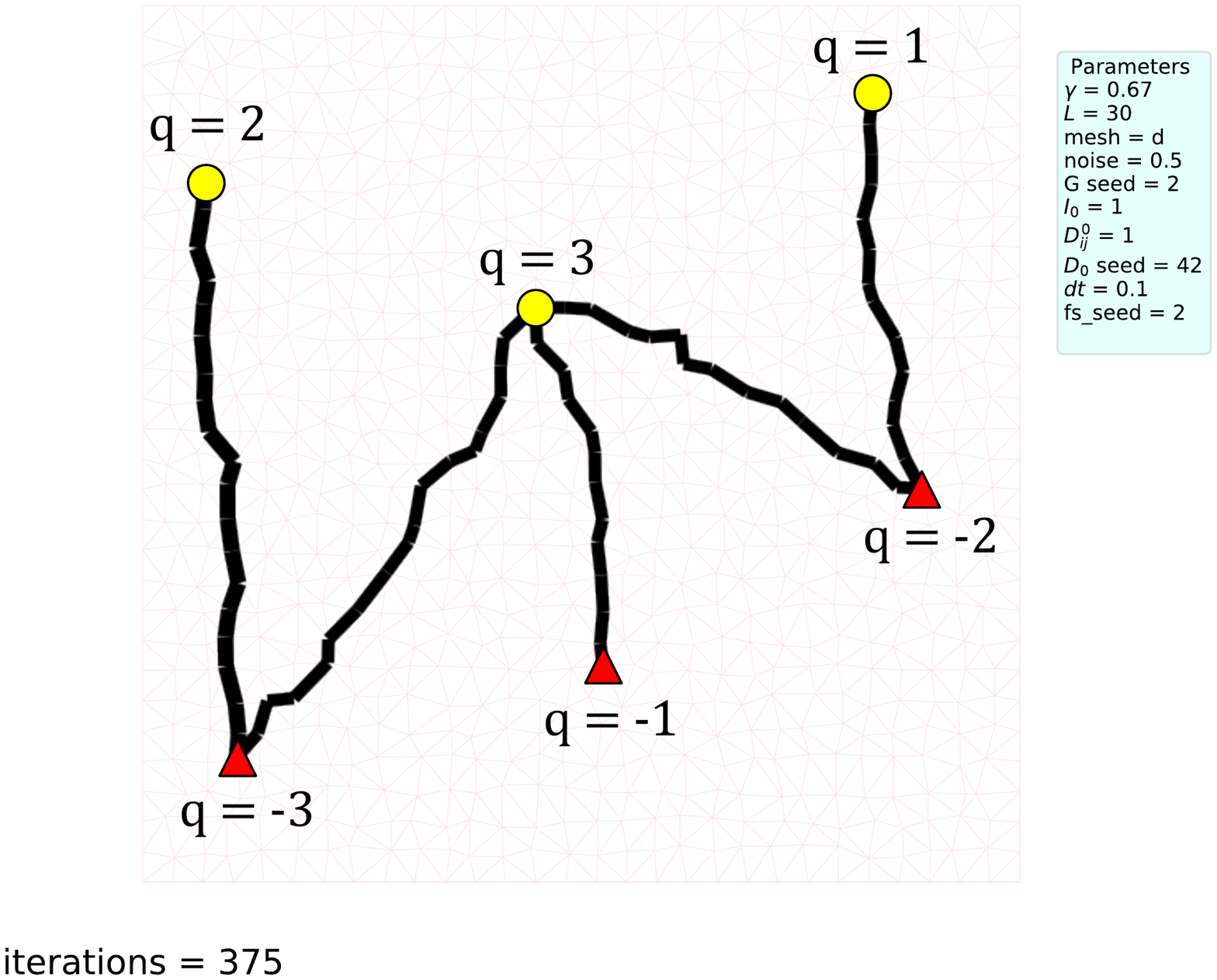} 

\caption{\textbf{(a-b)} Examples of (apparently) disconnected steady-state  networks of the dynamics \eqref{eq:new_adapt_rule_minE} for two different choices of the nodes' net flux distribution. The labels near each terminal represent their net flux, while the remaining nodes have $q=0$. \textbf{(c-d)} Examples of connected steady states for the same configurations of \textbf{(a)} and \textbf{(b)}, although disconnected solutions would be possible in principle. Note that the nodes fluxes are identical in cases \textbf{(a)} and \textbf{(c)}, apart from a permutation of the sources, and the same goes for the cases \textbf{(b)} and \textbf{(d)}, apart from a permutation of the sinks. All the simulations were done considering initial conductivities $D_{ij}(0)=1$. The thickness of the black lines is proportional to the radius of the channels. 
}
\label{fig:q_disconnected_connected} 
\end{figure}

Figures \ref{fig:mult_term_j} to \ref{fig:mult_term_o} also reveal that for certain choices of sources and sinks the adaptation mechanism can lead to apparently disconnected networks, where each ``disconnected component'' adapts almost independently from the others. Note, however, given the nature of adaptation dynamics \eqref{eq:new_adapt_rule_minE}, technically the components are not truly disconnected from each other, but are weakly coupled through edges with very small conductivities which asymptotically approach zero.Further simulations showed that, in general, this may  happen for distributions of the nodes' fluxes, $\vb*{q}$, which allow for a partition  of the graph $\G$ such that the sum of the nodes' fluxes 
of every subgraph is (non-trivially) zero. For instance, when the number of sources and sinks is equal and for each source there is a sink with a symmetric intensity, allowing the terminals to be grouped in pairs. Although this is a necessary condition, it's not sufficient to guarantee that the  steady-state  graph is apparently disconnected, otherwise, it would be the case of all the steady-state networks of Figure \ref{fig:q_disconnected_connected}. In the context of \Phy{}, these apparently disconnected solutions are unrealistic, since the organism grows and adapts as a single contiguous network. 

It's not clear what is the most appropriate choice of sources and sinks in the case of \textit{Physarum}. Naturally, food sources are sources of nutrients, but other stimulated regions can also be sources of signalling molecules \cite{Alim2017}.
However, it's certain that they aren't static as we are assuming here, since the flow is not steady, and periodically changes direction. Only fluctuations of the sources and sinks may replicate this shuttle streaming behaviour.

\FloatBarrier

\FloatBarrier

\subsection{Initial Conditions}

We now study the uniqueness of the steady-state  solutions of the dynamics \eqref{eq:new_adapt_rule_minE}. If the steady state  is unique, given a distribution of sources and sinks, $\vb*{q}$, and an initial mesh geometry, the system should converge to the same final network regardless of the initial conductivities, $D_{ij}(0)$. Note that until now we have only considered homogeneous initial conditions, $D_{ij}(0)=1$ for all $(i,j)\in E$. 

\begin{figure}[bht!] 

\newcommand{\mysub}[2][]{%
    \subfloat[#1]{\includegraphics[trim={2.9cm 3cm 3.5cm 3.1cm}, clip, width=0.249\textwidth]{#2}}%
}

\newcommand{\myfig}[1]{%
    \includegraphics[trim={2.9cm 3cm 3.5cm 3.1cm}, clip, width=0.25\textwidth]{#1}%
}

\begin{subfigure}{\textwidth}   
    \centering
    \begin{tabular}{@{}c@{}c@{}c@{}c@{}}
    \myfig{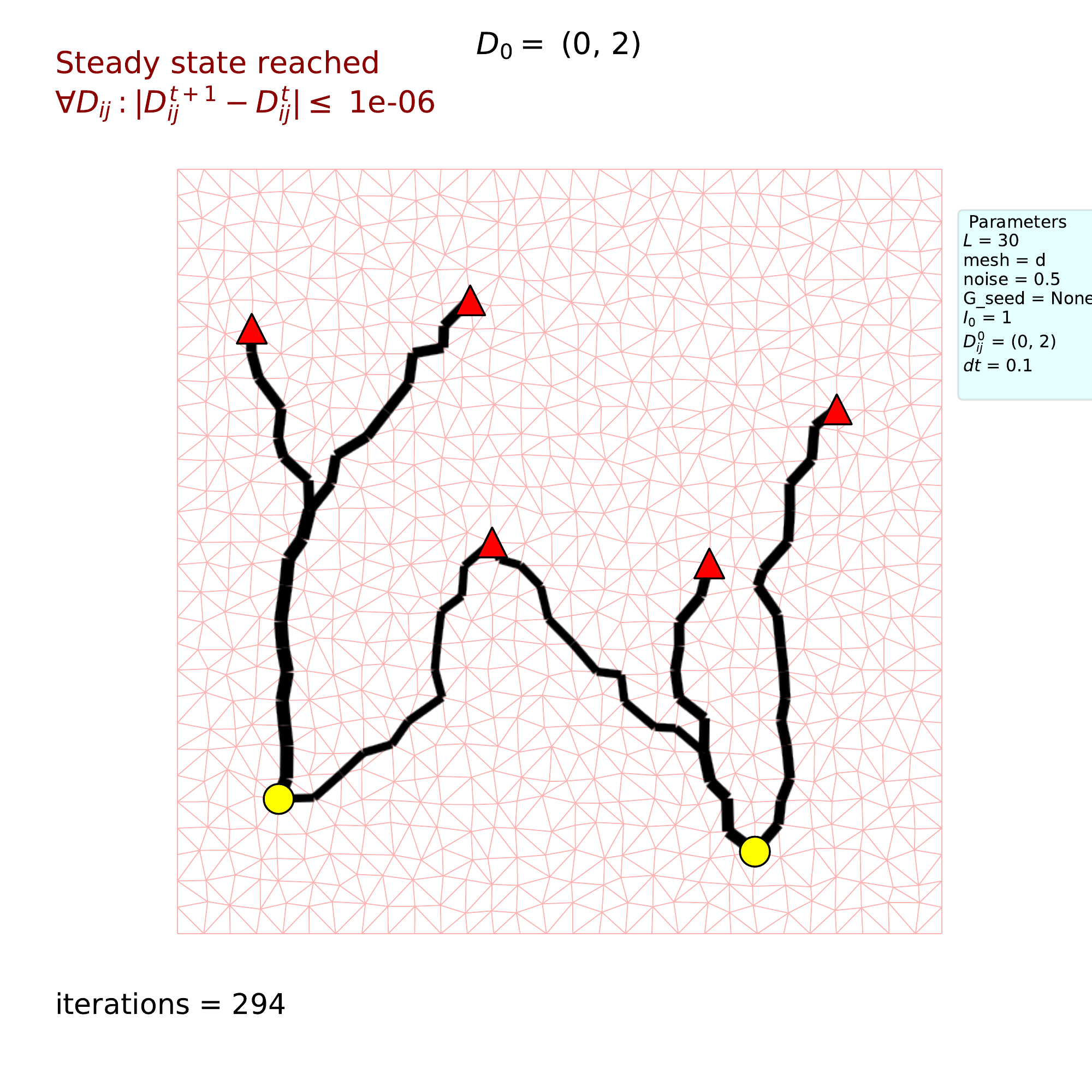} &
    \myfig{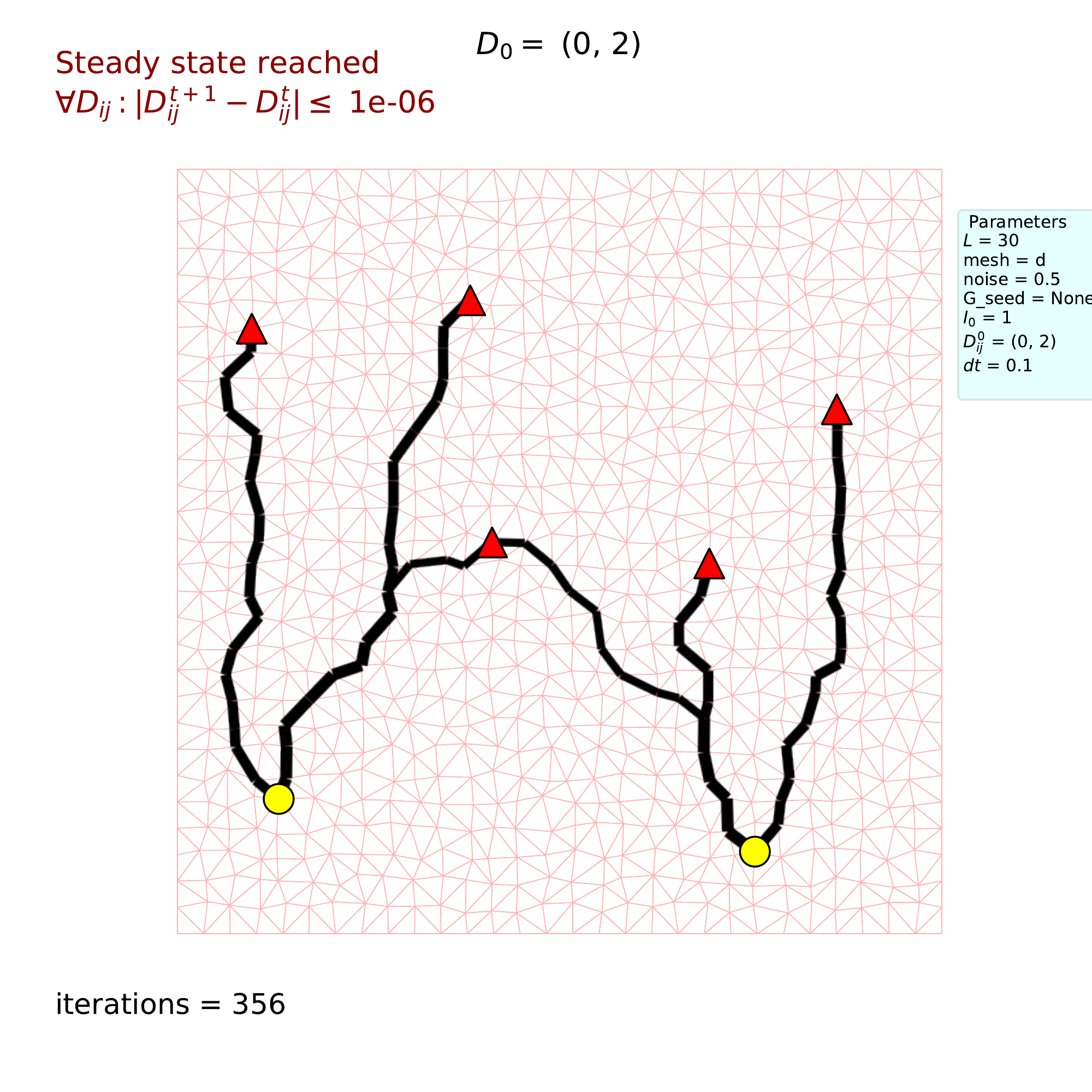} &
    \myfig{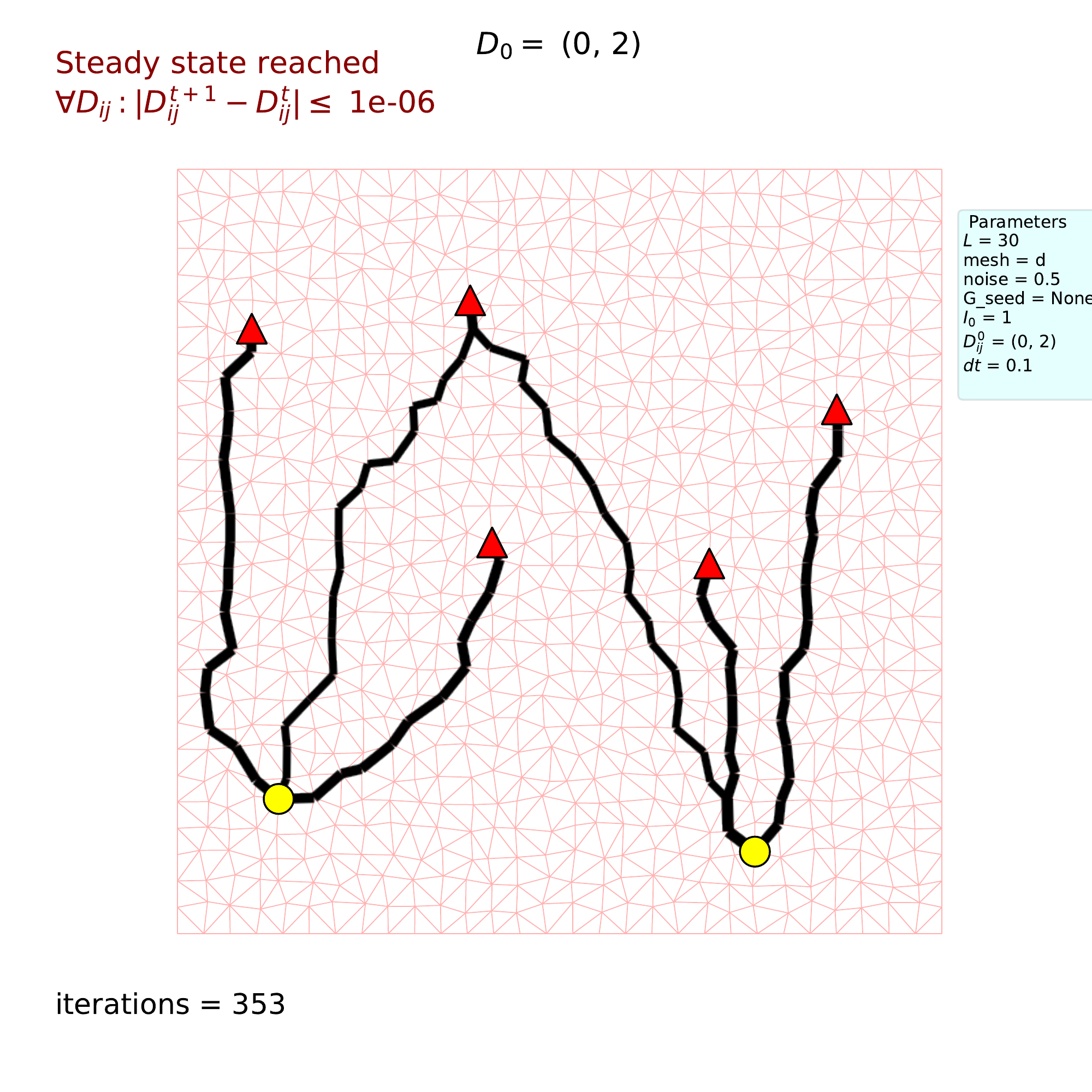} &
    \myfig{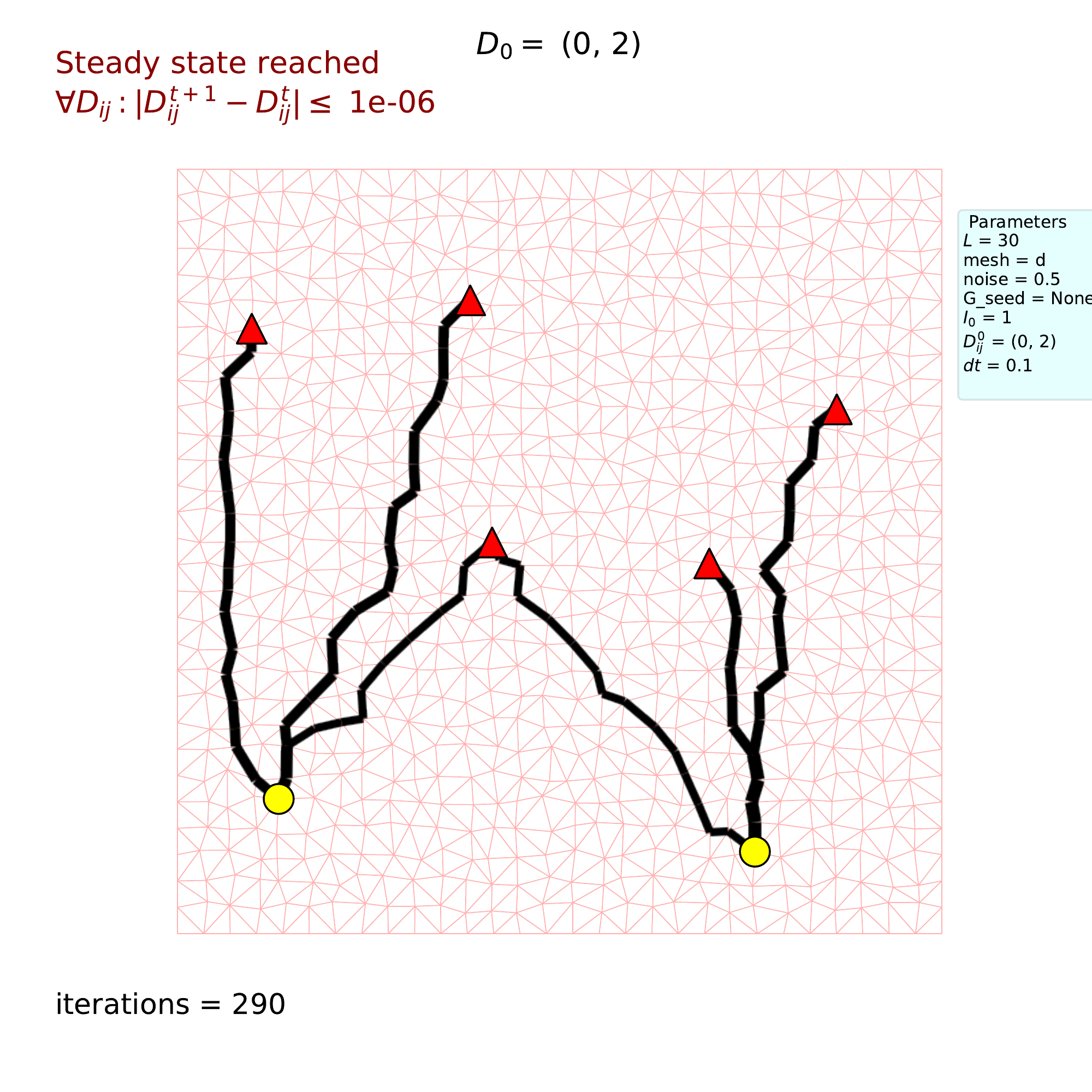} 
    \end{tabular}
    \caption{2 sources (yellow) with intensities $q_{source}=1/2$, and 5 sinks (red) with intensities $q_{sink}=-1/5$.}
\end{subfigure}
\\[0.2cm]
\begin{subfigure}{\textwidth}   
    \centering
    \begin{tabular}{@{}c@{}c@{}c@{}c@{}}
    \myfig{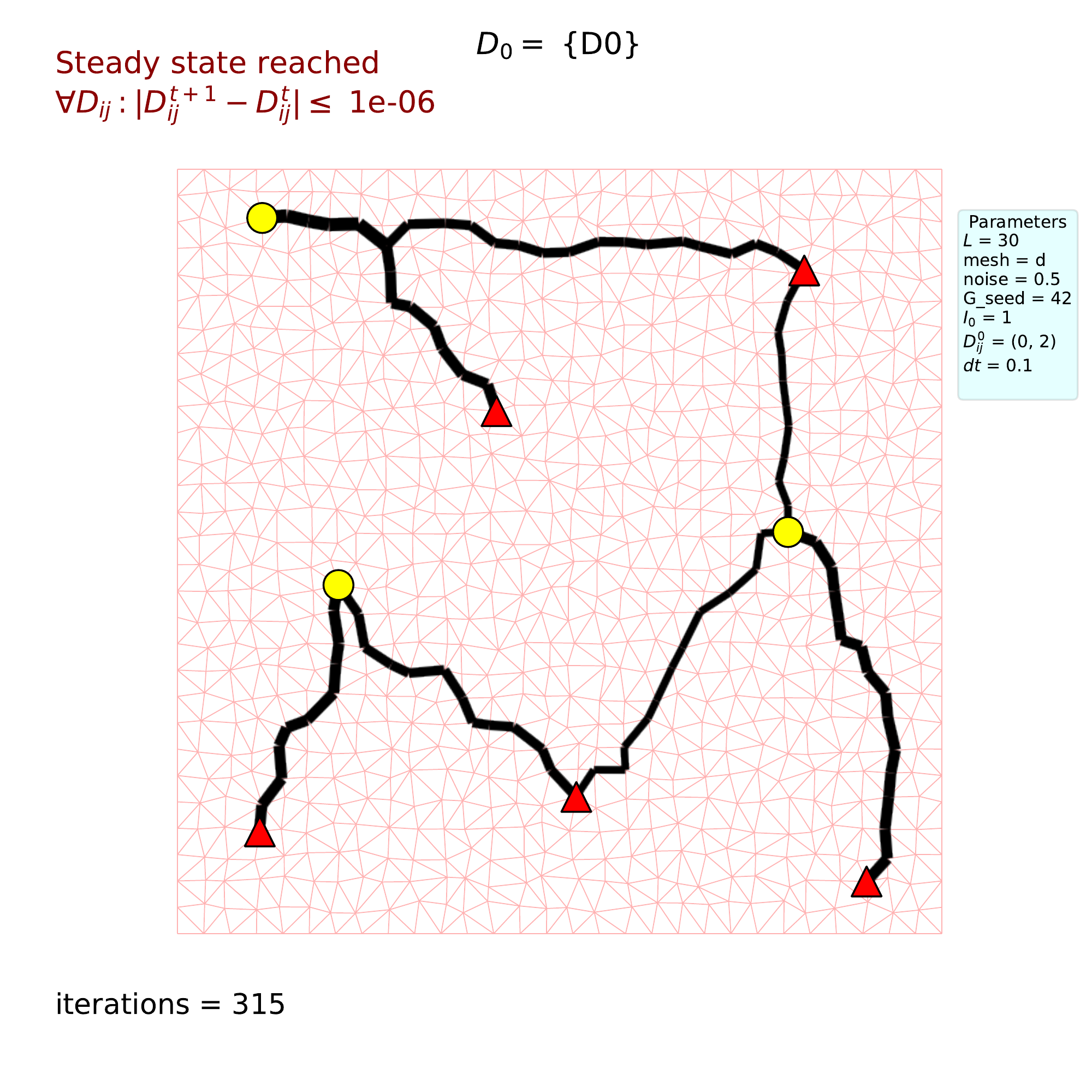} &
    \myfig{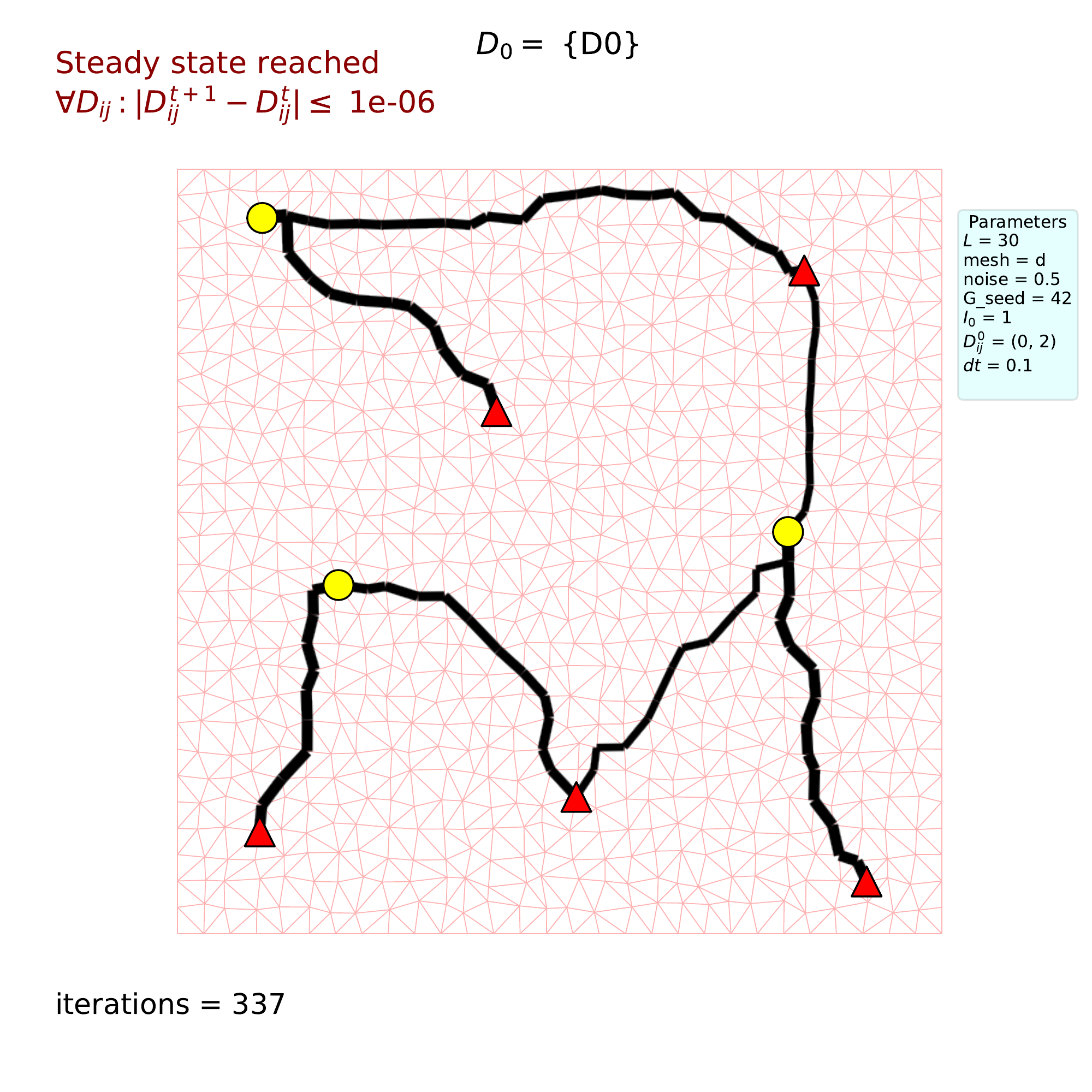} &
    \myfig{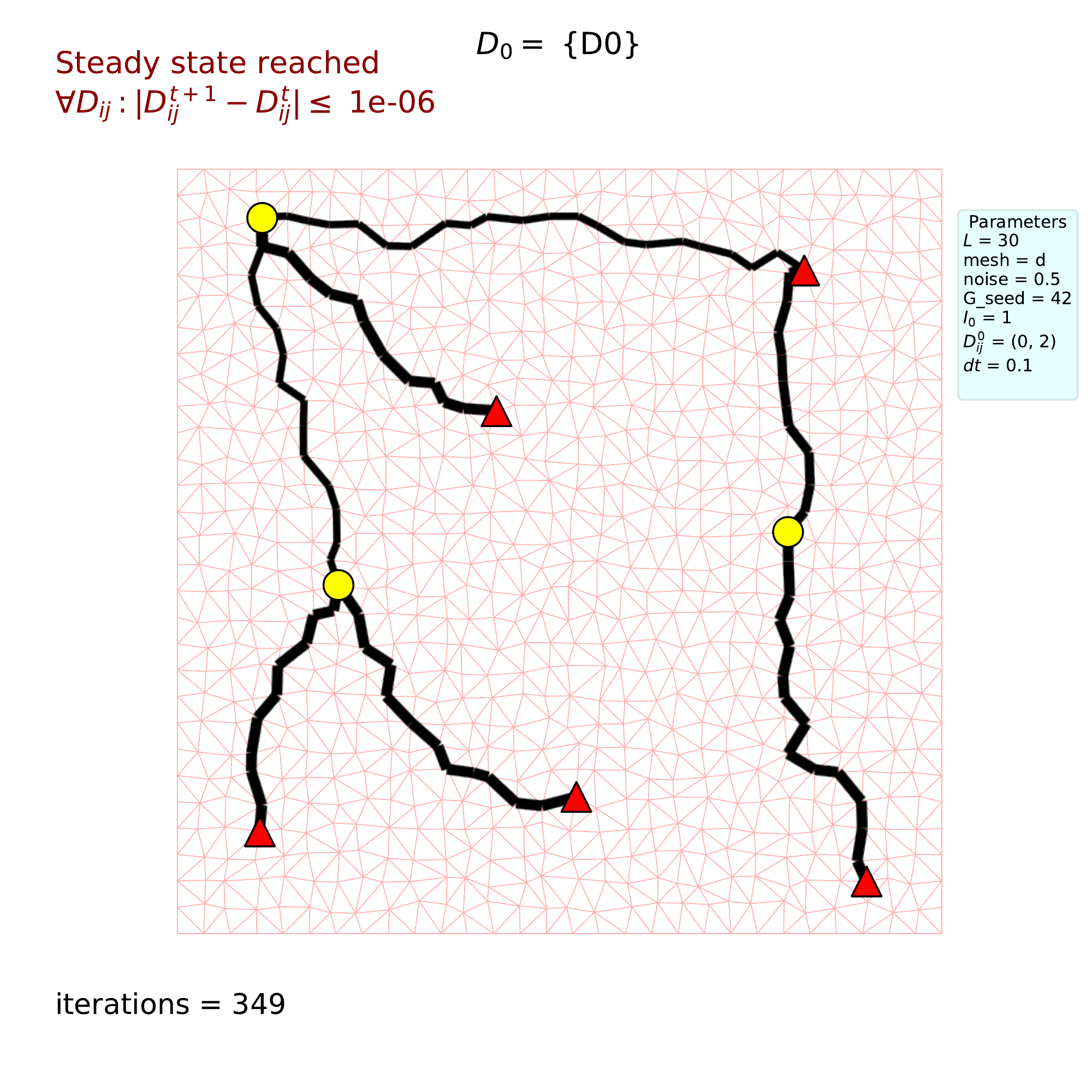} &
    \myfig{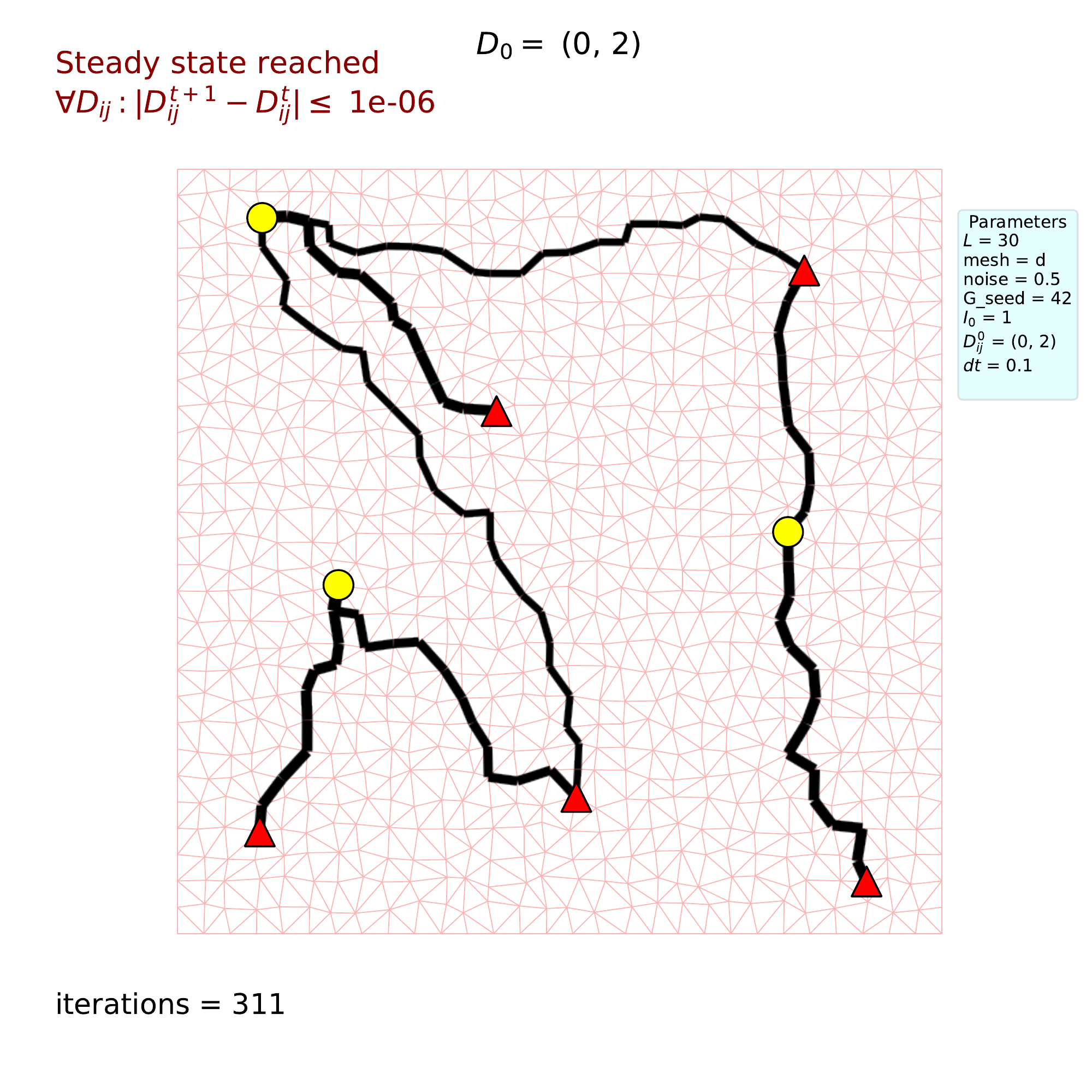}
    \end{tabular}
    \caption{3 sources (yellow) with intensities $q_{source}=1/3$, and 5 sinks (red) with intensities $q_{sink}=-1/5$.}
\end{subfigure}  

\caption{Dependency of the geometry of the optimal network on the initial conductivities of the channels. Each group of simulations, \textbf{(a)} and \textbf{(b)}, contains  examples of the steady-state  networks of \eqref{eq:new_adapt_rule_minE} for a given configuration of terminals and same initial mesh (red lines). Each steady state  was obtained by considering a different set of initial random conductivities uniformly distributed in the interval $[0,2]$. In both cases, the geometry of the final networks is highly sensitive to the initial conditions, which implies that the system may have multiple steady states. The thickness of the black lines is proportional to the radius of the channels.
}
\label{fig:Dij_random} 
\end{figure}

To check if that is the case, several simulations were repeated on the same configuration, but considering different sets of initial conductivities, which were drawn each time from a uniform distribution in the interval [0,2], i.e., $D_{ij}(0) \sim U(0,2)$. Figure \ref{fig:Dij_random} represents the steady states reached for two different configurations: one considering 2 sources and 5 sinks, and the other considering 3 sources and 5 sinks. The results show that for both cases, the geometry of the final network is highly sensitive to the initial conditions, and therefore we conclude that for a given setting the dynamics \eqref{eq:new_adapt_rule_minE} may have multiple steady-state  solutions, as one could have expected. Further simulations suggested that, in general, the higher the number of terminals, the easier it is for the system to converge to a different steady state. 

The multiplicity of the steady states can be explained through the
positive feedback loop between the flux and veins thickness inherent to the model's dynamics \eqref{eq:new_adapt_rule_minE}. Edges that start with higher conductivities in principle will carry more flow, and therefore are thickened at the expense of channels with initial lower conductivities, where the flow rate is likely lower, due to the volume conservation. This idea is repeated in the following time steps, which implies that the initially thinner channels are likely to shrink faster, while initially thicker channels are likely to thrive. Therefore, if the initial discrepancy of conductivities is considerable, some connections are initially favoured, which may affect the ultimate fate of the system.

Another interesting thing to test is how the system behaves under scaling transformations of the initial conductivities. If all the conductivities are scaled by a factor $\alpha>0$, i.e., $D_{ij}\rightarrow D_{ij}^\prime = \alpha D_{ij}$, the channel fluxes are scaled by the same amount ($Q_{ij} \rightarrow Q^\prime_{ij}=\alpha Q_{ij}$) due to the linearity of the \HP{} flow \eqref{eq:HP_flow}, while by \eqref{eq:volume_conservation} the volume of the network scales as $\V\rightarrow \V^\prime = \sqrt{\alpha}V$. Consequently, the adaptation dynamics (\ref{eq:new_adapt_rule_minE}) is transformed as

\begin{nalignat}[3]
&& \frac{d}{d\tau} \sqrt{D_{ij}^\prime} &= 
\frac{\V^\prime}{\beta}\frac{(Q^\prime_{ij})^{2/3}}{\sum\limits_{(k,m)\in E}L_{km} (Q_{km}^\prime)^{2/3}} - \sqrt{D^\prime_{ij}} 
&\quad \iff \\
\iff \quad &&
\bcancel{\sqrt{\alpha}}\frac{d}{d\tau} \sqrt{D_{ij}} &=
\bcancel{\sqrt{\alpha}}\frac{\V}{\beta}\frac{\cancel{\alpha^{2/3}}Q_{ij}^{2/3}}{\sum\limits_{(k,m)\in E}L_{km}\cancel{\alpha^{2/3}}Q_{km}^{2/3}} - \bcancel{\sqrt{\alpha}}\sqrt{D_{ij}}
&\quad \iff \\
\iff \quad &&
\frac{d}{d\tau} \sqrt{D_{ij}} &=
\frac{\V}{\beta}\frac{Q_{ij}^{2/3}}{\sum\limits_{(k,m)\in E}L_{km} Q_{km}^{2/3}} - \sqrt{D_{ij}} \;,
\end{nalignat}

\noindent implying that it remains invariant under the transformation  $D_{ij}\rightarrow \alpha D_{ij}$ with $\alpha > 0$. This means that if $\vb{D}(\tau)$ is solution of \eqref{eq:new_adapt_rule_minE}, so is 
$\vb{D}^\prime(\tau)\equiv \alpha \vb{D}(\tau)$. As the solution of \eqref{eq:new_adapt_rule_minE} is unique given an initial condition, this further implies that the only effect on the system of scaling the initial conductivities by a global factor ($\vb*{D}(0)\rightarrow \alpha \vb*{D}(0)$) is that the conductivities of the steady state  are also scaled by the same factor, but the geometry of the network shouldn't change ($\lim\limits_{\tau\rightarrow \infty}\vb*{D}(\tau)\rightarrow \alpha \lim\limits_{\tau\rightarrow \infty}\vb*{D}(\tau)$). Note that these observations can be generalised for the generic adaptation dynamics \eqref{eq:new_adapt_rule}, as long as the function $g$ satisfies $g(|\alpha Q_{ij}|)=\delta g(|Q_{ij}|)$ for some constants $\alpha>0$ and $\delta>0$, which is the case of \eqref{eq:new_adapt_rule_minE}, as $|\alpha Q_{ij}|^{2/3}=\alpha^{2/3}|Q_{ij}|^{2/3}$.

\begin{figure}[hbt!] 

\newcommand{\mysub}[2][]{%
    \subfloat[#1]{\includegraphics[trim={2.9cm 3cm 3.5cm 3.1cm}, clip, width=0.249\textwidth]{#2}}%
}

\newcommand{\myfig}[1]{%
    \includegraphics[trim={2.9cm 3cm 3.5cm 3.1cm}, clip, width=0.25\textwidth]{#1}%
}

\begin{subfigure}{\textwidth}    
    \centering
    \begin{tabular}{@{}c@{}c@{}c@{}c@{}}
    $D_{ij}(0)=0.1$ & $D_{ij}(0)=1$ & $D_{ij}(0)=5$ & $D_{ij}(0)=15$ \\[0.1cm]
    \myfig{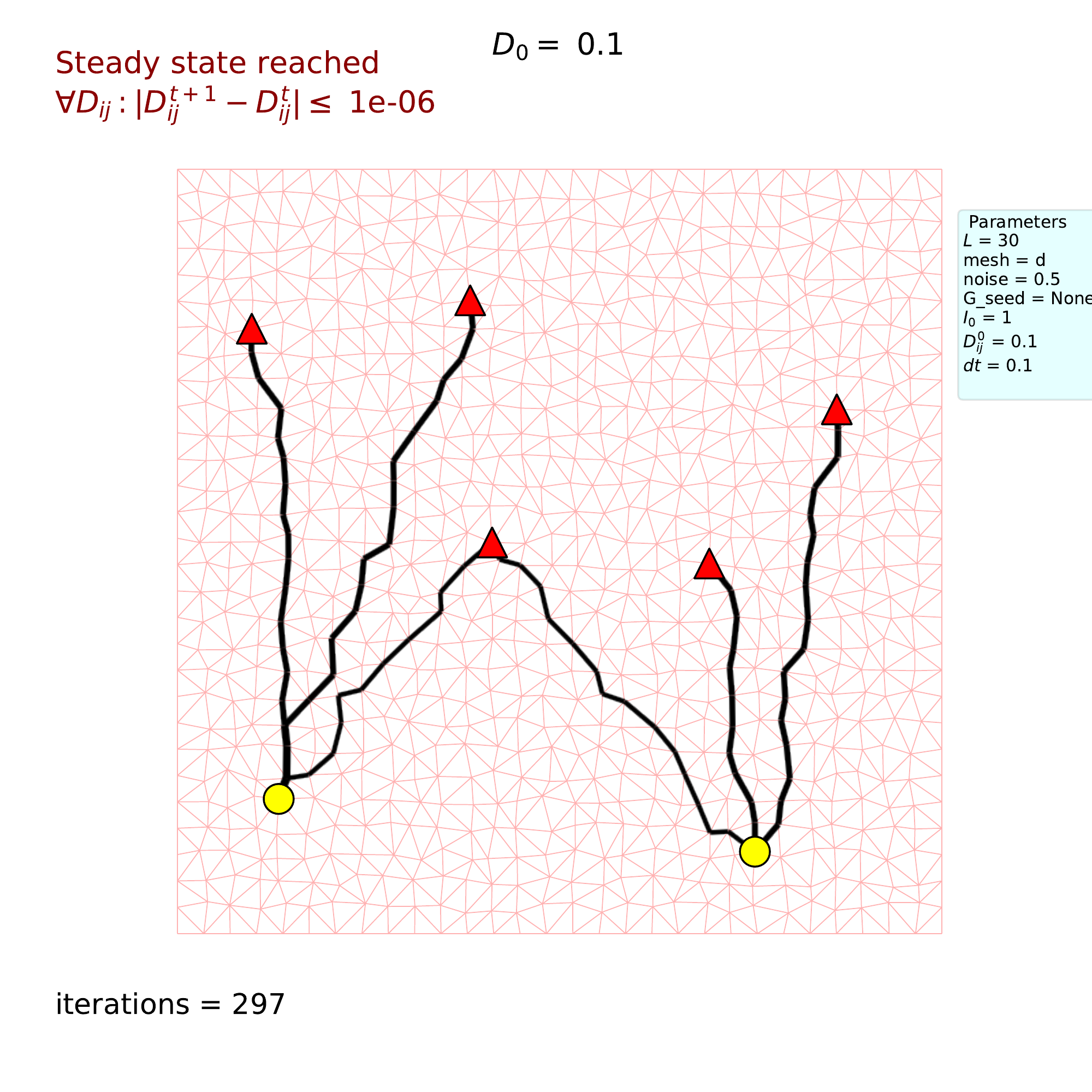} &
    \myfig{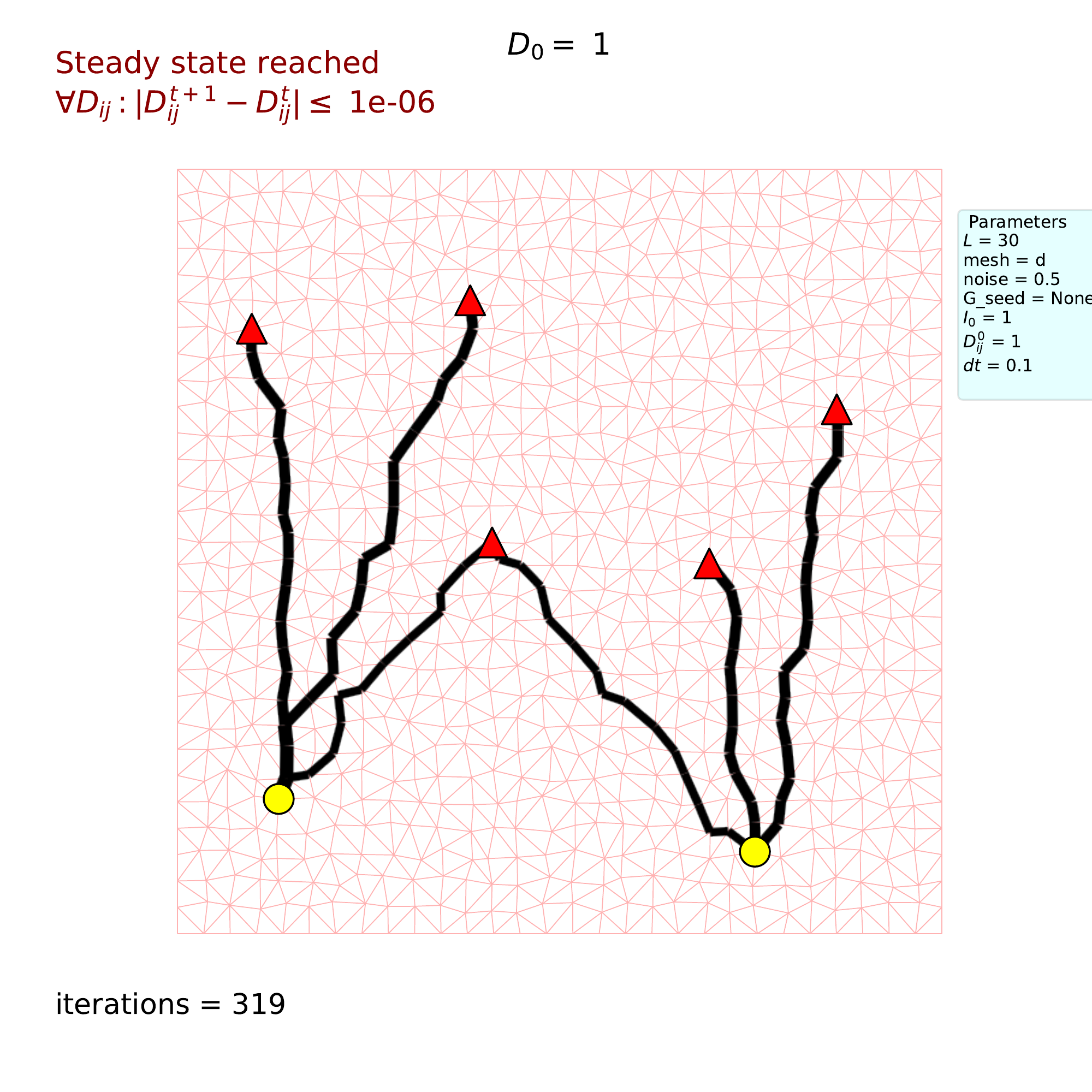} & 
    \myfig{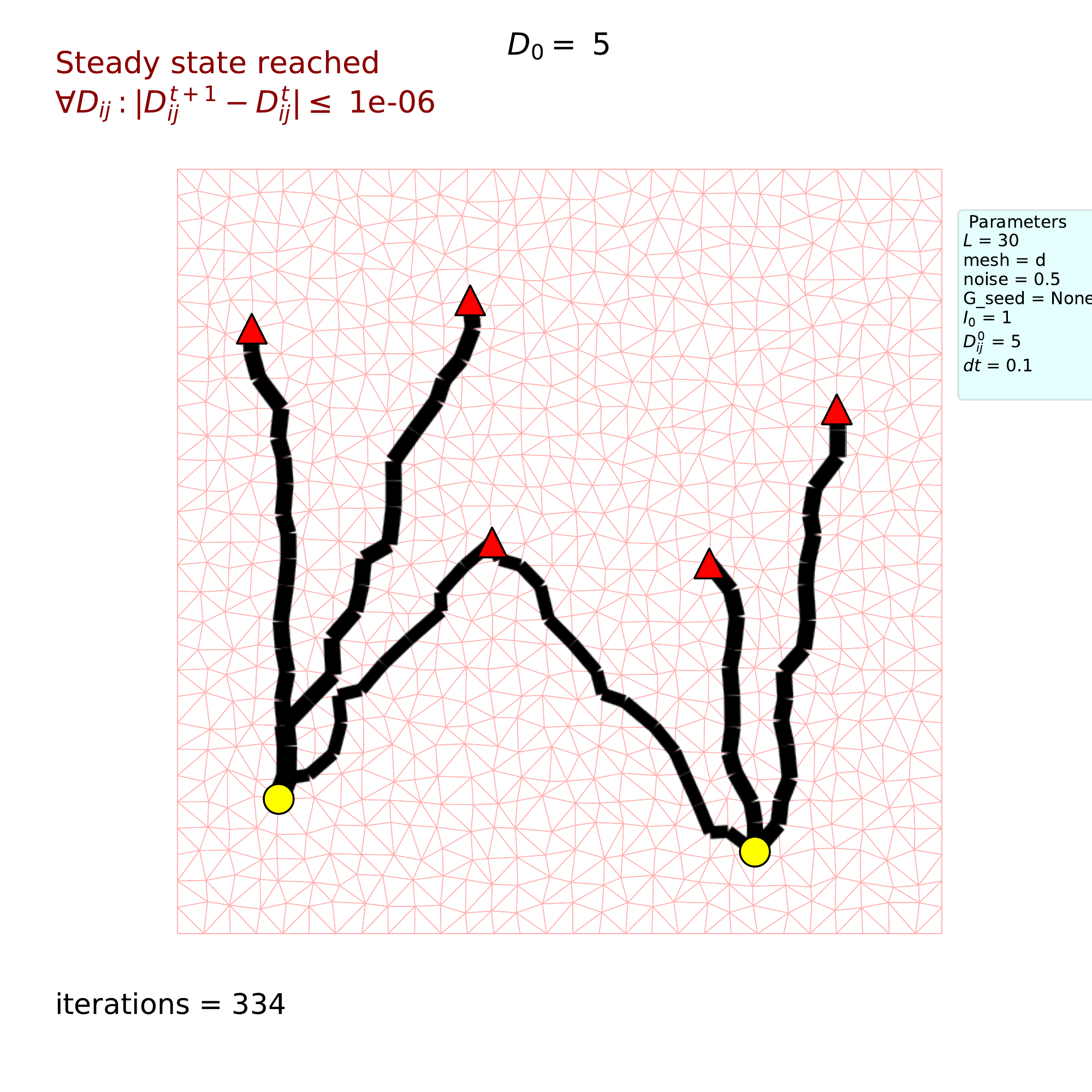} & 
    \myfig{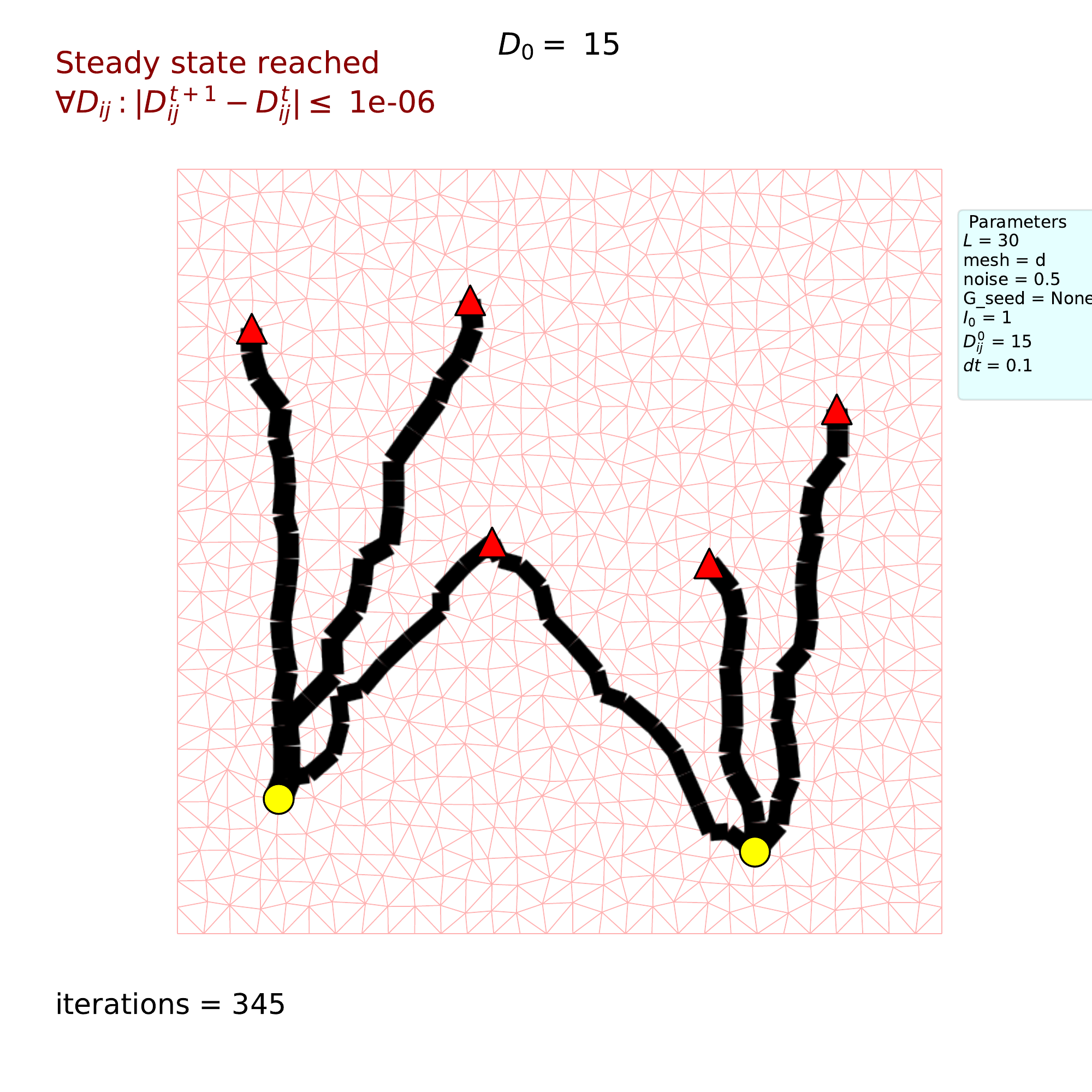}
    \end{tabular}
    \caption{2 sources (yellow) with intensities $q_{source}=1/2$, and 5 sinks (red) with intensities $q_{sink}=-1/5$.}
\end{subfigure}  
\\[0.2cm]
\begin{subfigure}{\textwidth}    
    \centering
    \begin{tabular}{@{}c@{}c@{}c@{}c@{}}
    \myfig{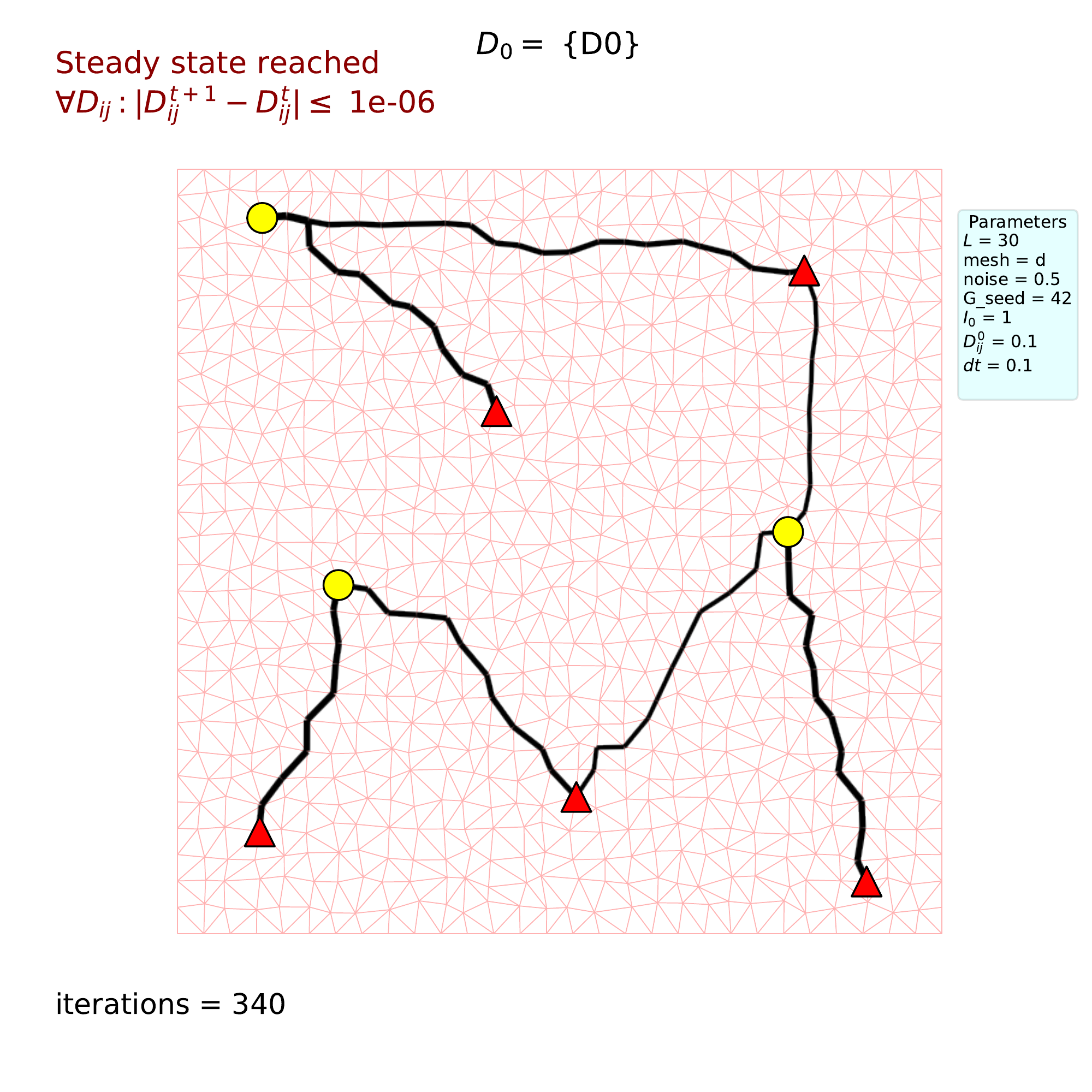} &
    \myfig{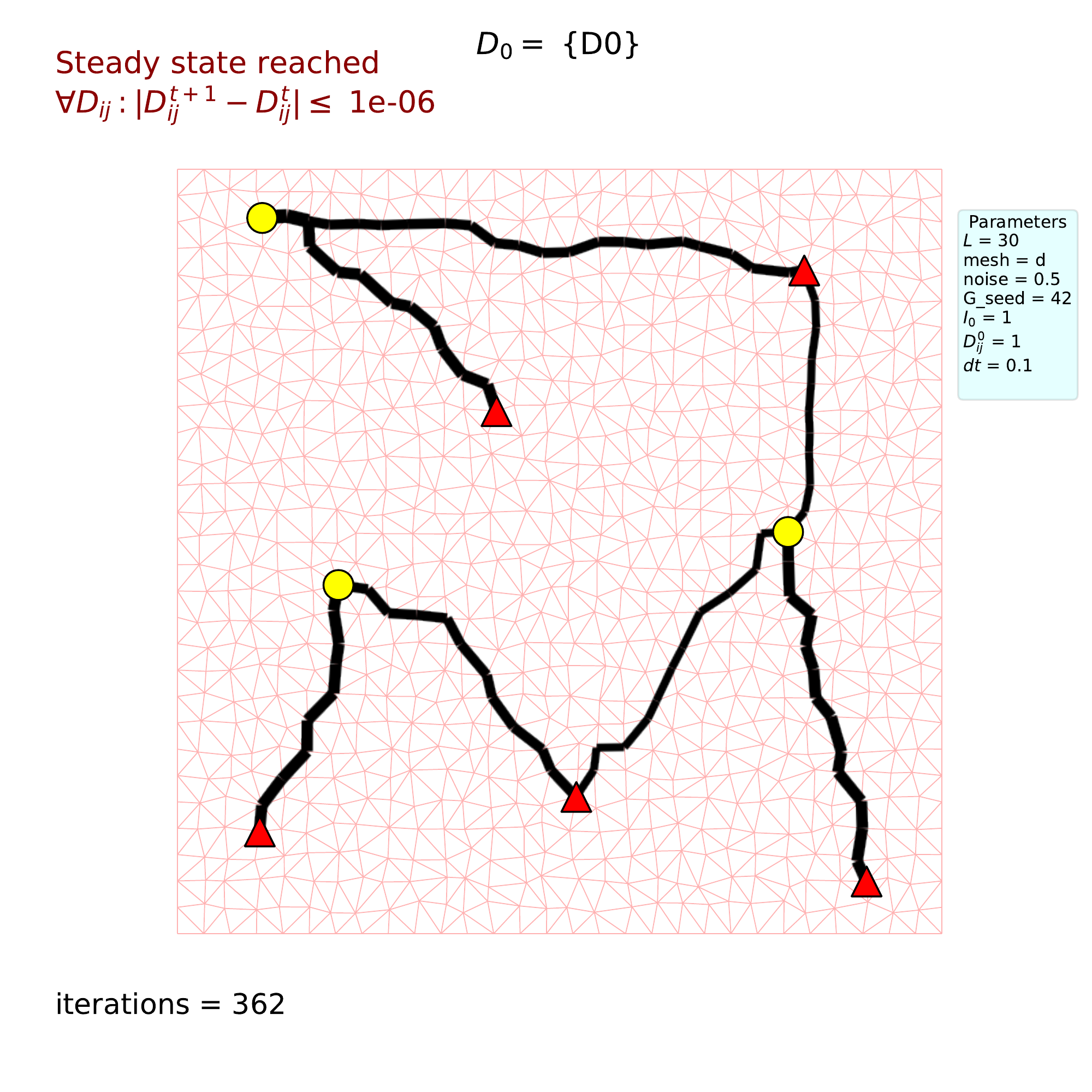} & 
    \myfig{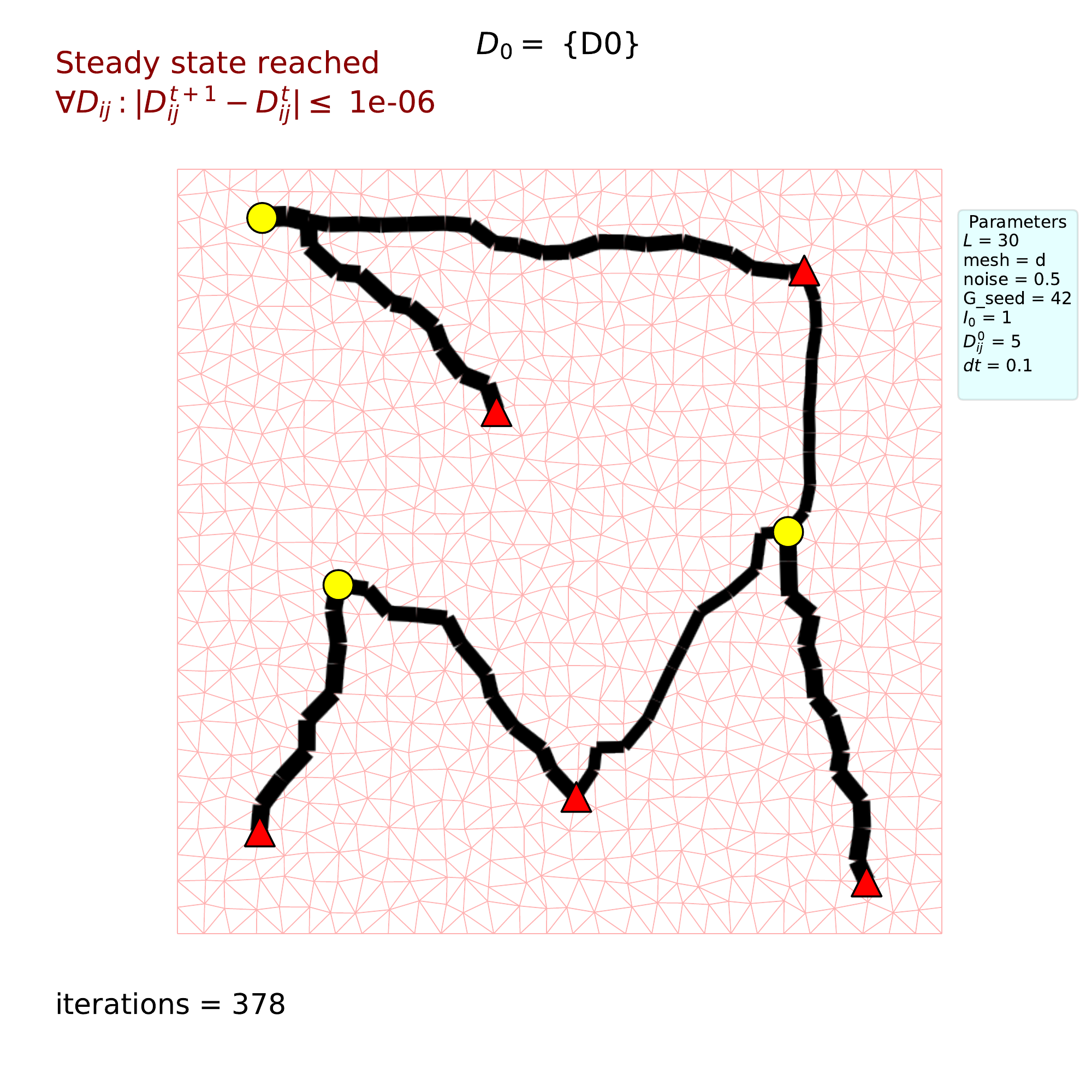} & 
    \myfig{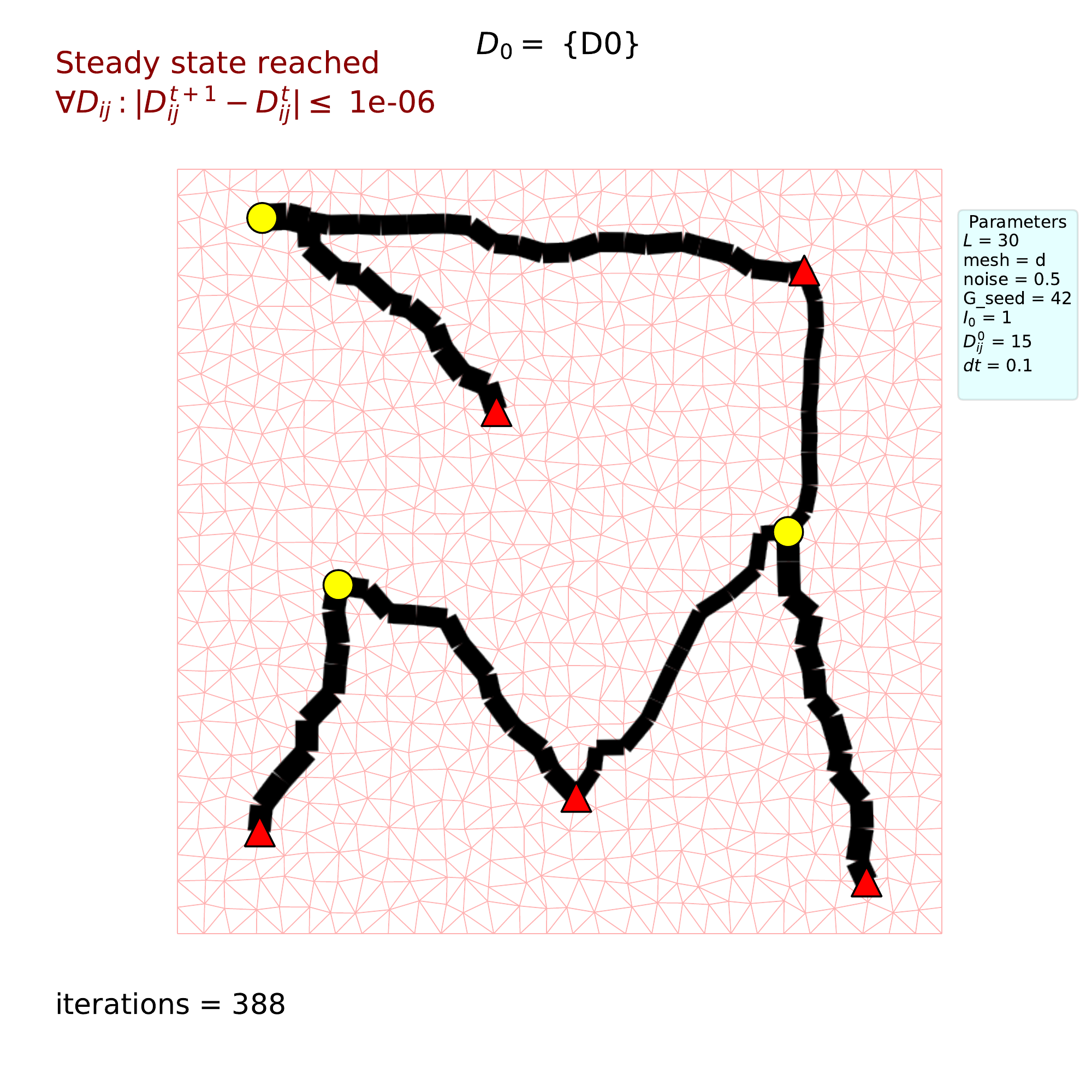}
    \end{tabular}
    \caption{3 sources (yellow) with intensities $q_{source}=1/3$, and 5 sinks (red) with intensities $q_{sink}=-1/5$.}
\end{subfigure}  

\caption{Dependency of the optimal network geometry on the scaling of the initial conductivities. The images correspond to the steady states of \eqref{eq:new_adapt_rule_minE} obtained for a different set of  homogeneous initial conductivities,  $D_{ij}(0)=D_0$, with $D_0 \in \{0.1,1,5,15\}$, considering the same configurations of Figure \ref{fig:Dij_random}. A scaling of the initial conductivities is reflected in a scaling of the steady-state conductivities by the same factor.
In each group of simulations, only the thickness of the channels ($\sim D_0^{1/4}$) is increased as $D_0$ increases, but the geometry of the network is left unchanged.   
}
\label{fig:Dij_initial_scale} 
\end{figure}

The scaling of the initial conditions was tested by performing repeated simulations on the same configuration, starting from different sets of initial homogeneous conductivities, $D_{ij}(0)=D_0$, where $D_0>0$ is 
the only parameter which was varied between runs. In Figure \ref{fig:Dij_initial_scale} are depicted the steady states obtained for different values of $D_0$, considering the same configurations of Figure \ref{fig:Dij_random}. The results confirm the observations, since for both cases, only the thickness of the channels of the steady-state networks changes when the initial conductivities are scaled by a given global factor, but the networks' geometry remains the same.

\FloatBarrier

\subsection{Phase Transition}

The choice of $g(|Q_{ij}|):= |Q_{ij}|^{2/3}$ in \eqref{eq:new_adapt_rule} used until now led us to study the more general class of polynomial functions, $g_\gamma(|Q_{ij}|):= |Q_{ij}|^\gamma$, where $\gamma > 0$ is a new parameter. In the following, we analyse the evolution of the system subject to the adaptation dynamics 

\begin{equation}
\frac{d}{d\tau} \sqrt{D_{ij}} =
\V
\frac{|Q_{ij}|^\gamma}{\sum\limits_{(k,m)\in E}L_{km} |Q_{km}|^\gamma} - \sqrt{D_{ij}}
\label{eq:new_adapt_rule_gamma}
\end{equation}

\noindent as a function of the parameter $\gamma$, assuming $\beta=1$ once more. Note that the steady state  of \eqref{eq:new_adapt_rule_gamma} for $\gamma=0$ is always  homogeneous, $\sqrt{D_{ij}^*} = \V / \sum_E L_{km}$, regardless of the initial conductivities, as $g_0(|Q_{ij}|) = 1$. Furthermore, if the initial conductivities are also homogeneous, $D_{ij}(0)=D_0$, the steady state  corresponds to the initial network, since $\V=D_0\sum_E L_{km}$, and therefore, $\sqrt{D_{ij}^*}=D_0=D_{ij}(t=0)$.

\begin{figure}[hbt!] 

\newcommand{\mysub}[2][]{%
    \subfloat[#1]{\includegraphics[trim={2.8cm 2.8cm 2.6cm 2.8cm}, clip, width=0.331\textwidth]{#2}}%
}

\centering

\mysub[$\gamma = 0.0$]{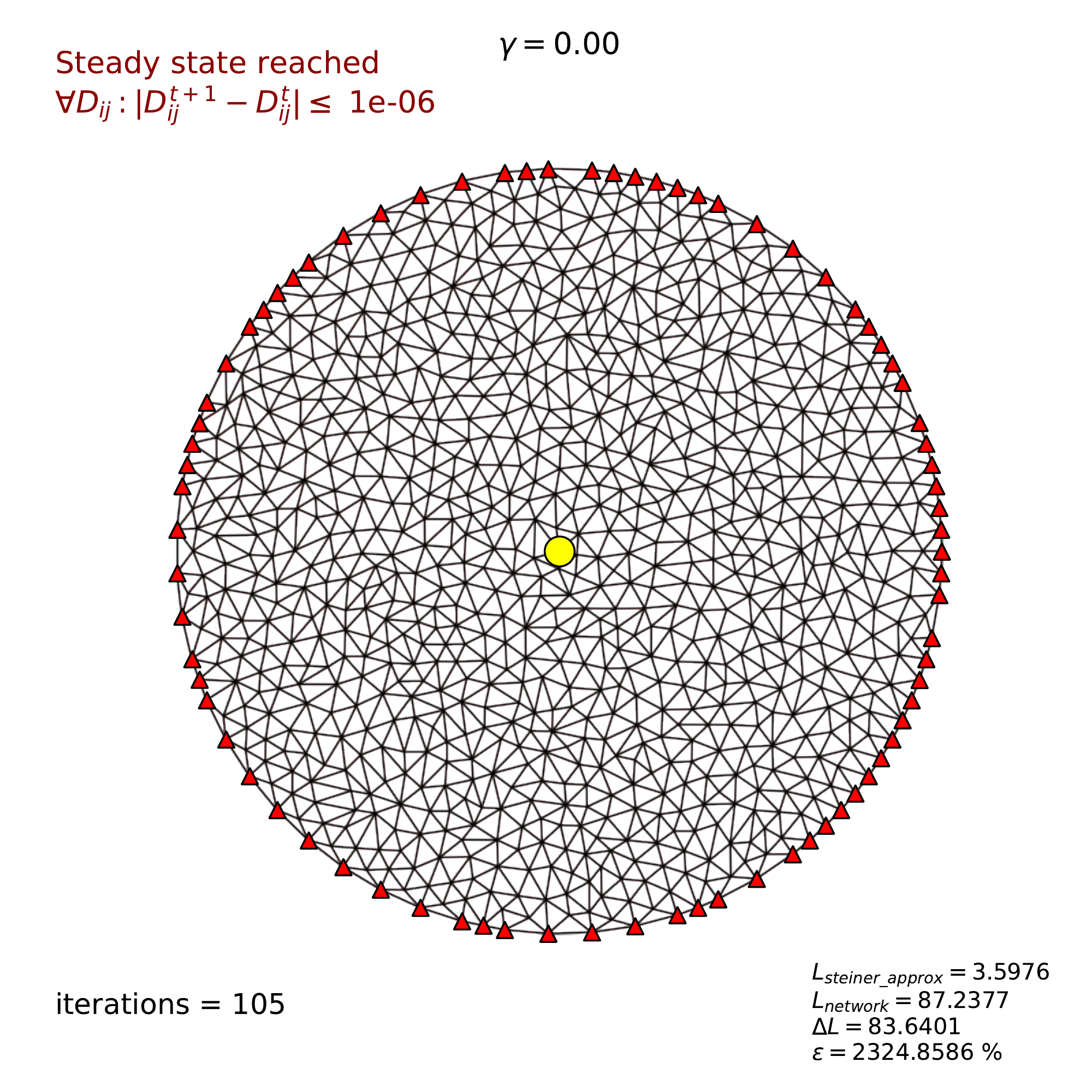} \hfill
\mysub[$\gamma = 0.39$]{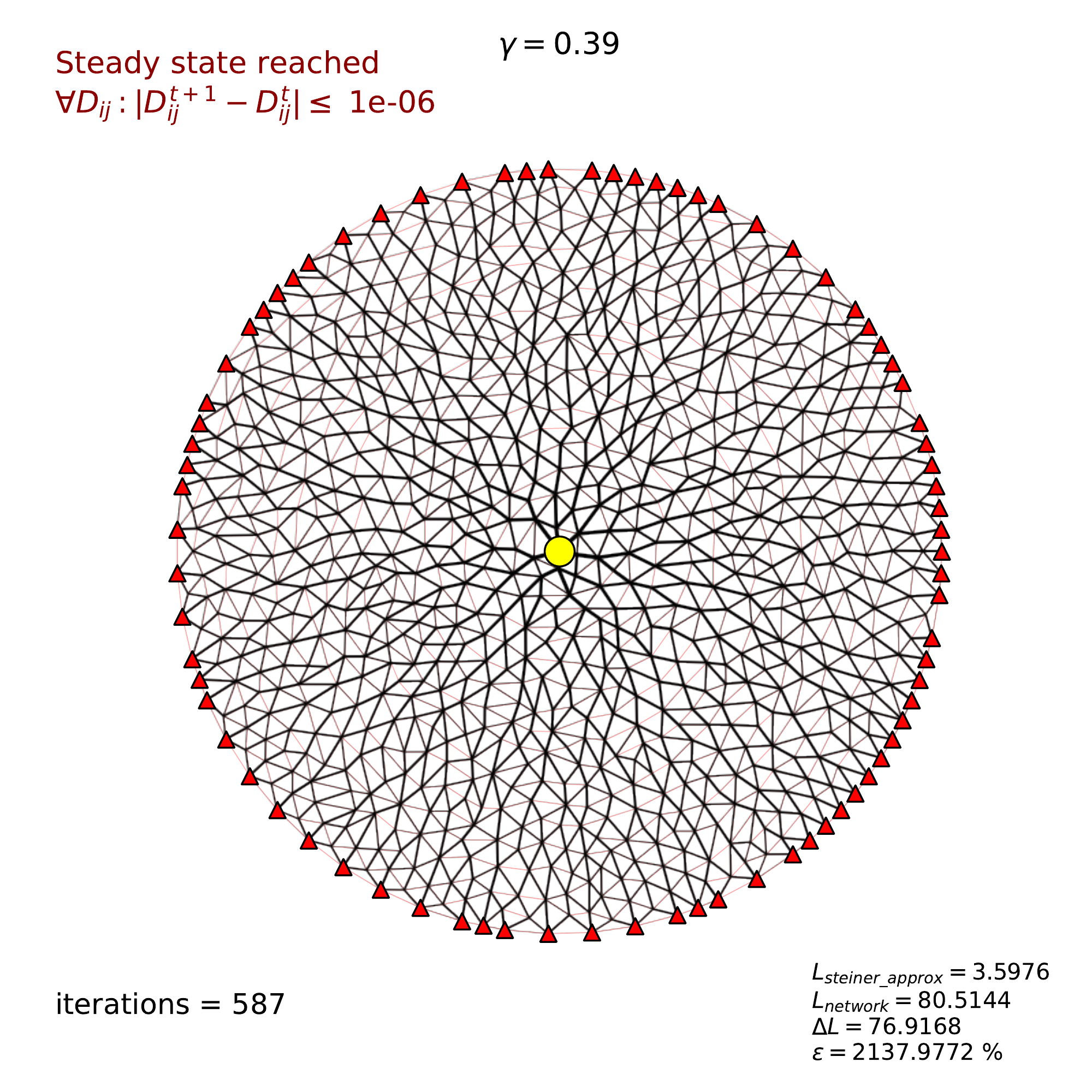} \hfill
\mysub[$\gamma = 0.47$]{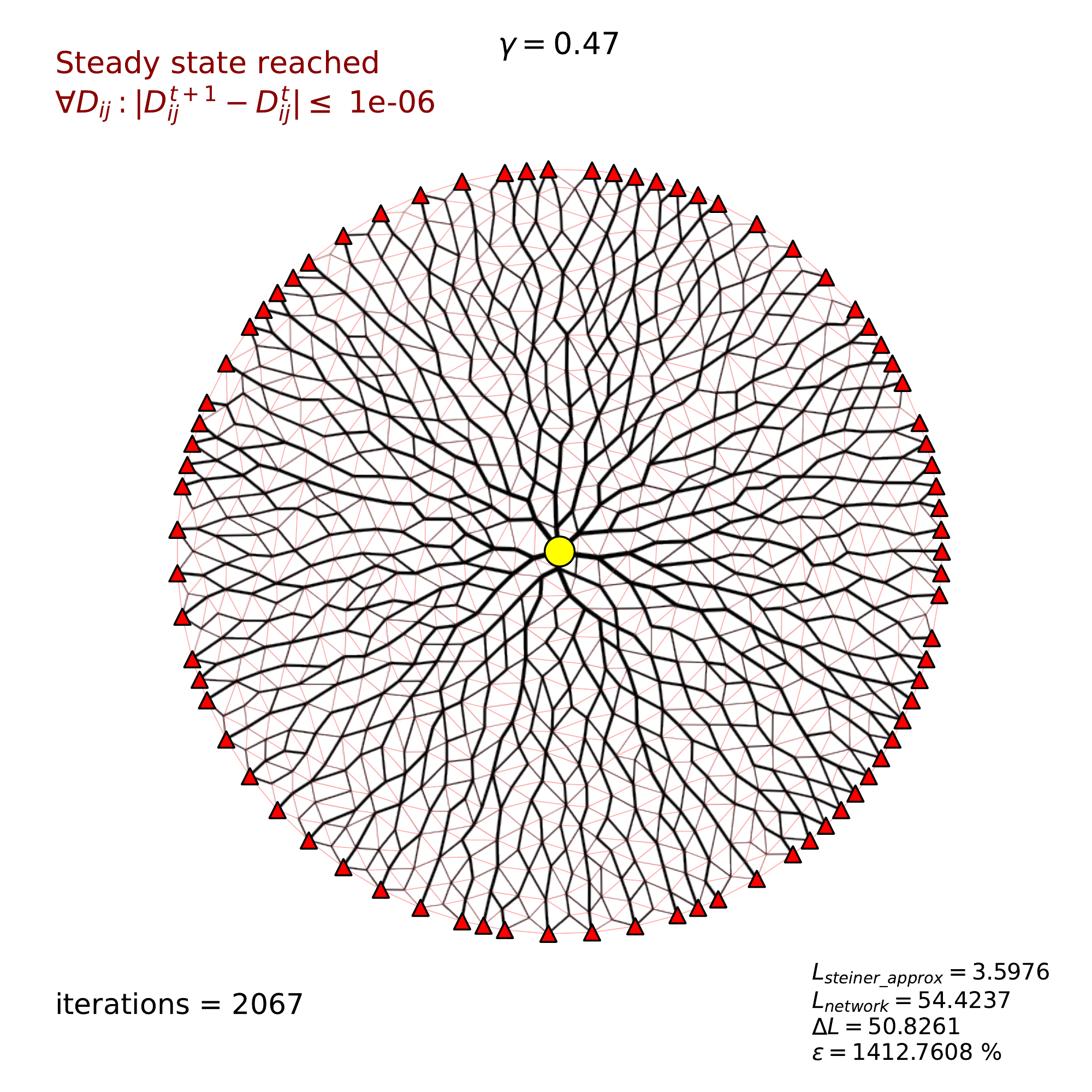} \\

\mysub[$\gamma = 0.51$]{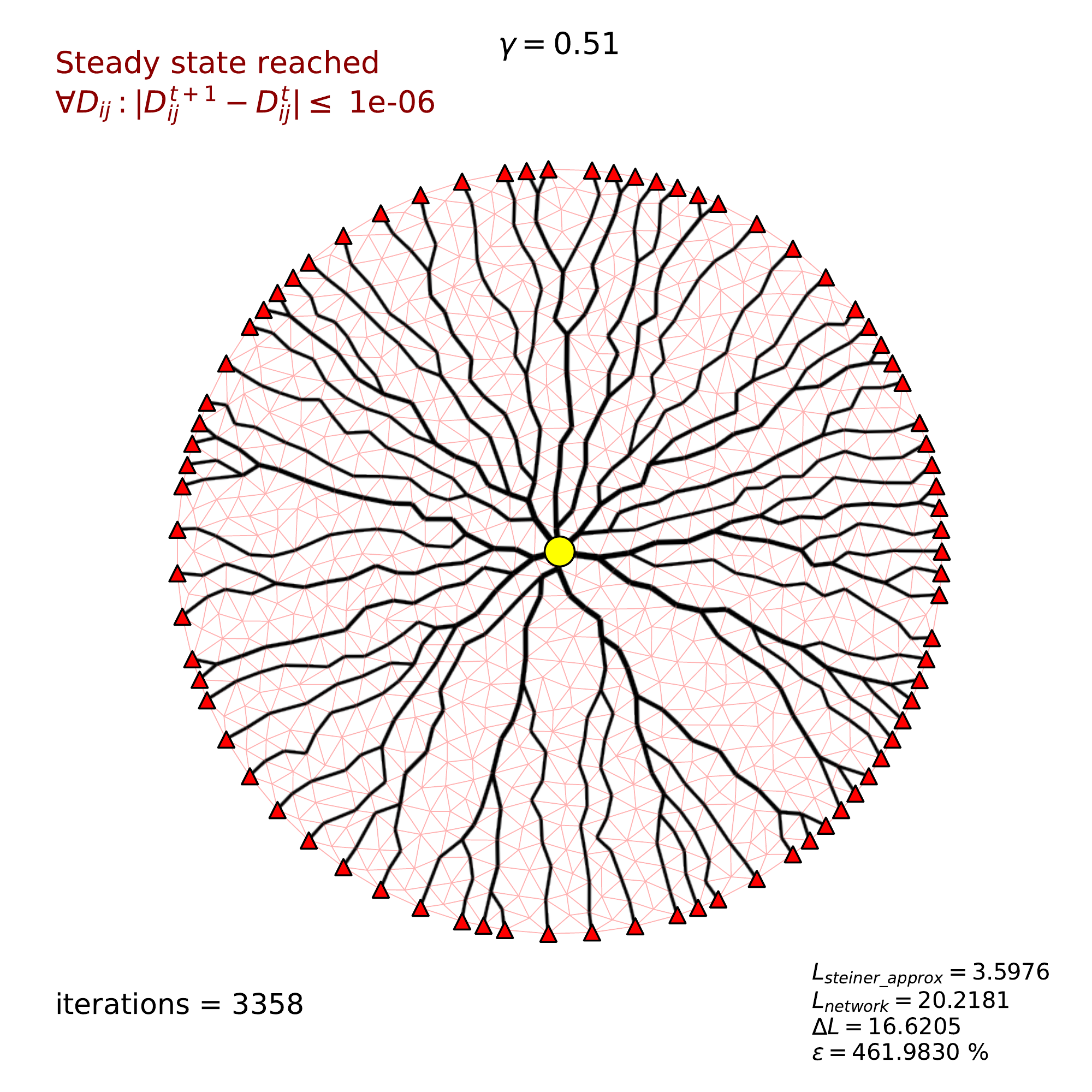} \hfill
\mysub[$\gamma = 0.59$]{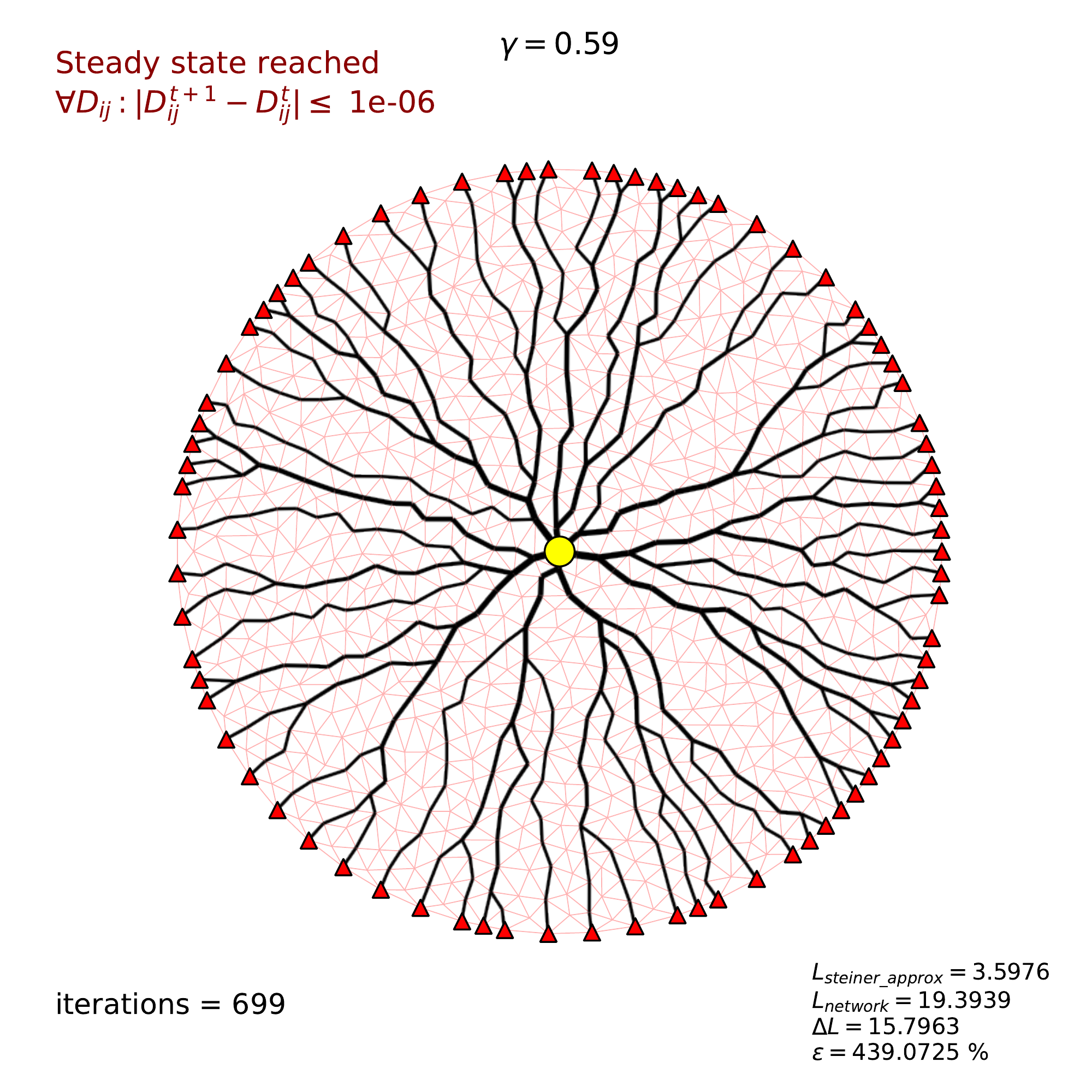} \hfill
\mysub[$\gamma = 0.67$]{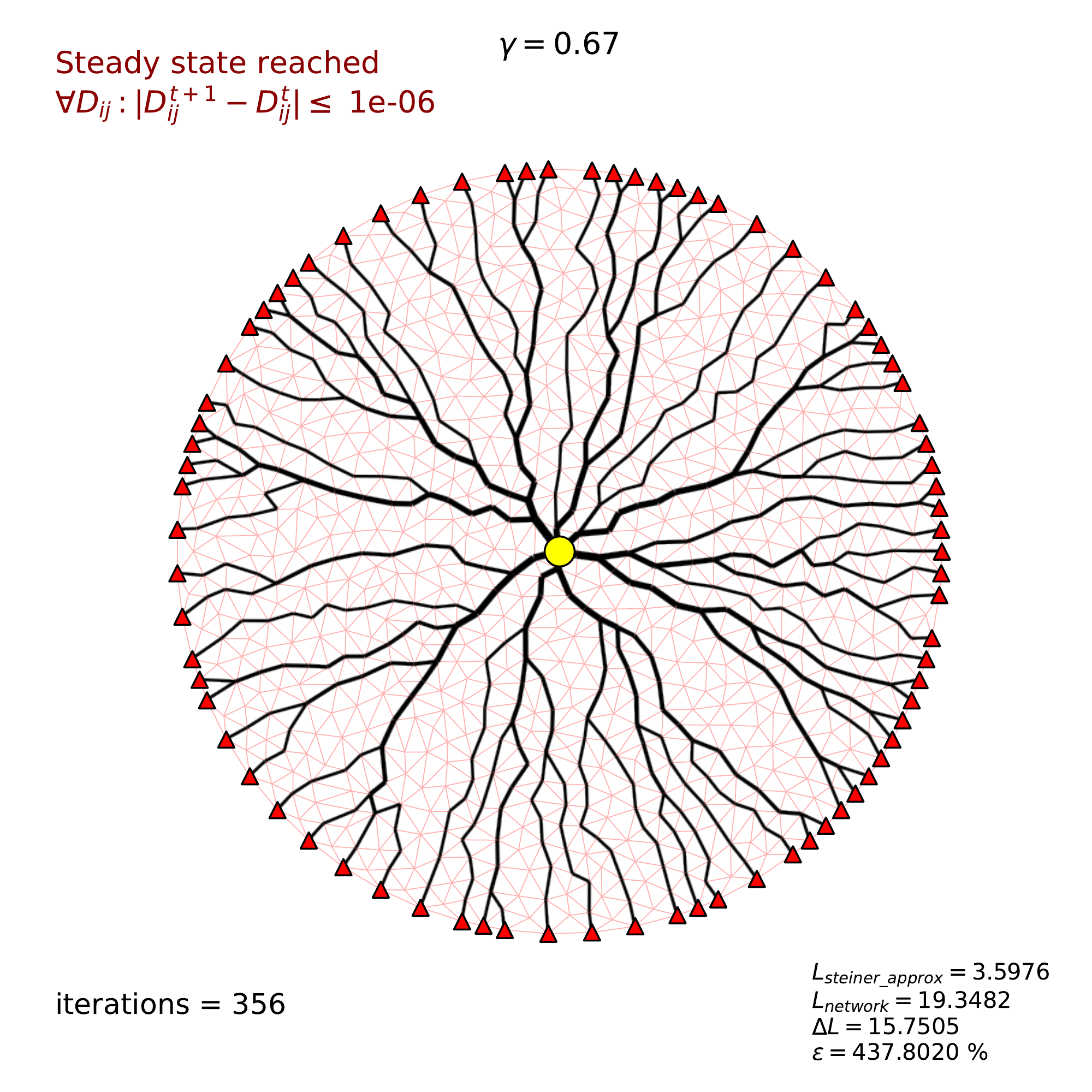}  \\

\mysub[$\gamma = 0.78$]{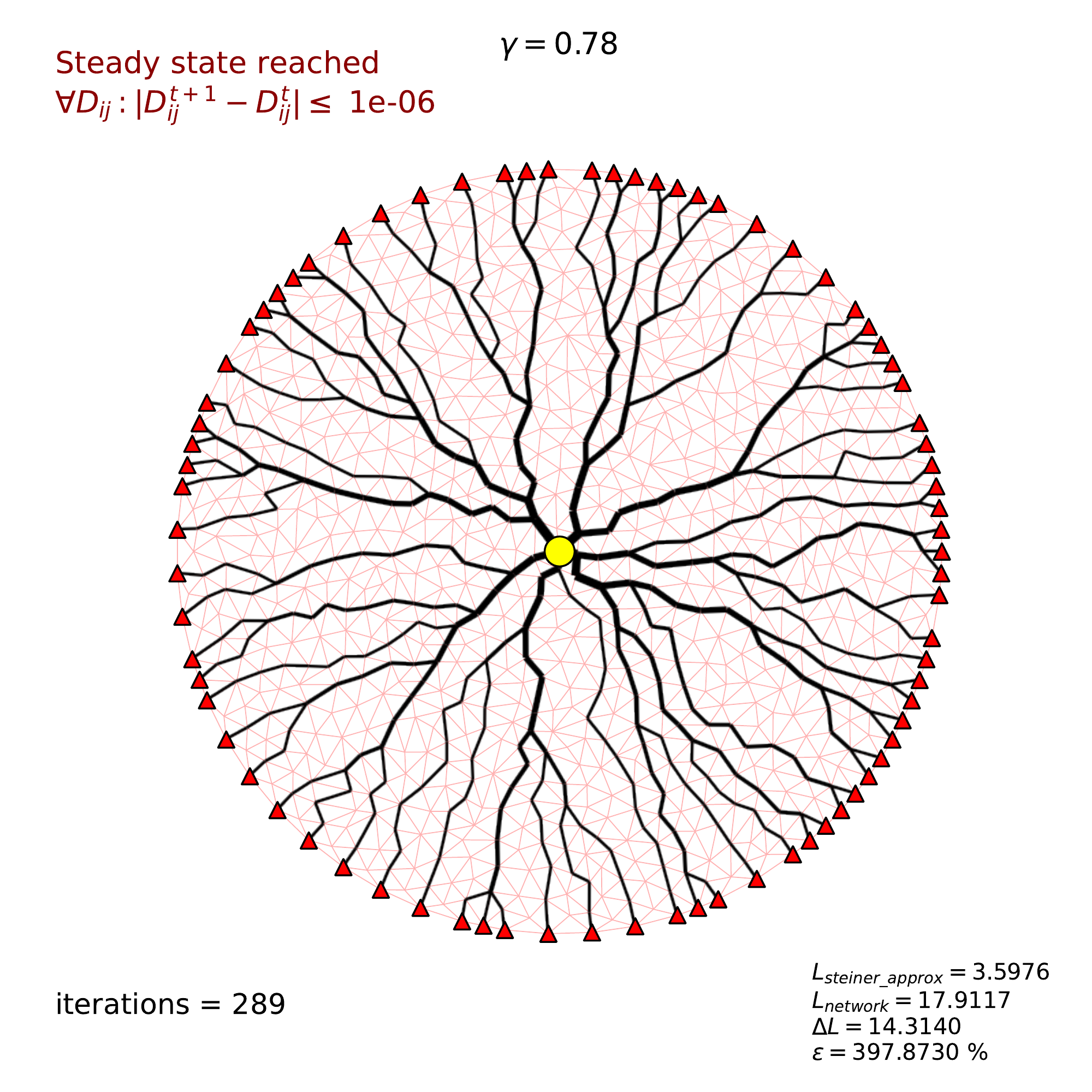} \hfill
\mysub[$\gamma = 0.98$]{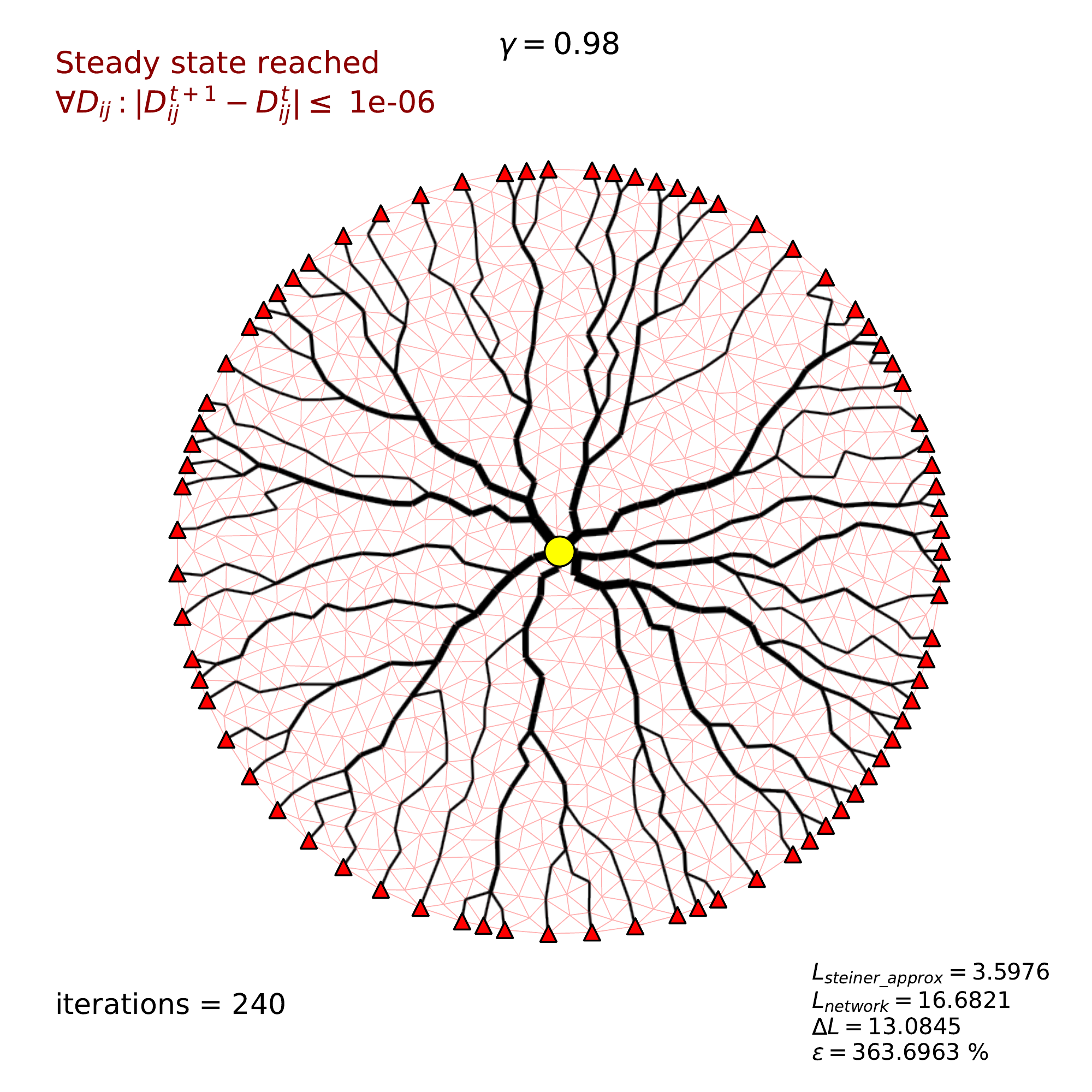} \hfill
\mysub[$\gamma = 1.18$]{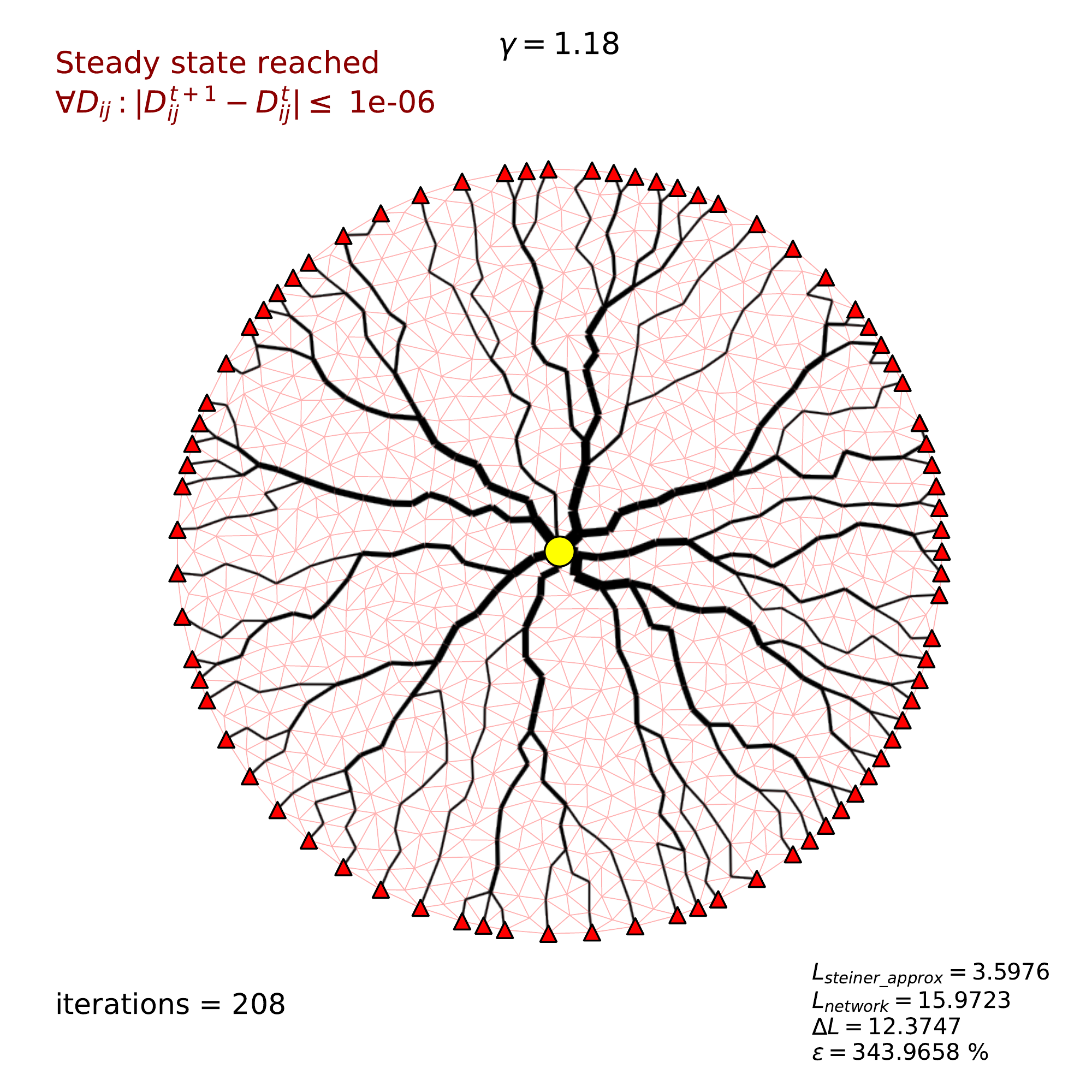}  \\

\mysub[$\gamma = 1.41$]{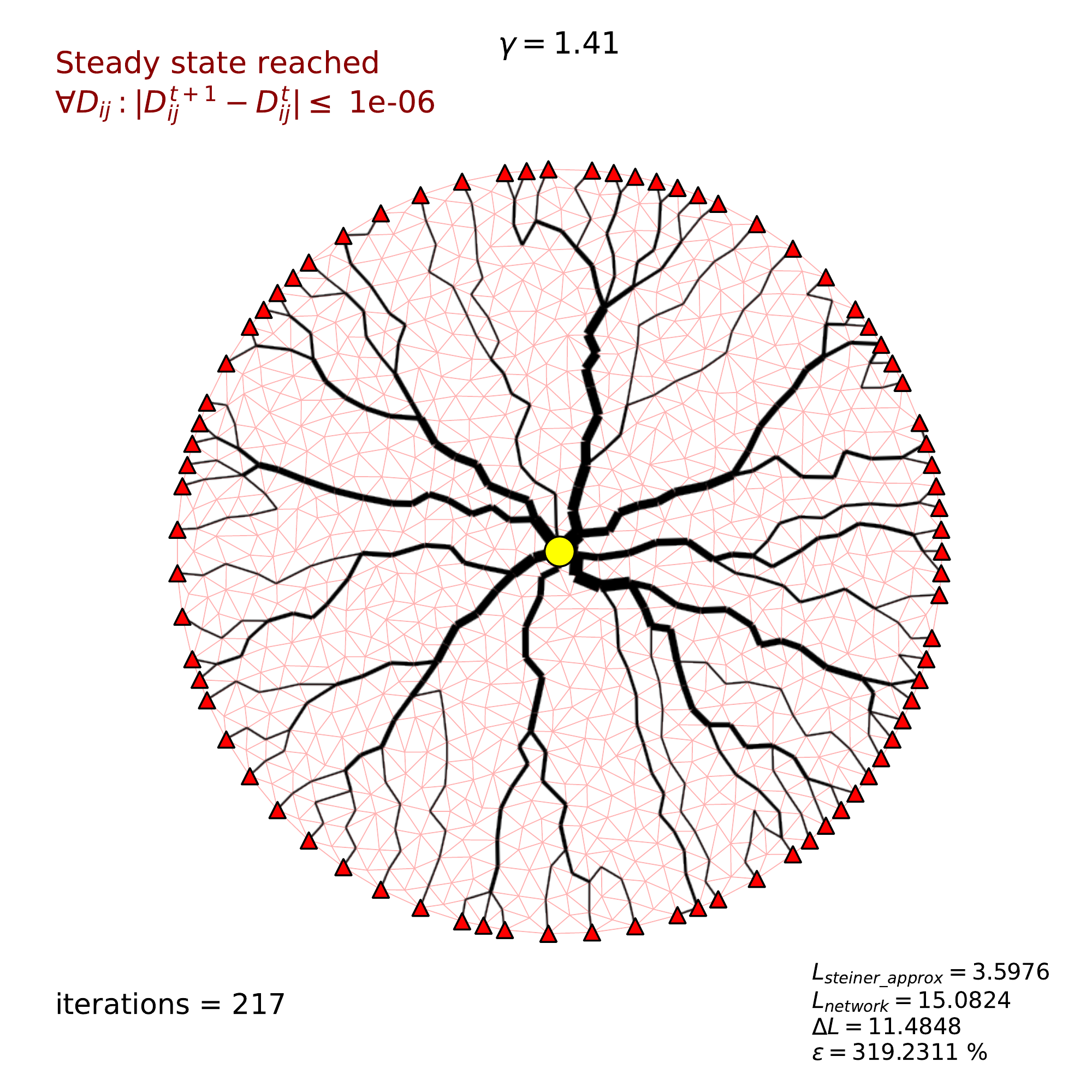} \hfill
\mysub[$\gamma = 1.76$]{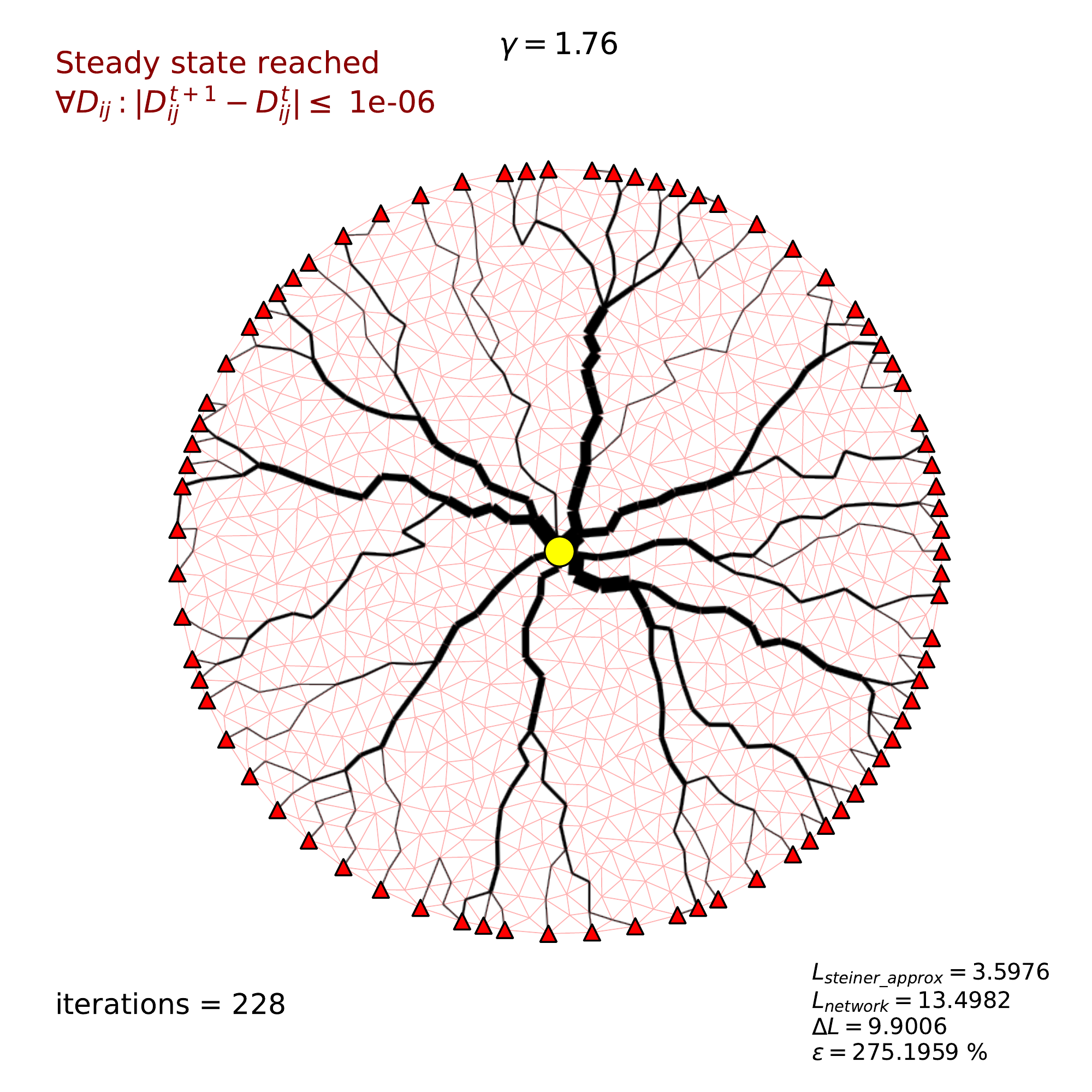} \hfill
\mysub[$\gamma = 2$]{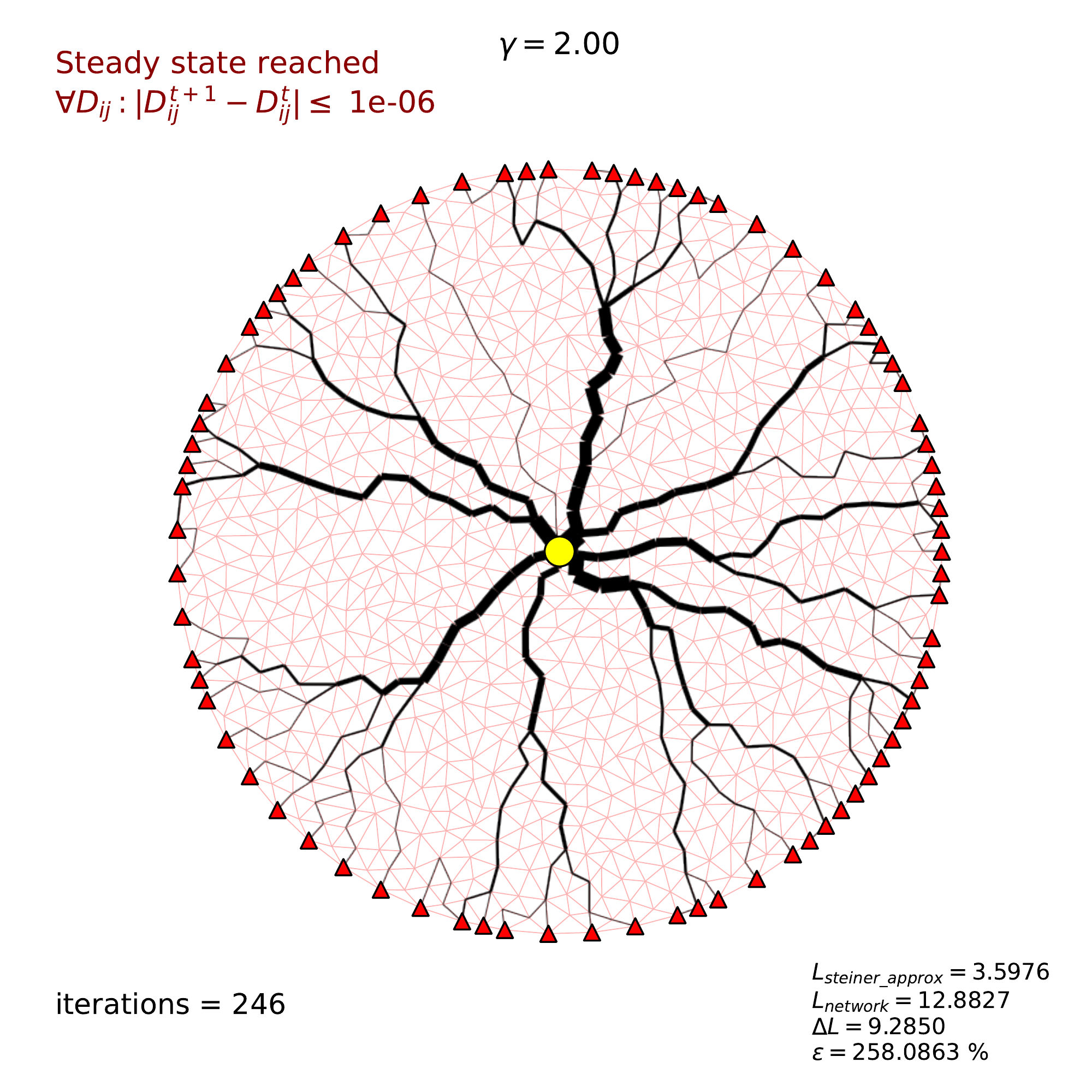}  \\

\caption{Dependency of the steady-state  networks of the dynamics \eqref{eq:new_adapt_rule_gamma} on $\gamma$, for the same configuration of Figure \ref{fig:phy_simulation_network}. Note the transition in the network topology from $\gamma=0.47$ to $\gamma=0.51$. For $\gamma < 1/2$ the networks have a redundant structure with loops and no hierarchical organisation, while for $\gamma > 1/2$ the networks have a tree-like structure with a clear hierarchy of tubes.}
\label{fig:phy_gamma_network} 
\end{figure}

Considering the same \textit{Physarum} scenario of the simulation in Figure \ref{fig:phy_simulation_network}, we have simulated the dynamics \eqref{eq:new_adapt_rule_gamma} for different values of $\gamma\in [0,2]$ starting from random initial conductivities, $D_{ij}(0)\sim U(1/2,3/2)$. Some examples of the steady-state networks reached by the system for a given $\gamma$ are shown in Figure \ref{fig:phy_gamma_network}.

The drastic change in the topology of the network suggests the existence of a phase transition in the system near $\gamma = 1/2$. For $\gamma < 1/2$, the steady-state  networks have a very redundant structure, and are characterised by having a high density of loops and many veins with similar radius. By contrast, for $\gamma > 1/2$, no loops remain as the conductivities of most vessels converge to zero. The steady states are trees spanned by the central source and the boundary sinks, with a clear hierarchy of vein thickness. As $\gamma$ increases, fewer primary veins remain, and it's more evident the hierarchy of the veins, with thinner veins as we move away from the central source. However, once more  we haven't found steady-state  networks with a reticulated hierarchical structure  for any value of $\gamma$, as observed in the real networks produced by \Phy{} (Figure \ref{fig:phy_network}), or in leaf venation networks. This is related to the choice of fixed sources and sinks previously discussed \cite{Corson2010}. 

\begin{figure}[hbt!] 

\captionsetup[subfloat]{position=top,skip=0.5pt}
\newcommand{\mysub}[2][]{%
    \subfloat[#1]{\includegraphics[trim={0cm 0cm 0cm 2.5cm}, clip, width=0.47\textwidth]{#2}}%
}
\centering

\mysub[\label{fig:D_ijhist_g0.1} $\gamma = 0.1$]{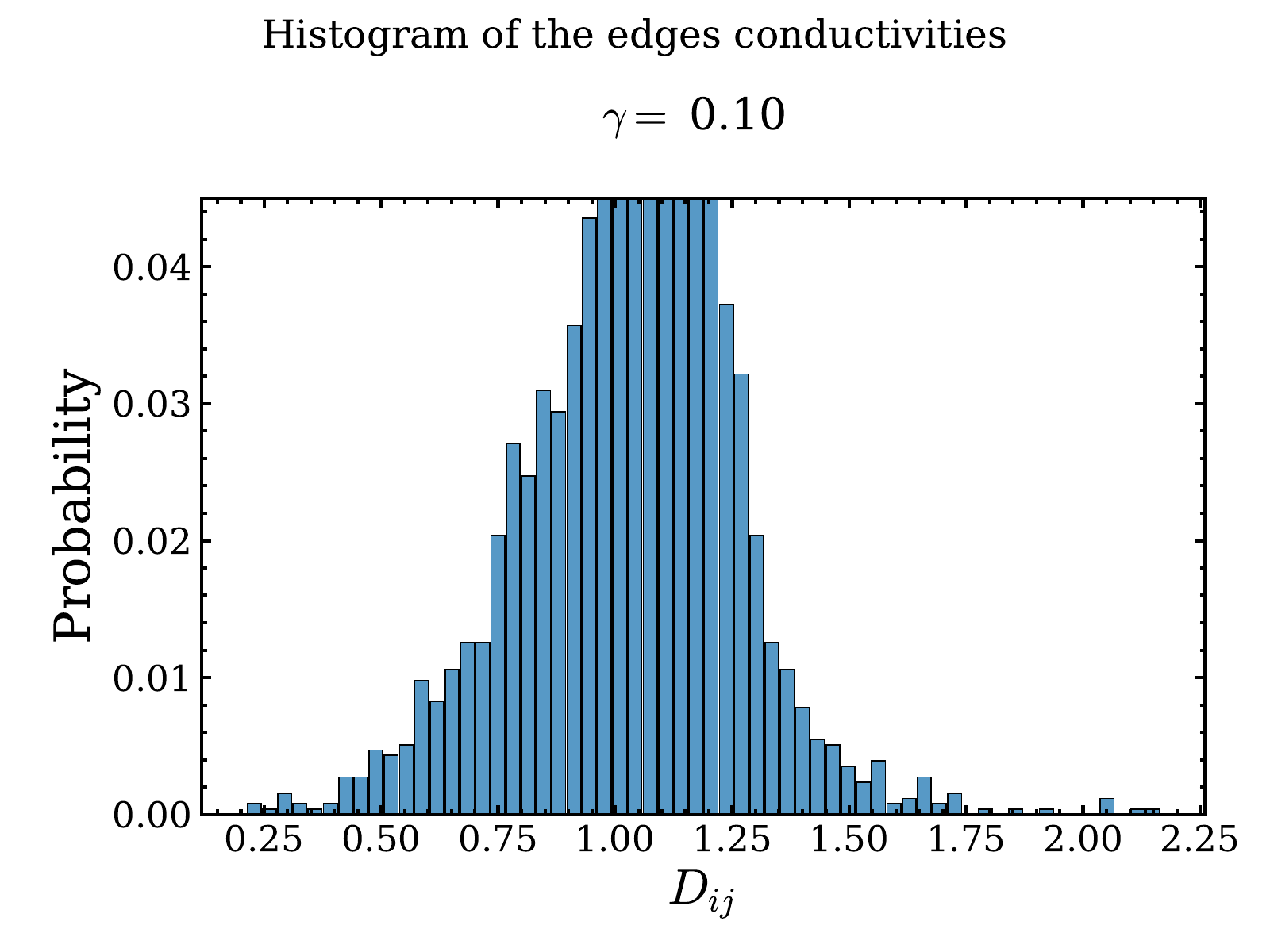}  \hfill
\mysub[\label{fig:D_ijhist_g0.47} $\gamma = 0.47$]{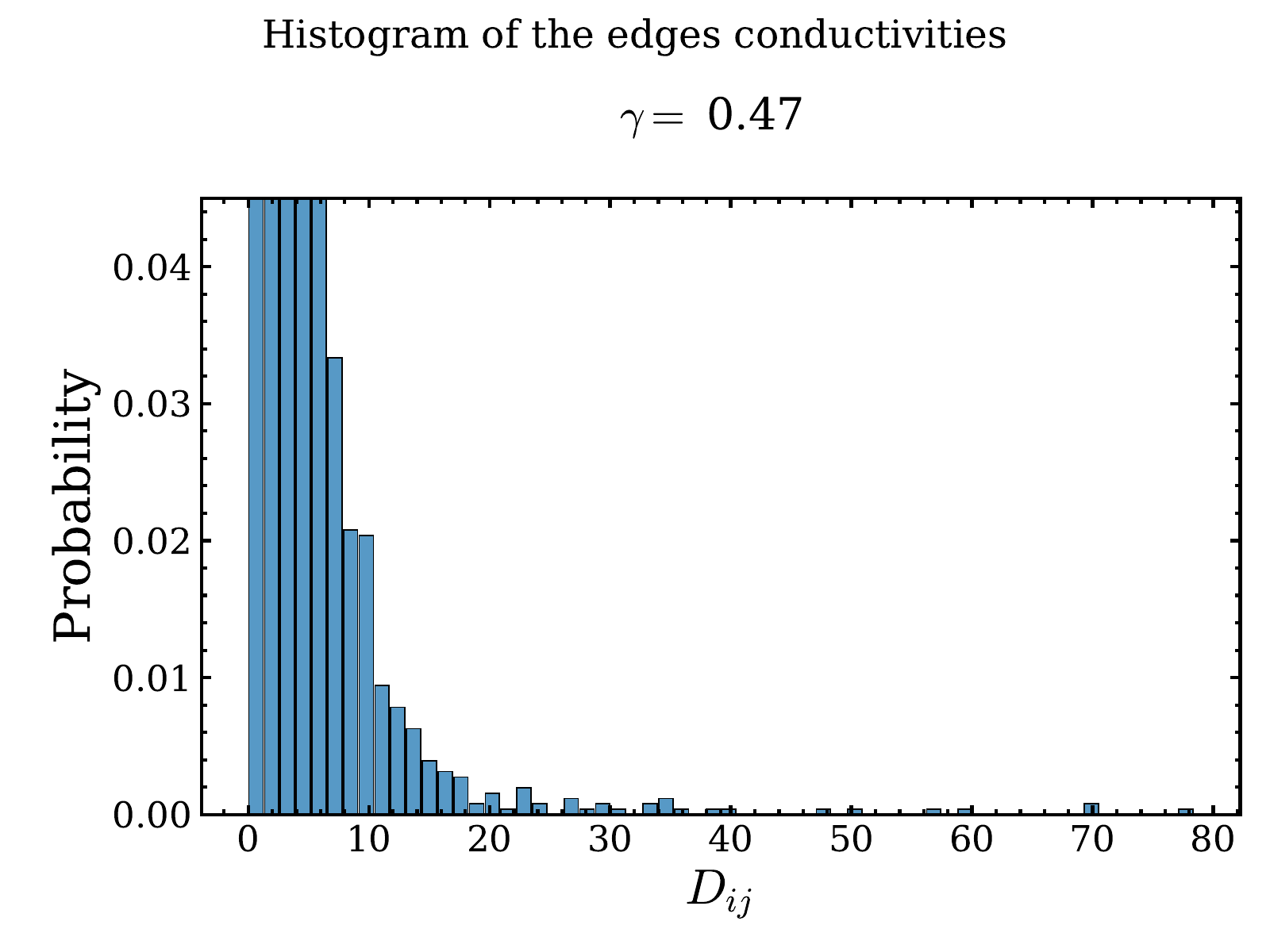} \\
\mysub[\label{fig:D_ijhist_0.51} $\gamma = 0.51$]{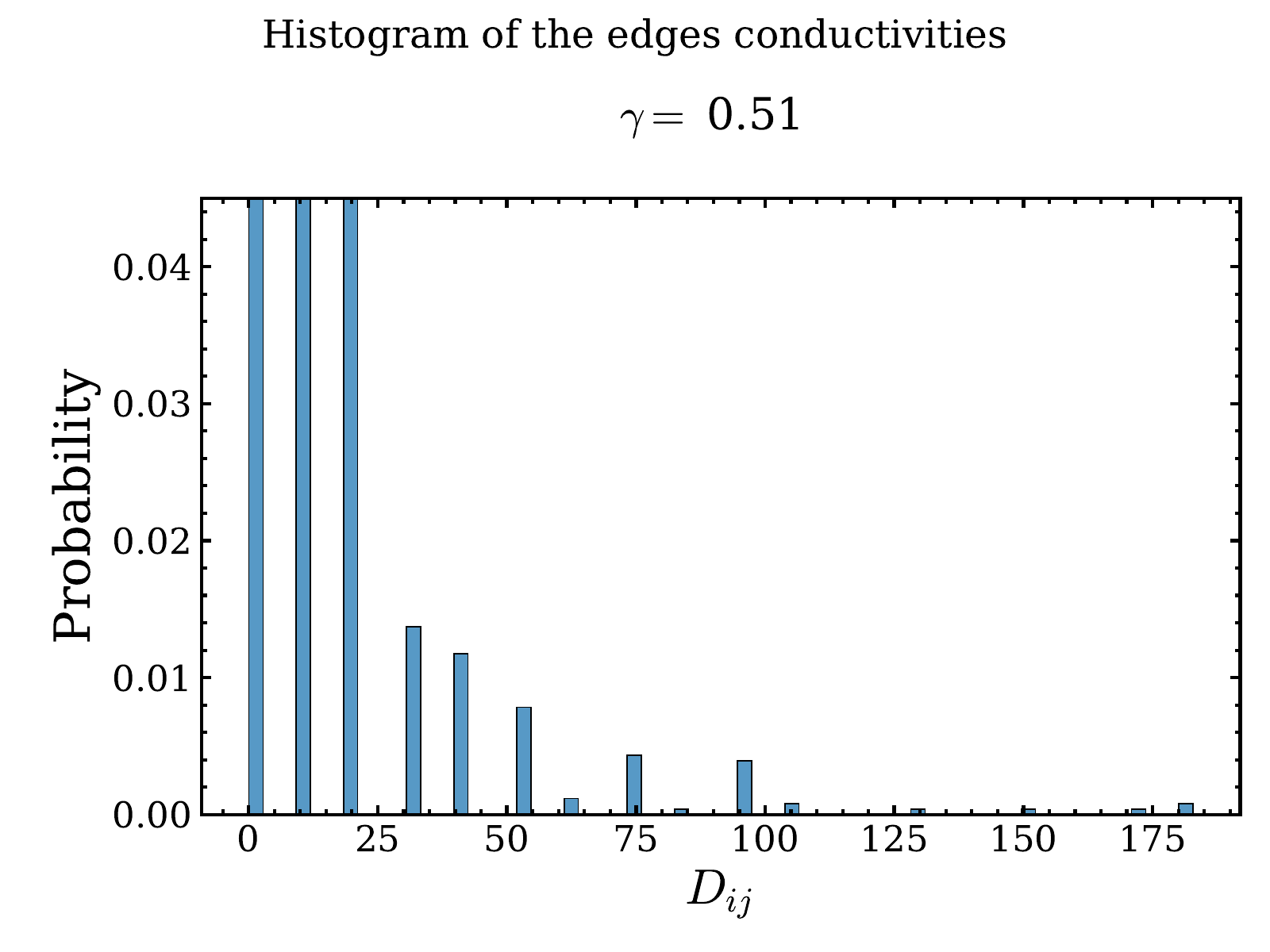} \hfill
\mysub[\label{fig:D_ijhist_0.67} $\gamma = 2/3$]{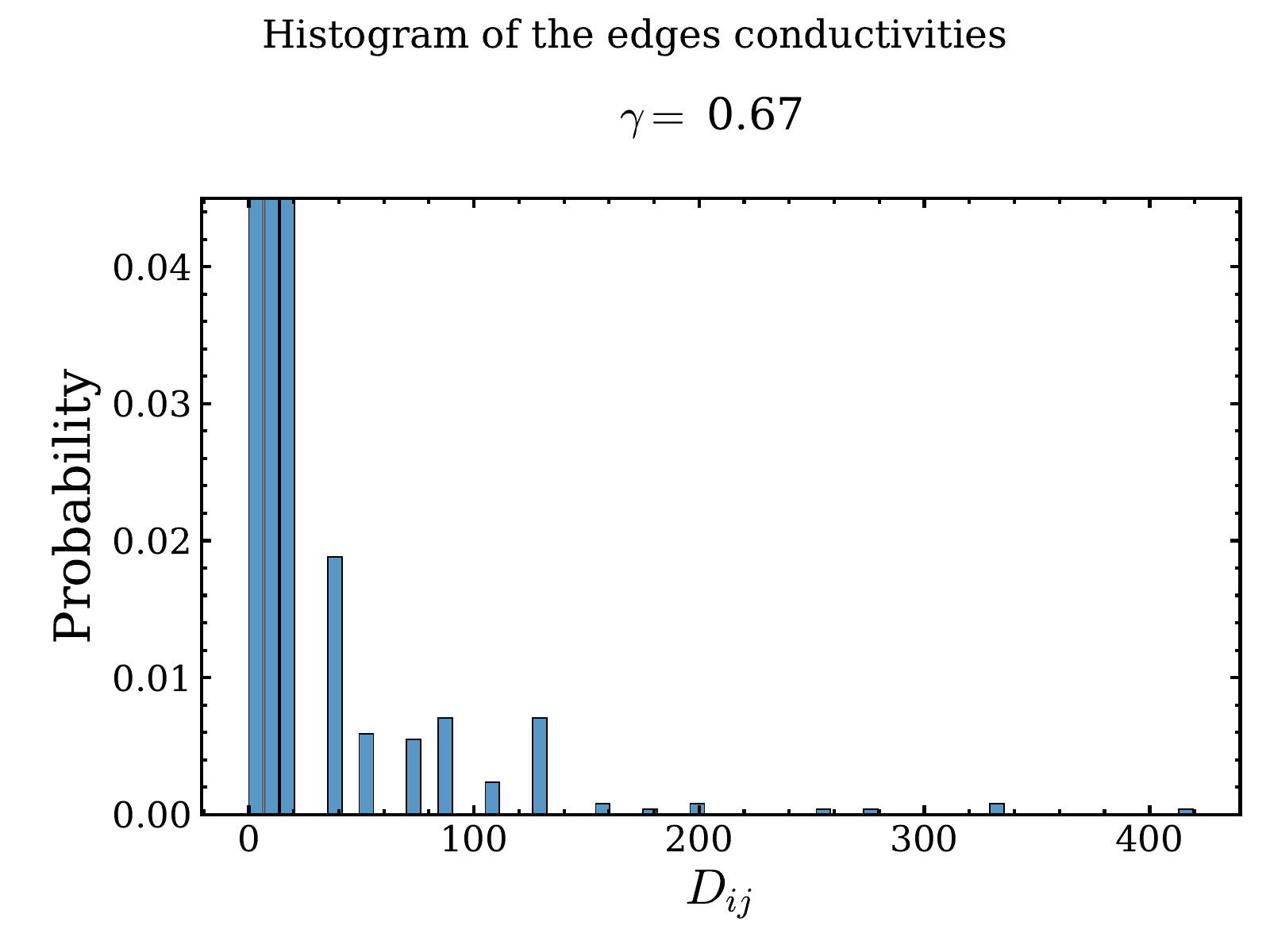}  \\

\caption{Histogram of frequencies of the steady-state  conductivities, 
for different values of $\gamma$, considering the same settings as in Figure \ref{fig:phy_gamma_network}. The distributions are different from each other. For $\gamma=0.1$, the distribution is very similar to the initial one. In the remaining cases, most of the conductivities are concentrated near zero, as most of the edges disappear. For clarity, the histograms focus on lower probabilities.
}
\label{fig:Dij_hist} 
\end{figure}

The differences in the topology of the networks can be better quantified by the histograms of the steady-state conductivities $D_{ij}$ for 4 different values of $\gamma$, presented in Figure \ref{fig:Dij_hist}. Each one has a very unique profile.

The phase transition was quantified through the evaluation of four metrics at the steady-state of each $\gamma$. The first two are the total power dissipation of the network \eqref{eq:energy_dissipation} and its total length,

\begin{equation}
    L = \sum_{(i,j)\in E} L_{ij} = 
    \sum_{(i,j)\in E} \sqrt{(x_i-x_j)^2 + (y_i-y_j)^2} \;,
    \label{eq:total_length}
\end{equation}

\noindent  where $(x_i,y_i)$ are the coordinates of the node $i$. The third is the flux coupling factor of the adaptation dynamics  \eqref{eq:new_adapt_rule_gamma}

\begin{equation}
    Z = \sum_{(i,j)\in E}L_{ij} |Q_{ij}|^\gamma \;,
    \label{eq:Z_gamma} 
\end{equation}

\noindent which can be seen as a measure of an effective network length, where the contribution of each channel is weighted by a function of its flux, $|Q_{km}|^\gamma$. Finally, we have considered the loop density, LD, as a simple measure of the network's redundancy, defined as the number of independent loops \footnote{By number independent loops it's meant the number of graph faces. Faces of a planar graph are regions enclosed by a set of edges that don't contain any other node or edge.}
of the steady-state  network, normalised by the number of independent loops of the initial network, which in our case, corresponds to the number of triangles of the initial Delaunay triangulation. Since the networks are connected and planar, the number of loops or faces, $f$, can be determined using Euler's formula $f = 1 + \edgesnum - \nodesnum$, where $\edgesnum=|E|$ is the number of edges, and $\nodesnum=|V|$ the number of vertices, yielding \cite{Corson2010} 
\begin{equation}
    \text{LD} = 
    \frac{1 + \edgesnum_{final} - \nodesnum_{final}}{1 + \edgesnum_{initial} - \nodesnum_{initial}} \;.
    \label{eq:loop_density}
\end{equation}

\noindent Note that only edges with conductivities above a threshold ($D_{thr}=5\times 10^{-4}$) were considered in the computation of the total length and the number of loops of the final networks.

\begin{figure}[hbt!] 

\newcommand{\mysub}[2][]{%
    \subfloat[#1]{\includegraphics[trim={0cm 0cm 0cm 1.3cm}, clip, width=0.47\textwidth]{#2}}%
}
\centering

\mysub[\label{fig:gamma_energy}Power dissipated by the steady-state networks, $\E(\gamma)$, normalised by $\E_0 \equiv \E(\gamma=0)$.]{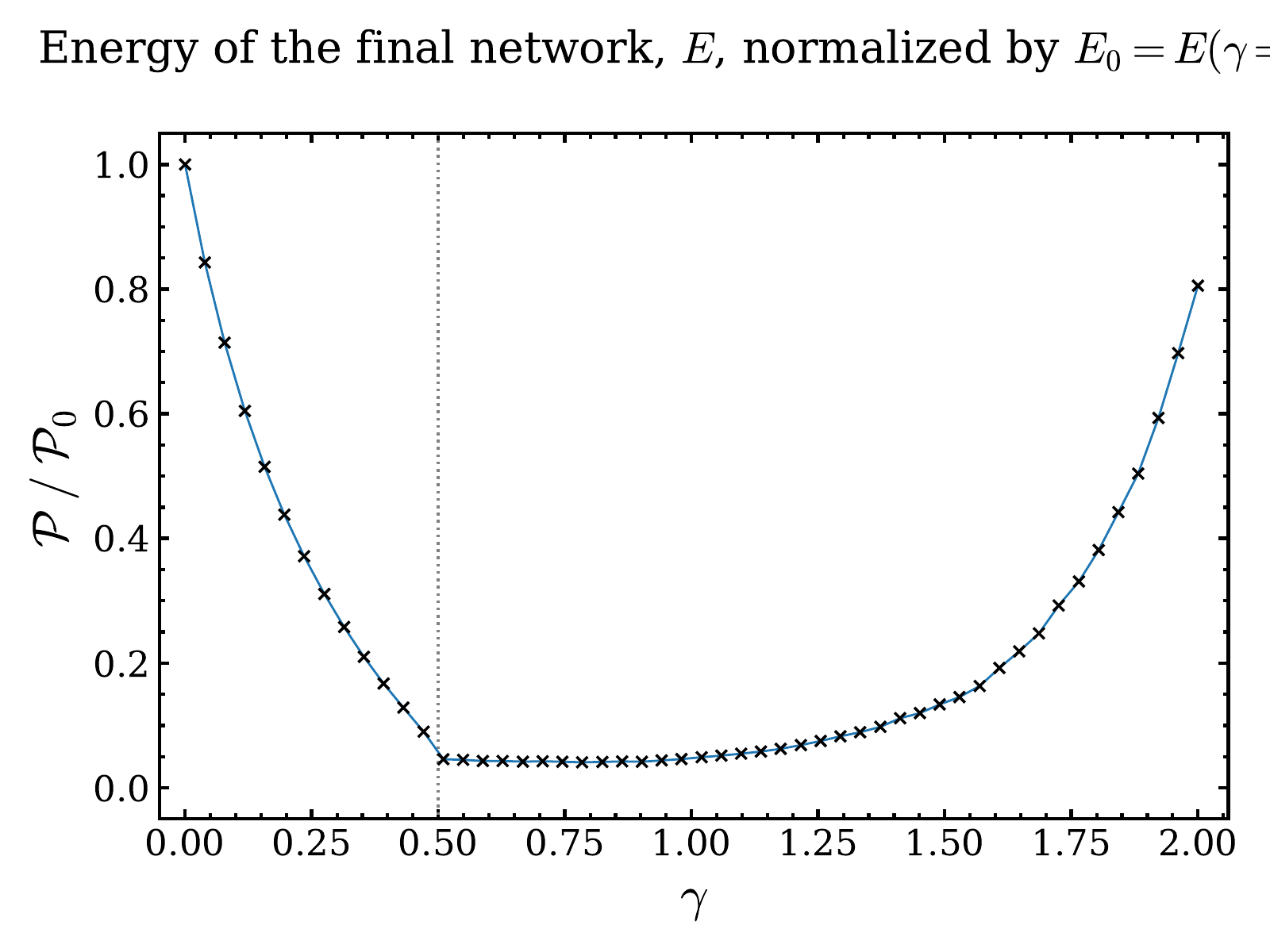} \hfill
\mysub[\label{fig:gamma_zoom_energy}Detailed view of the normalised power dissipated, $\E/\E_0$, near $\gamma=2/3$.]
{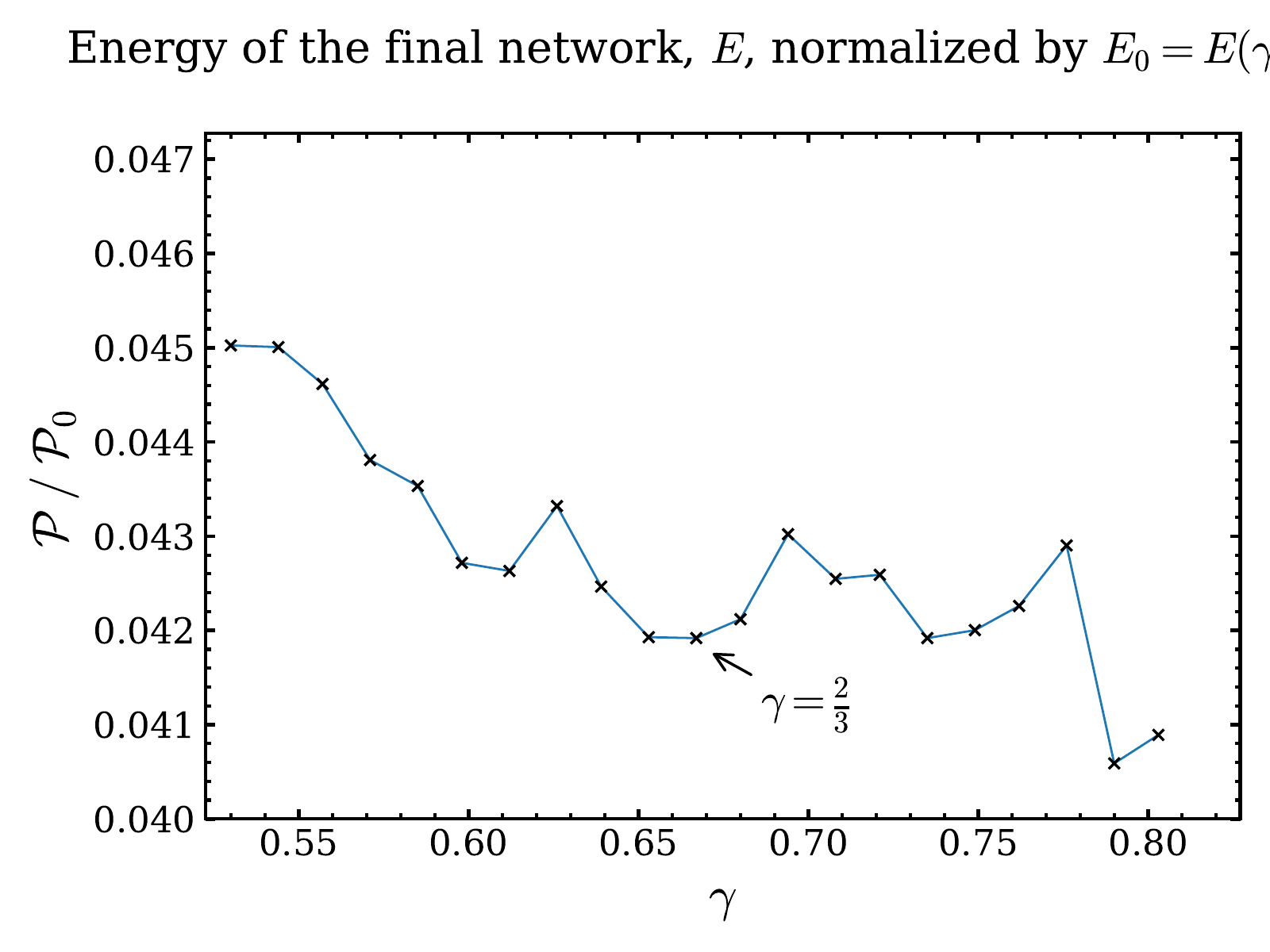} \\
\mysub[\label{fig:gamma_length} Total Length of the steady-state networks, $L$, normalised by $L_0 \equiv L(\gamma=0)$.]{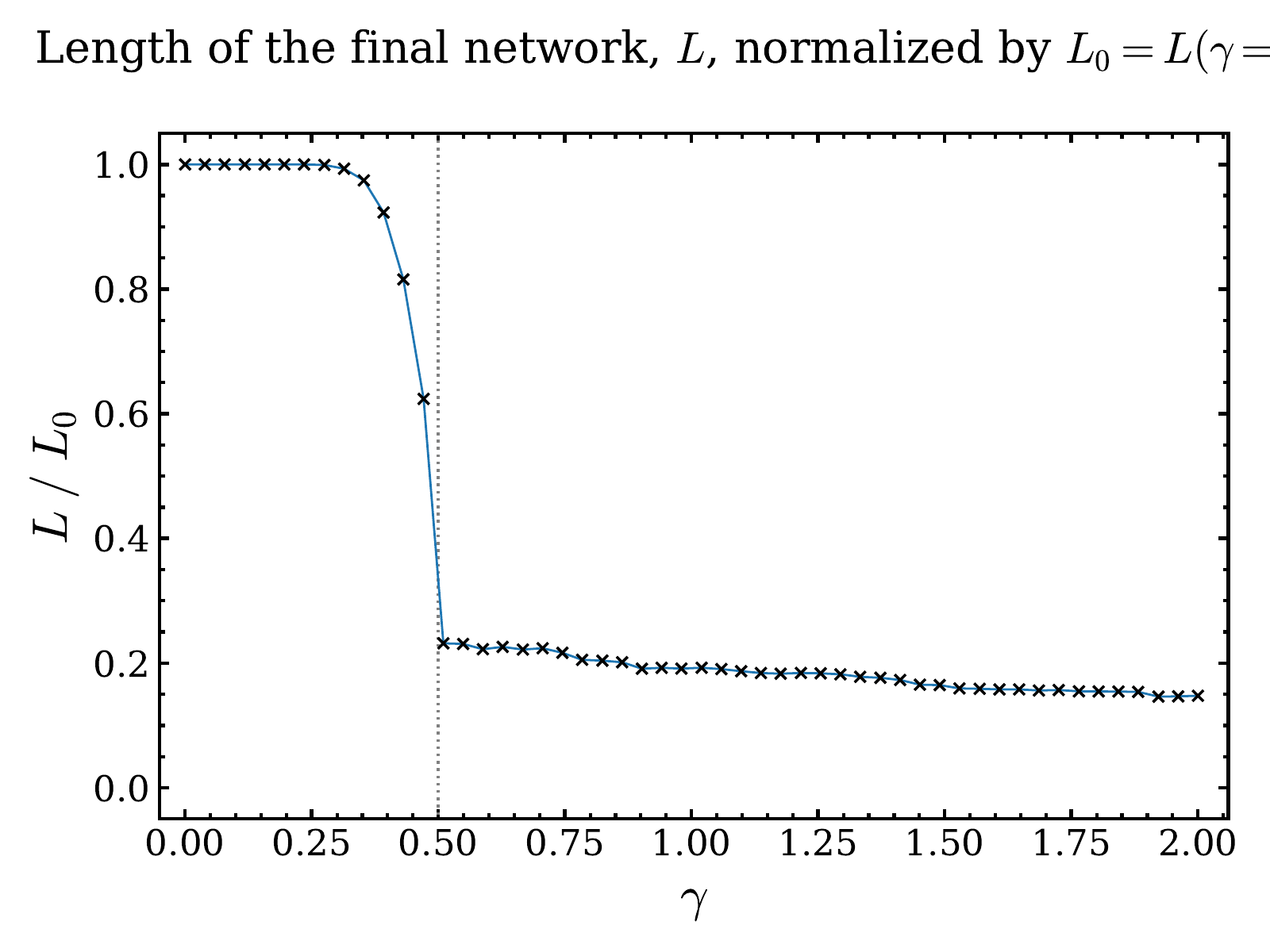} \hfill
\mysub[\label{fig:gamma_loop}Loop density of the steady-state networks defined as \eqref{eq:loop_density}.]{images/Gamma/disk/loops.pdf}  \\
\mysub[\label{fig:gamma_Z}Flux coupling factor of the steady-state networks, $Z(\gamma)$, defined as \eqref{eq:Z_gamma}, normalised by $Z_0 \equiv Z(\gamma=0)$.]{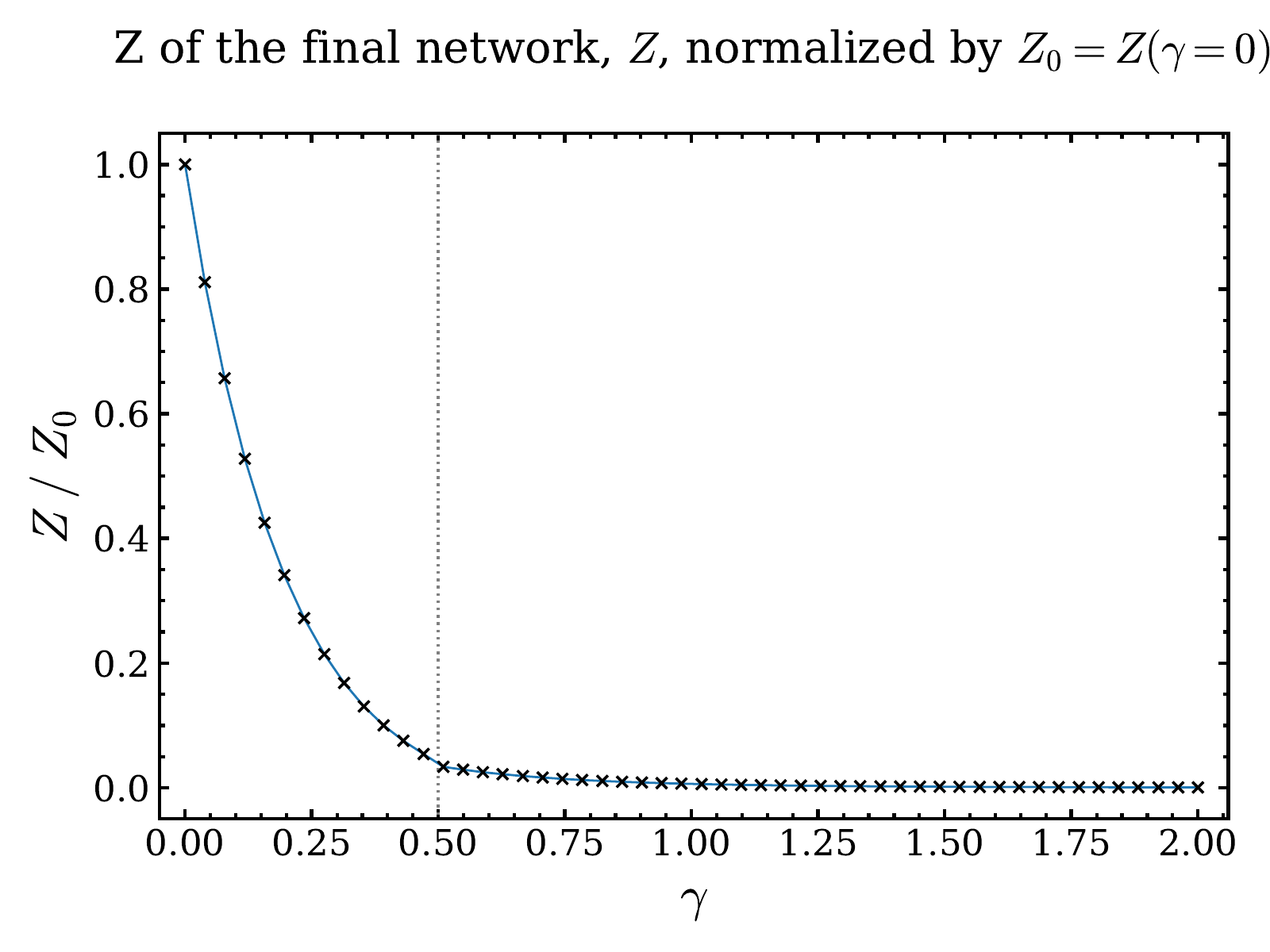} \hfill
\mysub[\label{fig:gamma_volume}Volume of the steady-state networks, $V(\gamma)$, normalised by the volume of the initial network, $V_0\equiv V(t=0)$.]{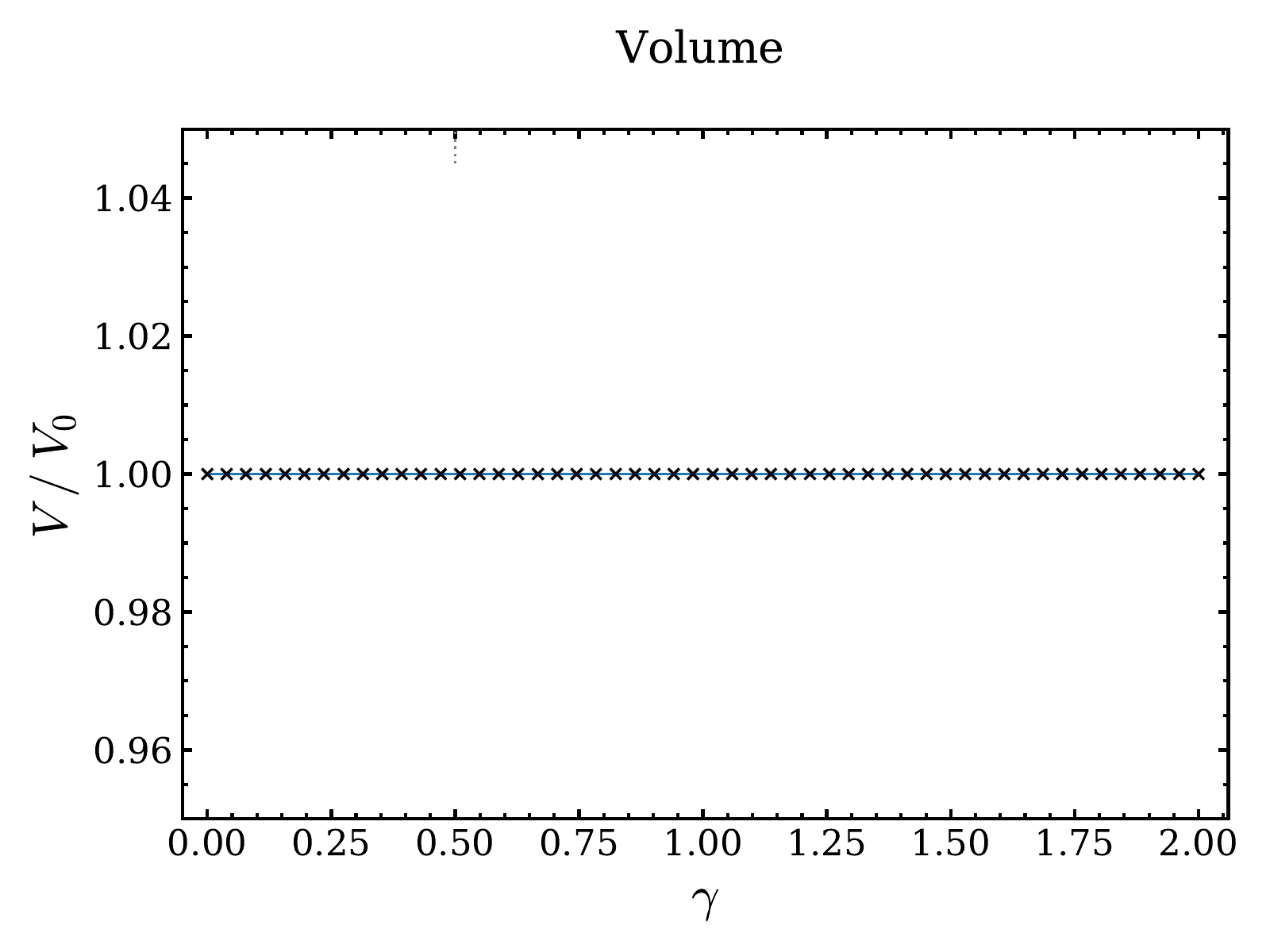}

\caption{Plots of different metrics as functions of gamma $\gamma$, evaluated at the steady state  of the simulations presented in Figure \ref{fig:phy_gamma_network}. The vertical dashed line marks the transition at $\gamma = 1/2$. Note the discontinuity of the slopes of the network's dissipation, total length and loop density. The correspondent graphs for other choices of initial conductivities and distribution of terminals showed the same tendency, thus supporting the universality of the phase transition. 
}
\label{fig:gamma_plots} 
\end{figure}

The dependency of these quantities on $\gamma$ is plotted in Figure \ref{fig:gamma_plots}. The change in the slope of the total network's dissipation, $\E(\gamma)$, observed at $\gamma = 1/2$, suggests the existence of a discontinuity on its first derivative with respect to  $\gamma$, which may correspond to a first-order phase transition, according to the Ehrenfest classification. For $\gamma \gtrsim 1$ the power dissipated at the steady state starts to notably increase, which can be understood based on the results of  Figure \ref{fig:phy_gamma_network}. As $\gamma$ increases, the number of channels connecting the sinks to the main branches decreases, and the channels are very thin, meaning they have low conductivities. This implies that they are associated with a large pressure drop $\Delta p_{ij}$ to reach the significantly high channel flux (the flux from the source is split between few channels). Therefore the power dissipated by each one, $\E_{ij}=\Delta p_{ij} Q_{ij}$ is very large. 

Figure \ref{fig:gamma_zoom_energy} represents a closer look at the  plot of the power dissipated  near $\gamma = 2/3$. Contrary to expectations, the minimum of the power dissipated is not reached for $\gamma=2/3$, as derived in \eqref{eq:minEDij}. This is most likely explained by the error propagation in the computation of the dissipation. By \eqref{eq:energy_dissipation}, the uncertainty of the dissipation, $\Delta \E_{ij}$ is related to the uncertainties of the conductivities, $\Delta D_{ij}$  and  the fluxes $\Delta Q_{ij}$ by 

\begin{equation}
    \frac{\Delta \E}{\E} =
    \sum_{(i,j)\in E} \left( 2 \frac{\Delta Q_{ij}}{Q_{ij}}
    - \frac{\Delta D_{ij}}{D_{ij}} \right) \;.
\end{equation}

Therefore, the sum of all the edge contributions with low conductivities and low fluxes can be considerably large. Other configurations of terminals were also tested, and in most cases, the minimum was indeed reached at $\gamma=2/3$.

The transition is even more clear in the graphs of the total length (Figure \ref{fig:gamma_length}) and loop density (Figure \ref{fig:gamma_loop}). In particular, the loop density starts and stays at its maximum for $\gamma \lesssim 0.3$, meaning that the adaptation doesn't change the topology of the network for low values of $\gamma$, and progressively decreases as $\gamma$ increases, becoming  zero for $\gamma > 0.5$, implying the absence of loops. This confirms that the steady-state  networks for $\gamma > 0.5$ are spanning trees assuming the threshold $D_{thr}=5\times 10^{-4}$. Conversely, as Figure \ref{fig:gamma_Z} shows, the quantity $Z$ decreases with $\gamma$ and displays a smooth transition near $\gamma=1/2$.

In Figure \ref{fig:gamma_volume} it's plotted the volume of the steady-state network normalised to the volume of the initial one as a function of $\gamma$, showing that the volume is always  conserved. Further simulations revealed that the phase transition is independent of the initial distribution of conductivities and of the distribution of sources and sinks.

\cleardoublepage

\chapter{Applications}
\label{chapter: Applications}

In this chapter, we explore some applications of the model, namely, maze solving and optimal network design. In the first case, we replicate the maze setting considered in \cite{Nakagaki2000}, where it was reported for the first time the \textit{Physarum} maze-solving abilities (Figure \ref{fig:maze}). In the second case, inspired by  \textit{Physarum} Tokyo experiments \cite{Tero1} (Figure \ref{fig:tokyo}), we simulate the model in an arena mimicking the Portuguese mainland. In this regard, we study the importance of flux fluctuations to build efficient and resilient networks, by introducing time-dependent distributions of sources and sinks in the model. The performance and topology of the resulting networks are then compared with those of the real Portuguese railway system.

\section{Maze Solving}

The famous maze experiments performed by Nakagaki et al. \cite{Nakagaki2000} showed that \textit{Physarum} can solve mazes when two food sources are placed at both ends (Figure \ref{fig:maze}). We now demonstrate that our model can reproduce this phenomenon. 

We have simulated the adaptation dynamics \ref{eq:new_adapt_rule_minE} over a graph describing  approximately the same maze used in \cite{Nakagaki2000}. The maze was generated from a $15\times 16$ square lattice, where some edges were removed and only the edges which outline the possible routes were left out. Like in the experiments, four possible routes connect the maze entrances (Figure \ref{maze_a}). The only difference is that all the possible paths have the same length due to the symmetry of the underlying square lattice, while in \cite{Nakagaki2000} the lengths of the $\alpha_1$, $\alpha_2$, $\beta_1$, $\beta_2$  segments which made up the paths were slightly different. The initial state represents the \textit{Physarum} plasmodium filling the entire maze. The food source stimuli which drive the optimisation in the experiments is mimicked by simulating the  model considering a source and a sink placed at the entrances of the maze. 

\begin{figure}[hbt!] 

\newcommand{\myfig}[1]{%
    \includegraphics[trim={2.8cm 2.8cm 3.2cm 2.8cm}, clip, width=0.235\textwidth]{#1}%
}
\begin{subfigure}{\textwidth}    
    \centering
    \begin{tabular}{ccc} %
    $t=0$ & $t=20$ & $t=130$ \\ %
    \myfig{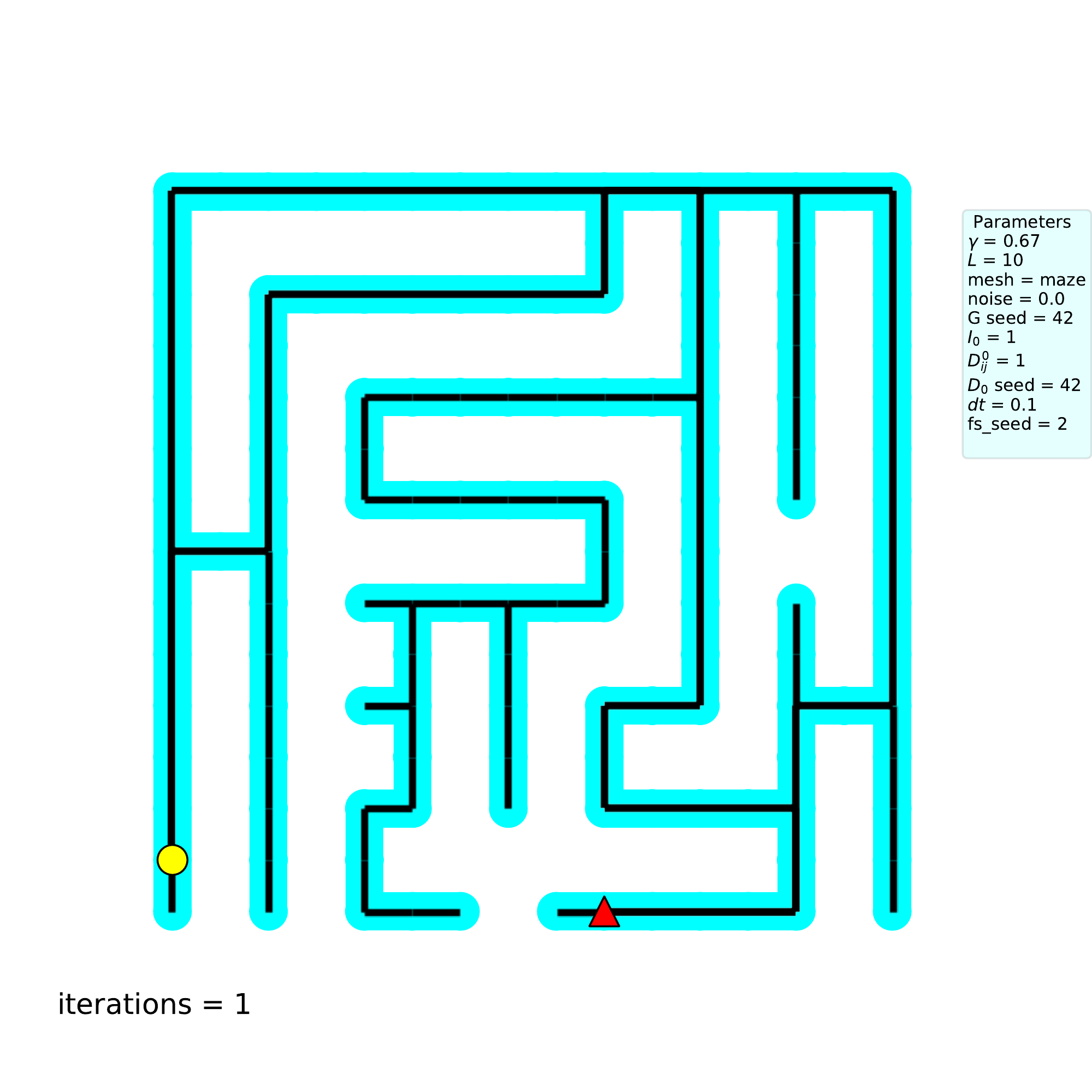} &
    \myfig{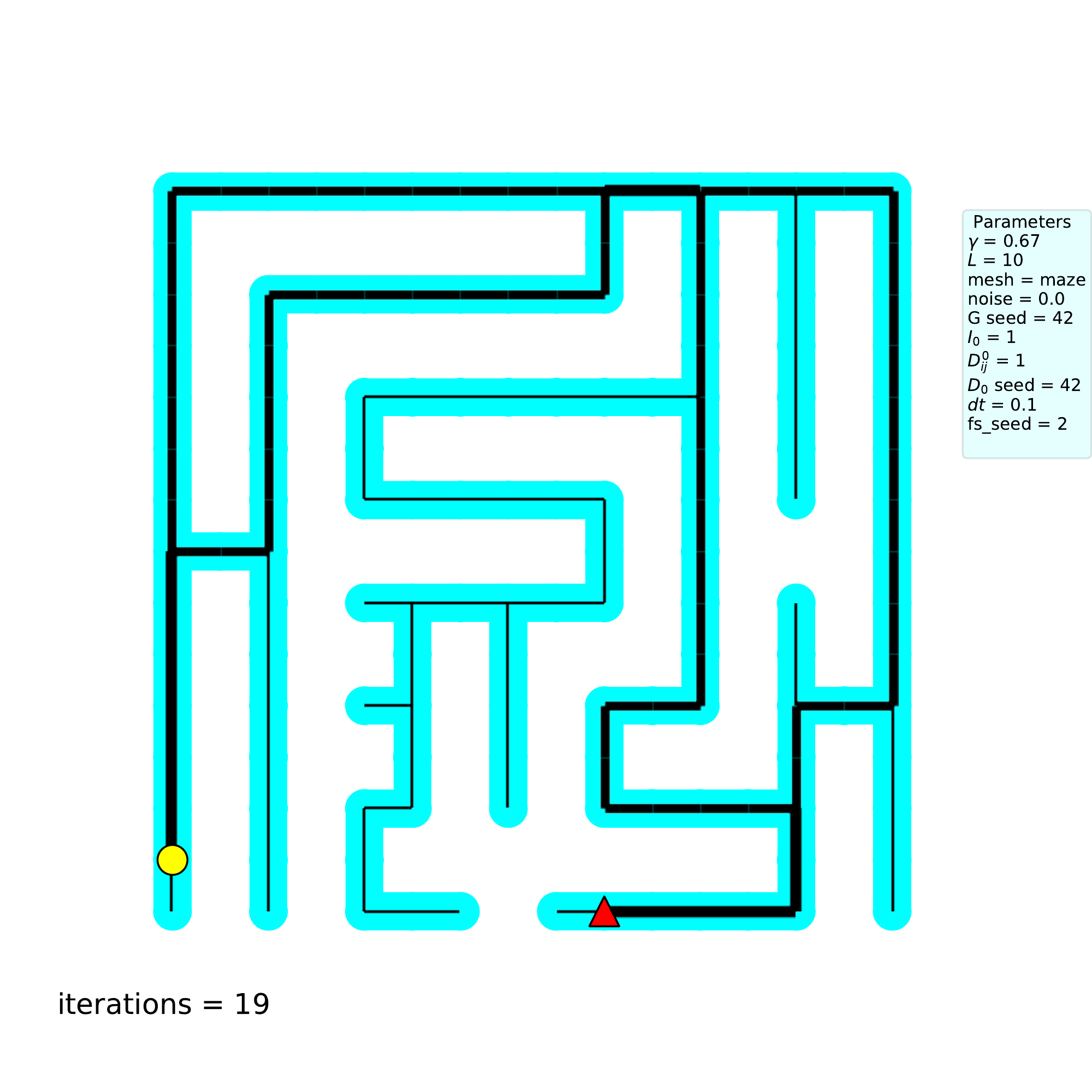} &
    \myfig{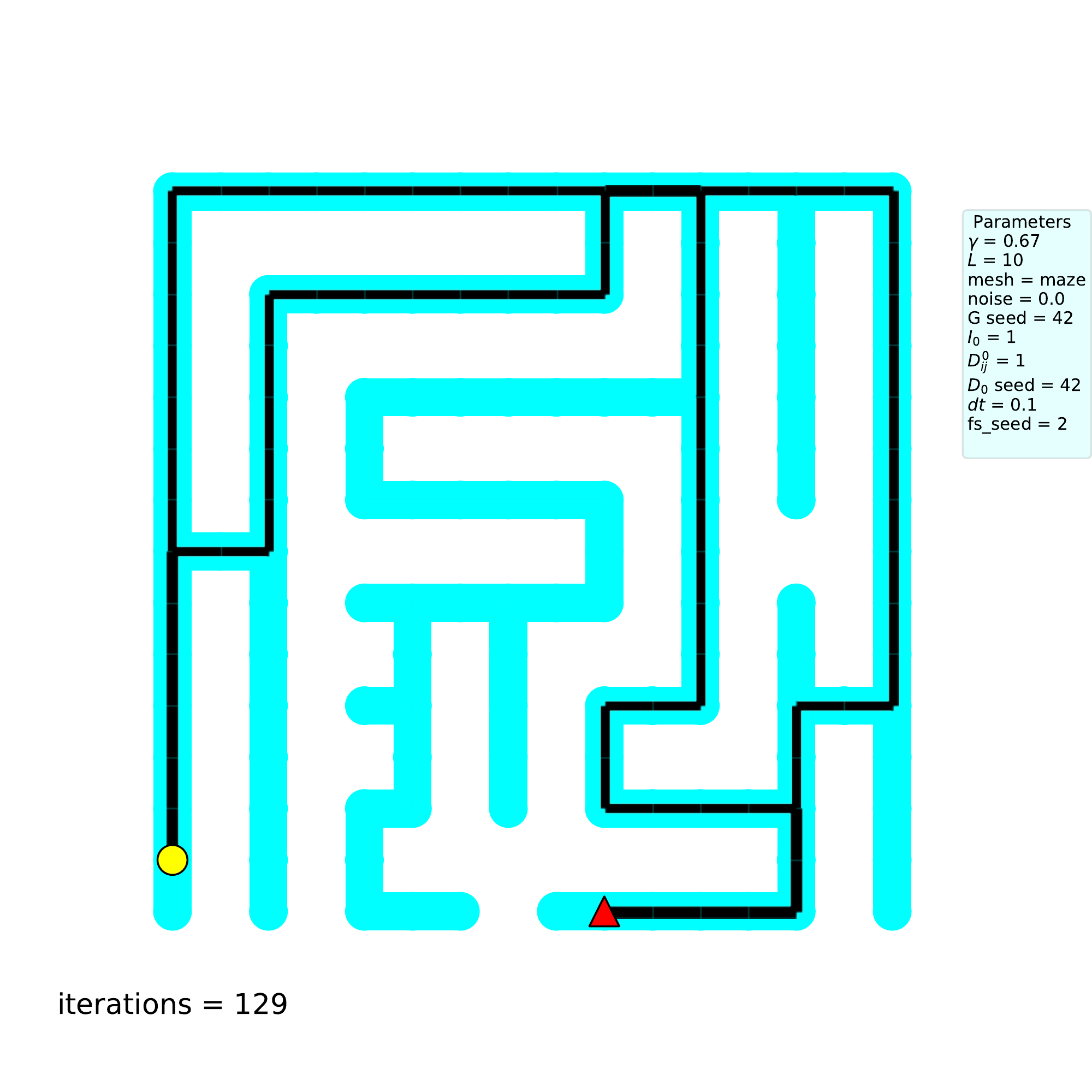} 
    \end{tabular}
    \caption{Simulation with initial homogeneous conductivities, $D_{ij}(0)=1$.}
    \label{fig:maze_homogeneous}
\end{subfigure}  
\\[0.2cm]
\setlength\tabcolsep{4pt}%
\begin{subfigure}{\textwidth}    
    \centering
    \begin{tabular}{@{}cccc@{}}
    $t=0$ & $t=20$ & $t=130$ & $t=250$ \\ %
    \myfig{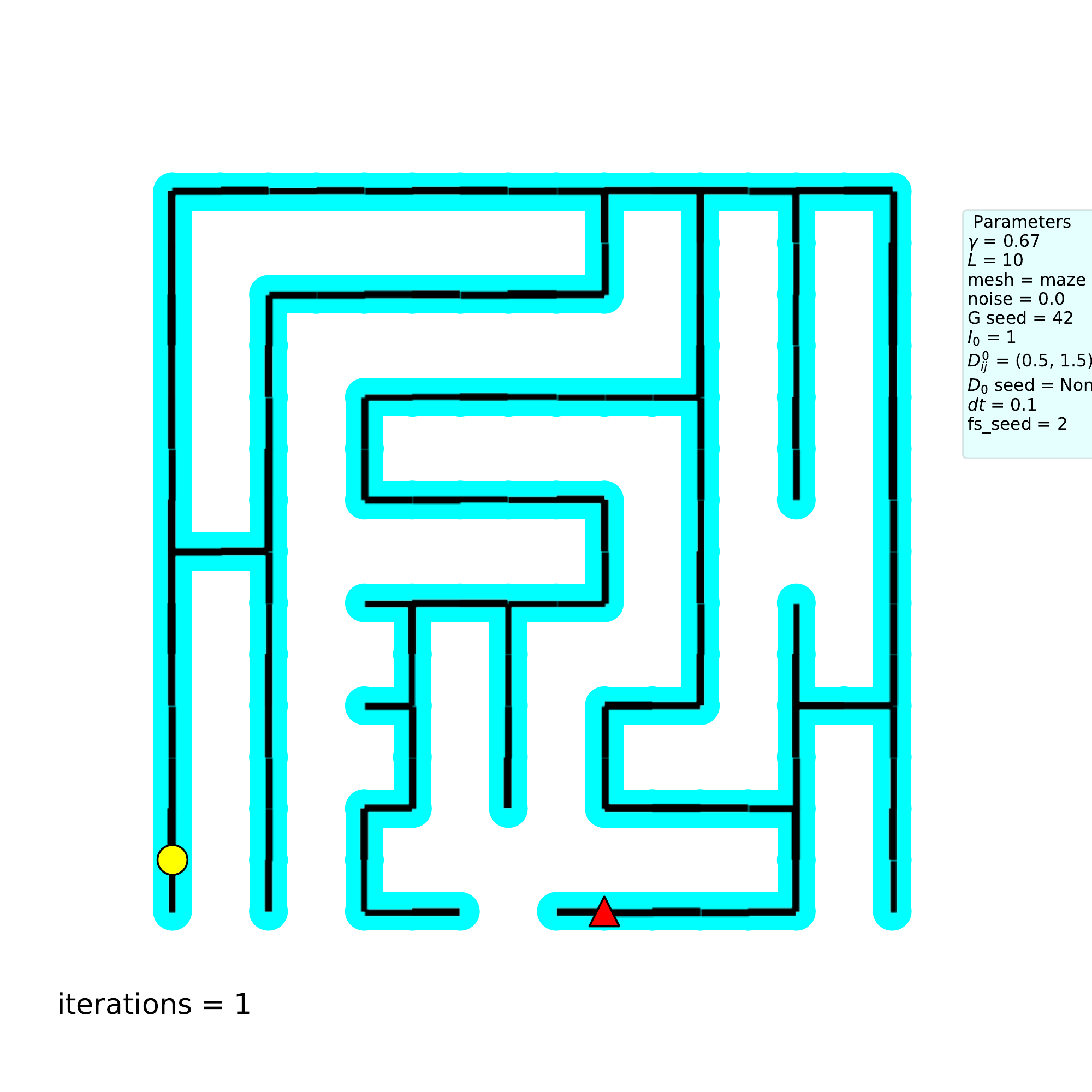} &
    \myfig{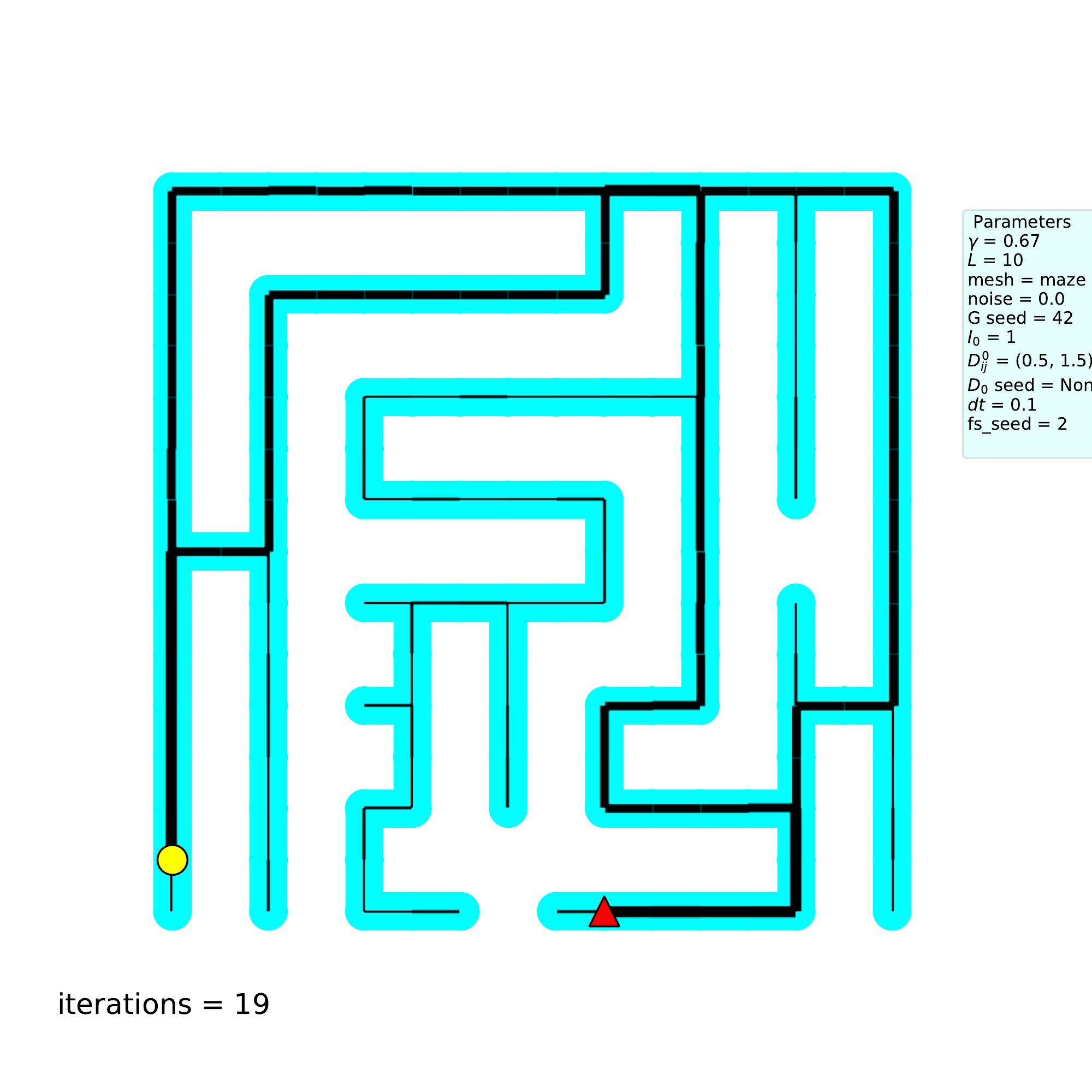} &
    \myfig{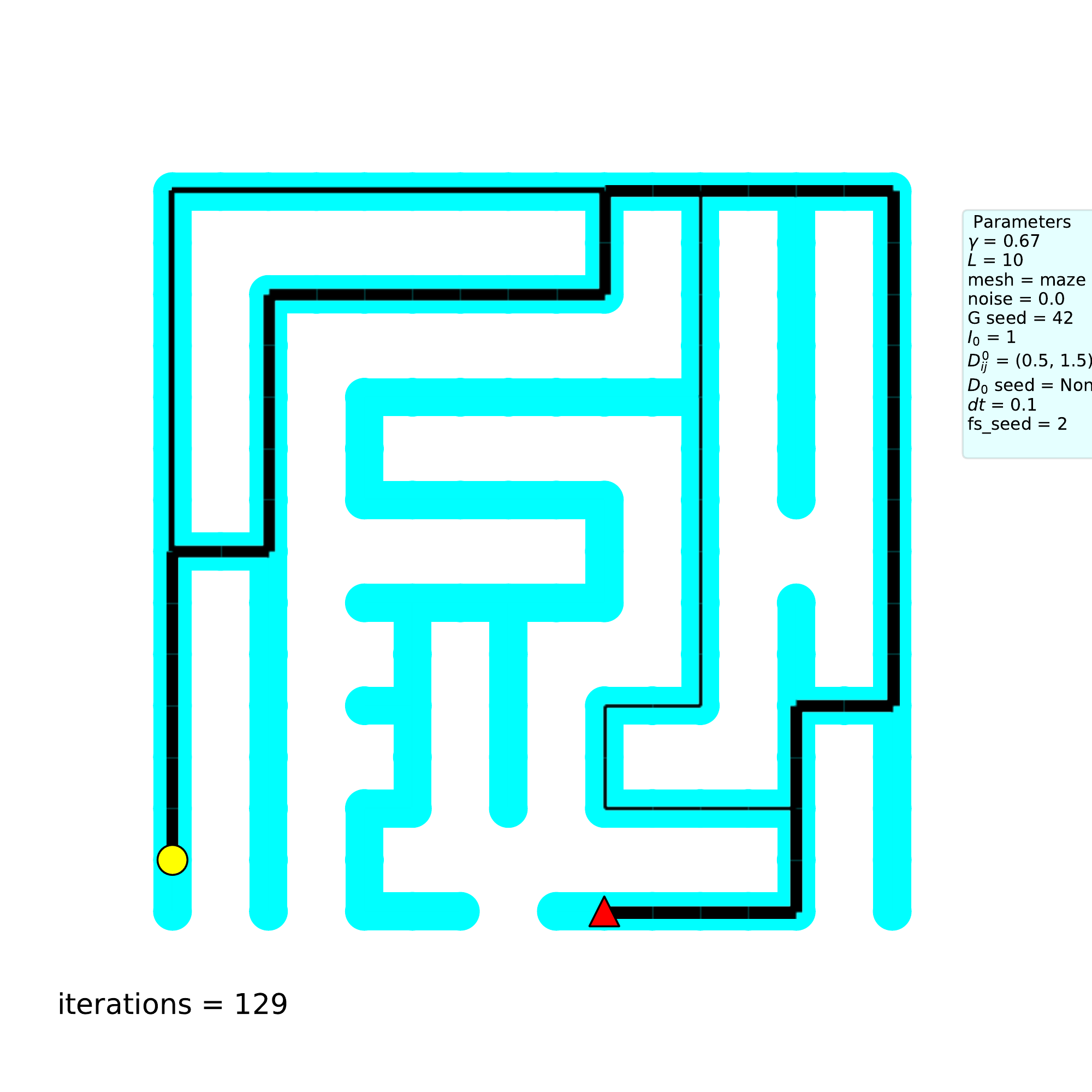} &
    \myfig{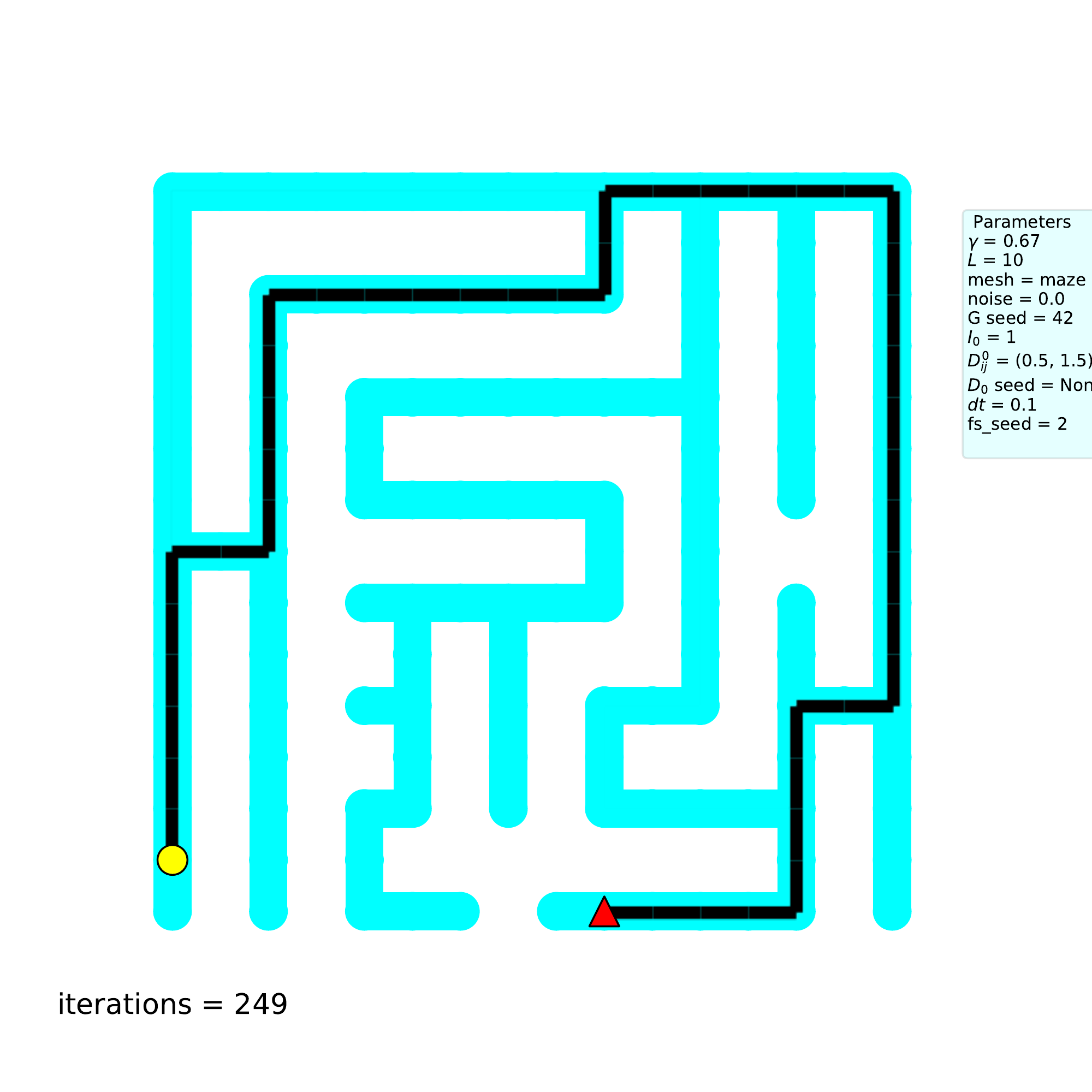} 
    \end{tabular}
    \caption{Simulation with initial random conductivities, $D_{ij}(0)\sim U(1/2,3/2)$.}
    \label{fig:maze_random}
\end{subfigure}  

\caption{Simulation of the model \eqref{eq:new_adapt_rule_minE} in a maze similar to that of Figure \ref{fig:maze}, starting from different sets of conductivities. The snapshots were taken at different iterations $t$ of the algorithm. \textbf{(a)} All possible solution paths are selected since they have the same total length and the initial conditions are the same. \textbf{(b)} First the dead ends disappear like in \textbf{(a)}, but due to the initial uneven distribution of conductivities, only the path which is initially favoured remains.
}
\label{fig:maze_simulation} 
\end{figure}

\begin{figure}[hbt!] 

\newcommand{\mysub}[2][]{%
    \subfloat[#1]{\includegraphics[trim={2.8cm 2.8cm 3.2cm 2.8cm}, clip, width=0.235\textwidth]{#2}}%
}

\centering

\mysub[]{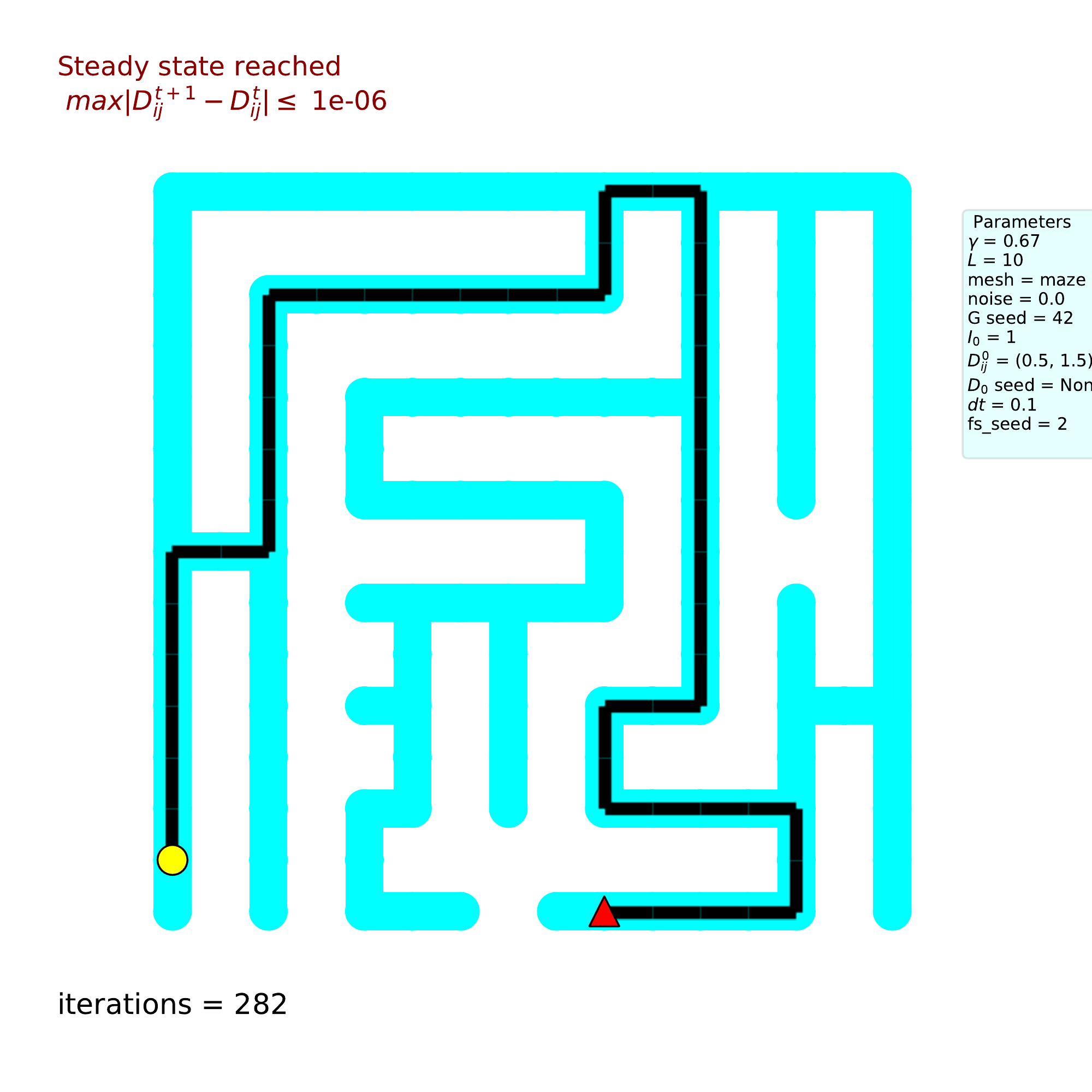} \hfill
\mysub[]{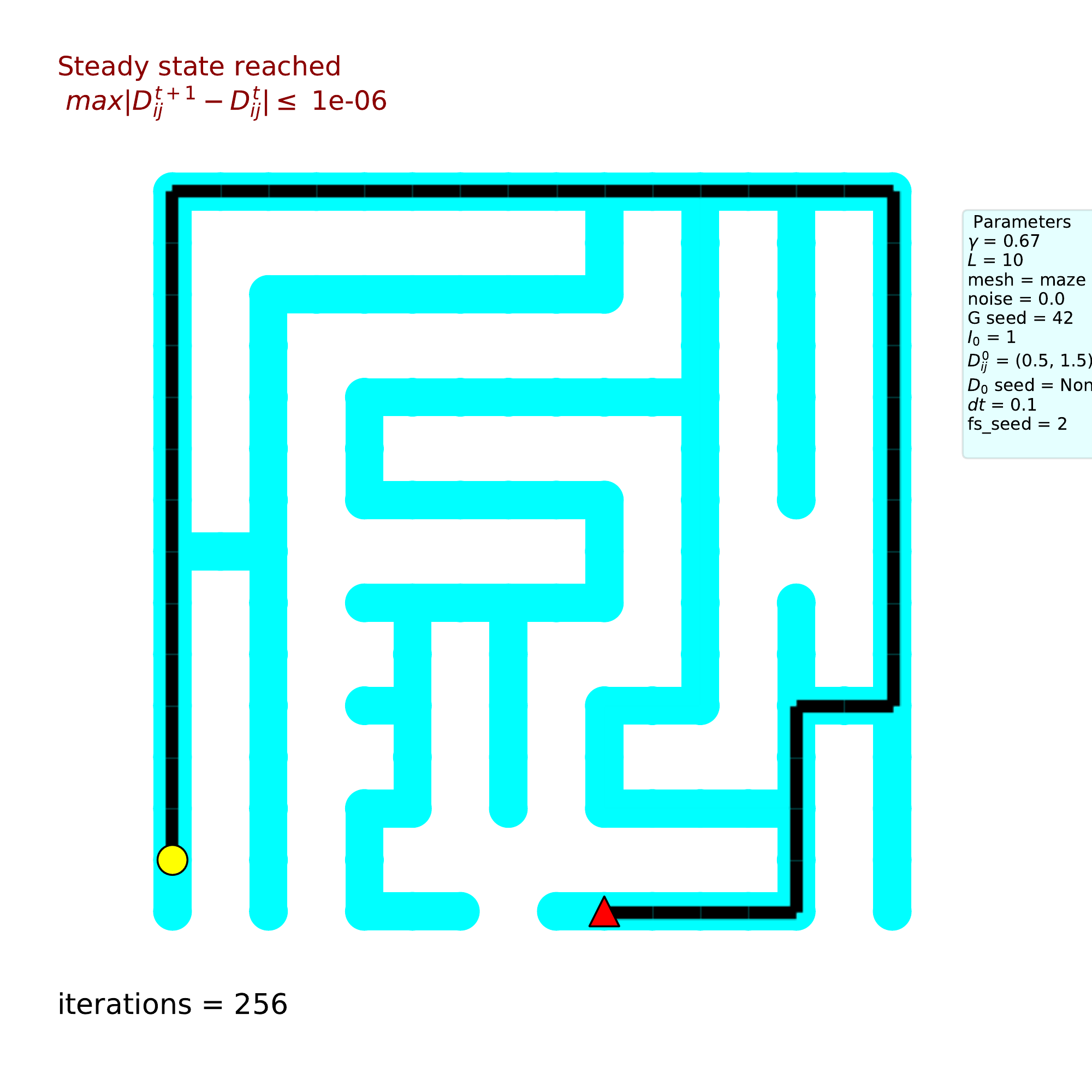} \hfill
\mysub[]{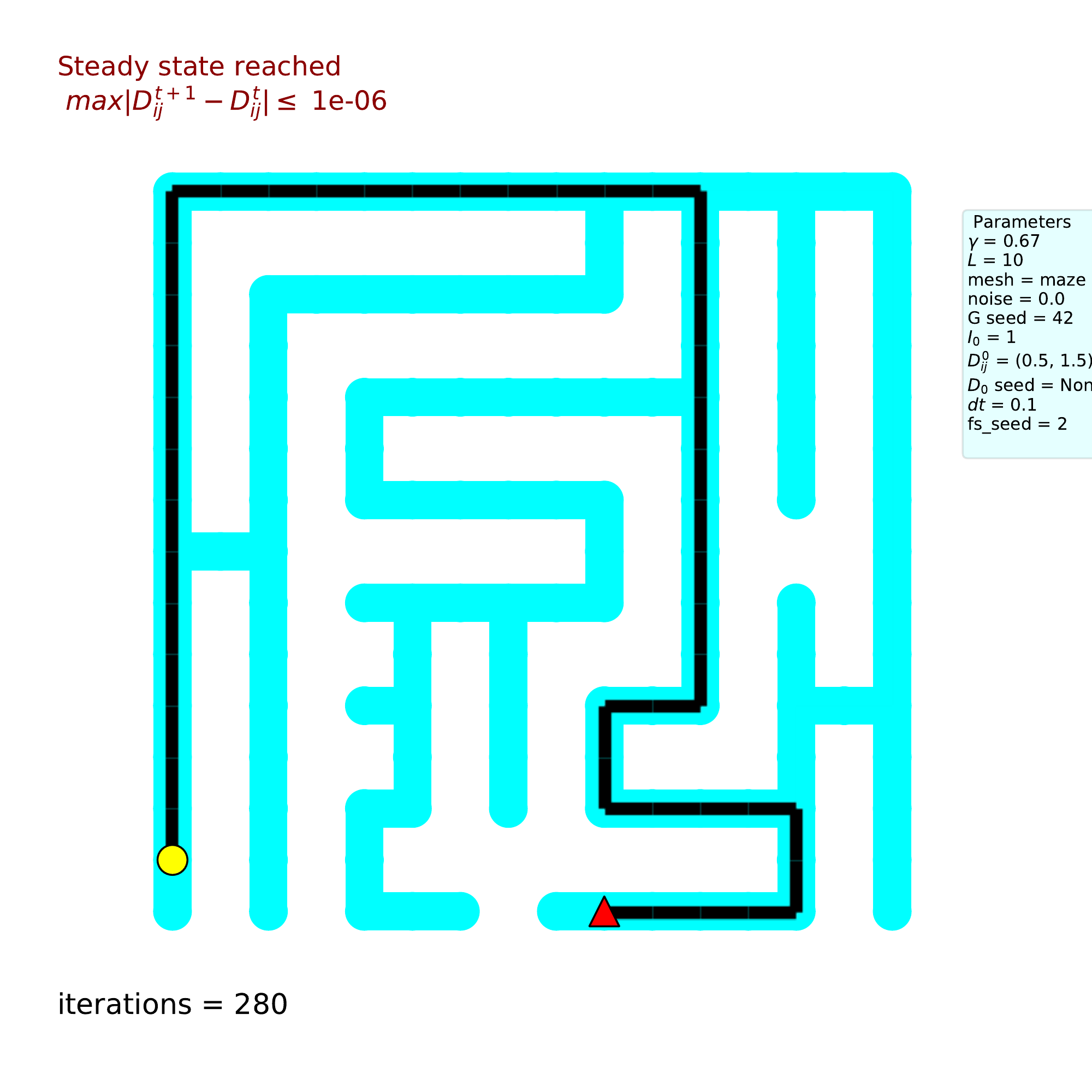} \hfill
\mysub[]{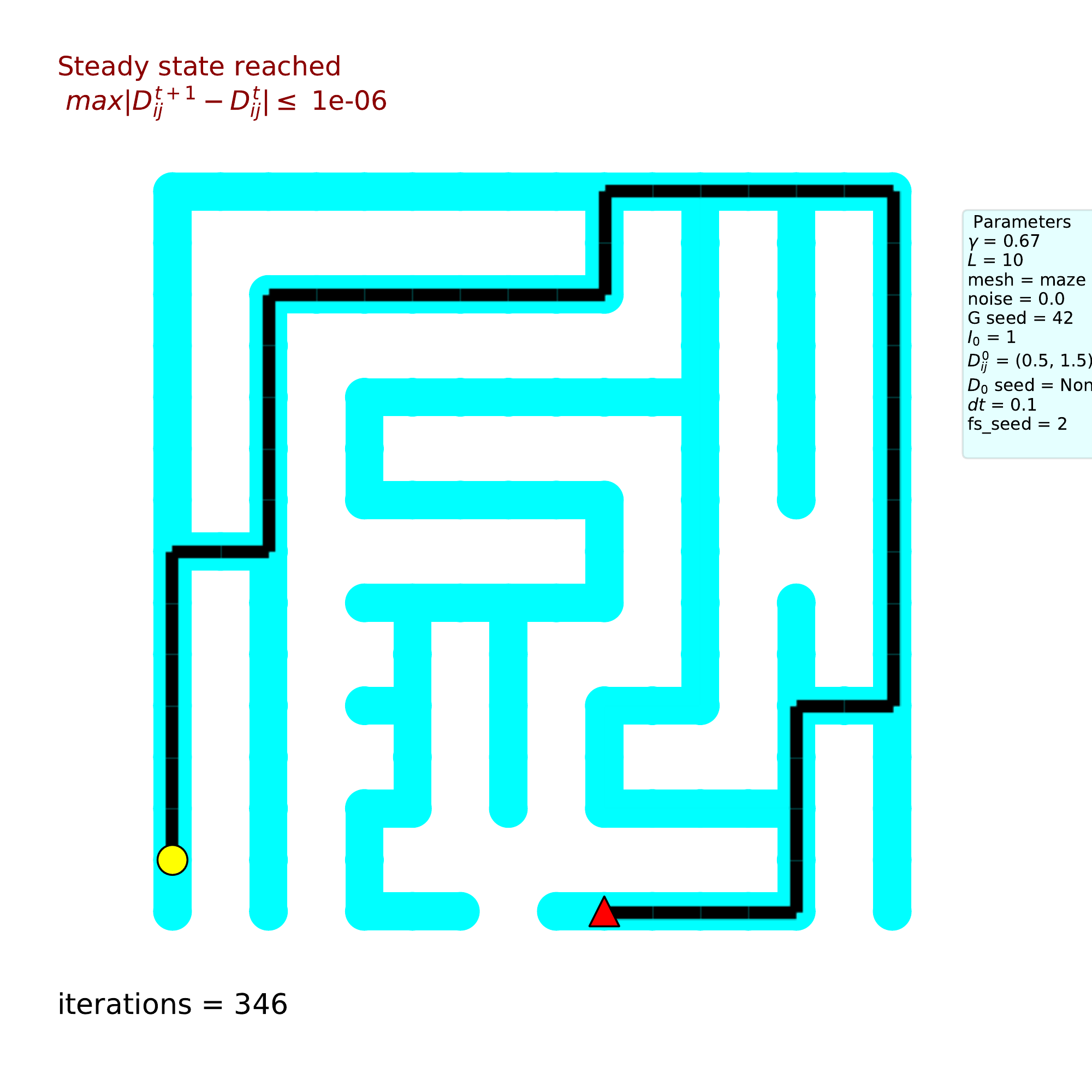} 

\caption{Maze solving considering non-homogeneous initial conductivities, $D_{ij}(0)\sim U(1/2,3/2)$. The images represent the final networks for different initial random states. The system converges to the initially favoured path, i.e., the path which has the highest average conductivity. All four possible paths can be reached by changing the random seed. 
}
\label{fig:maze_all_paths} 
\end{figure}

In Figure \ref{fig:maze_simulation}, it's shown snapshots of two simulations considering homogeneous and non-homogeneous initial conductivities. In both simulations, the paths which lead to dead ends are the first to vanish since there is no flux along those channels, and only paths connecting the source and sink survive. For homogeneous initial conditions (Figure \ref{fig:maze_homogeneous}), the steady state is a compromise between the four possible solutions. The adaptation mechanism can't select only one of the paths, as they both have the same length, and no channel conductivities are initially favoured. In contrast, when considering random initial conductivities (Figure \ref{fig:maze_random}), only one path remains at the end, since the symmetry of the system is broken. In this case, the simulations showed that after the dead-end cutting, the system seems to converge to the same degenerated steady state, but soon two of the redundant branches starts to disappear, and only one path survives. The path which is selected depends exclusively on the initial distribution of conductivities (Figure \ref{fig:maze_all_paths}).

Further simulations showed that if the square symmetry is broken by perturbing the position of the nodes, only one path remains regardless of the initial conditions since the path lengths are different. The dynamics of the model can thus account for the \textit{Physarum} maze solving ability. Another interesting thing to test would be to simulate the model on a 2D discretisation of the maze, by not ignoring the width of the maze segments. This would introduce more variability in the possible solutions and could accommodate the natural meandering of the \textit{Physarum} networks observed in Figure \ref{fig:maze}.

\section{Finding the Shortest Path}

Miyaji and Ohnishi \cite{Miyagi2008} proved mathematically that for any initial planar graph embedded in a two-dimensional surface,
the \textit{Physarum Solver} model \eqref{eq:Tero_update_rule_1} with the particular choice of $f(|Q_{ij}|) = |Q_{ij}|$ and $\mu=1$ always converges to the shortest-path connecting the two terminals regardless of the initial conductivities. This result was later generalised by Bonifaci et al. \cite{Bonifaci2013} for any graph topology, considering the same choice of parameters. We investigated if this holds for our model. For that, we have considered the adaptation dynamics with the choice of $g_\gamma(|Q_{ij}|) = |Q_{ij}|^\gamma$, i.e., the class of models \eqref{eq:new_adapt_rule_gamma}, and tested if, in the case of only two terminals, the system converges always to the shortest path for a certain value of $\gamma$. 

Extensive simulations were carried in planar graphs with 300 nodes resulting from a triangulation of a square region, for 20 different values of $\gamma$ in the range $[0.55,2]$. No values below $\gamma=1/2$ were considered since the previous analysis of the phase transition showed that for those values the steady states have redundant connections, and therefore can never correspond to the shortest path solution. 

In each case, the source and the sink were placed in the diagonally opposite corners of the square bounding region, to maximise the variability of paths, the dynamics could choose from. To prevent that any path is initially favoured, the simulations started always from an initial state of homogeneous conductivities, $D_{ij}(0)=1$. 

\begin{figure}[!hbt] 

\newcommand{\mysub}[2][]{%
    \subfloat[#1]{\includegraphics[trim={0cm 0cm 0cm 0cm}, clip, width=0.47\textwidth]{#2}}%
}

\centering

\mysub[\label{fig:SP_prob}]{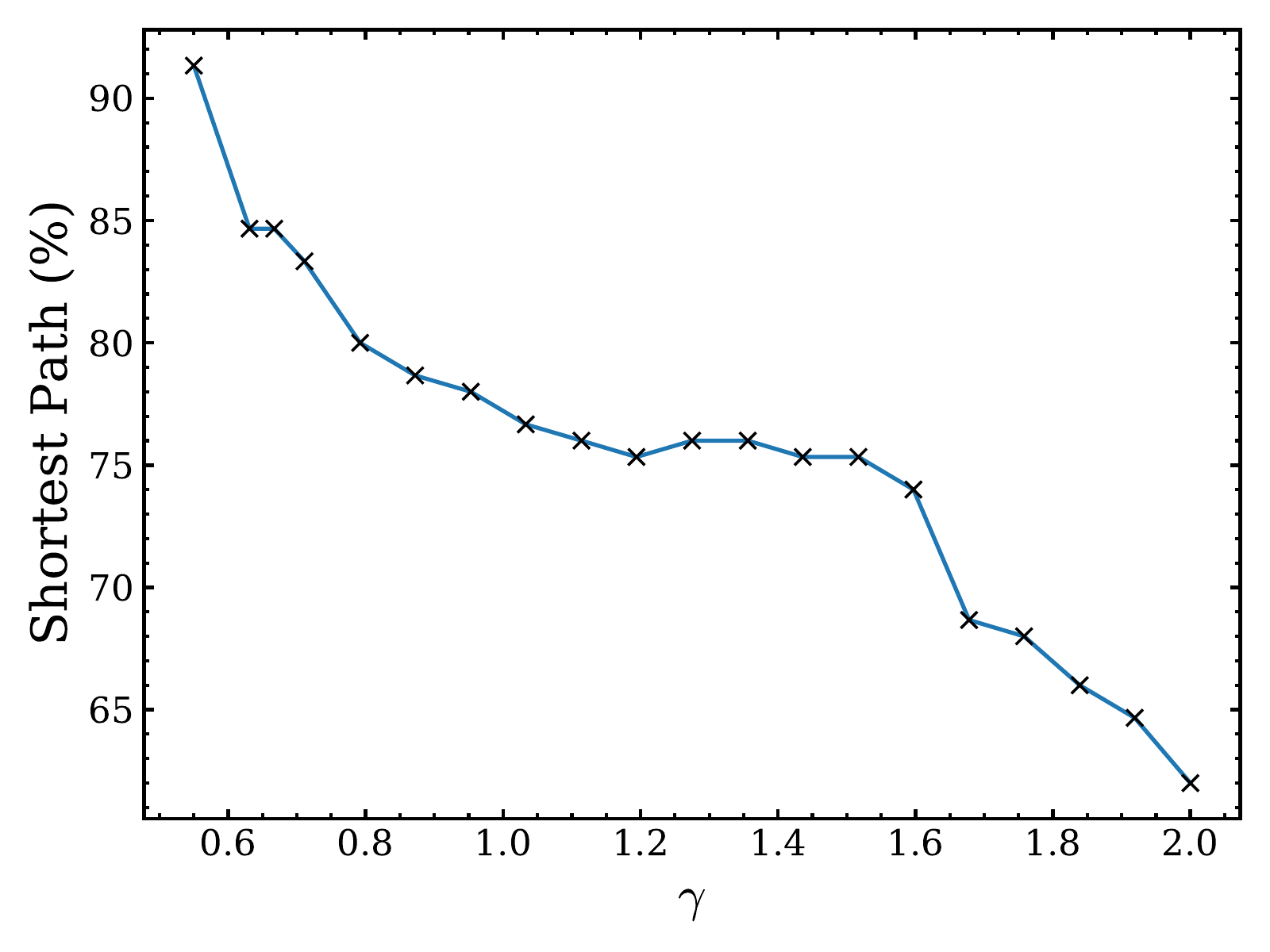} \hfill
\mysub[\label{fig:SP_L_error}]{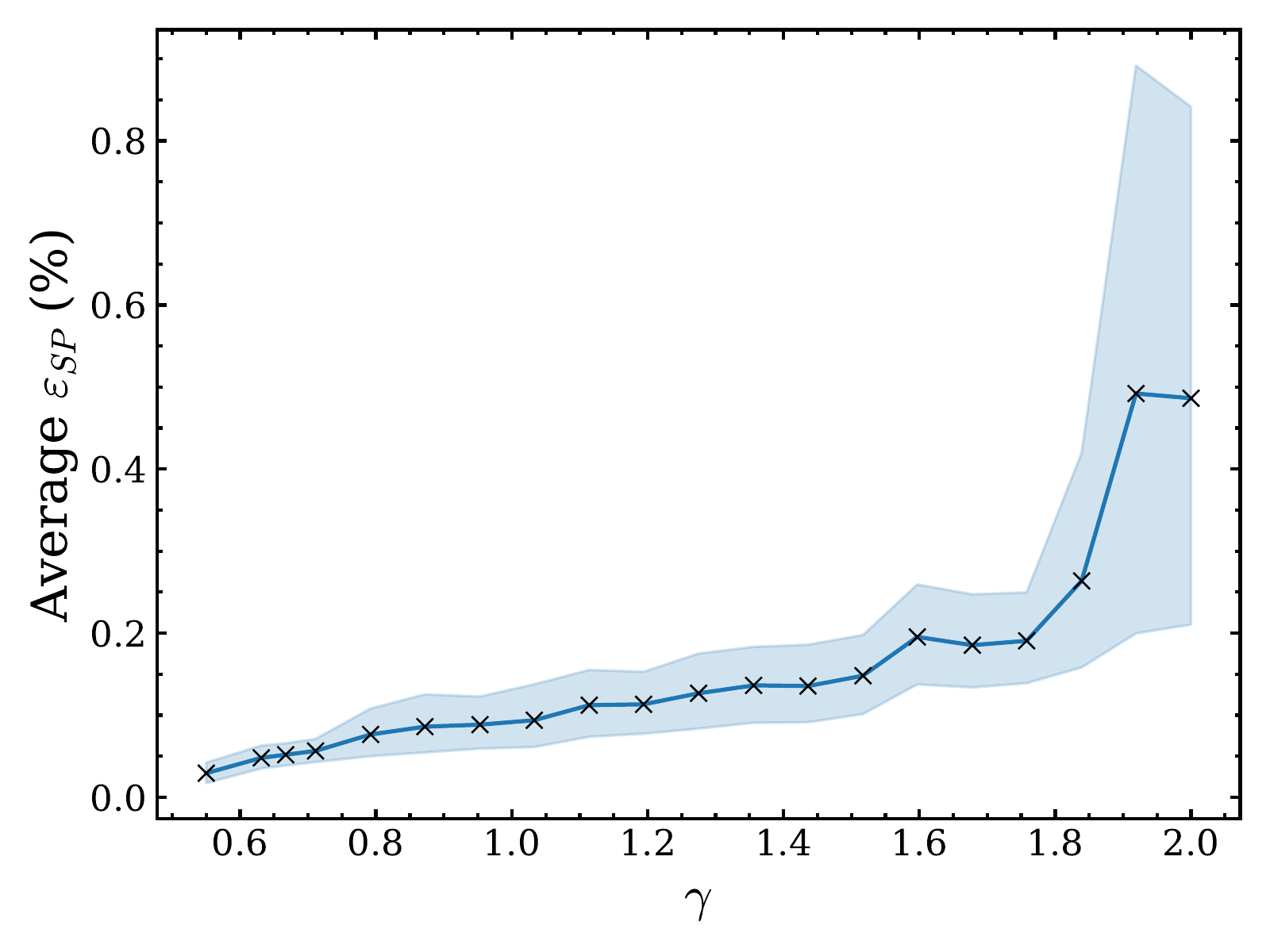} 

\caption{ 
\textbf{(a)} Probability of the system converging to the shortest path that connects two terminals as a function of the parameter $\gamma$, considering the dynamics \eqref{eq:new_adapt_rule_gamma}. \textbf{(b)} Average percentage error of the steady-state total path length, $L$, relatively to the length of the shortest path solution, $L_{SP}$, as a function of $\gamma$, i.e., $\varepsilon_{SP}=(L - L_{SP})/L_{SP}$. The computation of the average error only includes the final states different from the shortest path. The blue shaded region corresponds to the $95\%$ confidence interval of the mean error.  
The results are based on 150 realisations for each $\gamma$ value, each realisation corresponding to a different initial mesh with 300 nodes. The source and the sink were always placed in diagonally opposite corners of the square. In all the cases, the system starts from homogeneous distribution of conductivities, $D_{ij}(0)=1$.
}
\label{fig:shortest_path_gamma} 
\end{figure}

In Figure \ref{fig:SP_prob}, it's plotted the probability of the system to converge for the shortest path as a function of $\gamma$,  based on 150 realisations for each $\gamma$, each corresponding to a  different initial mesh. The results show that the probability tends to decrease as $\gamma$ increases. In particular, for $\gamma = 2/3$ the probability is near $85\%$. However, there isn't a single value of  $\gamma$ which guarantees that the final solution is always the shortest path. Nevertheless, for all values of $\gamma$ the deviations of the total length of the steady state from the length of the shortest path solution are on average very small. This is shown by the results of Figure \ref{fig:SP_L_error}, where it's plotted the relative error of the total length as a function of $\gamma$, averaged between the simulations where the system didn't converge to the shortest path. One can observe that the deviations tend to increase with $\gamma$, but are always below $1\%$, showing the solution is on average very close to the shortest path.  More tests  should be performed to validate the high values of probabilities and small deviations obtained, by considering larger graphs, smaller time steps and a higher number of realisations. In particular, it would be interesting to test values near $\gamma = 0.55$, where the probability is above $90\%$.

\section{Approximating Portugal's Rail System}

The topology of a biological network has a great impact on its performance and vulnerability, and thus on the chances of survival of the organism. The network structure should provide an effective distribution of the resources (transport efficiency) by ensuring a small average path length between any two points. However, the total cost to build the network, which is proportional to its overall length, should be minimised due to the limited amount of available resources. Thus, the transport efficiency should be maximised under the constraint of low cost. On the other hand, the network must be robust enough, by providing secondary paths that ensure its normal functioning in case of damage or random failure of some links. Once again, the number of redundant pathways have to be balanced with the additional cost of producing them. Therefore, the design of an optimal network requires a complex trade-off between the production cost, transport efficiency and fault tolerance. 

As seen in the section \ref{section:Tokyo_experiment}, a famous experiment carried by Tero et al. in 2010 showed that \textit{Physarum} builds networks with a good compromise between these three metrics and whose values are comparable to those of real-world infrastructure  networks, in particular, to the Tokyo railway system. Motivated by this experiment, we have simulated the adaptation dynamics \eqref{eq:new_adapt_rule_gamma} considering a mesh with the shape of mainland Portugal, for different values of $\gamma$ and other model parameters,and compared the results with the Portuguese railway system. 

\begin{figure}[hbt] 
    \centering
    \includegraphics[width=0.33\textwidth]{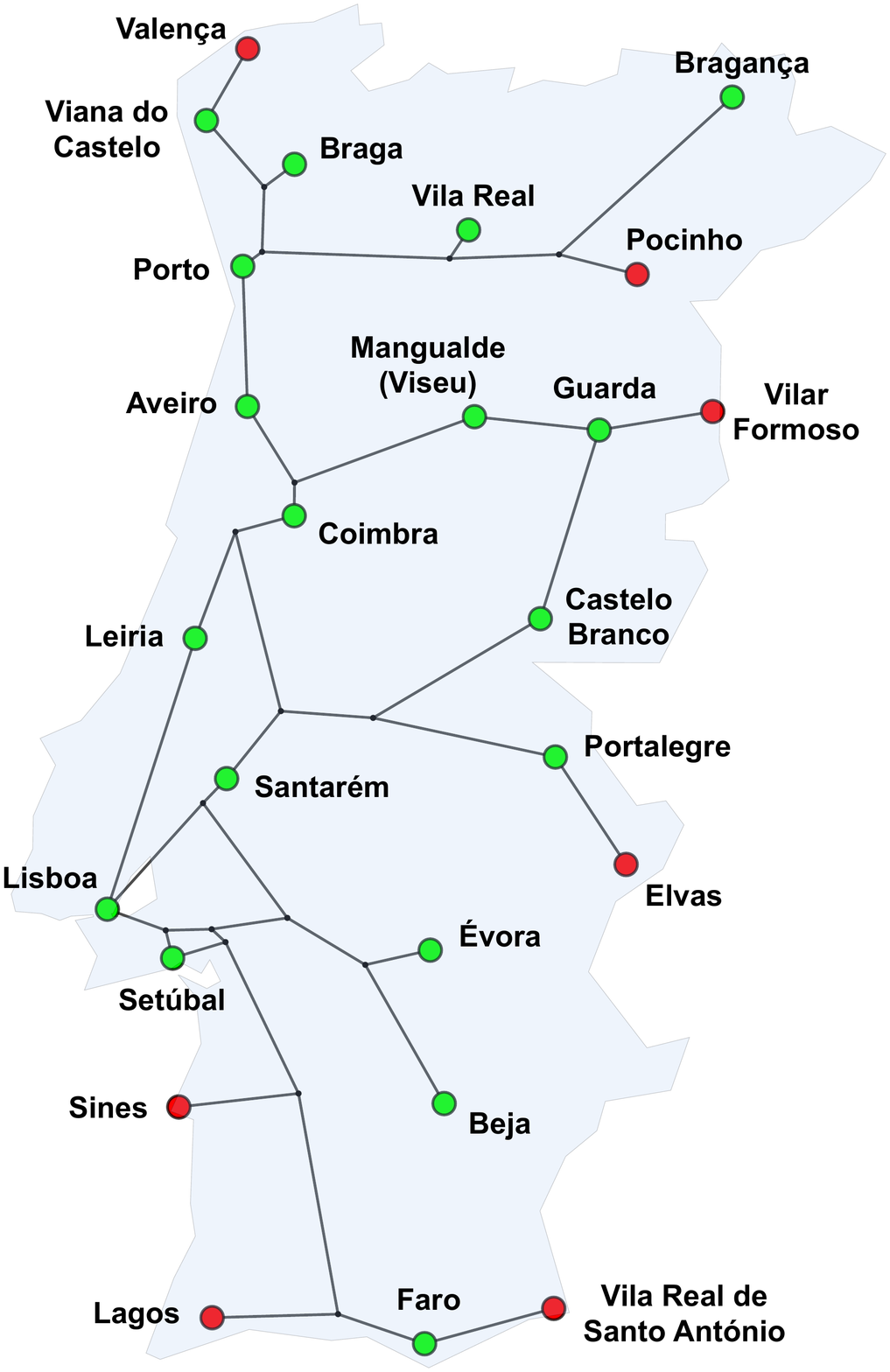}
    \caption{Approximation of the part of the real Portuguese rail network used in this study, which connects all the 18 district capitals (green nodes) and 5 additional terminal cities (red nodes). Note that Mangualde was used to represent the capital district Viseu. Some rail lines are no longer active but were included for consistency.}
    \label{fig:railway_cities_labels}
\end{figure}

The boundary of Portugal was approximated by a polygon, followed by a triangulation of the enclosed region which resulted in an initial network with 1005 nodes and 2817 edges. We have considered a configuration of 25 terminals, representing the geographical locations of the 18 Portuguese  district capitals and 7 additional major cities, except for the Viseu district which was represented by the city Mangualde for convenience. 
The optimised networks resulting from the dynamics \eqref{eq:new_adapt_rule_gamma} were compared with the section of the Portuguese rail network which connects those cities  \cite{rede_CP} as depicted in Figure \ref{fig:railway_cities_labels}. Note that the rail lines connecting Bragança to Tua (\textit{Linha do Tua}, 1887–2008) and Vila Real to Régua (\textit{Linha do Corgo}, 1906-2009) are no longer active, but were included for comparison purposes.

The initial graph is assumed to be embedded in the Earth's surface, meaning that the position of the node  $i$ is defined by its geographical coordinates: latitude $\lat_i$ and longitude $\lon_i$. Consequently, the length of the edge $(i,j)$, $L_{ij}$, is given by the geodesic distance between the two end-node positions, which can be approximated using the Haversine formula, yielding

\begin{equation}
    L_{ij} = 2 R_E \arcsin{\left(\sqrt{
    \sin^2\left(\frac{\lat_i - \lat_j}{2}\right) +
    \cos(\lat_i)\cos(\lat_j)\sin^2\left(\frac{\lon_i - \lon_j}{2}\right) }\right)} \;,
    \label{eq:haversine}
\end{equation}

\noindent where $R_E =\SI{6371}{km}$ is the Earth's radius. 

The performance of the optimised networks was evaluated in terms of cost, transport efficiency and fault tolerance. The total cost of producing the network is measured by its total length (TL) 

\begin{equation}
    \text{TL} = \sum_{(i,j)\in E} L_{ij}
    \label{eq:total_length} \;,
\end{equation}

\noindent where the lengths of the edges, $L_{ij}$, are given by \eqref{eq:haversine}. Note that in the case of biological networks the cost of producing the networks should be defined more accurately by the total section area of the channels. However, here we are  interested in applications to road and rail transportation, where the costs for the construction company are proportional to the total network length. 

On the other hand, from the perspective of a traveller, an optimal network should ensure a fast travel between two destinations by minimising the distance between them. Thus, the transport efficiency (TE) is defined as the inverse of the average minimum distance (MD) between all the distinct pairs of terminal cities. If $d_{SP}(i,j)$ denotes the length of the shortest path in the graph connecting the nodes $i$ and $j$, and $T$ is the set of the $|T|$ terminals, then 

\begin{equation}
    \text{TE} = \text{MD}^{-1} =
    \left( \frac{2}{|T|(|T|-1)} 
    \sum_{\substack{i < j \\i,j\in T}}
    d_{SP}(i,j) \right)^{-1} \;.
    \label{eq:transport_efficiency}
\end{equation}

Lastly, the robustness of the networks is measured by its fault tolerance (FT), which is defined as the probability of the network remaining connected after a single edge is removed. The probability of disconnecting the network equals its fraction of edges which are \textit{bridges}\footnote{In graph theory, a \textit{bridge} is any edge whose removal increases the number of connected components of the graph.}. Thus, FT is given by

\begin{equation}
    \text{FT} = 1 - \frac{b}{|E|} \;,
    \label{eq:fault_tolerance}
\end{equation}

\noindent where $b$ is the number of \textit{bridges} of the final network. Note that, as before, only the edges with conductivities above the threshold, $D_{thr}=5\times 10^{-4}$ are considered in the computation of the three metrics of the steady-state networks.

\begin{figure}[hbt] 
    \centering
    \captionsetup[subfloat]{justification=centering}
    \subfloat[\label{fig:railway_graph}Railway \\
    \textbf{(1384, 313, 0.45)}
    ]{\includegraphics[width=0.2\textwidth]{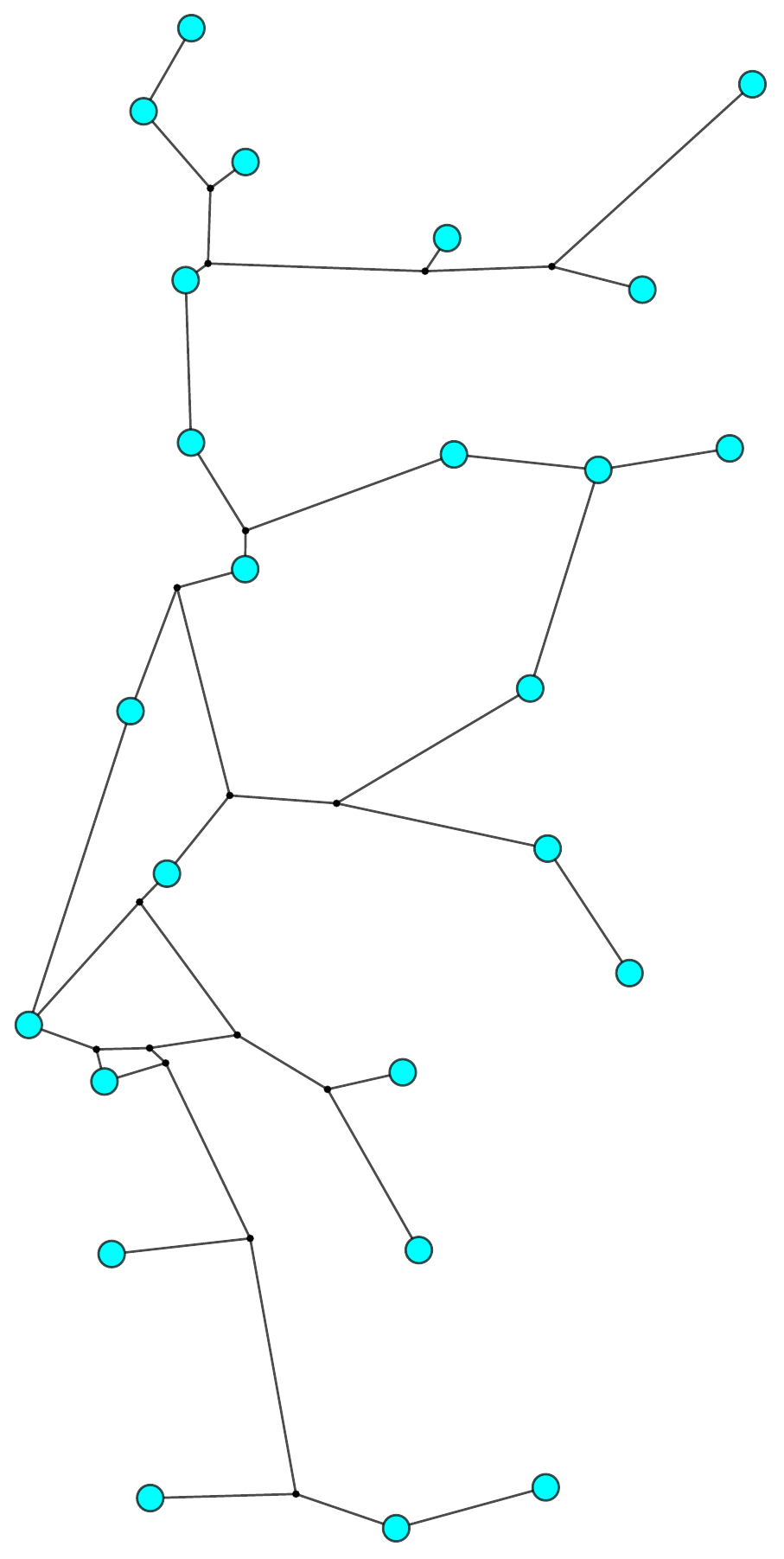}} 
    \hspace{0.1\textwidth}%
    \subfloat[\label{fig:MST_graph}MST \\
    \textbf{(1848, 417, 0.00)}
    ]{\includegraphics[width=0.2\textwidth]{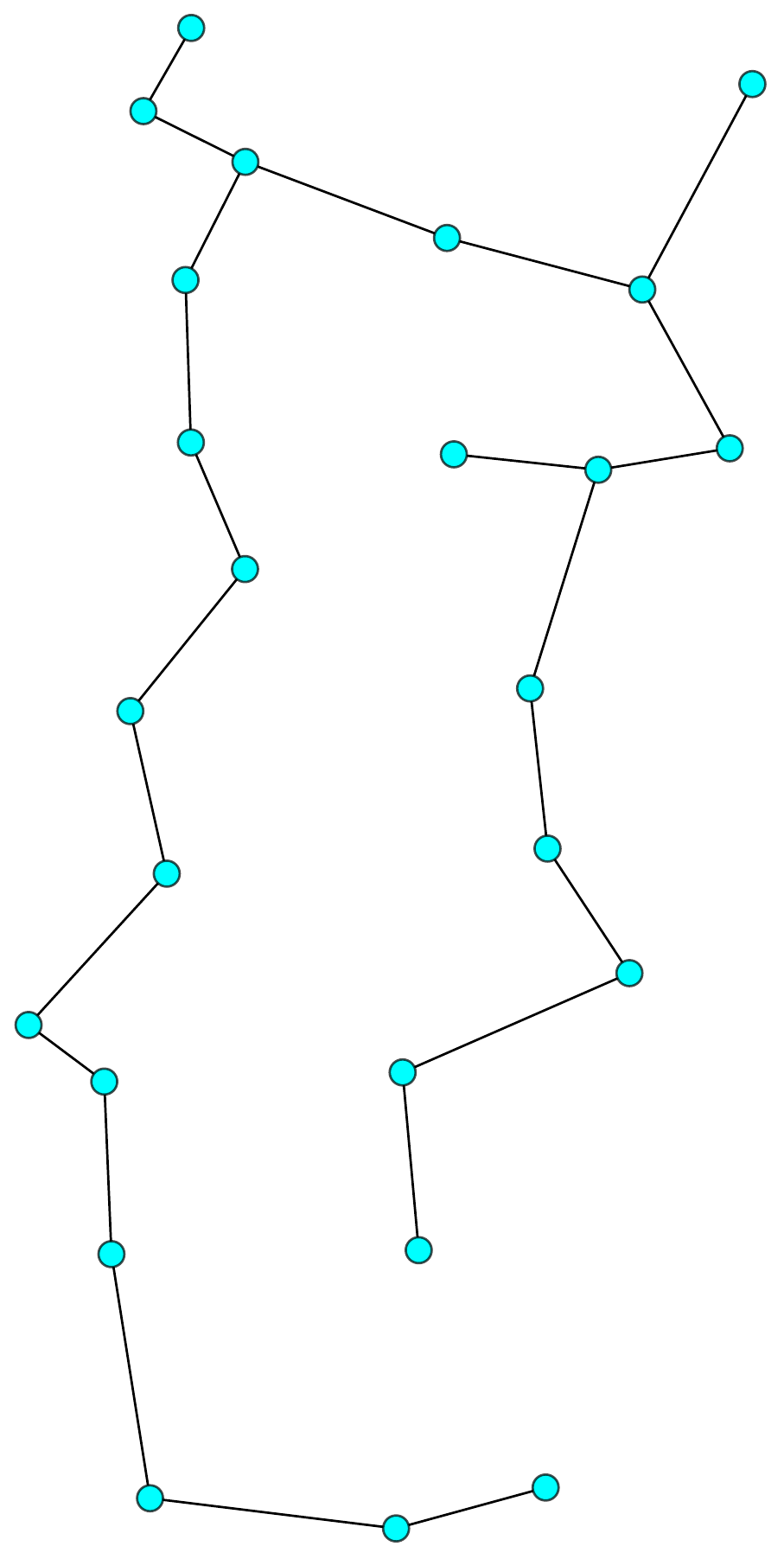}} 
    \hspace{0.1\textwidth}%
    \subfloat[\label{fig:CG_graph}CG \\
    \textbf{(68009, 227, 1.00)}
    ]{\includegraphics[width=0.2\textwidth]{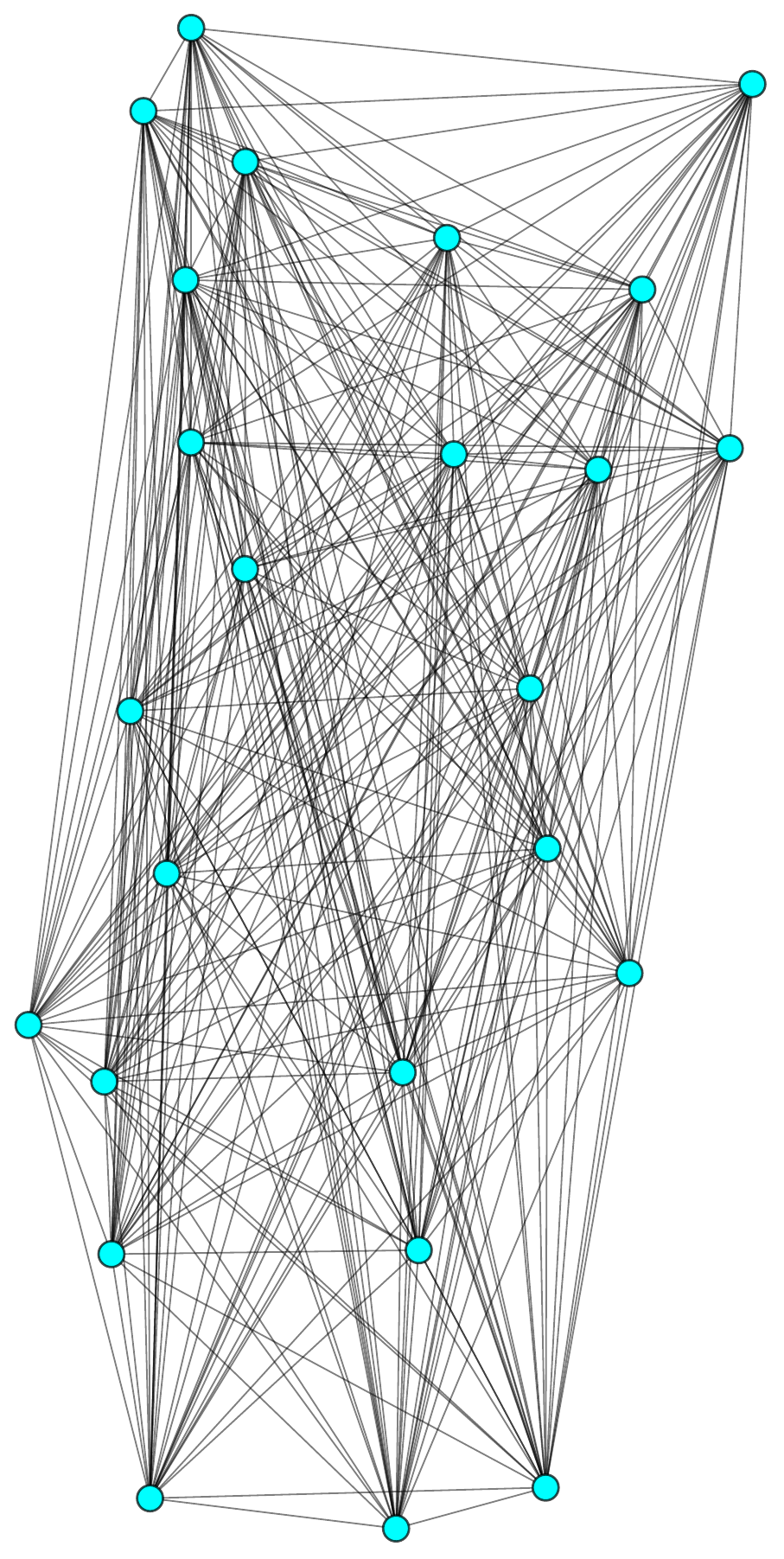}} 
    \caption{ 
    \textbf{(a)} Graph of Portugal's rail system (Figure \ref{fig:railway_cities_labels}) with all the city nodes (terminals) marked in blue. \textbf{(b)}  The minimum spanning tree (MST) connecting the same set of cities, i.e., the \textit{tree} graph spanned by the city nodes with the minimal possible total length \eqref{eq:total_length}.
    \textbf{(c)} The complete graph (CG) connecting every pair of cities by a distinct edge. The legend of each graph refers to the metrics \textbf{(TL, MD, FT)}, where TL and MD are given in kilometres.}
    \label{fig:Portugal_graphs}
\end{figure}

Besides the actual railway system (Figure \ref{fig:railway_graph}), the performance of the final networks was also compared with those of the minimal spanning tree (MST)  and the complete graph (CG) spanned by the city nodes. The MST\footnote{The (geometric) minimum spanning tree (MST) spanned by a set of points (terminals) embedded in a manifold is the graph that connects all the terminals together by geodesic lines, without any cycles, such that the total length of the lines is minimised. Note, however, if other additional nodes are allowed (Steiner points), the graph with the minimum total length which connects all the terminals, either directly or via the Steiner points, is in general different, and it's known as the (geometric) minimum Steiner Tree.} (Figure \ref{fig:MST_graph}) is by definition the graph that connects all the city positions with minimal possible total cost \eqref{eq:total_length}, while the CG (Figure \ref{fig:CG_graph}) is the graph that connects every pair of cities by a distinct edge, maximising, therefore, the transport efficiency \eqref{eq:transport_efficiency} and the fault tolerance (FT$=1$) at the expense of a tremendous cost. The cost (TL), transport efficiency (TE) and fault tolerance (FT) of the final networks were normalised to the corresponding values for the CG, yielding \CG{TL}, \CG{TE}, \CG{FT}. A comparison between the metrics of these graphs is presented in Table \ref{tab:graph_metrics}. To compare the overall performance, the trade-off between the transport efficiency, fault tolerance and the cost was captured by two benefit-cost measures, defined as the ratios BCR\textsubscript{TE}$=\CG{TE}/\CG{TL}$ and BCR\textsubscript{FT}$=\CG{FT}/\CG{TL}$.

\begin{table}[bht!]
    \centering
\begin{tabular}{r|cccc|cccc|c}
\toprule
   Graph &  TL (km) &  TL\textsubscript{CG} &  TL\textsubscript{MST} &  TL\textsubscript{railway} &  TE (km$^{-1}$) &  TE\textsubscript{CG} &  TE\textsubscript{MST} &  TE\textsubscript{railway} &    FT \\
\midrule
    mesh &    31482 &                  0.46 &                  22.75 &                      17.04 &          0.0042 &                  0.95 &                   1.75 &                       1.31 &  1.00 \\
 railway &     1848 &                  0.03 &                   1.34 &                       1.00 &          0.0032 &                  0.73 &                   1.33 &                       1.00 &  0.45 \\
     MST &     1384 &                  0.02 &                   1.00 &                       0.75 &          0.0024 &                  0.55 &                   1.00 &                       0.75 &  0.00 \\
      CG &    68009 &                  1.00 &                  49.14 &                      36.80 &          0.0044 &                  1.00 &                   1.83 &                       1.38 &  1.00 \\
\bottomrule
\end{tabular}
    \caption{Comparison between the initial mesh (mesh), real railway (railway), minimum spanning tree (MST) and complete graph (CG) in terms of the total length (TL), transport efficiency (TE) and fault tolerance (FT). The columns $X_{G}$ denotes the metric $X$ of the graph of each row  normalised by the one of the graph $G$. 
    }
    \label{tab:graph_metrics}
\end{table}

Until now we have been always considering the adaptation under a fixed set of sources and sinks.The previous analysis revealed that this either resulted in tree-like networks ($\gamma > 1/2$) with zero fault tolerance, or poorly-optimised networks similar to the initial mesh, ($\gamma < 1/2 $) which aren't cost-efficient. In both cases, this means that networks have an overall low performance, conversely to the ones built by \textit{Physarum}. To tackle this issue, we now introduce fluctuations in the flux  by considering time-dependent sources and sinks, similar to the original model \cite{Tero1}. At each step of the algorithm, two nodes are randomly selected from the set of terminals to drive the flow: one acts as a source with intensity $q_{source}=I_0$ ($I_0>0$) and the other as a sink $q_{sink}=-I_0$, while the remaining terminals have $q=0$. This emulates 
more closely the shuttle streaming characteristic of \textit{Physarum} networks, by changing the flux direction in each vessel over time, although not in an exactly periodic way.

Due to the high fluctuations of the channel fluxes induced by the stochastic choice of the terminals, the stopping condition considered in the case of fixed terminals \eqref{eq:stop_condition} is hardly ever met. Simulations showed that although the system converges to a stable network topology, the channel conductivities display unceasing oscillations which are larger the greater is the time step used, $\dt$, being hard to establish their bounds beforehand. Therefore for the stochastic case, we have considered a different stopping criterion. As we are mostly concerned with the topology of the steady state, we consider that the algorithm converges when the topology of the network remains unchanged in a period of 500 iterations, meaning that the set of edges with conductivities above a given threshold ($D_{thr}=5\times10^{-4}$) doesn't change over that period.
To regularise the effect of the fluctuations on the adaptation, a smaller time step was used, $\dt = 0.02$, corresponding therefore to a period of 10 time units without any changes in topology. This stopping criterion has some limitations since it depends on the choice of the number of iterations, which was carefully chosen according to the $\dt$ used and the overall the time scale of the adaptation mechanism.

\newpage

\subsection{Dependence of the performance on $\gamma$}

It was first analysed the dependency of the topology and performance of the final networks on the parameter $\gamma$ of the adaptation dynamics \eqref{eq:new_adapt_rule_gamma}. Numerous simulations were carried out for different values of $\gamma$, considering the same mesh, initial conditions ($D_{ij}(0)=1$) and seed. 

\begin{figure}[H] 

\captionsetup[subfloat]{labelformat=empty,position=bottom,skip=3pt}

\newcommand{\mysub}[2][]{%
    \subfloat[#1]{\includegraphics[trim={4.8cm 2cm 4.8cm 1.8cm}, clip, width=0.26\textwidth]{#2}}%
}

\setlength{\fboxrule}{1.5pt}
\newcommand{\mysubf}[2][]{%
    \subfloat[#1]{\includegraphics[trim={4.8cm 2cm 4.8cm 1.8cm}, clip, width=0.26\textwidth,valign=t,cfbox=green!80!black]{#2}}
}

\centering

\mysub[\textbf{\normalsize (0.45, 30031, 237, 1.00)}]{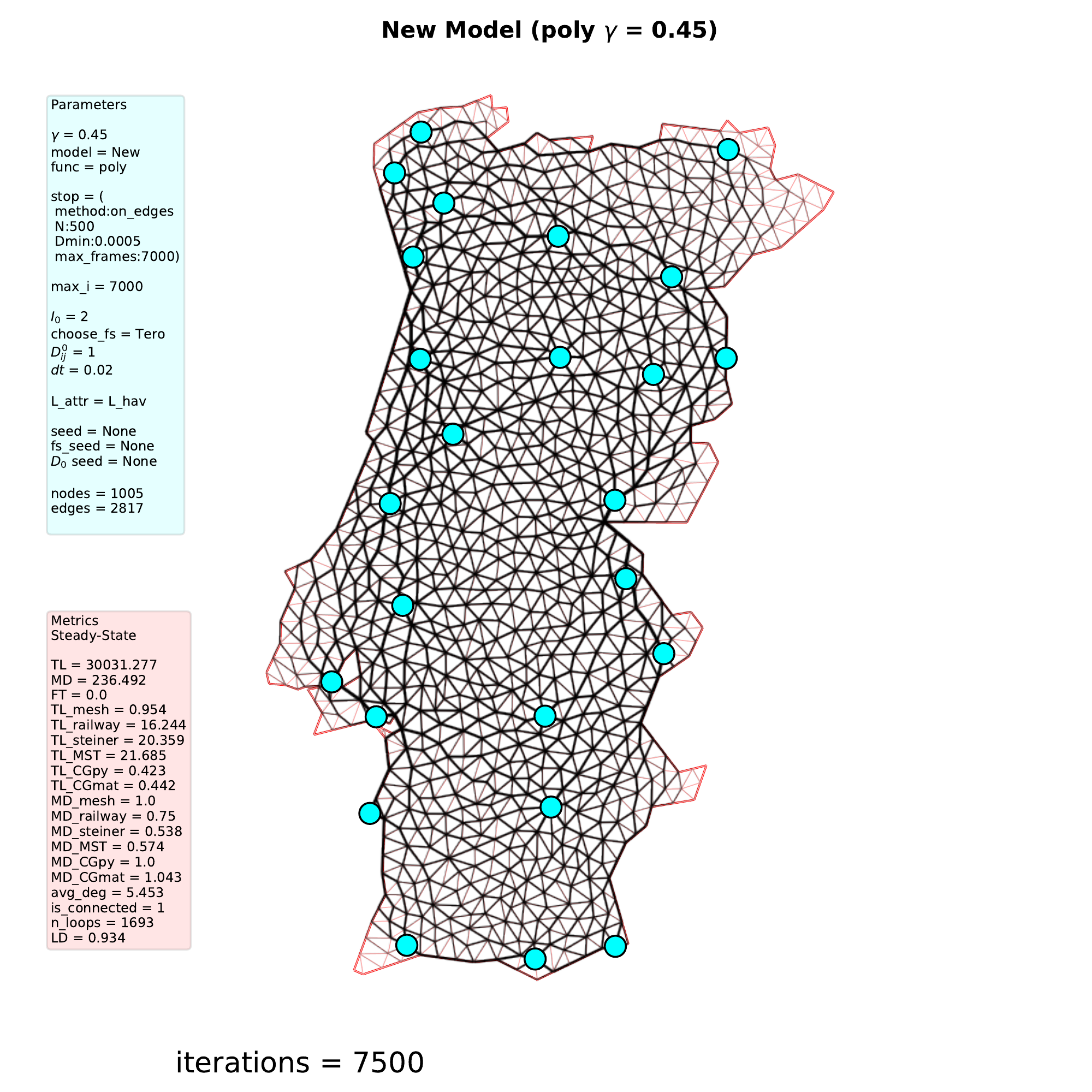} 
\hfill
\mysub[\textbf{\normalsize (0.55, 2803, 275, 1.00)}]{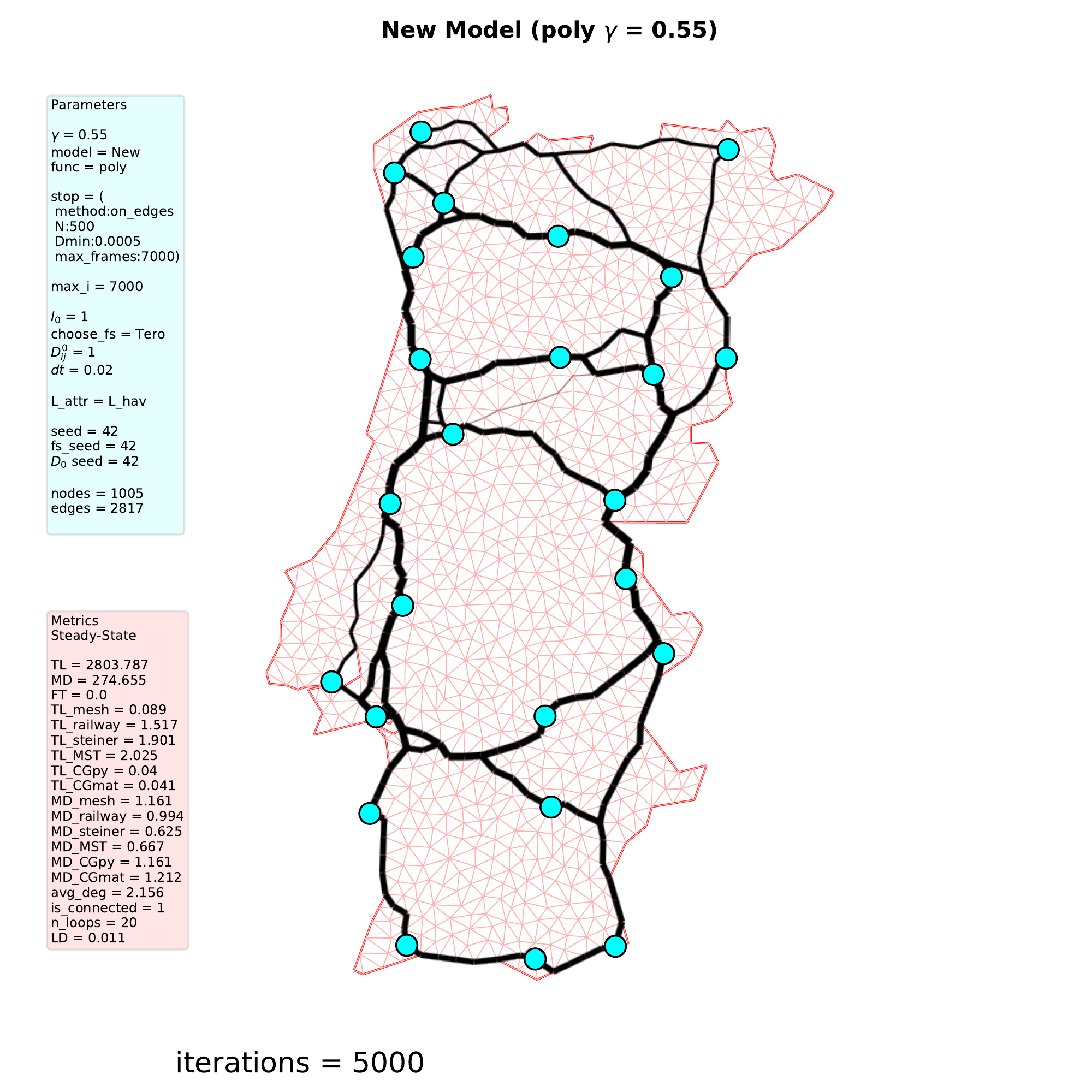} \hfill
\mysubf[\textbf{\normalsize ($\bm{2/3}$, 2124, 293, 0.98)}]{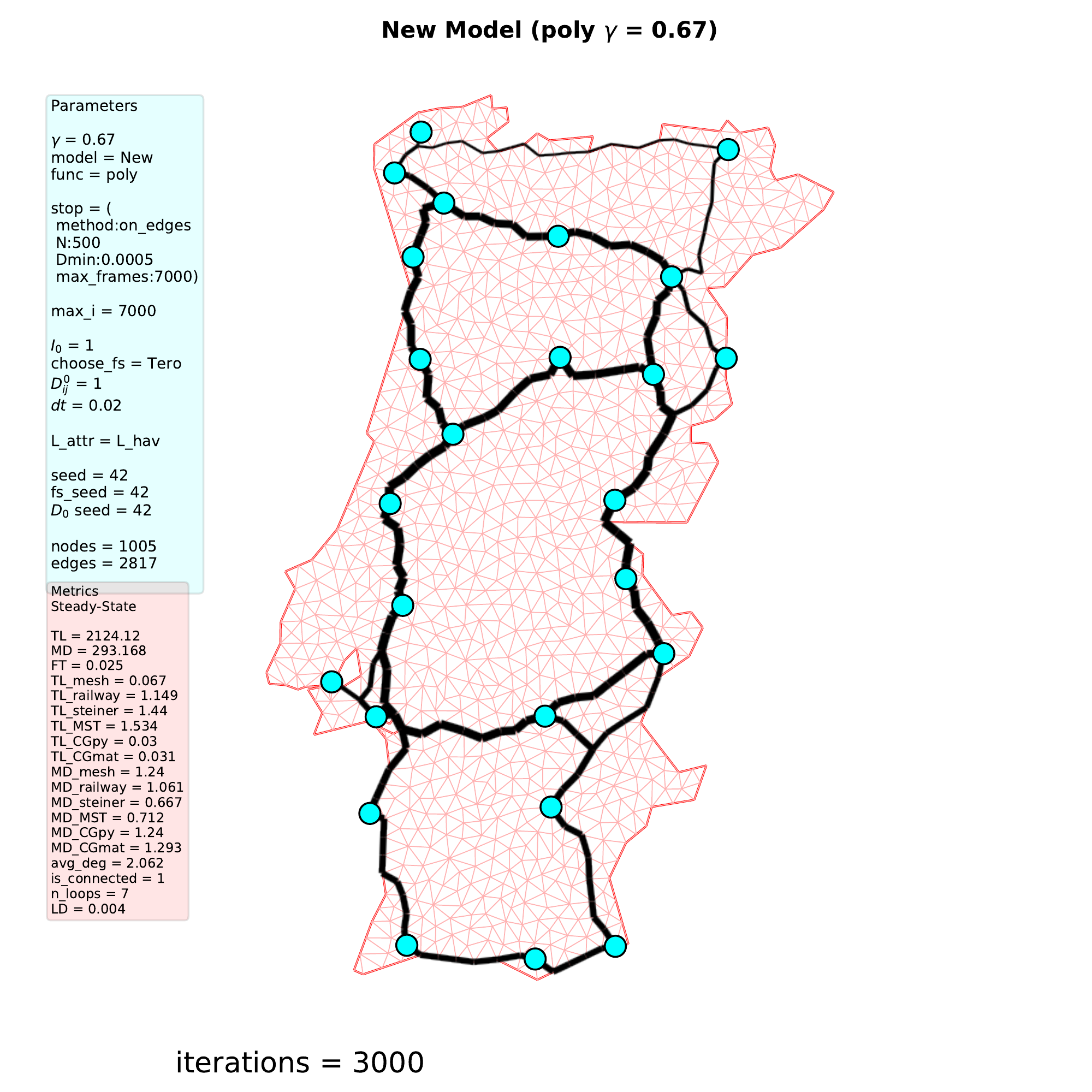} \\
\mysubf[\textbf{\normalsize (0.75, 1817, 305, 0.88)}]{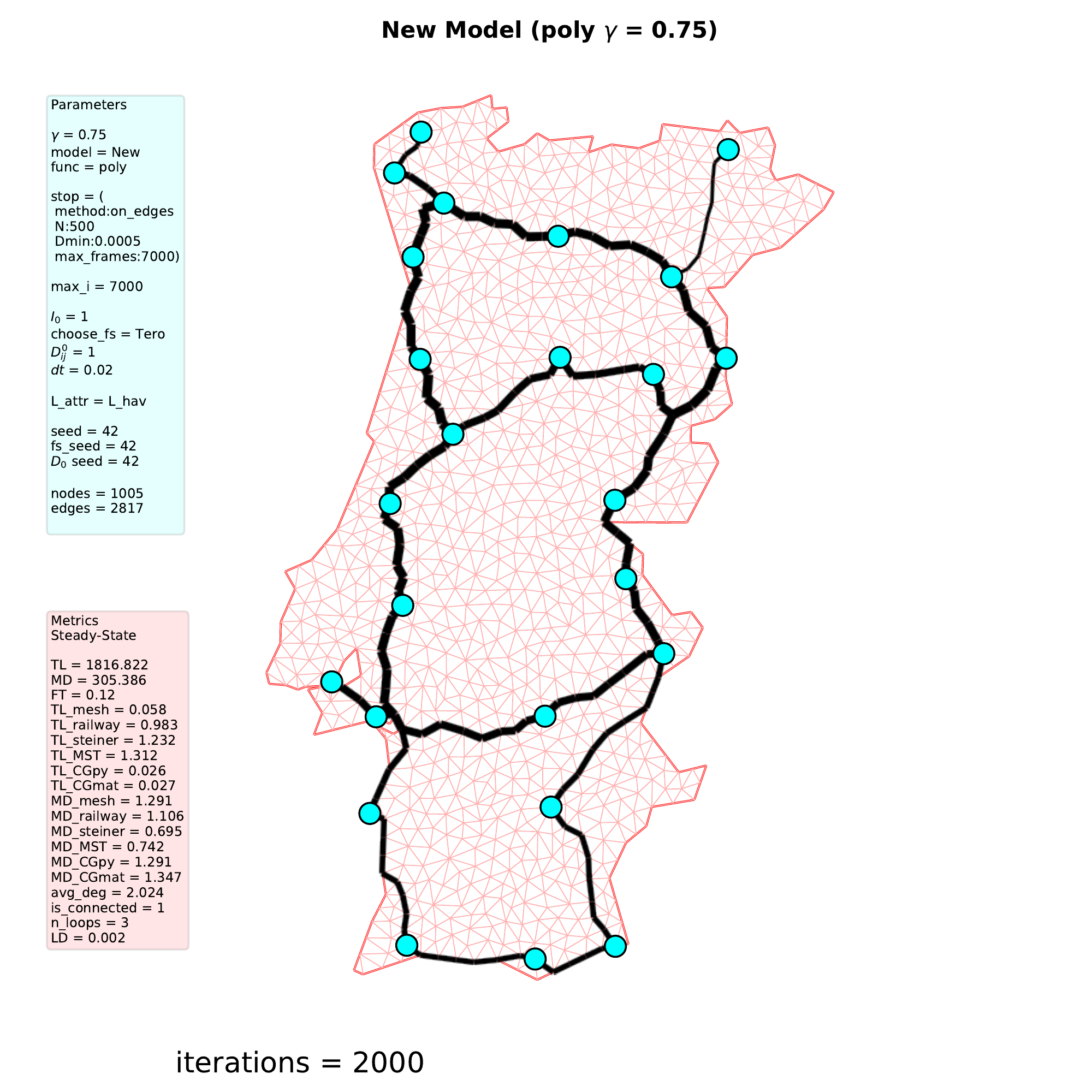} \hfill
\mysubf[\textbf{\normalsize (0.95, 1667, 323, 0.87)}]{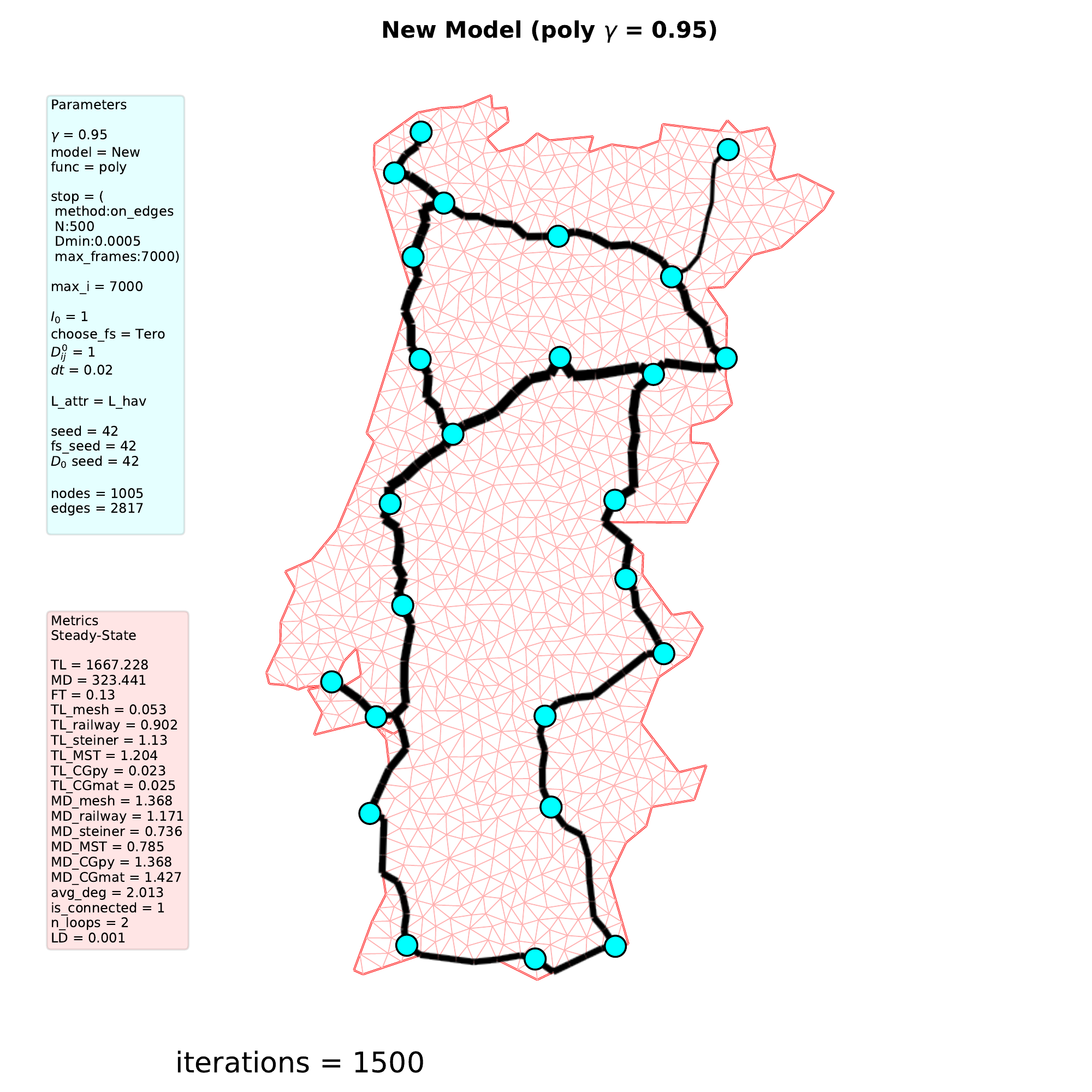} \hfill
\mysub[\textbf{\normalsize (1.05, 1566, 359, 0.35)}]{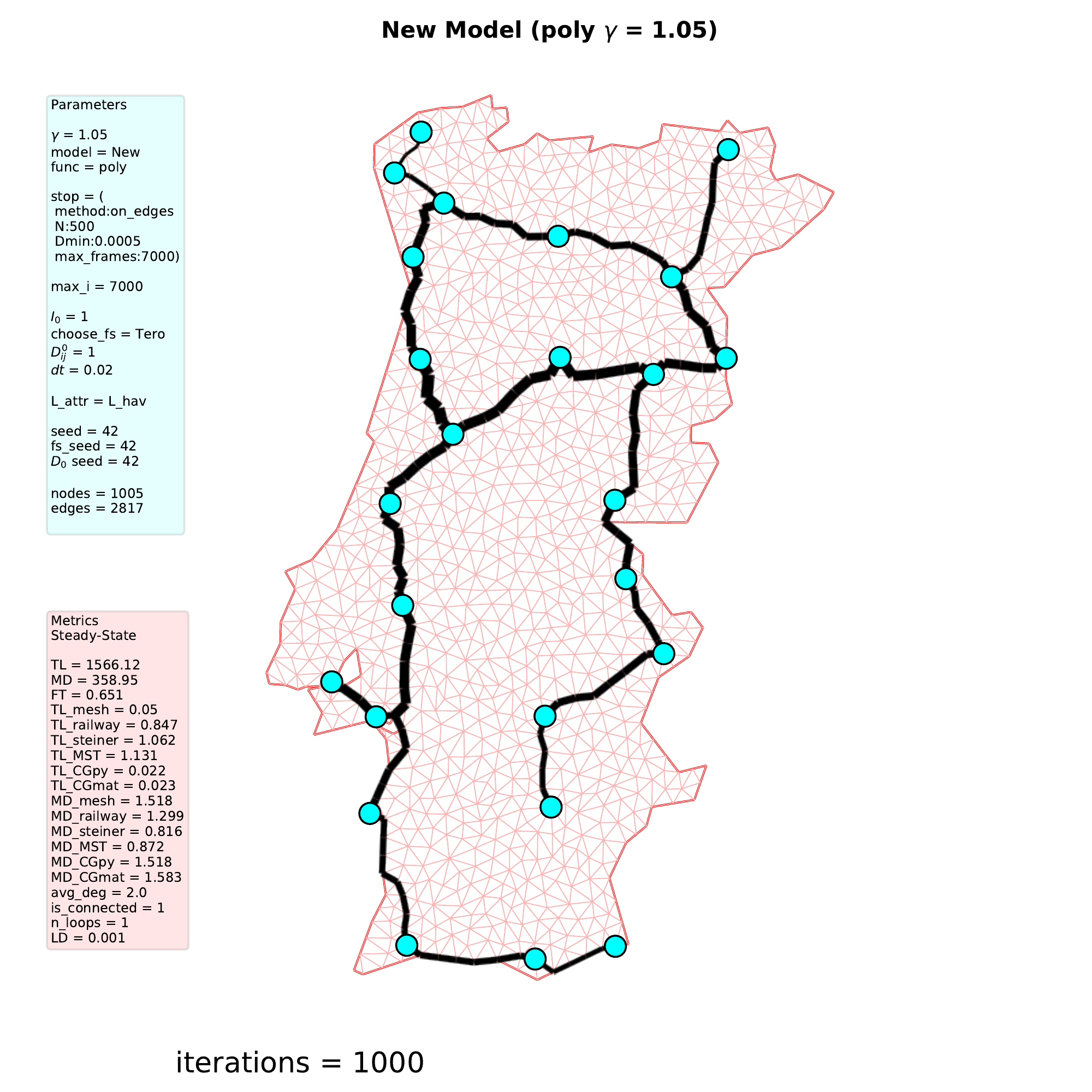}  \\
\mysub[\textbf{\normalsize (1.35, 1500, 379, 0.00)}]{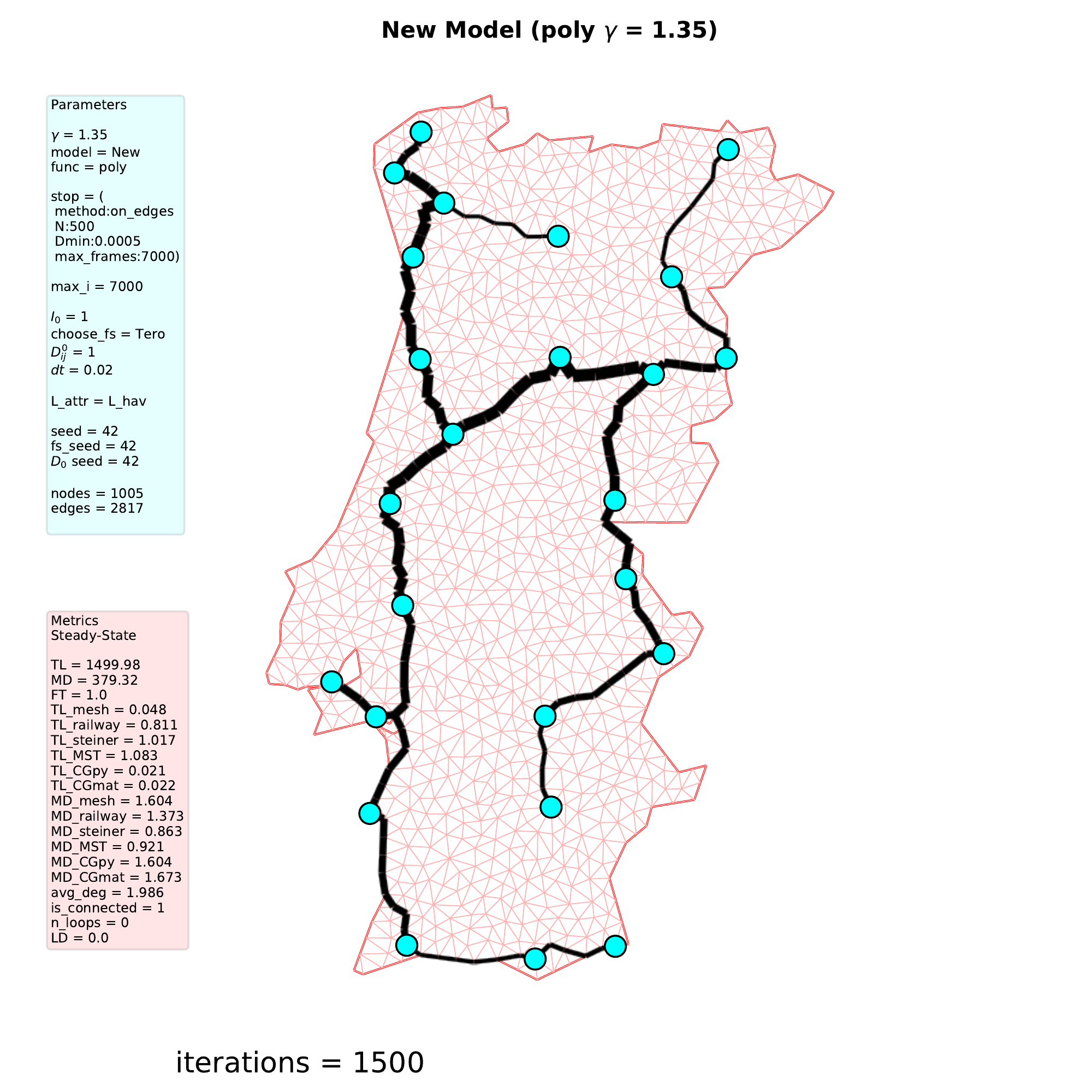} \hfill
\mysub[\textbf{\normalsize (1.55, 1500, 379, 0.00)}]{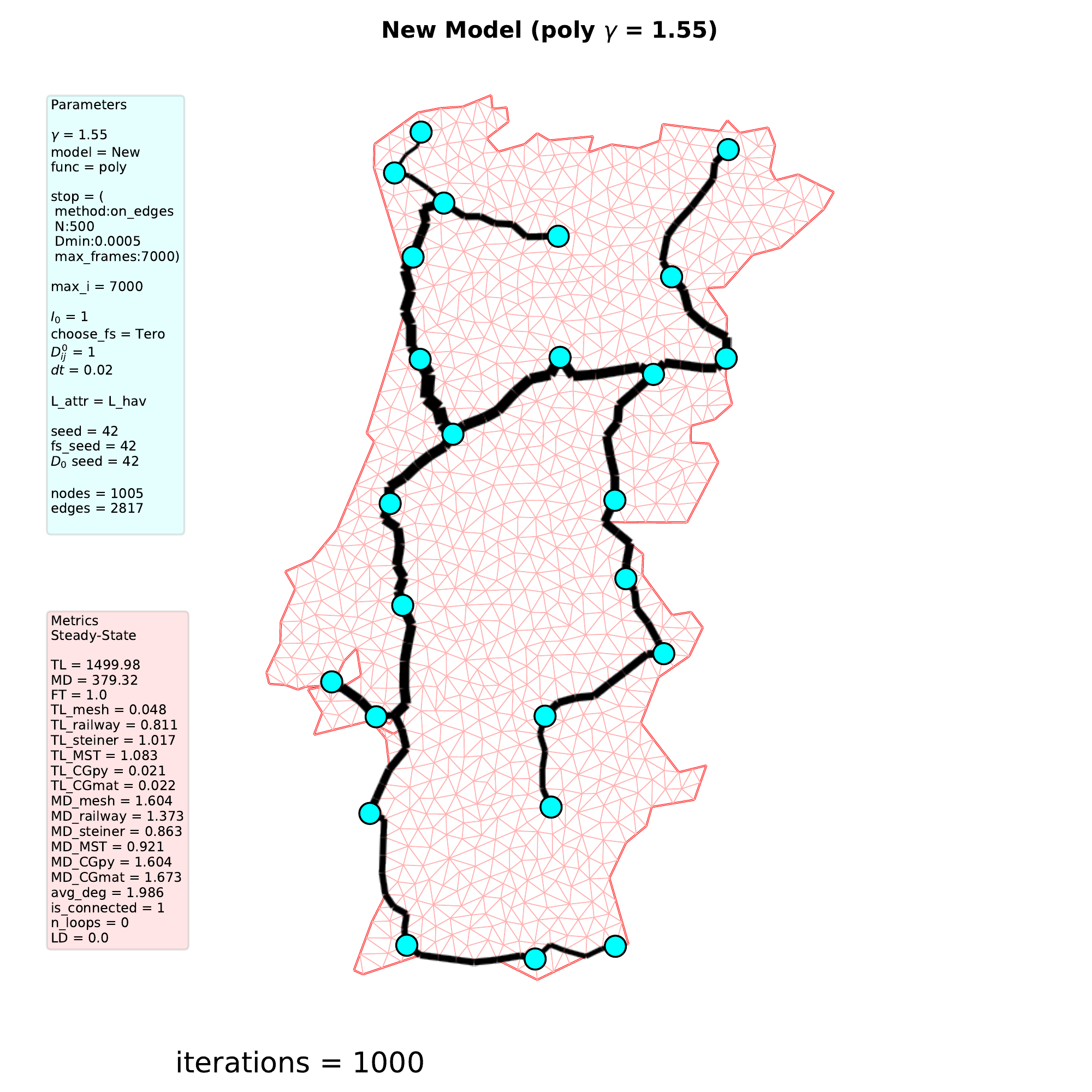} \hfill
\mysub[\textbf{\normalsize (1.95, 1489, 385, 0.00)}]{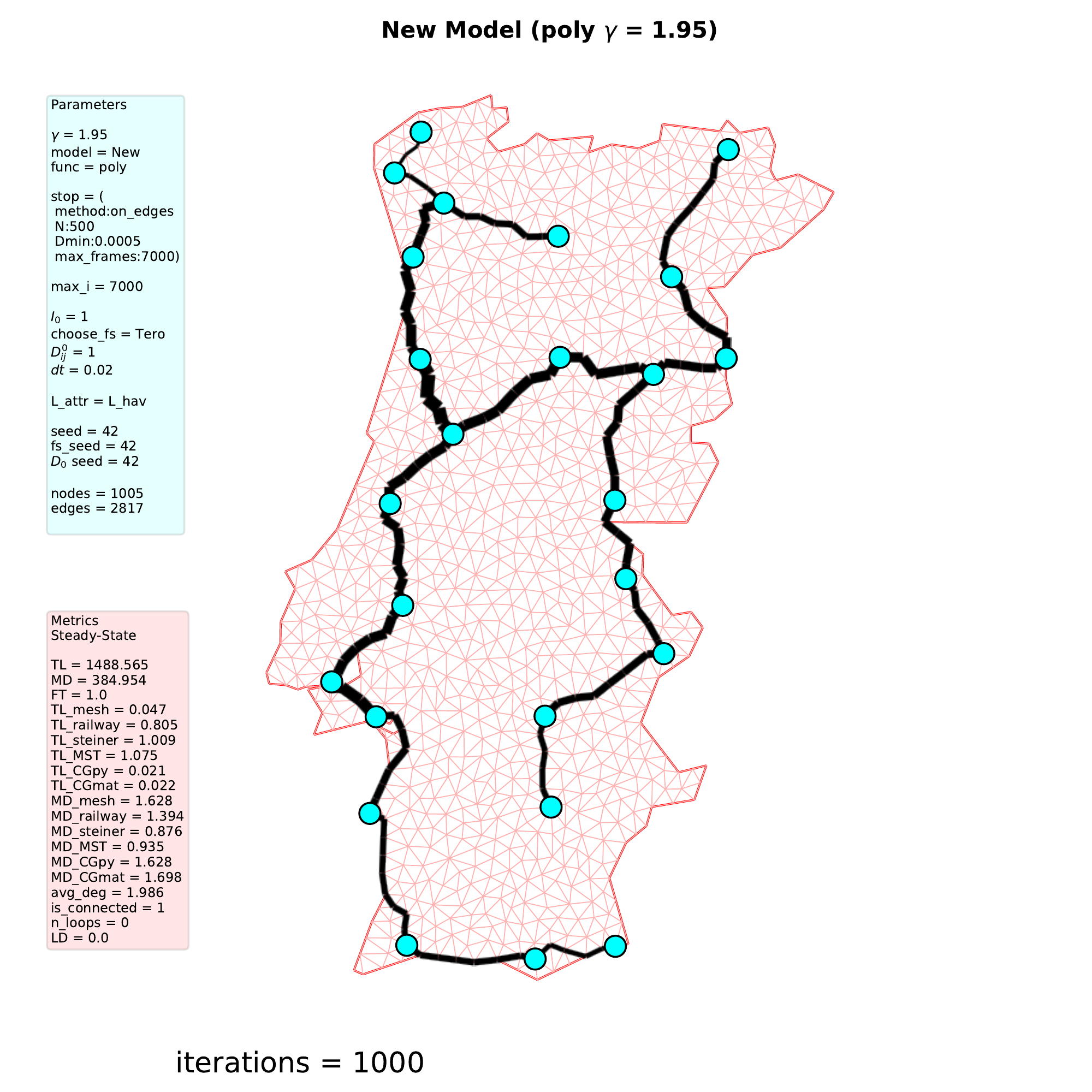}  \\

\caption{Topology of the networks resulting from the adaptation dynamics \eqref{eq:new_adapt_rule_gamma} as a function of the parameter $\gamma$, considering a stochastic choice of the source-sink pair from the set of terminals in each step of the algorithm ($I_0=1$ and $D_{ij}(0)=1$). For $\gamma < 1/2$ the dynamics result in poorly-optimised networks very close to the initial mesh, similar to the case of fixed terminals. As $\gamma$ is increased, the networks slowly evolve towards the minimum spanning tree (Figure \ref{fig:MST_graph}), losing all the redundant paths which provide robustness to the network. The legend of each image refers to the network metrics \textbf{($\bm{\gamma}$, TL, MD, FT)}, where TL and MD are given in kilometres. The top 3 networks with the best overall performance are highlighted in green. 
} 

\label{fig:portugal_gamma} 
\end{figure}

Some examples of the different networks reached by the system are given in Figure \ref{fig:portugal_gamma}. For low values of $\gamma$, the flux fluctuations result in the formation of stable redundant 
paths that improve the network's robustness. In particular, for $\gamma < 1/2$, similarly to the case of fixed terminals, most of the initial mesh remains and very few preferential pathways are formed, resulting in networks with a huge cost. A drastic change in the topology is observed near $\gamma = 1/2 $, characterised by a substantial reduction in the network's total length yet some alternative routes remain, leading to 
a better compromise between cost and fault tolerance. A good trade-off between the cost and the transport efficiency is also reached for low values of $\gamma$  due to the formation of additional bifurcation points, which resemble Steiner tree type of connections. As $\gamma$ is increased the redundant paths progressively disappear, and the system slowly converges towards the MST solution (Figure \ref{fig:MST_graph}). The minimisation of the cost is achieved with the inevitable complete loss of the network's robustness. Simulations also showed that for a given $\gamma$, the structure of the final network depended slightly on the random seed used, however these observations remain always valid. 

\begin{figure}[H] 

\newcommand{\mysub}[2][]{%
    \subfloat[#1]{\includegraphics[trim={0cm 0cm 0cm 0cm}, clip, width=0.495\textwidth]{#2}}%
}

\centering

\subfloat[\label{fig:gamma_TE_TL}]{\includegraphics[
trim={0cm 7cm 3.5cm 0cm}, clip, height=0.57\textwidth]{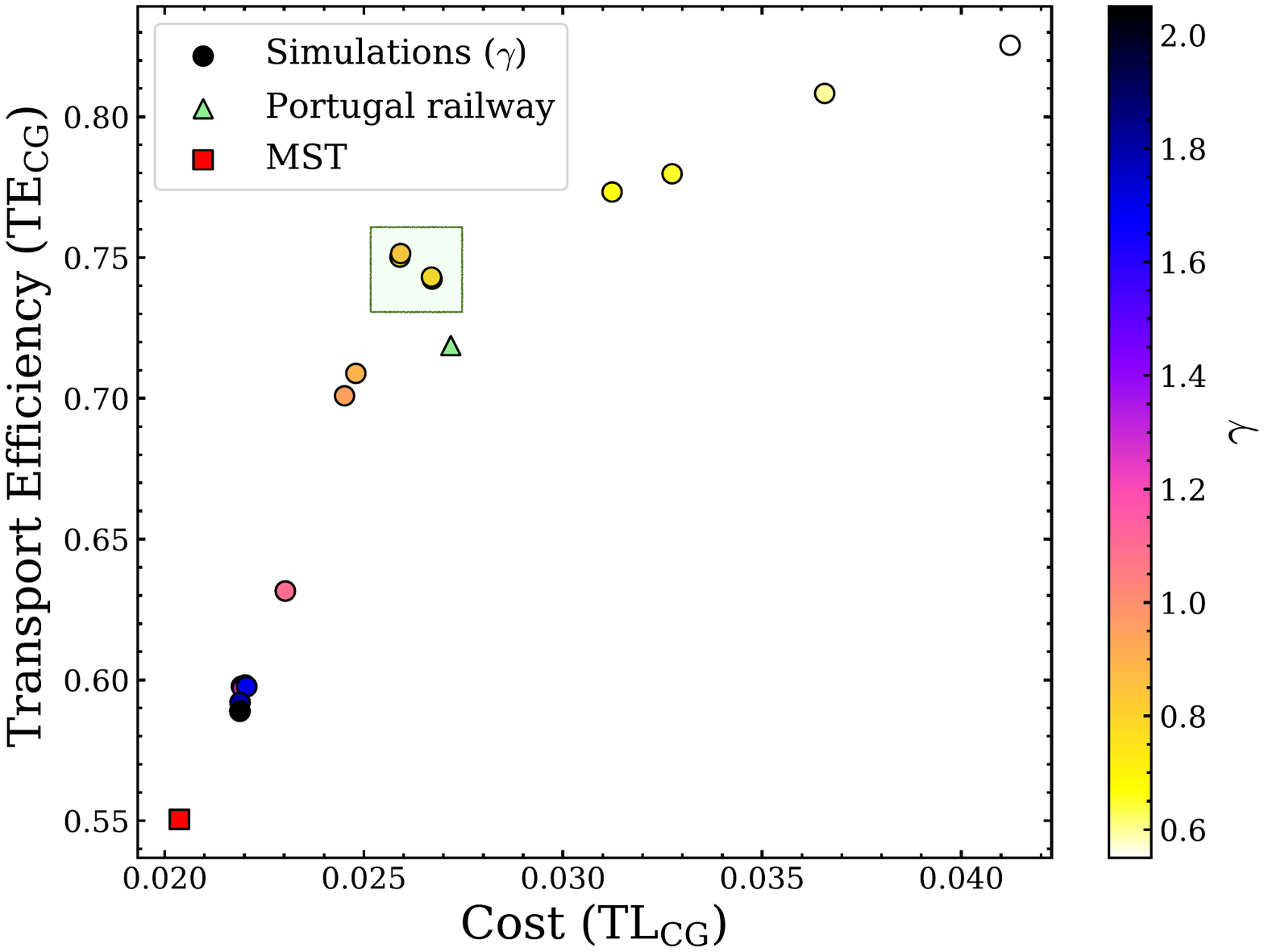}}  \hspace{0.3cm}%
\subfloat[\label{fig:gamma_FT_TL}]{\includegraphics[
trim={0cm 7cm 0cm 0cm}, clip, height=0.57\textwidth]{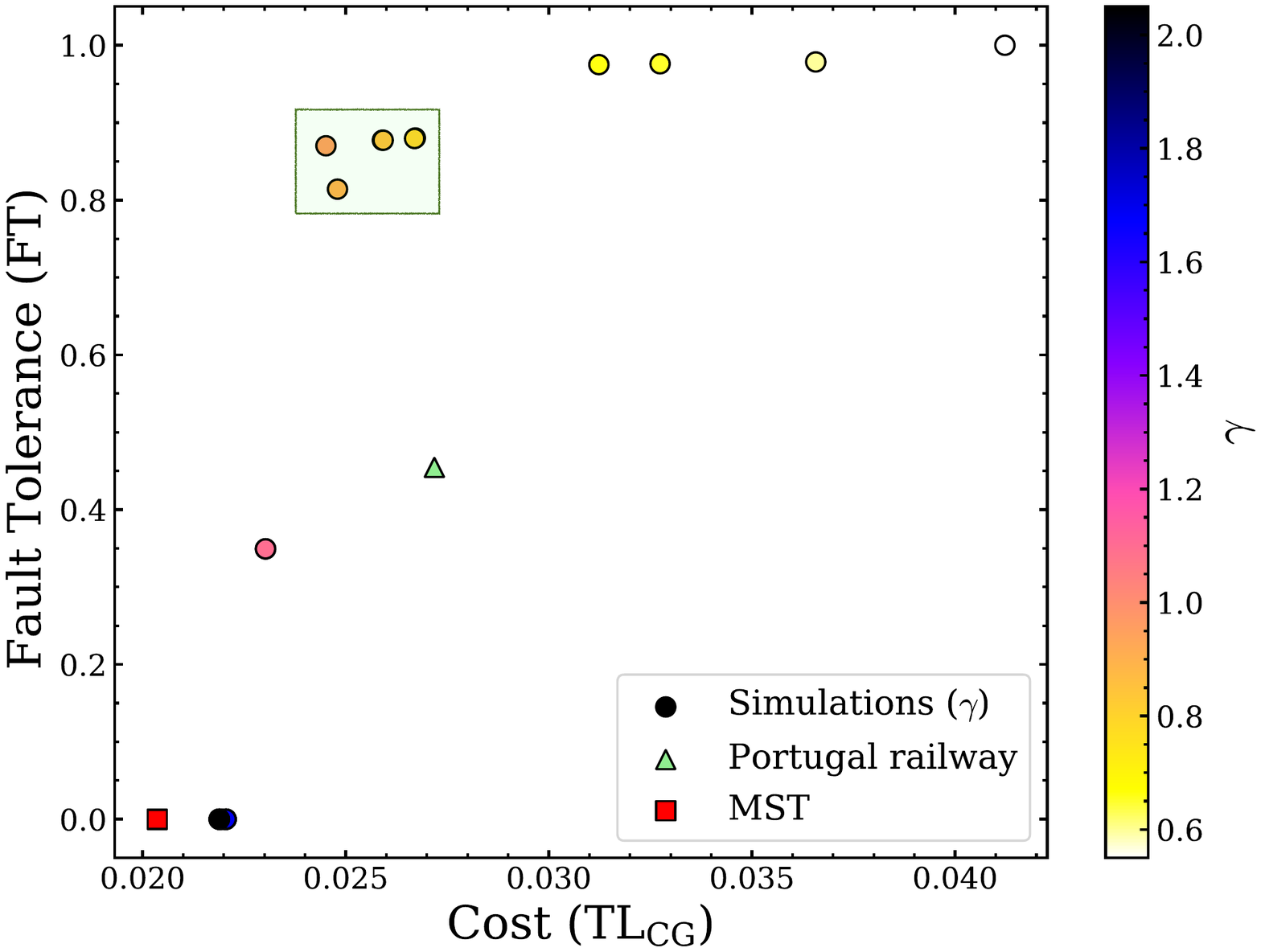}} \\
\subfloat[\label{fig:gamma_BCR}]{\includegraphics[
trim={0cm 2.7cm 0cm 2.3cm}, clip,
height=0.45\textwidth]{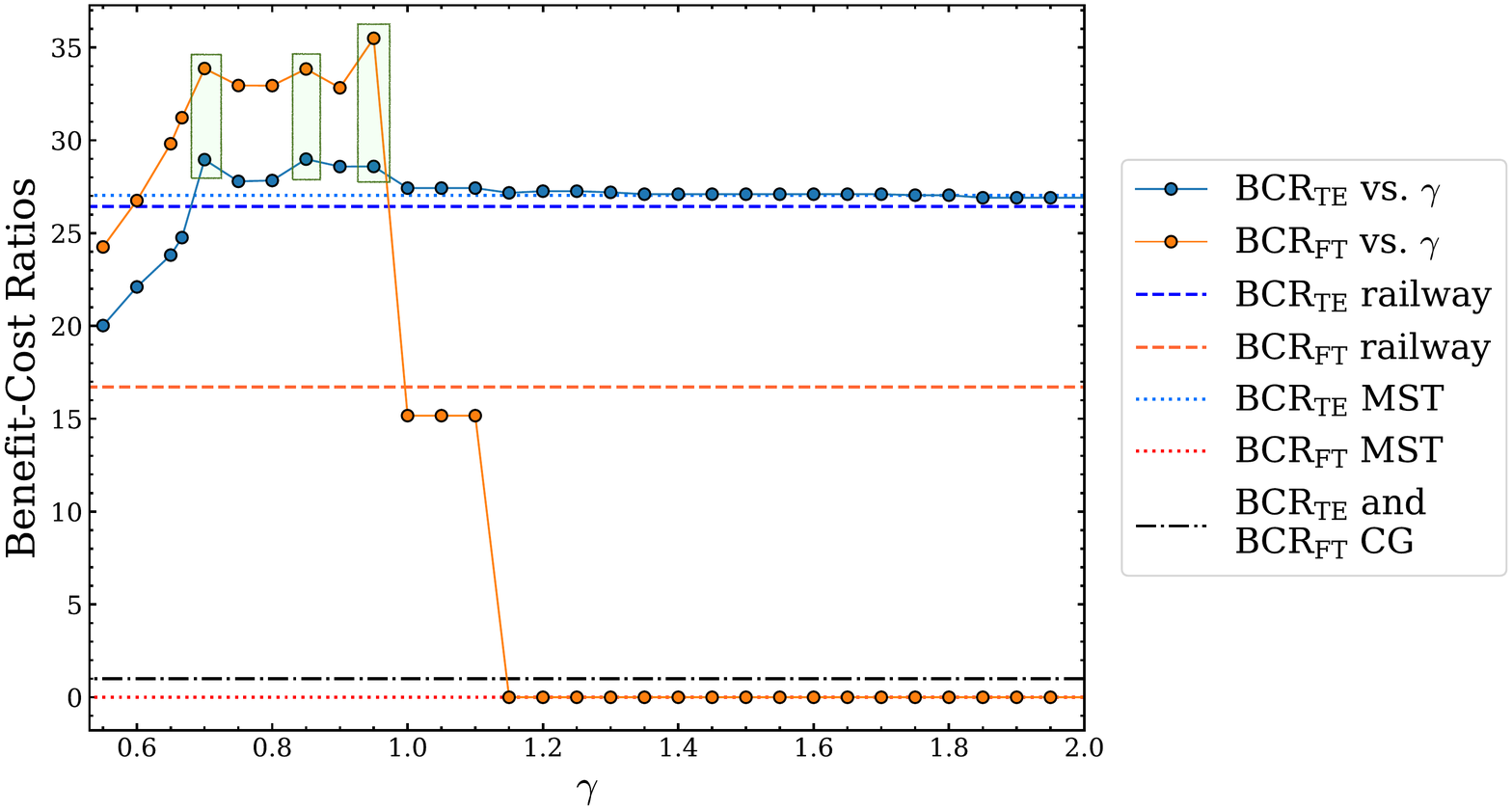}}

\caption{Network performance of the adaptation dynamics \eqref{eq:new_adapt_rule_gamma} as a function of the parameter $\gamma$, with the other parameters fixed, including the random seed. \textbf{(a-b)} Transport efficiency \eqref{eq:transport_efficiency} and fault tolerance \eqref{eq:fault_tolerance}  plotted against the total length of the network \eqref{eq:total_length} (cost). The metrics are normalised to those of the complete graph (CG) connecting the city nodes, yielding \CG{TL}, \CG{TE}, \CG{FT}.
The coloured circles represent the simulation results as $\gamma$ was varied from 0.55 to 2.00, considering the stochastic choice of the source-sink pair with $I_0=1$, and initial conditions $D_{ij}(0)=1$. The results were compared with the same normalised metrics of the real railway (green triangles) and MST network (red squares). \textbf{(c)} Plots of the benefit-cost ratios, defined as  BCR\textsubscript{TE}$=\CG{TE}/\CG{TL}$ and BCR\textsubscript{FT}$=\CG{FT}/\CG{TL}$, as the function of $\gamma$, compared with the ones of the real railway, MST and CG. The proposed optimal models (i.e., which result in networks with the best performance trade-off) are highlighted in green.
}
\label{fig:portugal_gamma_metrics_plots} 
\end{figure}

The trade-off between the network's cost, transport efficiency and fault tolerance can be better quantified by the plots of Figure \ref{fig:portugal_gamma_metrics_plots}. As the Figures \ref{fig:gamma_TE_TL} and \ref{fig:gamma_FT_TL} confirm, the transport efficiency and fault tolerance tends to decrease as $\gamma$ increases. Interestingly, most of the simulation results of the first plot lie in a well-defined curve that resembles the Pareto front \cite{Dilao2009, Muraro2013,Katifori2019} associated with the compromise between maximising the efficiency while minimising the overall cost. By definition, the networks lying on the Pareto front can't achieve a better transport efficiency without an increase of the cost, neither can have a lower cost without a decrease in the efficiency. The real railway is quite far from this frontier. In particular, one can observe that for $\gamma \sim 0.8$, the simulations achieve significantly better fault tolerance and a higher transport efficiency comparing to the real railway, with a slightly smaller cost, implying  a much better benefit-cost trade-off. The overall performance, captured by the two benefit-cost ratios,  BCR\textsubscript{TE}$=\CG{TE}/\CG{TL}$ and BCR\textsubscript{FT}$=\CG{FT}/\CG{TL}$, is depicted in Figure \ref{fig:gamma_BCR}, as a function of $\gamma$. For $\gamma \leq 0.6$, the excessive cost doesn't compensate the increase of the network's robustness and transport efficiency, resulting in worse performance than the MST and the real railway. However, in the interval $\gamma \in [0.7, 1[$, the situation completely changes, and simulations result in networks with a much better compromise between the three metrics than any other graph. For higher values of $\gamma$, the networks still achieve a slightly better efficiency-cost trade-off than the real railway, although the network's resilience is completely lost. The graph with the worst compromise is naturally the complete graph, due to the tremendous cost (Table \ref{tab:graph_metrics}) of connecting all the pairs of cities individually. In conclusion, for $\gamma \in [0.7, 1[$, the model results in networks with the overall best performance and, in general, higher than the performance of the real railway, MST and CG.

\subsection{Dependence of the performance on the stochastic choice of terminals}

The introduction of the flux fluctuations in the system opens numerous possibilities for the criterion of selecting the sources and sinks in each step. One can consider a deterministic time-dependent distribution of the nodes flux, $\vb*{q}$, or a purely stochastic one based on some hypothesis, as we have considered in the previous analysis. 

The choice of a specific set of sources and sinks in a given step tends to reinforce preferentially 
the channels along the shortest paths connecting them at the expense of the remaining ones. Consequently, the final network results from a compromise of averaging out the selected routes in all time steps. Therefore, different methods of choosing the driving terminals in each step lead in principle to distinct network topologies, ultimately affecting their fitness. In the context of \textit{Physarum}, it's not straightforward to decide what is the more adequate criterion based on the available experimental data, 
and due to the limitations of the model in describing through  simplistic terms the complex and not-well-understood mechanism underlying the network optimisation. 

We now compare different stochastic methods of choosing the  sources and sinks in each step and study their impact on the topology and performance of the optimised networks. Five different cases were studied. The first one considered was the original method proposed by Tero et al. \cite{Tero1} of randomly choosing in each step one source-sink pair from the set of terminals, such that $q_{source}= - q_{sink}= I_0$. This method is referred to as the ``Random pair'' method.  

From the perspective of a traveller who wants to move from one city (the source) to any other the fastest way possible, the network should be a good compromise of all the shortest paths between the cities. A similar argument can be applied to \textit{Physarum}, which seeks to transport the nutrients throughout the network in a fast and efficient way, by establishing multiple short connections between the available food sources, enabling an effective  management of the food consumption and distribution. Therefore, it makes sense to consider that at each time step, one terminal is randomly assigned as the source while all the remaining terminals are sinks, receiving an equal amount of fluid, i.e., $q_{sink} = - q_{source} / (|T|-1)$ where $q_{source}=I_0$. This method is designated by ``Random source'', and should in principle maximise the transport efficiency. 

However, biologically speaking, there is no specific argument that sustains the hypothesis of  only one pair of food sources being ``activated'' at a given moment, or one food acting as a source while all the remaining act like sinks. At a given step, all the food sources can be actively pumping nutrients, so any possible source-sink state should be possible. For this reason, we consider the case where  the nodes' net fluxes of all the terminals ($q_i$ with  $i \in T$) are time-dependent random variables subject to the constraint 

\begin{equation}
    \sum_{i \in \text{sources}} q_i = - \sum_{i \in \text{sinks}} q_i = I_0
    \label{eq:sum_qi_I0} \;.
\end{equation}
In this way, at each step, a random combination of sources and sinks is generated. This method is referred to as the ``All random'' case.

We also compare with the case of fixed terminals, where some cities were assigned as sources and the others as sinks from the beginning (``Fixed Terminals'' method). Different combinations of sources and sinks were tested. As seen before, the choice of fixed terminals can lead to disconnected solutions. Thus, to ensure that the final network remained connected, in each case, we have considered a fixed random distribution of node fluxes which satisfies \eqref{eq:sum_qi_I0}. All the above methods were tested considering the parameterisation of the model which minimises the power dissipated, i.e., $\gamma = 2/3$ in \eqref{eq:new_adapt_rule_gamma}. 

Finally, we also studied the networks produced by the \textit{Physarum Solver} model \eqref{eq:Tero_update_rule_1}, with the choice of a sigmoidal response typically used in literature,  $f(|Q_{ij}|)=|Q_{ij}|^\gamma/(1+|Q_{ij}|^\gamma)$, and  considering the stochastic choice of the source-sink pair as in the ``Random pair'' case. This method is designated by ``PS - random pair''.
In this case, the simulations were performed using $\gamma = 1.8$ and $I_0=2$, which according to \cite{Tero1} are the parameters which yielded networks mimicking the Tokyo rail system with the best trade-off between cost, efficiency and fault tolerance. To establish an even comparison, all the remaining methods were also simulated considering the same total inlet flux, $I_0$.  

\begin{figure}[hbt!] 

\newcommand{\mysub}[2][]{%
    \subfloat[#1]{\includegraphics[trim={4.8cm 2cm 3.5cm 1.8cm}, clip, width=0.3\textwidth]{#2}}%
}

\centering

\captionsetup[subfloat]{justification=centering}
\mysub[``Random pair'' \\  
\textbf{\normalsize (2017, 292, 0.92)}
]{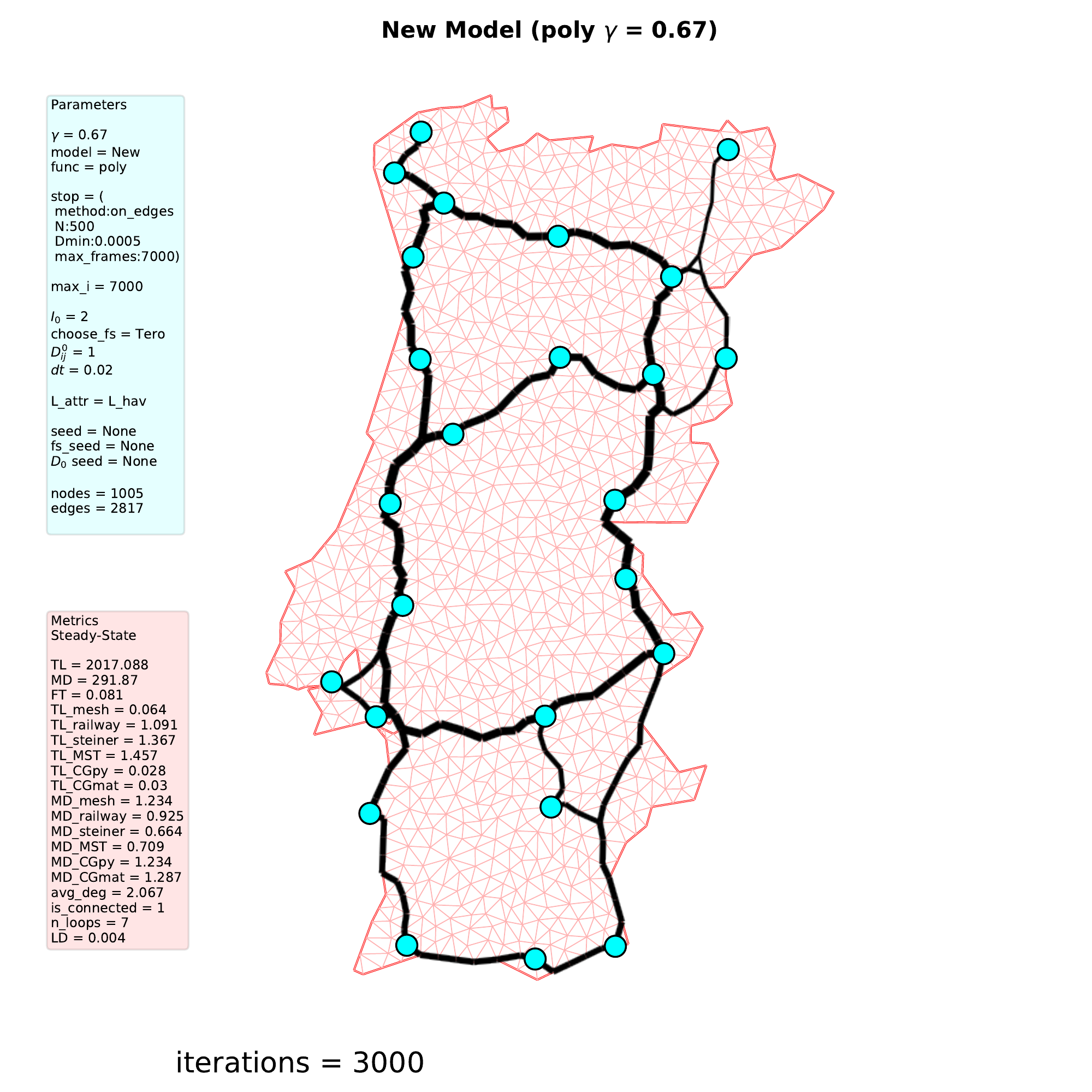}
\hfill
\mysub[``Random source'' \\
\textbf{\normalsize (1790, 290, 0.60)}
\label{fig:Random_source}]{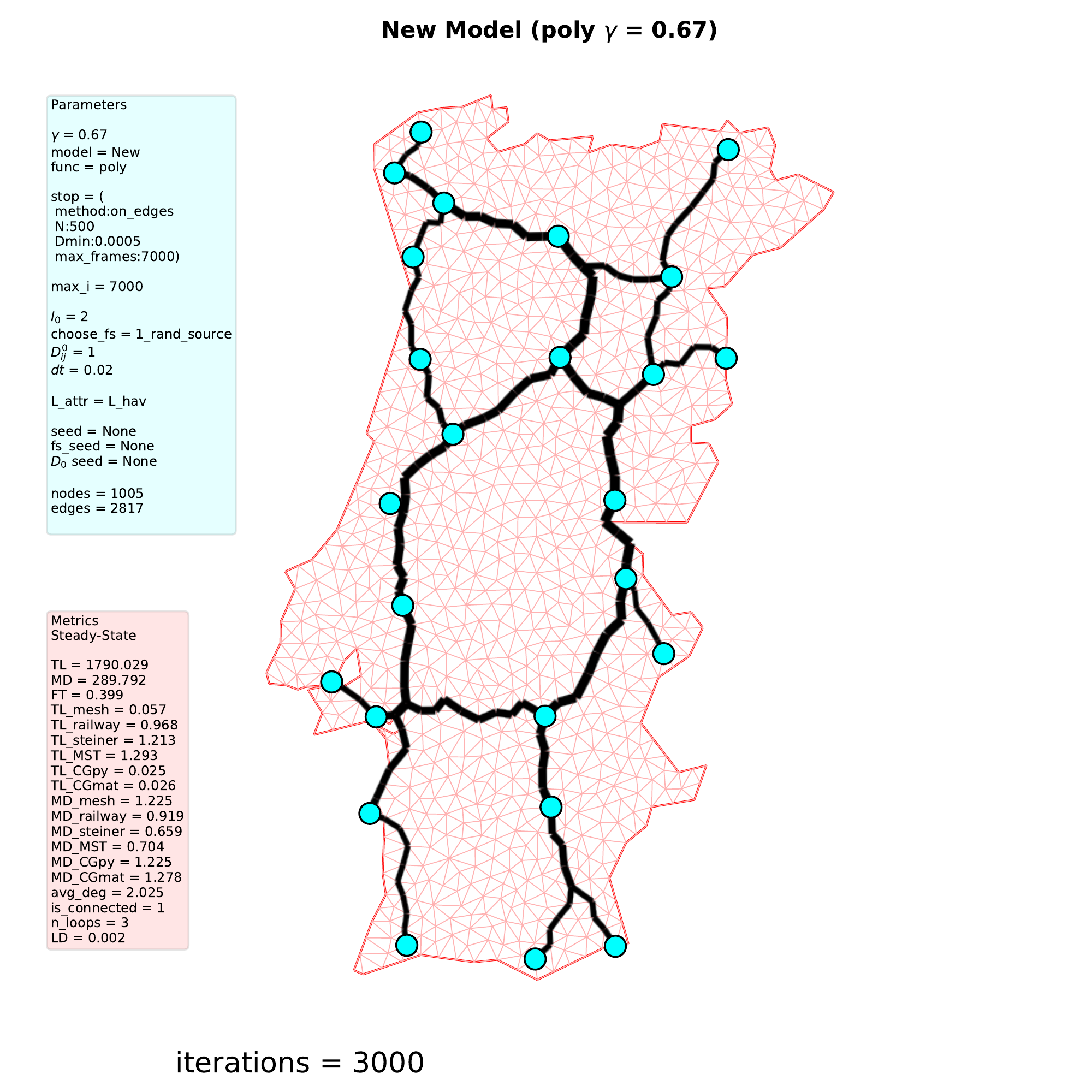} 
\hfill
\mysub[``All random'' \\
\textbf{\normalsize (1714, 304, 0.91)}
]{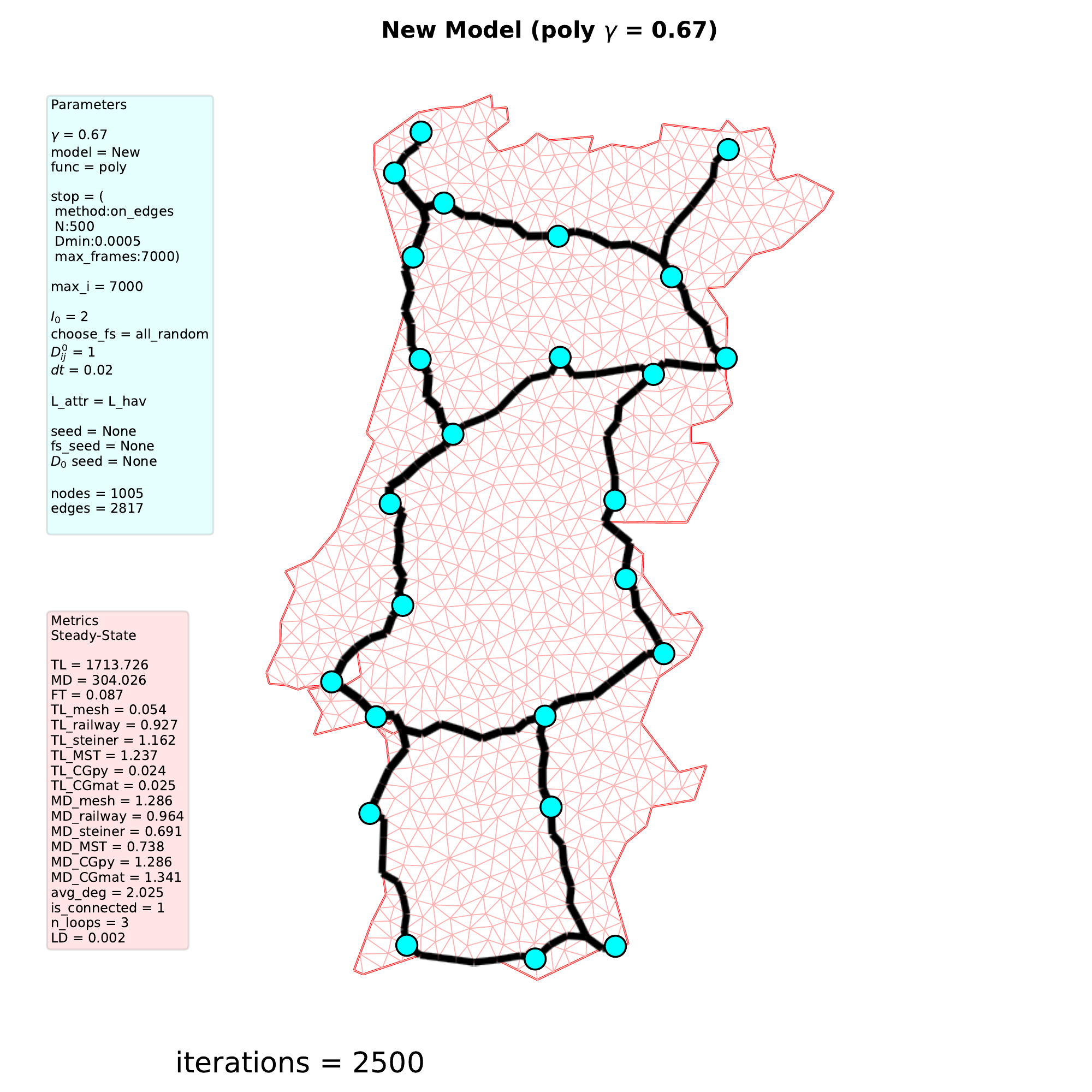}
\\
\mysub[``Fixed terminals'' \\
\textbf{\normalsize (3166, 391, 0.00)}
]{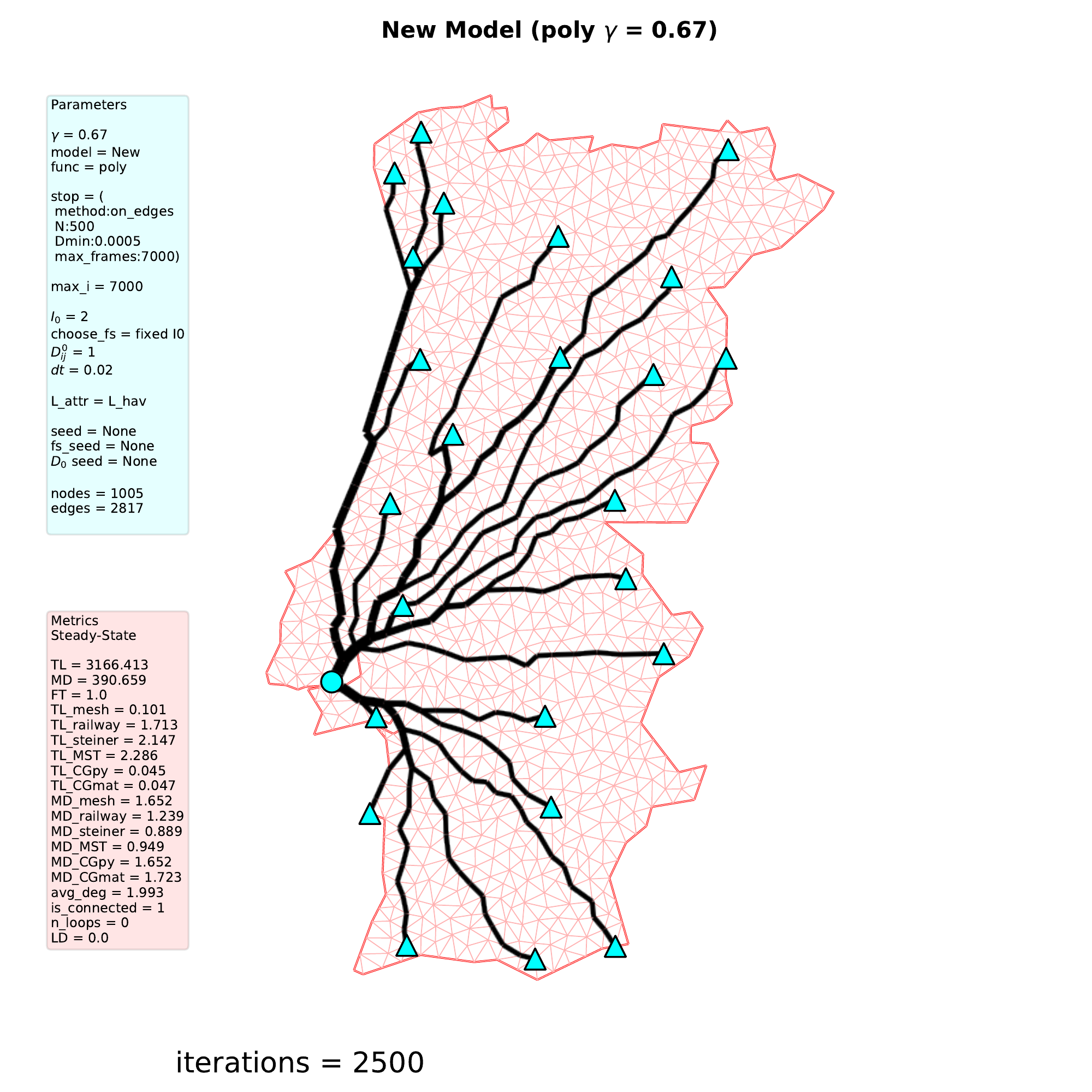} 
\hspace{0.1\textwidth}
\mysub[``PS - random pair'' \\
\textbf{\normalsize (1644, 326, 0.92)}
]{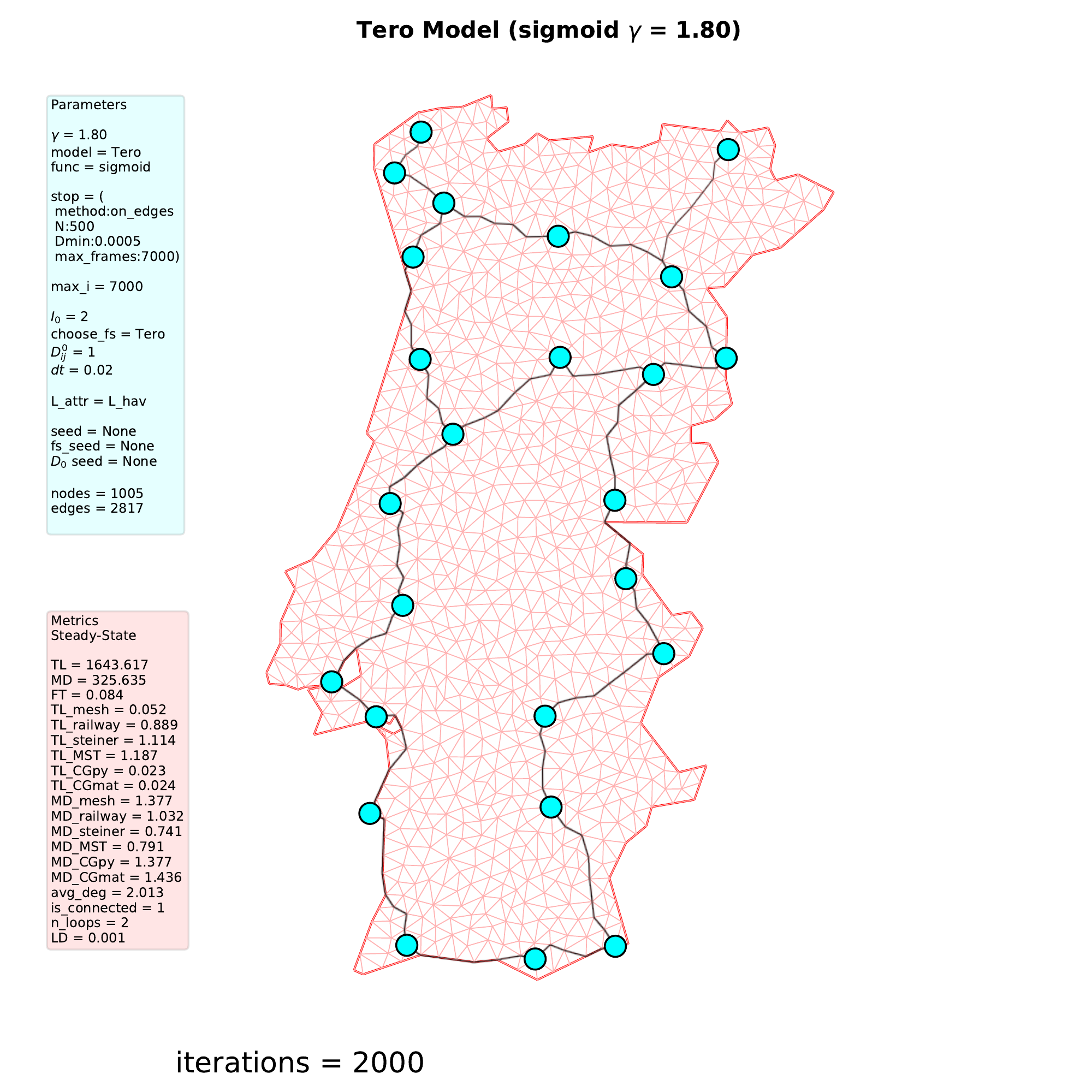} 

\caption{Topology of the optimised networks considering different methods of choosing the sources and sinks in each step. Note that the examples given are not fully representative, since the topology may change significantly due the stochastic nature of the algorithm. 
\textbf{(a)} \textbf{``Random pair'' -} Model \eqref{eq:new_adapt_rule_gamma} with $\gamma=2/3$, considering the choice of random source-sink pair from the set of terminals in each step. 
\textbf{(b)} \textbf{``Random source'' -} Model \eqref{eq:new_adapt_rule_gamma} with $\gamma=2/3$, where in each step a random terminal is chosen as the source, and the remaining terminals are sinks receiving the same amount of flux. 
\textbf{(c)} \textbf{``All random'' -} Model \eqref{eq:new_adapt_rule_gamma} with $\gamma=2/3$, where in each step all the terminals are randomly chosen either as sources or sinks. The terminals net flux are random variables subject to the constraint  \eqref{eq:sum_qi_I0}. 
\textbf{(d)} \textbf{``Fixed Terminals'' -} Model \eqref{eq:new_adapt_rule_gamma} with $\gamma=2/3$, considering fixed sources and sinks from the beginning. In this case,  the capital Lisbon acts like a source (blue circle) of flux, while all the remaining cities are sinks (blue triangles).
\textbf{(e)} \textbf{``PS - random pair'' -} \textit{Physarum Solver} model \eqref{eq:Tero_update_rule_1} with the choice of a sigmoid update function $f(|Q_{ij}|)=|Q_{ij}|^{1.8}/(|Q_{ij}|^{1.8}+1)$
typically used in literature and $\mu=1$. The choice of terminals in each step is the same as ``Random pair'' method. Note how much thinner are the channels of the final network comparing to the remaining cases, due to the total volume not being conserved in this case. In all the cases the  total inlet flux  is $I_0=2$, and the same initial conditions were used, $D_{ij}(0)=1$. The legend of each image refers to the network metrics \textbf{(TL, MD, FT)}, where TL and MD are given in kilometres.
}
\label{fig:portugal_method}
\end{figure}

Examples of the typical networks produced by the different methods are represented in Figure \ref{fig:portugal_method}. As the images show, the different choices lead to steady states with distinct topological features. It's also interesting to note how thinner are the selected channels by the \textit{Physarum Solver} model comparing to any other case, which is explained by the adaptation mechanism not conserving the network's volume in this case.

\begin{figure}[hbt!] 

\captionsetup[subfloat]{labelformat=empty,position=bottom,skip=3pt}

\newcommand{\mysub}[2][]{%
    \subfloat[#1]{\includegraphics[trim={4.8cm 2cm 3.5cm 1.8cm}, clip, width=0.3\textwidth]{#2}}%
}

\centering

\mysub[\textbf{\normalsize (2670, 523, 0.00)}]{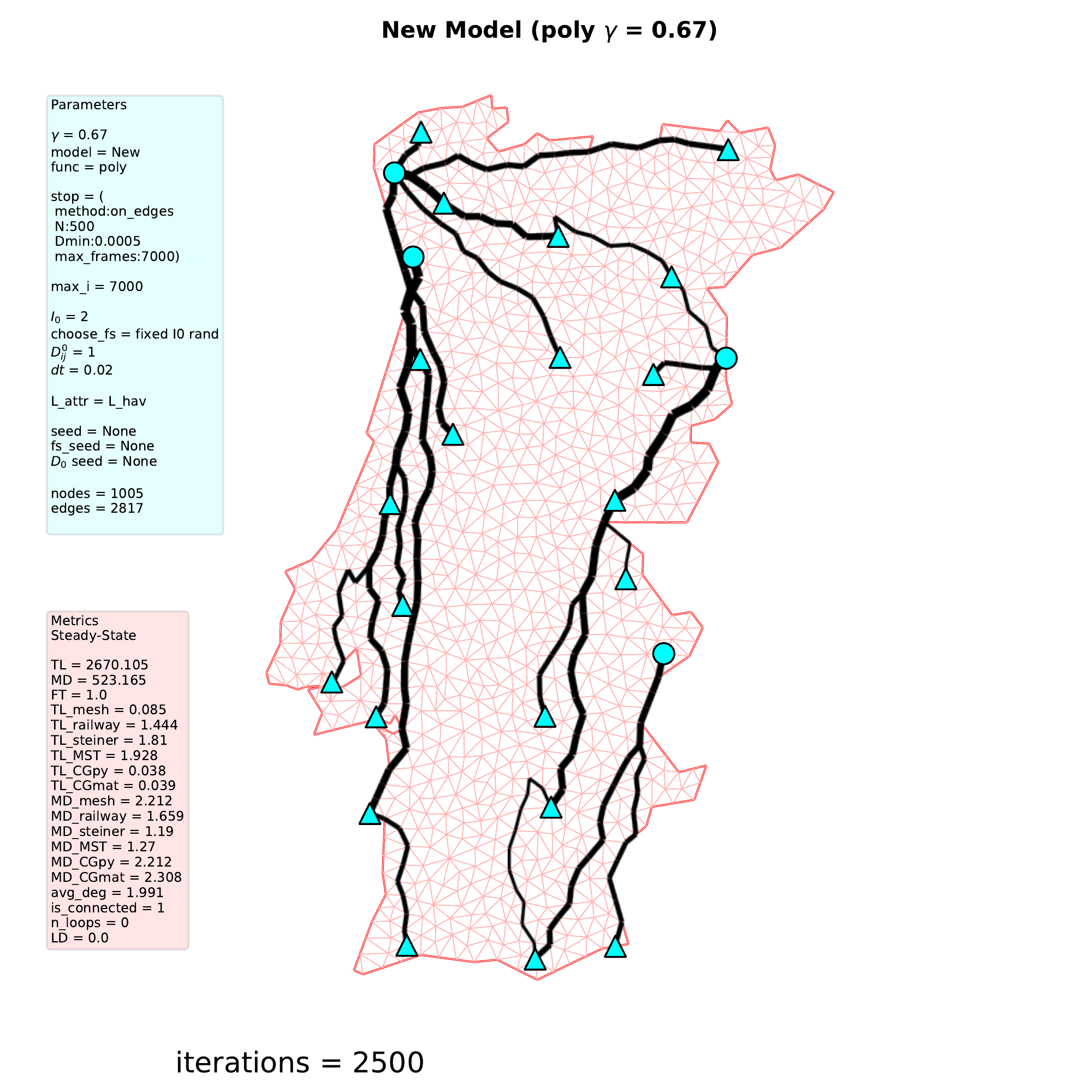}
\hfill
\mysub[\textbf{\normalsize (2748, 548, 0.00)}]{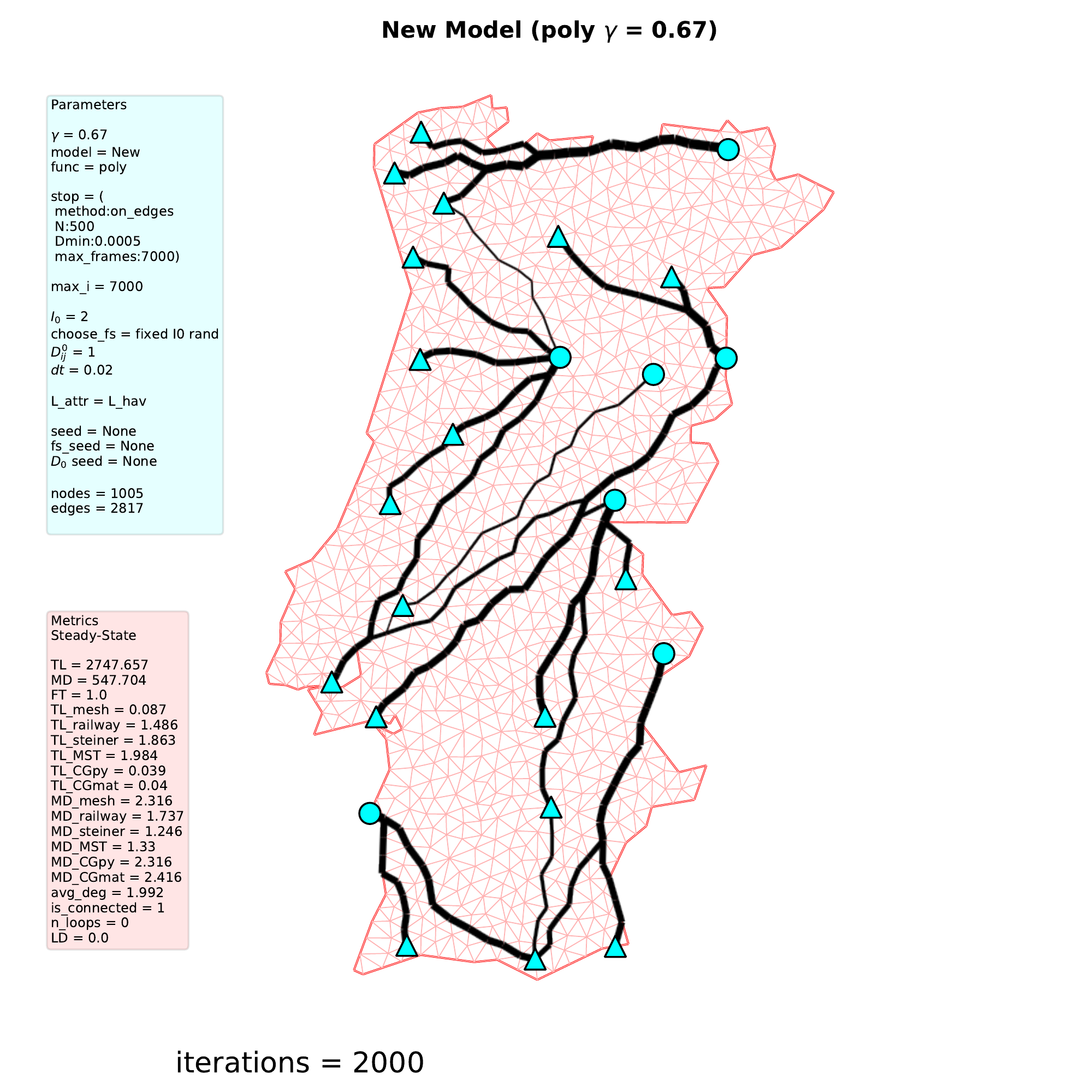} 
\hfill
\mysub[\textbf{\normalsize (2167, 459, 0.00)}]{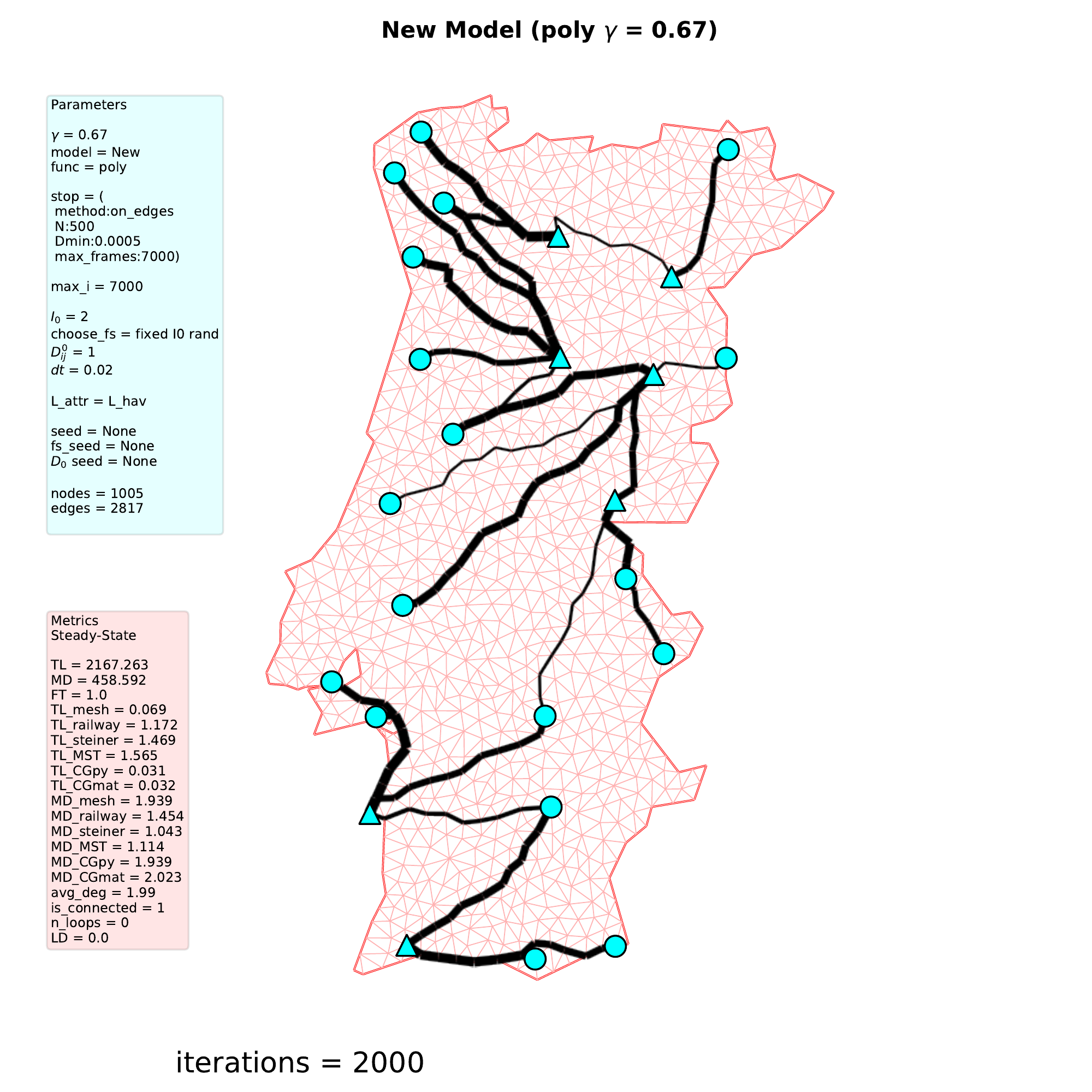}

\caption{Typical networks obtained when considering a static set of sources and sinks, for $\gamma = 2/3$. Different choices of sources and sinks lead to completely different networks, which are always \textit{trees}. Clearly, these topologies are far from being optimal due to excessive cost of connecting all the terminals and the zero fault tolerance, resulting in low transport efficiency. Only the adaptation subject to time-dependent sources and sinks can result in efficient and resilient networks. The legend of each image refers to the network metrics \textbf{(TL, MD, FT)}, where TL and MD are given in kilometres.
}
\label{fig:portugal_fixed_examples}
\end{figure}

\begin{figure}[hbt!] 

\newcommand{\mysub}[2][]{%
    \subfloat[#1]{\includegraphics[trim={0cm 0cm 0cm 0cm}, clip, width=0.495\textwidth]{#2}}%
}
\centering

\subfloat[]{\includegraphics[trim={0cm 7cm 0cm 7cm}, clip, width=0.495\textwidth]{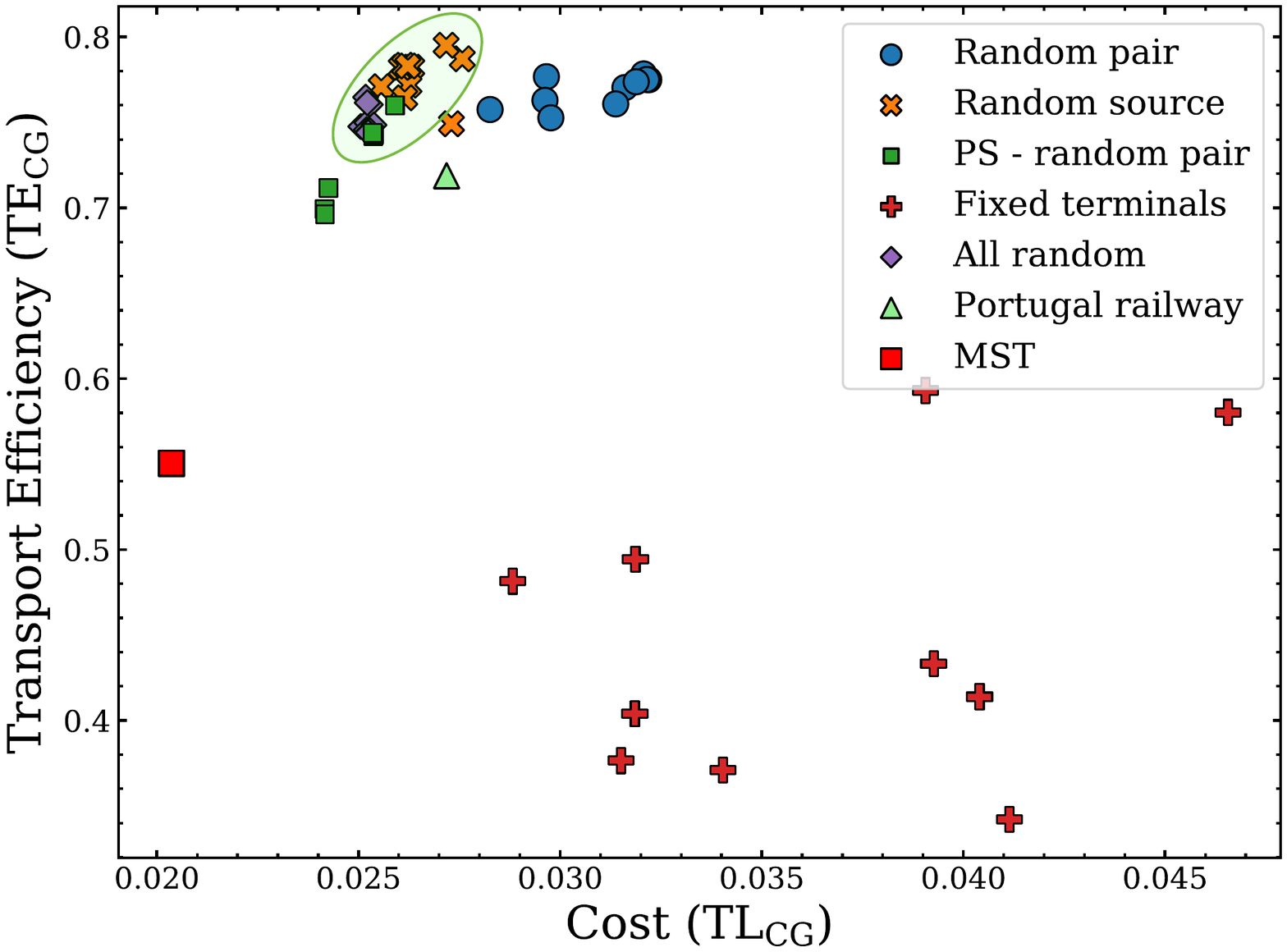}}  \hfill
\subfloat[]{\includegraphics[trim={0cm 7cm 0cm 7cm}, clip, width=0.495\textwidth]{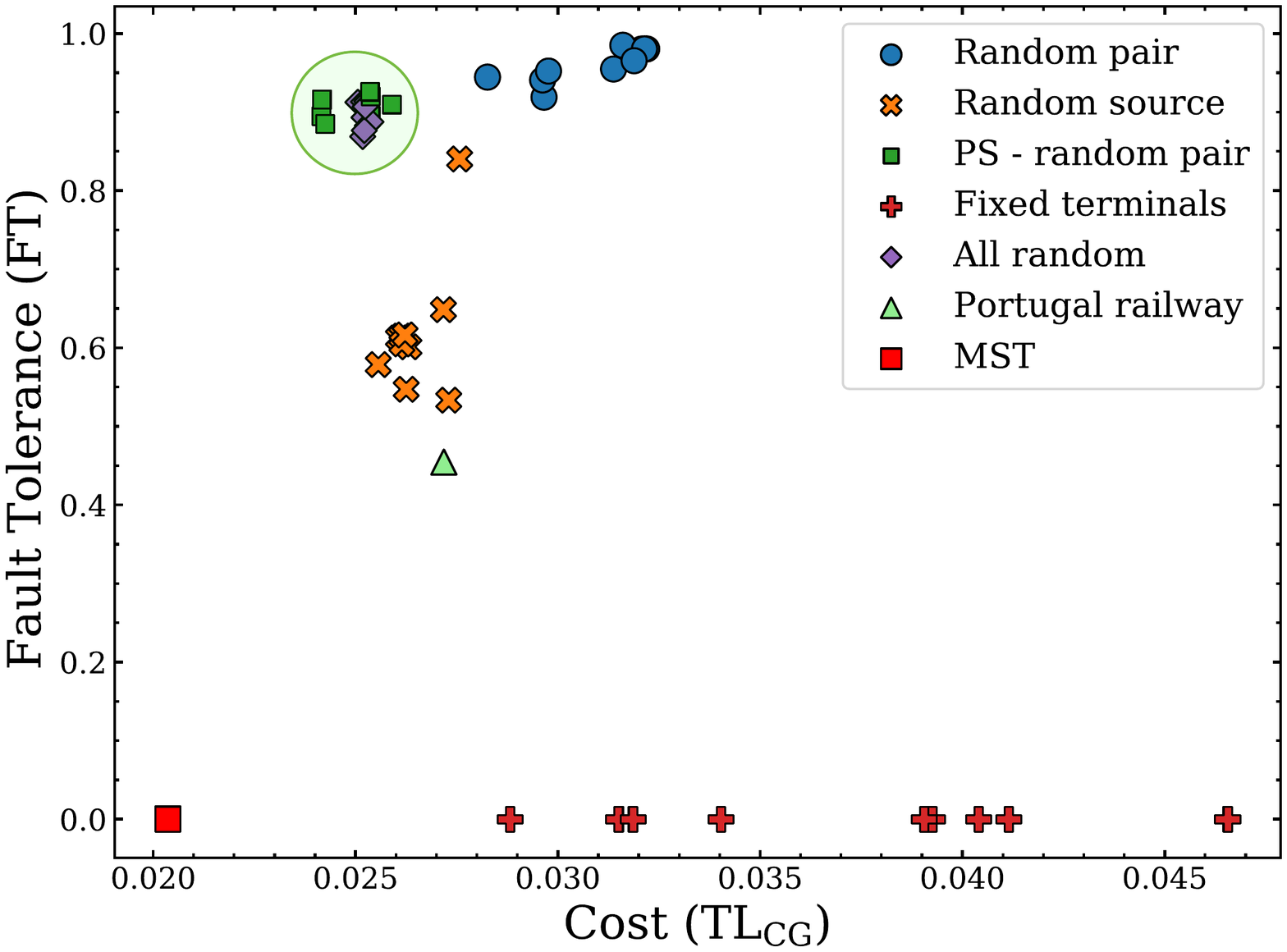}} \\
\subfloat[]{\includegraphics[trim={0cm 7cm 0cm 7cm}, clip,width=0.495\textwidth]{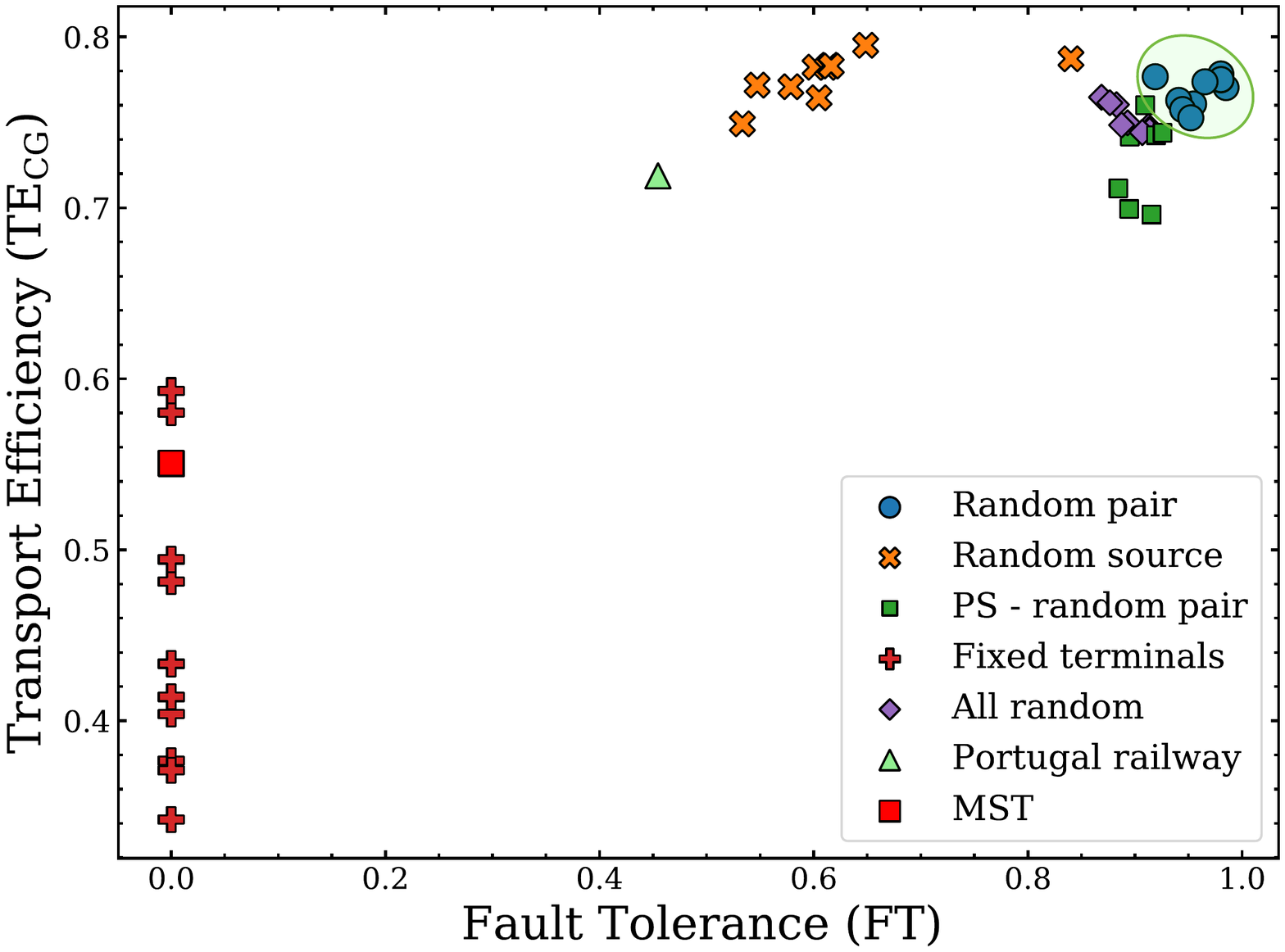}} \hfill
\subfloat[]{\includegraphics[trim={0cm 7cm 0cm 7cm}, clip, width=0.495\textwidth]{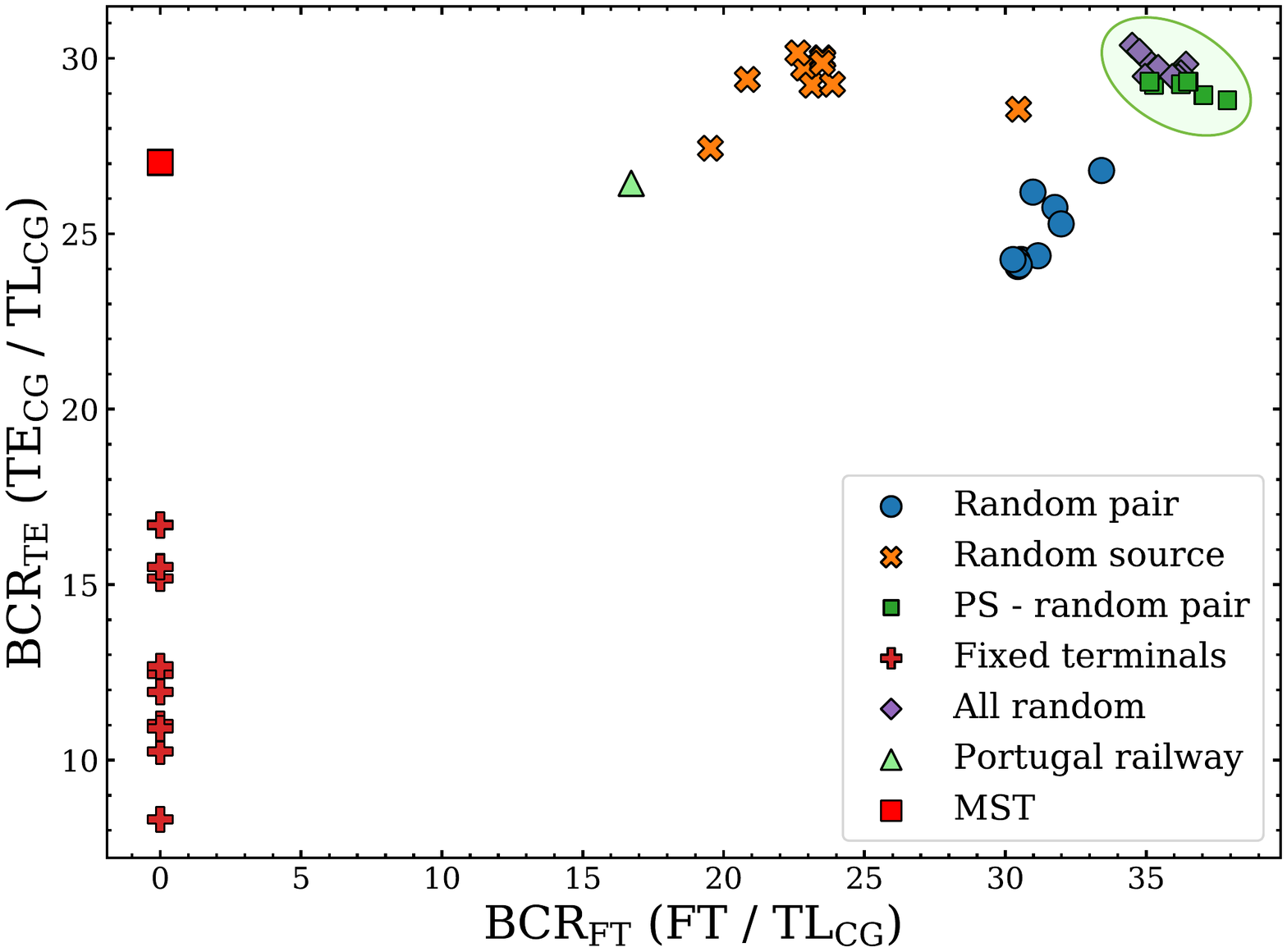}}

\caption{ Performance of the networks for different methods of choosing the sources and sinks over time, as described in Figure \ref{fig:portugal_method}. \textbf{(a-c)} Plots showing the trade-off  between the cost (total length), transport efficiency and fault tolerance of the networks. The metrics are normalised to those of the CG graph. Each type of marker corresponds to a different method. The results were compared with the same normalised metrics of the real railway (green triangles) and MST network (red squares) \textbf{(d)} The benefit-cost ratios, defined as  BCR\textsubscript{TE}$=\CG{TE}/\CG{TL}$ and BCR\textsubscript{FT}$=\CG{FT}/\CG{TL}$, plotted against each other, measuring the overall compromise between the three metrics for the different methods of choosing the terminals. The simulations which achieved the best trade-off between each pair of metrics are highlighted in green. Overall, the stochastic networks are more robust and efficient than any other network, especially comparing to the case of fixed terminals.
}
\label{fig:portugal_method_metrics_plots} 
\end{figure}

To establish a comparison between the methods, the overall  performance of the networks was again quantified in terms of cost, transport efficiency and fault tolerance. The average performance of the different methods, based on 10 realisations for each case, is summarised in Table \ref{tab:portugal_method_metrics}. A more visual quantification is given by the plots of Figure \ref{fig:portugal_method_metrics_plots}, depicting the trade-off between the different metrics of all simulations.

The choice of fixed terminals is definitely the method that leads to networks with the worst performance by far in every respect. In the plots of Figure \ref{fig:portugal_method_metrics_plots}, the points corresponding to this method are all scattered, reflecting the great variety of \textit{tree} topologies 
depending on the specific arrangement of sources and sinks (Figure \ref{fig:portugal_fixed_examples}), but in every case, the performance metrics are consistently low. The characteristic tree-like topology of the steady states entails a great cost without any benefit in terms of transport efficiency, as the terminals are on average very distant from each other, and in terms of tolerance to damage, as no redundant paths are formed ($\text{FT}=0$). As a result, the average values of the metrics are the lowest ones from all the methods by a large gap, in particular, the cost-benefit ratios. This justifies the importance of the flux fluctuations to build efficient and resilient networks. 

One could have expected that the ``All random'' method would lead to the highest diversity of topologies, due to being the most stochastic one. However, the proximity of the points in the plots of Figure \ref{fig:portugal_method_metrics_plots}suggests quite the opposite. Actually, this method imposes more restrictions on the topology of the final network, as all the states of sources and sinks are possible, meaning that the connections between the terminals must accommodate all these possibilities. However, there is still the possibility that the method used to generate the random distribution of node fluxes in each step might have introduced some bias in the states actually generated.

As expected, the ``Random source'' method is the one that results in the maximum transport efficiency on average. However, the ``All random'' method reaches a slightly better trade-off between the efficiency and the cost, achieving the highest BCR\textsubscript{TE}. By contrast, from all the stochastic methods, ``Random source'' is the one with the lowest fault tolerance and BCR\textsubscript{FT}, which is related with the characteristic topology of the resulting networks, as seen in Figure \ref{fig:Random_source} -- the cities closer to the border are connected by tree-like branches to a robust core connecting the interior cities. 

Although the \textit{Physarum Solver} has the worst transport efficiency from the stochastic methods, it achieves a similar BCR\textsubscript{TE} comparing to the ``All random'' and ``Random source'' cases, only because it produces networks with the lowest total length of all the methods. In compensation, it attains the highest fault tolerance, which together with this low-cost results in the best cost-robustness trade-off (highest BCR\textsubscript{TE}).

The ``All Random'', ``Random source'' and ``PS - random pair'' methods produced networks with better metrics than the railway graph (Figure \ref{fig:railway_graph}). This is almost true for ``Random pair'' except for the total length, which is considerably higher than that of the railway graph, leading to a slightly lower BCR\textsubscript{TE}. All the stochastic methods also achieved a better overall performance than the MST graph (Figure \ref{fig:MST_graph}), although the BCR\textsubscript{TE} of the latter is slightly higher than the ``Random pair'' case, only because it has the lowest cost by definition.  The discrepancy is even higher when comparing with the CG (Figure \ref{fig:CG_graph}). Although the CG reaches the maximum efficiency and fault tolerance, it has the highest cost by a very large margin, resulting in the worst cost-benefit ratios. In comparison, the stochastic methods achieve more than $70\%$ of the efficiency of the CG, for only less than $3\%$ of the cost, resulting in at least 25 times better cost-benefit relationship.

Note that, from the previous analysis, we can conclude $\gamma = 2/3$ is not the value of $\gamma$ which results in networks with the best overall performance. The plot of Figure \ref{fig:gamma_BCR} suggests that $\gamma=0.7$ could have been a better choice, since, from all the values of $\gamma$ tested, it's the one which achieves the highest BCR\textsubscript{TE} and the second-highest  BCR\textsubscript{FT}, reaching the best compromise between the three metrics. This would allow a more reliable and fair comparison with the \textit{Physarum Solver} method, whose parameters were cherry-picked to achieve the highest performance by extrapolating the results of the Tokyo experiments\cite{Tero1}. 

In conclusion, the results clearly show the importance of flux fluctuations in the design of low-cost, efficient and robust flow networks \cite{Corson2010,Katifori2010,Tero1,Katifori2019}. Due to the Pareto nature of the optimisation, the criteria of choosing the best stochastic method to generate the driving terminals depends on the relevance of these network features in a given context. Assuming we want the best overall trade-off between the three metrics (TL, TE, FT), the results suggest that the ``All Random'' method is the best one, given that is the method which achieved the highest BCR\textsubscript{TE} and highest BCR\textsubscript{FT} from all the tests of our model.

\begin{table}[hbt!]
    \centering
    \begin{adjustbox}{max width=\textwidth}
    \begin{tabular}{lccccc}
\toprule
           Method & TL\textsubscript{CG} ($\times 10^{-5}$) & TE\textsubscript{CG} ($\times 10^{-3}$) & FT ($\times 10^{-3}$) & BCR\textsubscript{TE} ($\times 10^{-1}$) & BCR\textsubscript{FT} ($\times 10^{-1}$) \\
\midrule
       All random &                            2523 $\pm$ 3 &                             751 $\pm$ 3 &           896 $\pm$ 5 &                              298 $\pm$ 1 &                              355 $\pm$ 2 \\
    Random source &                           2648 $\pm$ 20 &                             777 $\pm$ 4 &          620 $\pm$ 27 &                              294 $\pm$ 3 &                             234 $\pm$ 10 \\
 PS - random pair &                           2471 $\pm$ 22 &                             718 $\pm$ 8 &           909 $\pm$ 4 &                              291 $\pm$ 4 &                              368 $\pm$ 4 \\
      Random pair &                           3086 $\pm$ 44 &                             768 $\pm$ 3 &           960 $\pm$ 7 &                              249 $\pm$ 4 &                              312 $\pm$ 5 \\
  Fixed terminals &                          3645 $\pm$ 178 &                            449 $\pm$ 27 &             0 $\pm$ 0 &                             125 $\pm$ 10 &                                0 $\pm$ 0 \\
\midrule
     MST &                                    2036 &                                     551 &                     0 &                                      270 &                                        0 \\
 railway &                                    2718 &                                     719 &                   455 &                                      264 &                                      167 \\
      CG &                                  100000 &                                    1000 &                  1000 &                                       10 &                                       10 \\
\bottomrule
\end{tabular}

\end{adjustbox}
\caption{The mean and corresponding standard error of the performance metrics for different methods of choosing the sources and sinks in each step of the algorithm. The methods are described in Figure \ref{fig:portugal_method}. The metrics are normalised to the corresponding values of the CG (Figure \ref{fig:CG_graph}). The results are based on 10 runs for each method and are sorted by the benefit-cost ratio of the network's transport efficiency, BCR\textsubscript{TE}. At the bottom is presented the same metrics of the MST, railway graph and CG for comparison. The stochastic methods produce networks with much better performance than the case of fixed terminals. Overall, the stochastic networks also have a better benefit-cost relationship than the MST, CG and the real railway.
}
\label{tab:portugal_method_metrics}
\end{table}
\cleardoublepage

\chapter{Modelling \textit{Physarum}'s Growth}
\label{chapter:growth}

\textit{Physarum} grows and progressively rearranges its network structure as it forages. So far we have only considered the network optimisation of static organisms, and neglected  the growth mechanism. We have developed a generic model describing the adaptation dynamics of a static flow network of distensible channels filled with an incompressible fluid, and applied to the particular case of \textit{Physarum}. In this chapter, we extend the previous model to accommodate the formation of new channels, connecting the growth to the network optimisation, as an attempt to describe the foraging behaviour of \textit{Physarum}, neglecting its body mobility.

A simple stochastic model of cellular growth was proposed by Eden in 1961 \cite{Eden1961}, which is described as follows. The cells are represented by square lattice sites and growth can only occur at the boundary i.e., from one occupied cell to one adjacent free cell. Initially, only one cell is occupied. At each time step, a random boundary cell is selected to reproduce to a neighbour empty cell with a given probability that may depend on different factors, leading to different types of cluster formations. 

Different variations of the Eden model applied to the growth of \textit{Physarum} were recently studied by Ferreira and Dilão \cite{JFerreira_Tese}. In this case, the network development is simulated by tracing the linage that connects the starting cell to a given cell at the boundary. Each edge of the graph connects the father to the daughter cell. In particular, it was considered a flow-based growth version which shares some similarities with the one proposed here. The flow was driven by the starting cell (food source) and the outside perimeter cells which behave like sinks with fixed pressures, $p=0$. After computing the channel fluxes, the probability that a sink is occupied in the current step is proportional to the amount of flux received, $q$.  However no coupling between the growth and the adaptation mechanisms was explicitly considered, and the formation of channels is merely probabilistic. Here we propose a simple growth model which solves these two issues. 

\section{Growth Model}

To incorporate the growth into the previous adaptation model,  \textit{Physarum} is now represented by a dynamic graph. The growth is regarded as the formation of new channels at the boundary of the organism when there are enough nutrients in the neighbourhood to build them. The nutrients are supplied by active food sources, initially placed at certain nodes, and transported to the nodes at the boundary where they are stored until they are used in the veins' development. In this way, the boundary nodes behave like sinks of the nutrients flux, mimicking simulated regions where the growth occurs (growth fronts).  

For simplicity, it's assumed that the dynamics take place on top of a pre-existing mesh resulting from a Delaunay triangulation, which means that the \textit{Physarum}'s network at a given time is a subgraph of the underlying mesh. This implies that all the channels have already a pre-determined orientation and fixed length, $L_{ij}$, but ensures that the organism grows as a planar graph. The formation of the new channels is simulated by the  progressive activation of the edges of the underlying mesh when the cost of producing it is overcome. Each edge of the mesh is thus associated with a nutrient cost of activation, which in principle should depend on its length, and can have two possible states:

\begin{itemize}
    \item \textbf{Active:} the edge represents a vein of the \textit{Physarum's} network. 
    \item \textbf{Inactive:} the edge is not currently part of the network.  
\end{itemize}

Similarly to the previous model, each active channel is considered as a cylindrical elastic tube whose diameter can change in response to the flux flowing through it. They are characterised by a conductivity, $D_{ij}$, which is subject to the same  adaptation dynamics \eqref{eq:new_adapt_rule}. If the channel $(i,j)$ is inactive, $D_{ij}=0$, otherwise $D_{ij}>0$. A newly formed channel is initialised with an arbitrary conductivity value $D_{ij}(0)=D_0>0$.

On the other hand, each node $i$ is characterised by an amount of nutrients, $m_i$, and can have three possible states:

\begin{itemize}
    \item  \textbf{Empty state:} inactive nodes that are not yet part of the \textit{Physarum} network.
    \item  \textbf{Growing state:} nodes located at the growing margins of \textit{Physarum} (boundary)
    that store temporarily nutrients from the sources and  participate in channels formation. These nodes are referred to as ``boundary nodes''. Active nodes are in the growing state as long as they contain at least one inactive neighbour node. 
    \item \textbf{Transport state:} nodes that can't give rise to new channels anymore, and serve only as intermediaries to  transport nutrients to  the boundary, where the formation of new channels can occur. 
\end{itemize}

In addition, any given node can have a food source, which is ``activated'' from the moment that the node is activated i.e., it's changed to the growing state.

\subsection{Algorithm}

The algorithm compromises three main steps which are described as follows. Initially, the \textit{Physarum} is represented by a single node containing a food source.

\subsubsection{Computation of the fluxes}

In each time step, the nutrients flow from the active food sources to the current boundary sinks via the active channels, where they are accumulated. In the simplest case, there is only one food source with magnitude $q_{source}=I_0$, which supplies an equal amount of nutrients to all the boundary sinks, although other distributions of node fluxes can be considered as well. In any case, the conservation of the mass  \eqref{eq:sum_qi} requires that the input flux from the active food sources balances the flux of the boundary sinks. For simplicity, it's assumed that the food sources never run out. Given the current sets of boundary nodes and of active food sources, the fluxes of the active channels are computed through the conservation laws \eqref{eq:sum_Qij}, as described in section \ref{section:computation_Qij}.

Then, the nutrient reserves of the boundary nodes are increased according to the flux that each receives i.e., $|q_{i}|$ with $i \in \text{boundary}$. Assuming the fluid has a unitary density, the amount of nutrients that a given node accumulates per time step is

\begin{equation}
    dm_i = |q_{i}|\dt  \qquad  , i \in \text{boundary}
    \label{eq:dm_boundary}
\end{equation}

\noindent where $\dt$ is the duration of the time step. In the beginning, while \textit{Physarum} is a single node covering a food source and no channels are yet formed, only the last step is applied.

\subsubsection{Growth stage}

The next step is the network's growth. For each boundary node, it's computed the set of the incident edges which are currently inactive, i.e., edges where new channels can be formed. The nutrient reserves of the node are equally distributed among those neighbour inactive edges where they are accumulated, mimicking the formation of the channel tips. In practice, the production cost of each inactive edge is reduced by the amount of nutrients given. If the nutrients transferred to an edge exceeds its production cost, the excess is kept stored on the boundary node. When the cost of producing the channel reaches zero, the edge is activated, and a new channel is created with a conductivity $D_{ij}(0)=D_0$. If the other end node of the edge is inactive (empty state), the node is activated on the ``growing state'', and the boundary is extended with the new sink node. Note that every time a new channel with length $L_{ij}$ is formed, the total network's volume, $\V$, is increased by $\sqrt{D_0}L_{ij}$. It's also assumed that the cost of producing a channel is proportional to its initial volume.  

Finally, for every boundary node, it's checked if they contain any inactive neighbour node. In case they don't, the node state is changed to the ``transport state'', and growth can no longer occur starting from that node. The eventual nutrient reserves of the node are evenly distributed between the neighbours on the ``growing state'' which give continuity to the network development.

\subsubsection{Network adaptation}

Finally, the conductivities of the active channels are adapted according to the dynamics \eqref{eq:new_adapt_rule} for a given choice of the function $g$. In the following we consider the polynomial choice used previously, $g_\gamma(|Q_{ij}|)=|Q_{ij}|^\gamma$. Note that, although the total volume of the network, $V$, in \eqref{eq:new_adapt_rule} increases over time due to the continuous network growth, it's still conserved in the adaptation process.

\section{Results}

\subsection{One food source}

\begin{figure}[hbt!] 
\captionsetup[subfloat]{labelformat=empty,position=top,skip=0.5pt}

\newcommand{\mysub}[2][]{%
    \subfloat[#1]{\includegraphics[trim={2cm 2.7cm 2.5cm 3.1cm}, clip, width=0.31\textwidth]{#2}}%
}

\centering

\mysub[$t=300$]{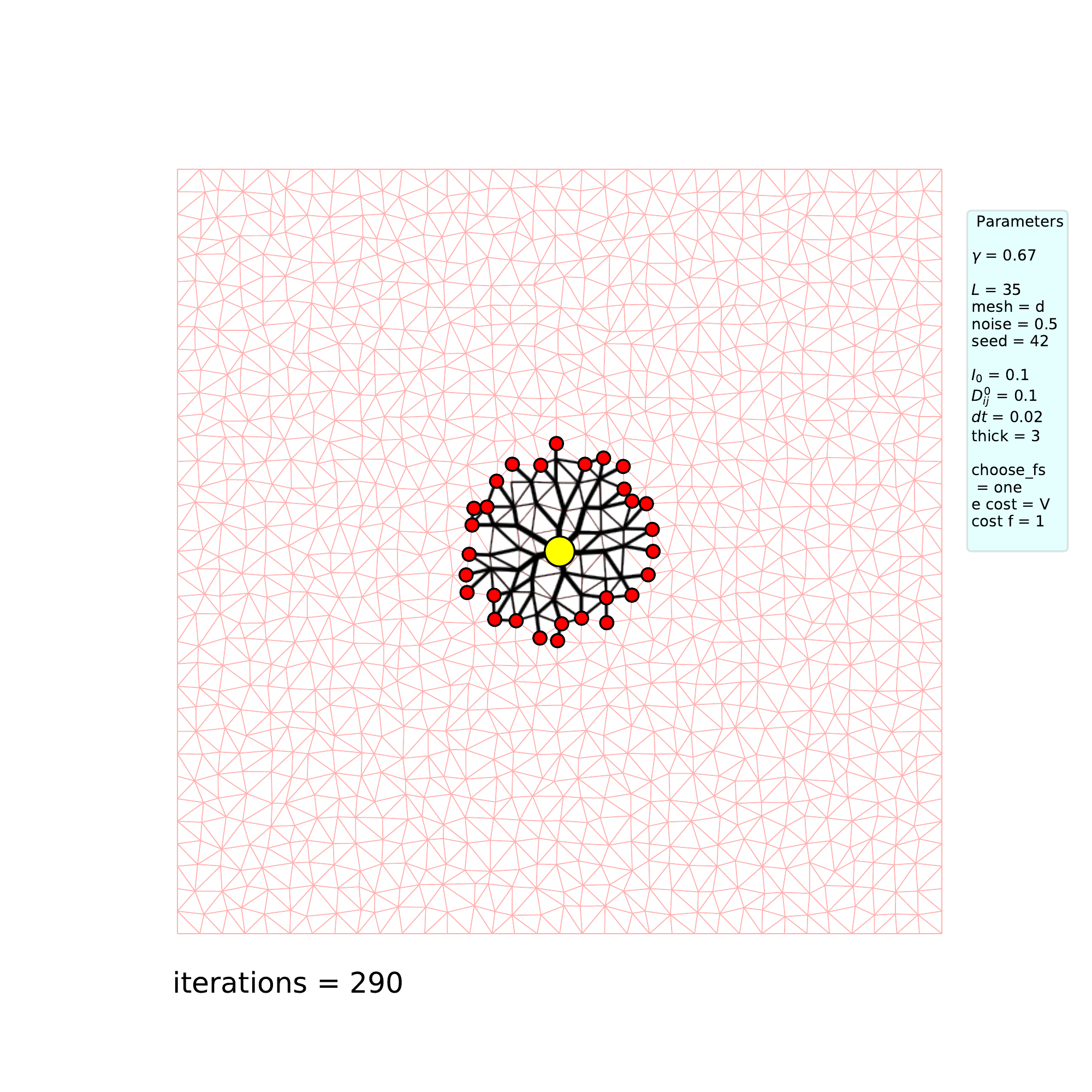} \hfill
\mysub[$t=900$]{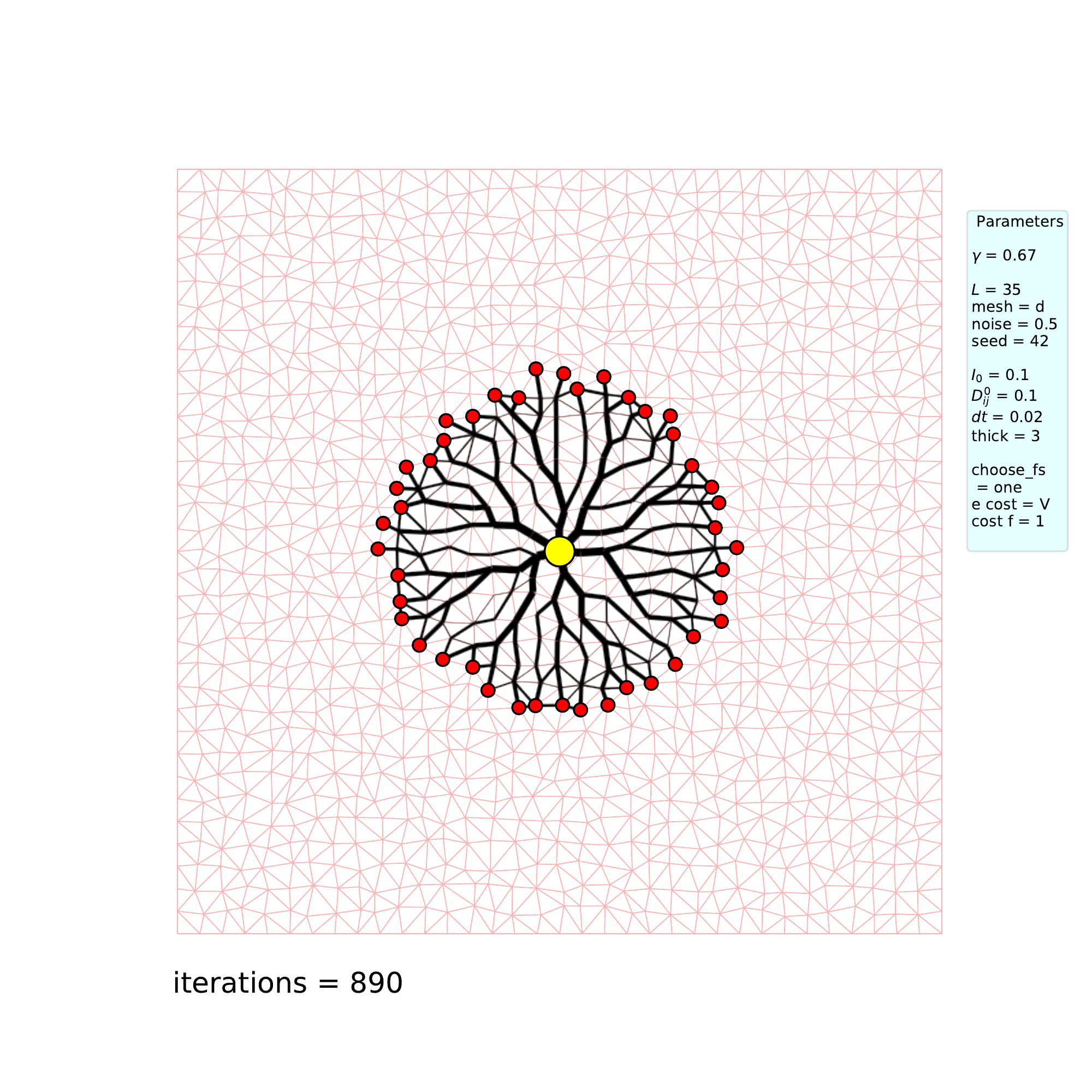} \hfill
\mysub[$t=1500$]{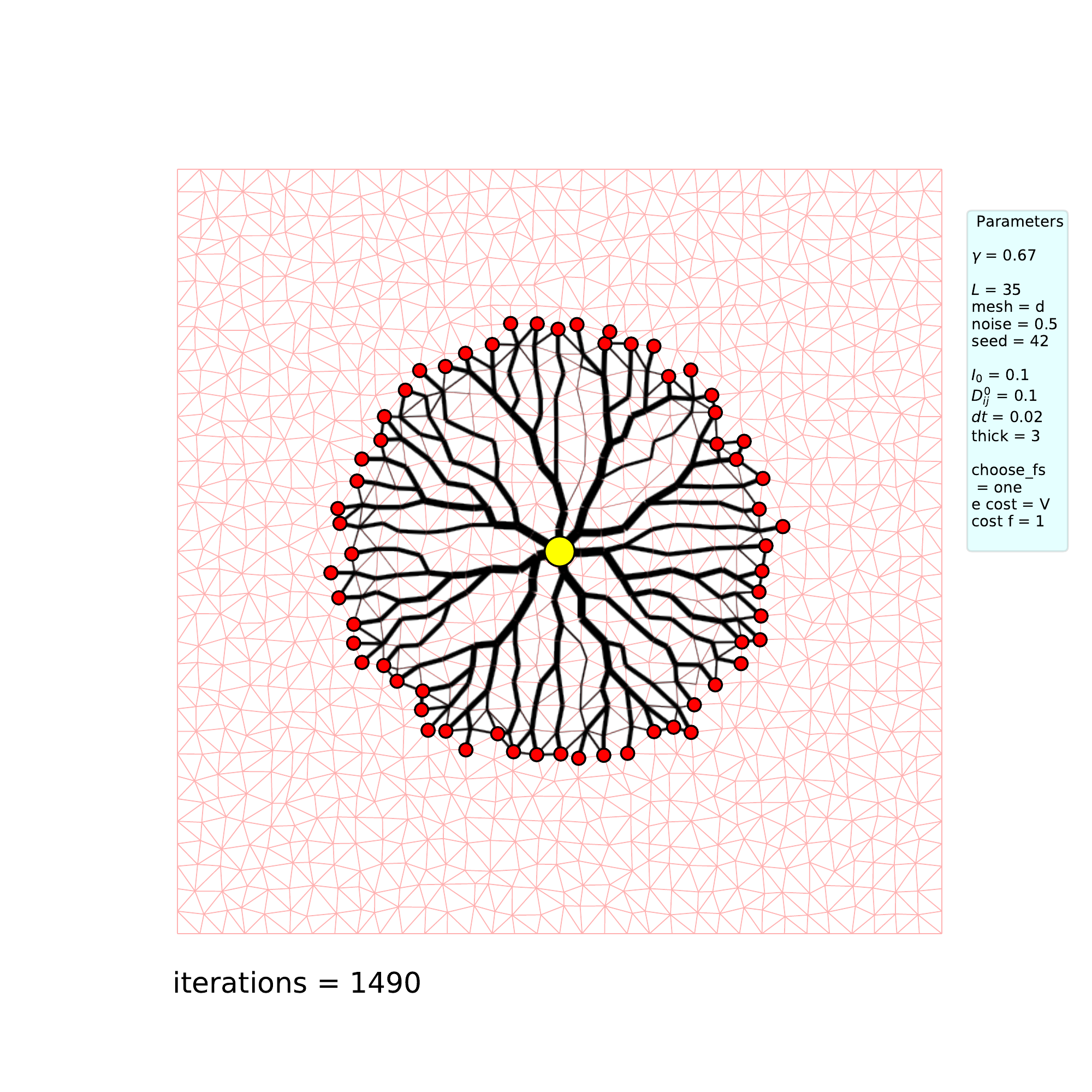} \\

\mysub[$t=2100$]{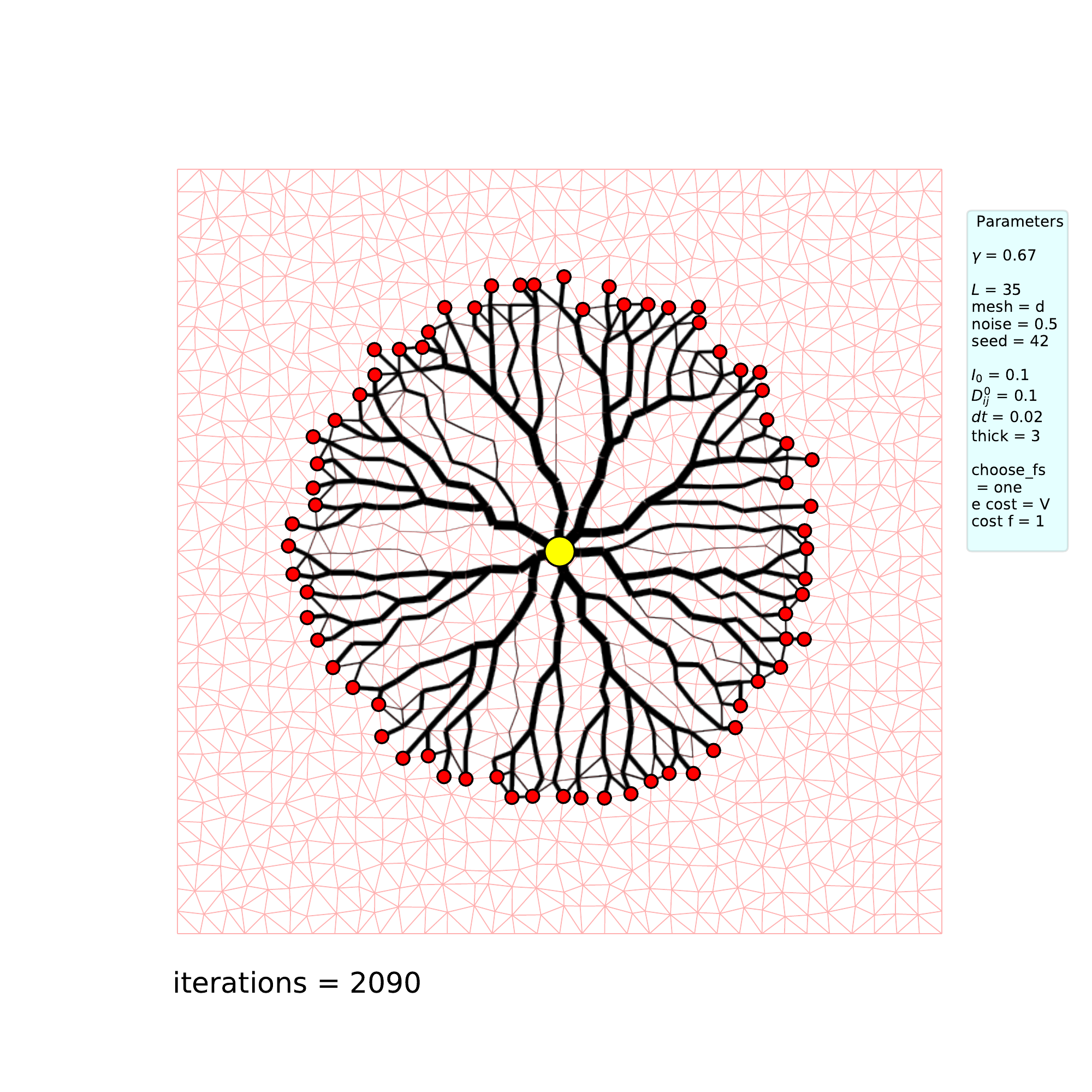} \hfill
\mysub[$t=2700$]{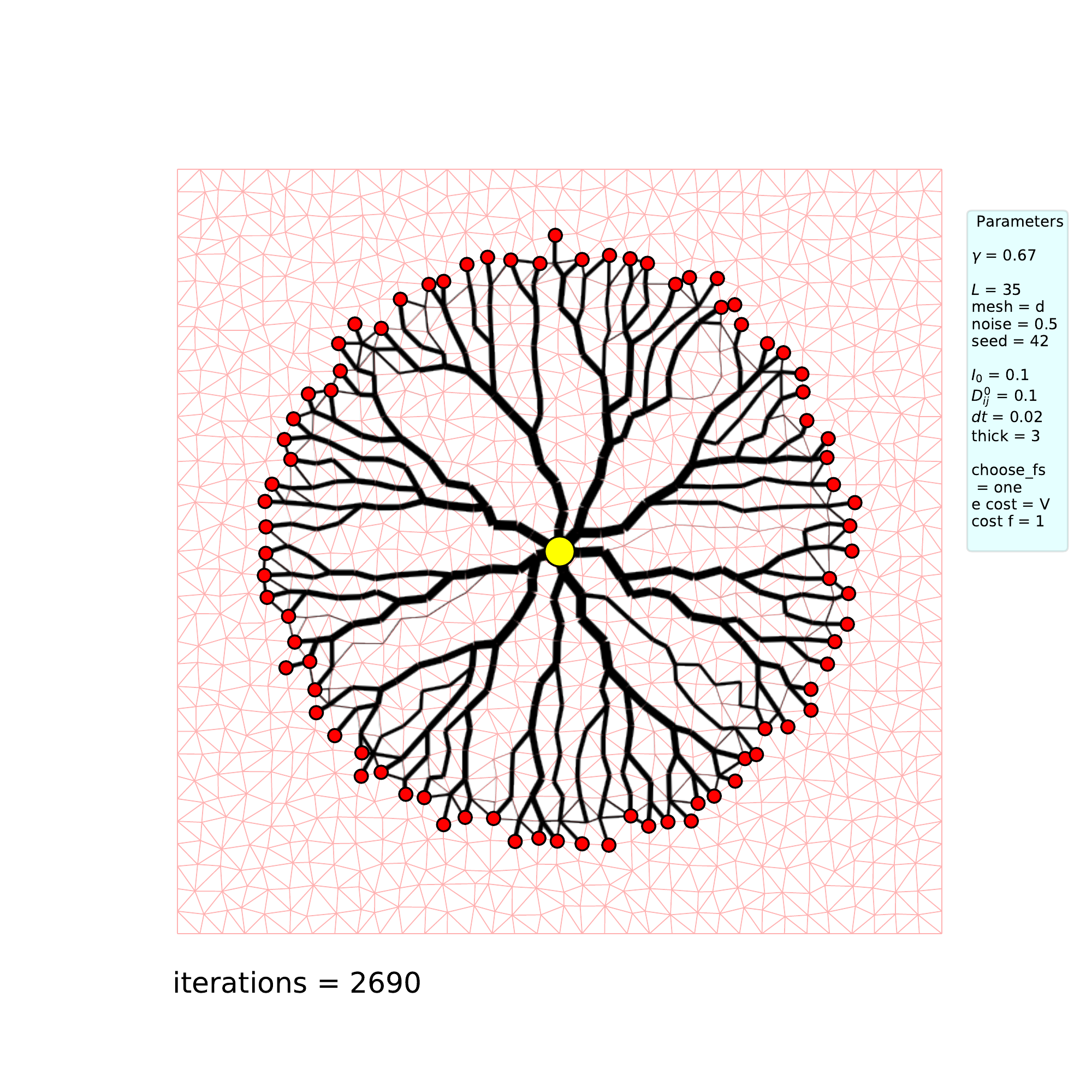} \hfill
\mysub[$t=3900$]{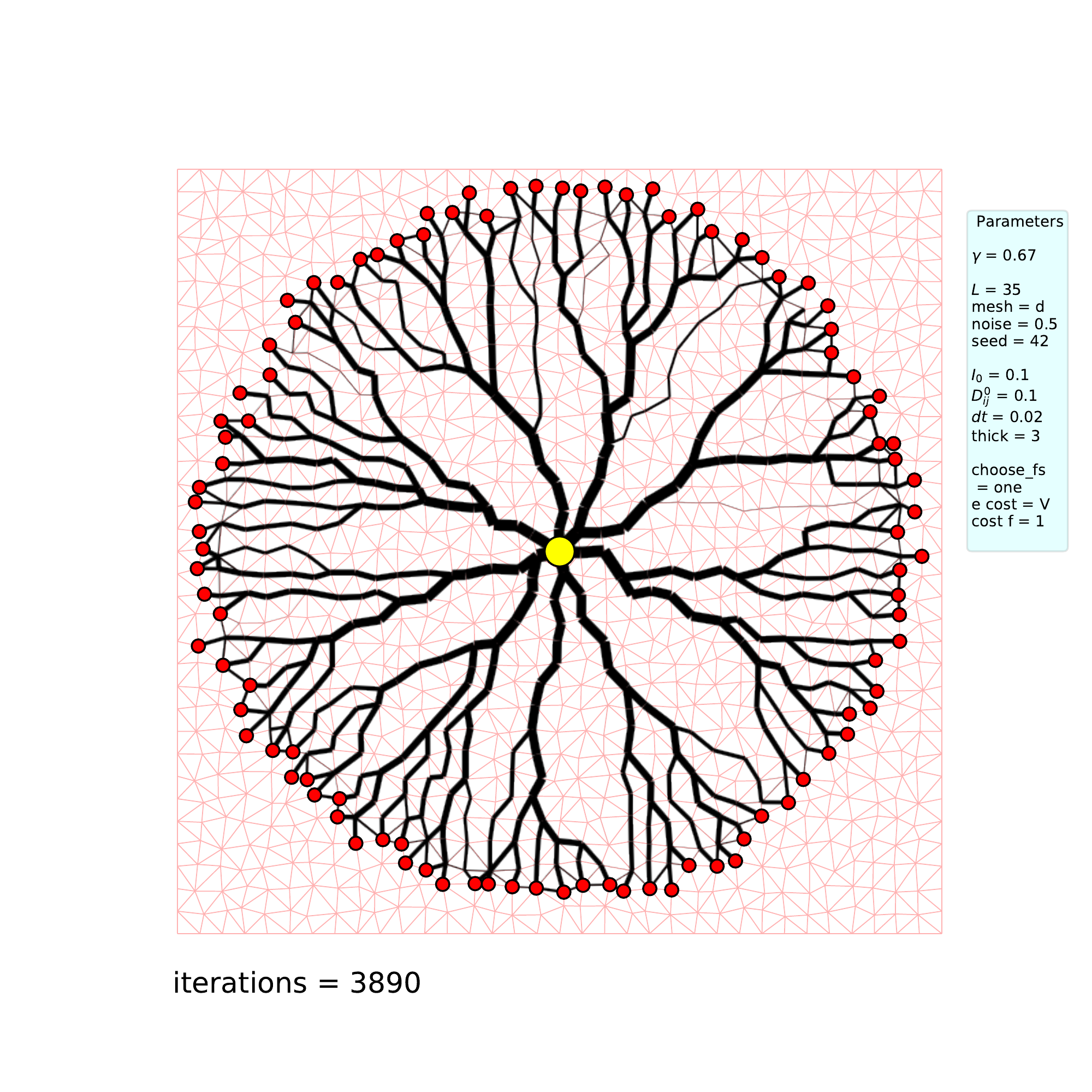} 

\caption{
Physarum growing from a food source considering the adaptation dynamics \eqref{eq:new_adapt_rule_gamma} with $\gamma=2/3$. The growth is driven by a central food source (yellow) which pumps nutrients to the moving boundary (red) where there is a constant uptake of nutrients to create new channels (with an initial conductivity $D_0$). In each time step, an amount of $I_0\dt$ nutrients given by the source is distributed evenly between the boundary sinks. As \textit{Physarum} grows, the channels adapt their thickness according to the flux of nutrients flowing through. Simulation carried out with $D_0=0.1$, $\dt=0.02$, $I_0=0.1$. The dynamics result in a tree-like network that resembles the networks produced by the real organism (Figure \ref{fig:phy_network}), although no stable cross-links between the main veins are formed. The labels $t$ designate the time step in which the snapshots were taken.
}
\label{fig:growth_g0.67} 
\end{figure}

We start by analysing the simplest case of the individual growing from a food source, without any more food sources available in the surroundings. In each step, the food source 
with intensity $q_{source}=I_0$ supplies an equal amount of nutrients to all the current boundary sinks. In Figure \ref{fig:growth_g0.67} it's shown snapshots of a simulation for the choice of parameters  $I_0=0.1$ and $\gamma=2/3$, and considering that channels are formed with a conductivity $D_0=0.1$.

As the images show, the growth occurs in a more or less isotropic fashion and the growth fronts are circular.
Furthermore, the growth slows down as time passes, since the boundary is progressively extended, and thus less food is supplied per time step to each boundary sink. These two features of the growth mechanism are biologically consistent. Note that the former is related to the underlying irregular mesh used, resulting from a Delaunay triangulation. Other types of meshes were tested as well, namely square, triangular and hexagonal lattice graphs. However, even when adding a small noise to the node positions to break the lattice symmetries, simulations have shown that the growth fronts weren't circular due to the regularity of the underlying graph, meaning the lattice symmetries were still quite noticeable. 

The dynamics resulted in tree-like networks which share some similarities with the networks produced by the real organisms (Figure \ref{fig:phy_network}). In particular, as the network grows the inner channels are highly optimised, leading to the formation of distinct fan-shaped growth fronts with a more dense branching as we move away from the source. This is especially noticeable as time increases.  However, an important feature is still missing, the interconnections between the main veins which result in a loopy architecture that provides robustness to the network. In the simulations for $\gamma = 2/3$, these redundant connections are not stable and ultimately disappear. 

\begin{figure}[hbt!]
\captionsetup[subfloat]{labelformat=empty,position=top,skip=0.5pt}

\newcommand{\mysub}[2][]{%
    \subfloat[#1]{\includegraphics[trim={2cm 2.7cm 2.5cm 3.1cm}, clip, width=0.31\textwidth]{#2}}%
}

\centering

\mysub[$t=300$]{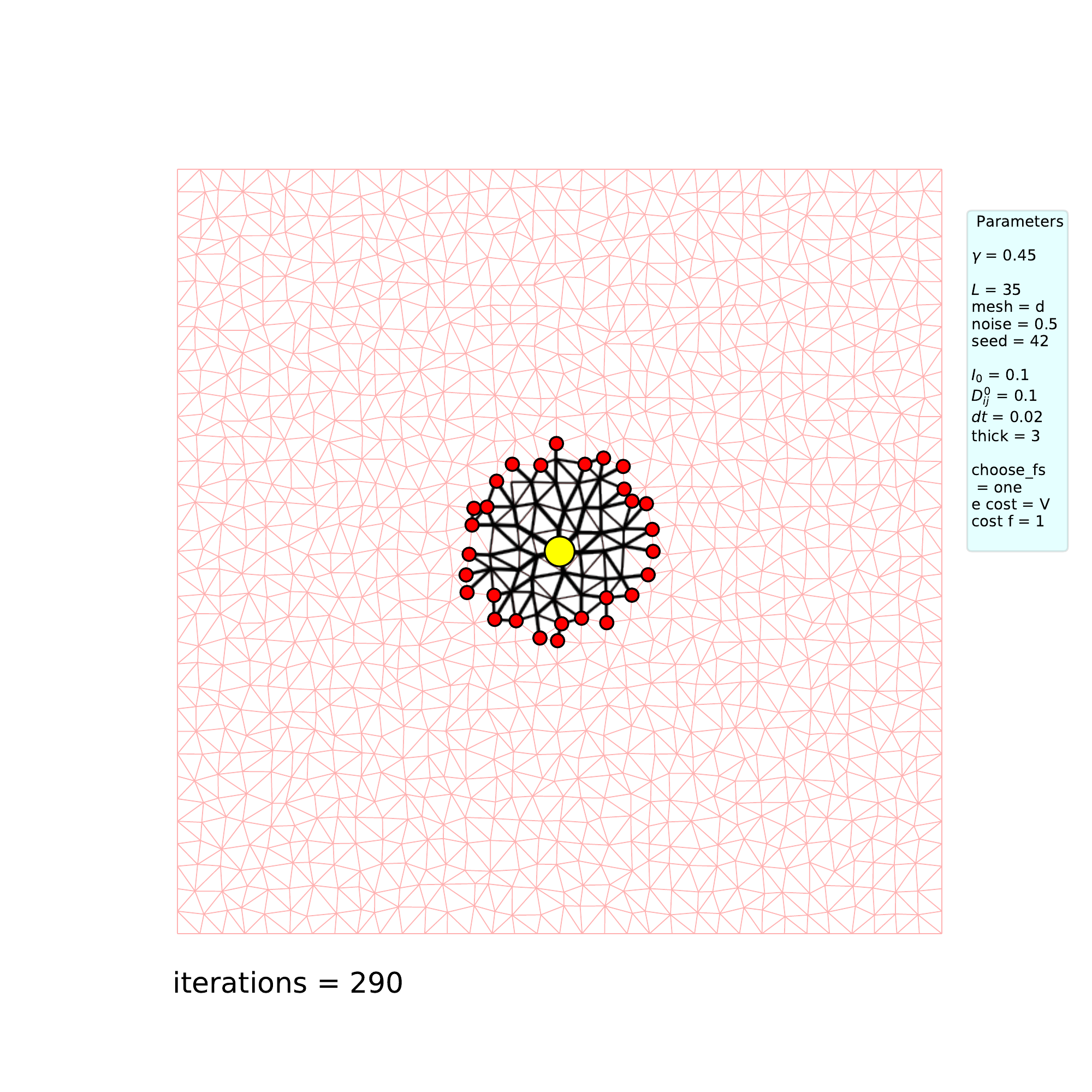} \hfill
\mysub[$t=900$]{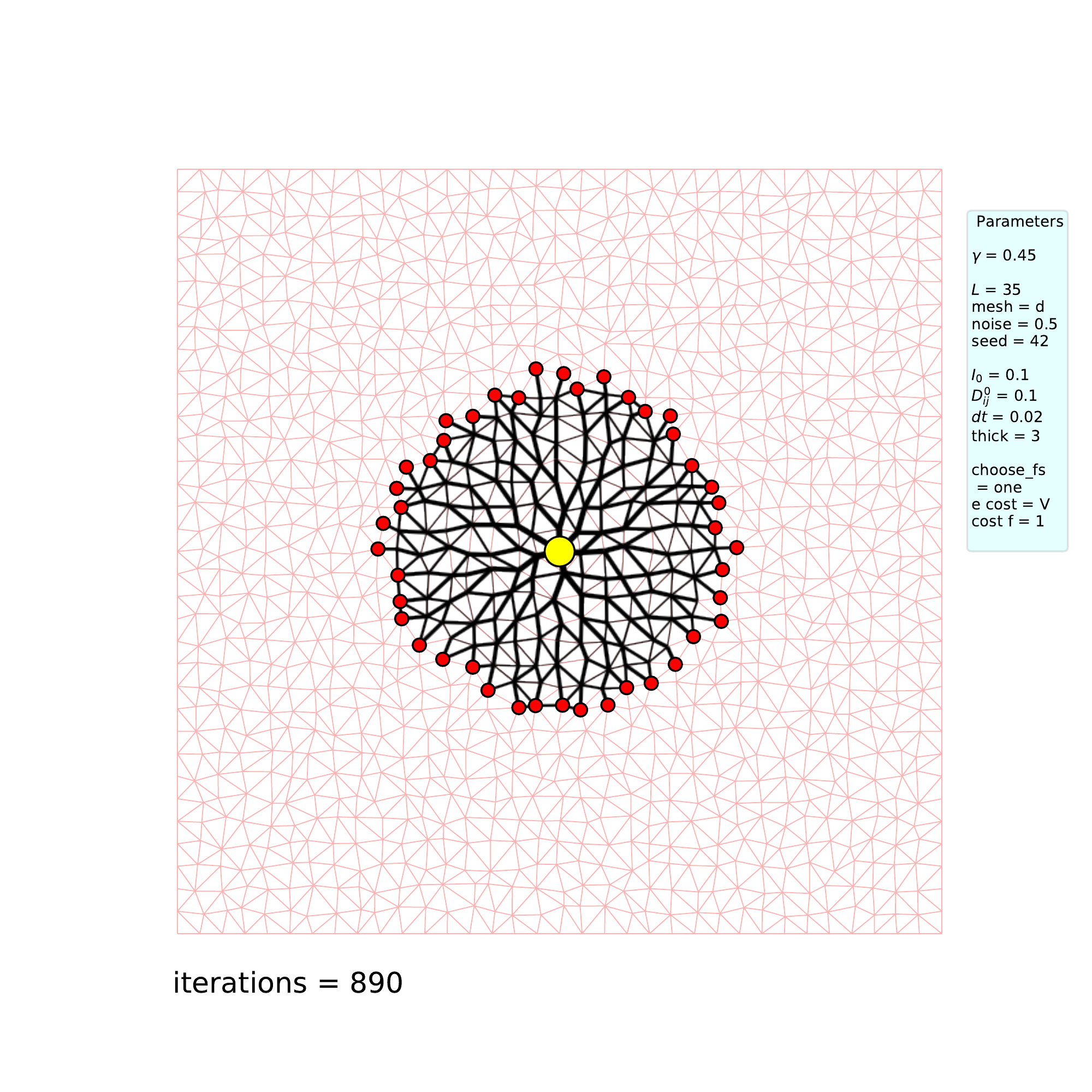} \hfill
\mysub[$t=1500$]{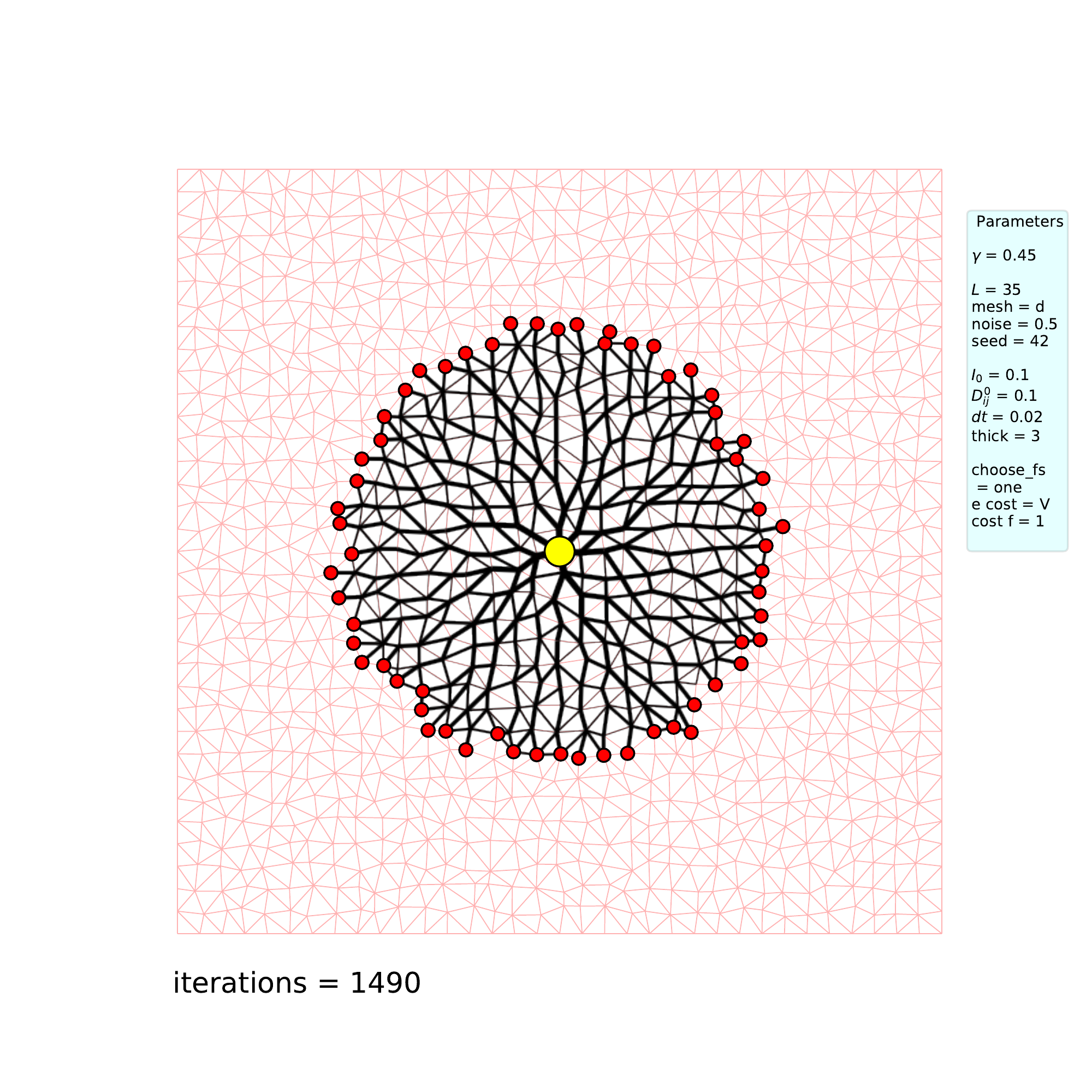} \\

\mysub[$t=2100$]{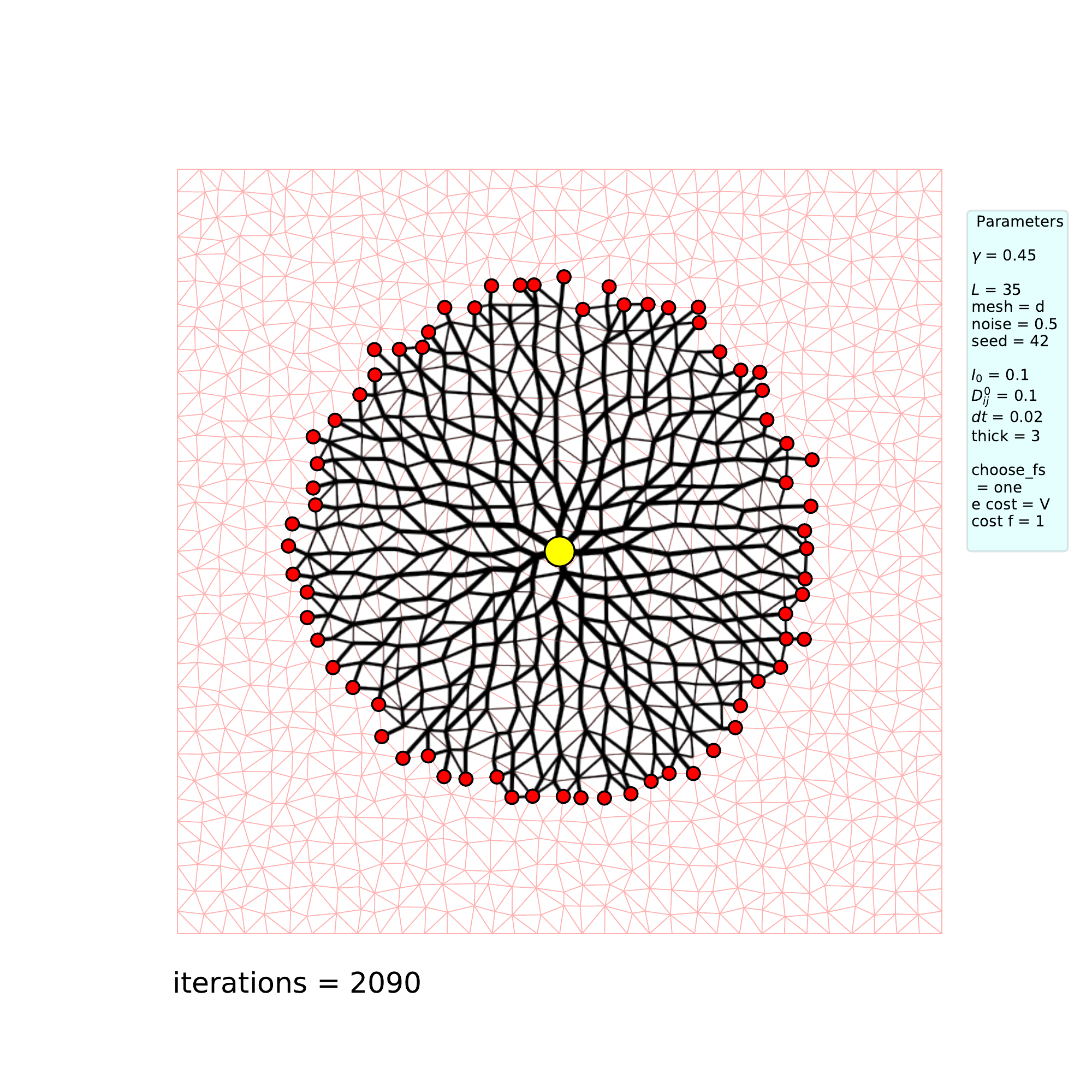} \hfill
\mysub[$t=2700$]{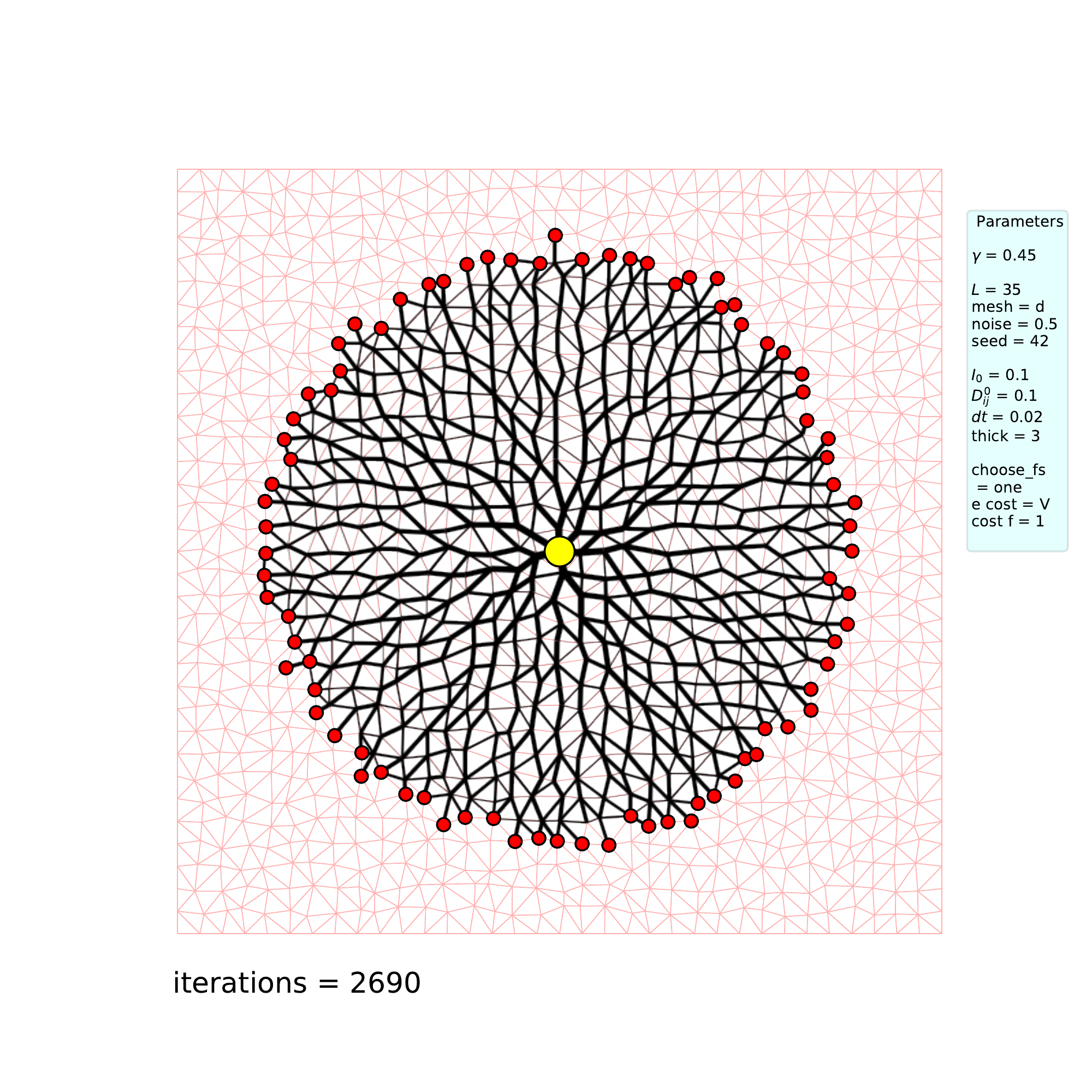} \hfill
\mysub[$t=3900$]{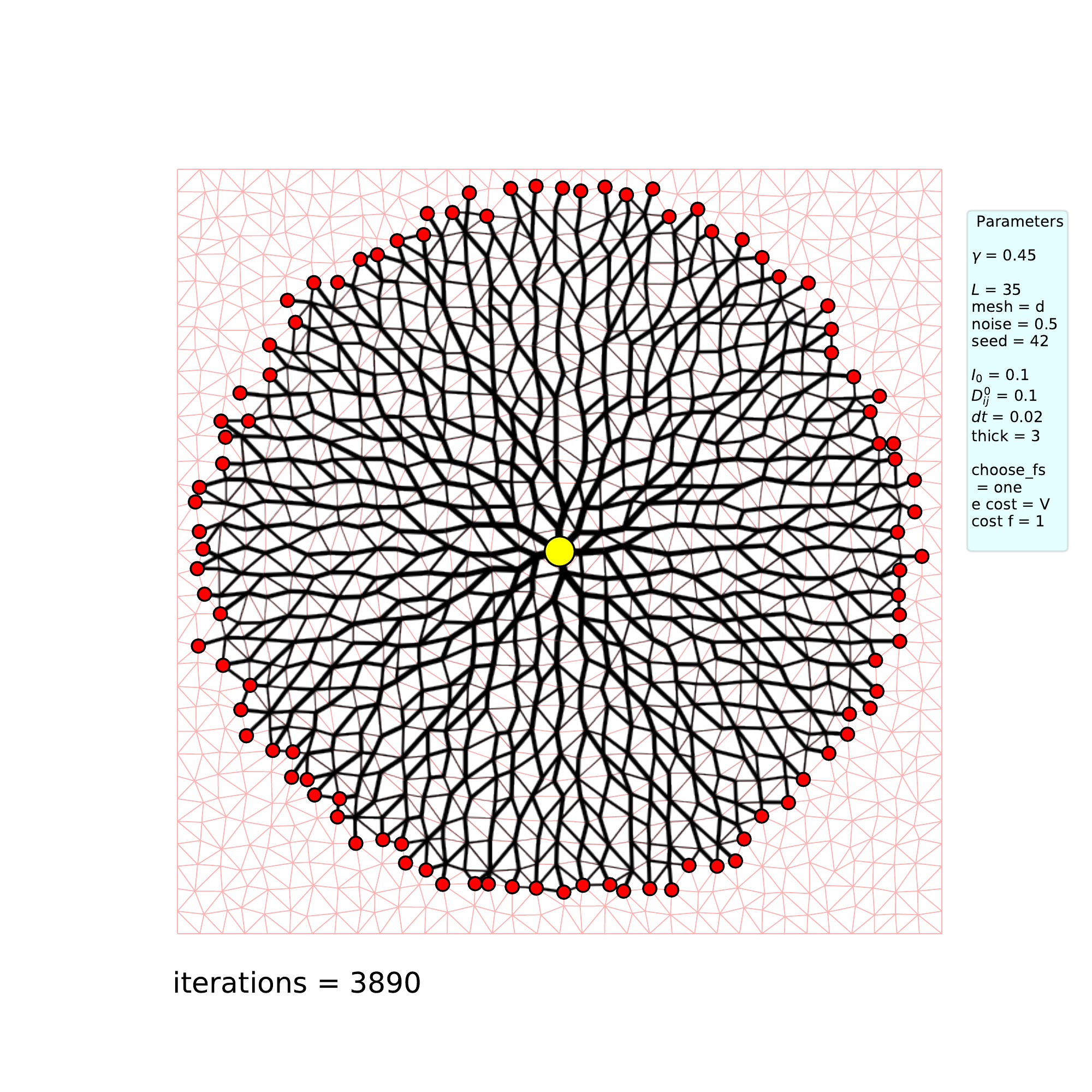} 

\caption{
Physarum growing from a food source considering the adaptation dynamics \eqref{eq:new_adapt_rule_gamma} with $\gamma=0.45$. Same settings as in Figure \ref{fig:growth_g0.67}. For $\gamma < 1/2$ the dynamics result in highly redundant networks very close to the underlying mesh. Despite the redundancy, due to the lack of optimisation and a clear hierarchy of vein thickness, the networks don't resemble the ones produced by the real organisms. The labels $t$ designate the time step in which the snapshots were taken.
}
\label{fig:growth_g0.45} 
\end{figure}

By contrast, similar to the static cases, for $\gamma < 1/2$ the networks end up being highly redundant, as Figure \ref{fig:growth_g0.45} shows.  However, they are  poorly optimised and there isn't a clear hierarchy of veins as observed in the real \textit{Physarum} networks, where the thicker main veins branch into progressively thinner ones.

\begin{figure}[hbt!] 

\captionsetup[subfloat]{labelformat=empty,position=top,skip=0.5pt}

\newcommand{\mysub}[2][]{%
    \subfloat[#1]{\includegraphics[trim={2cm 2.7cm 2.5cm 3.1cm}, clip, width=0.31\textwidth]{#2}}%
}

\centering

\mysub[$t=300$]{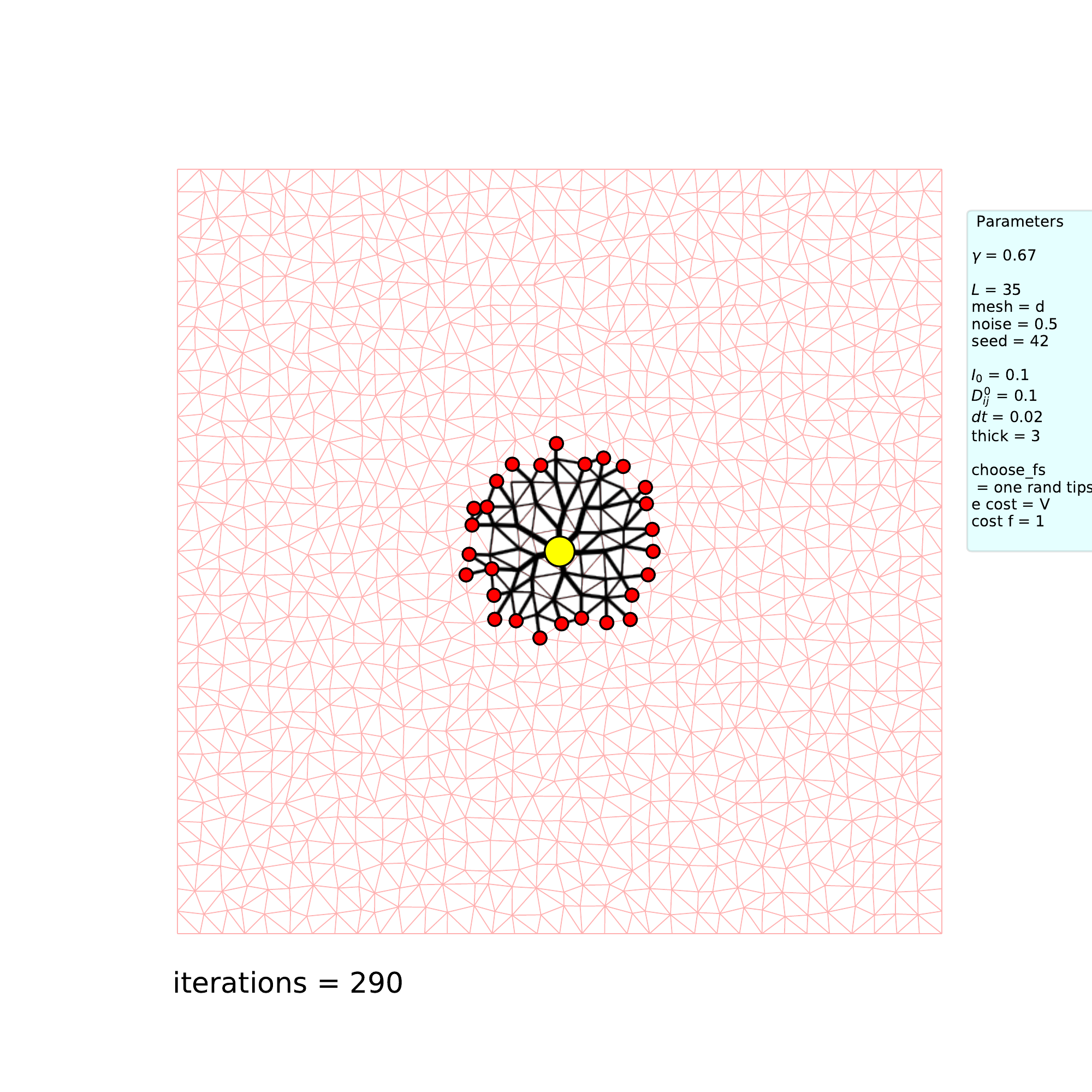} \hfill
\mysub[$t=900$]{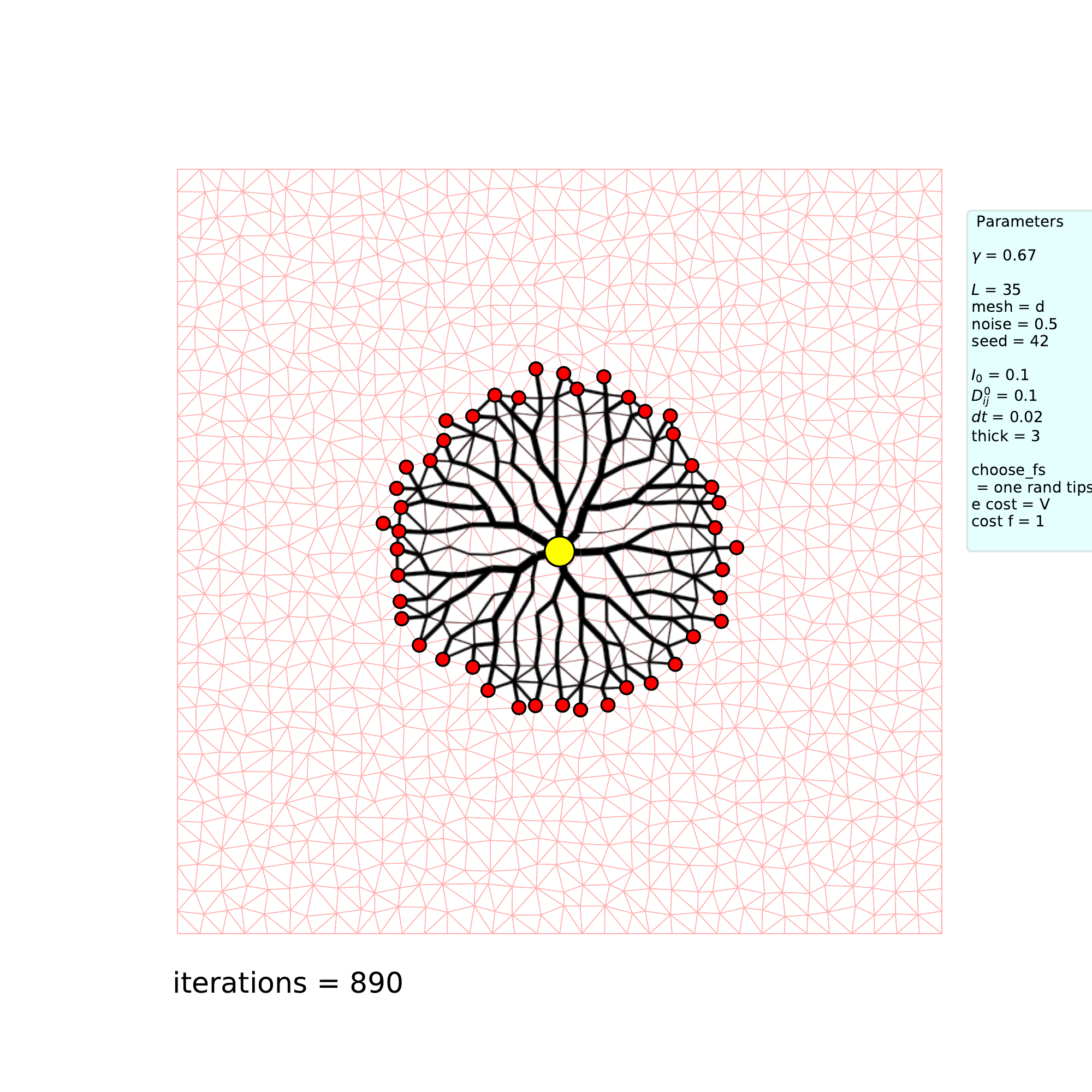} \hfill
\mysub[$t=1500$]{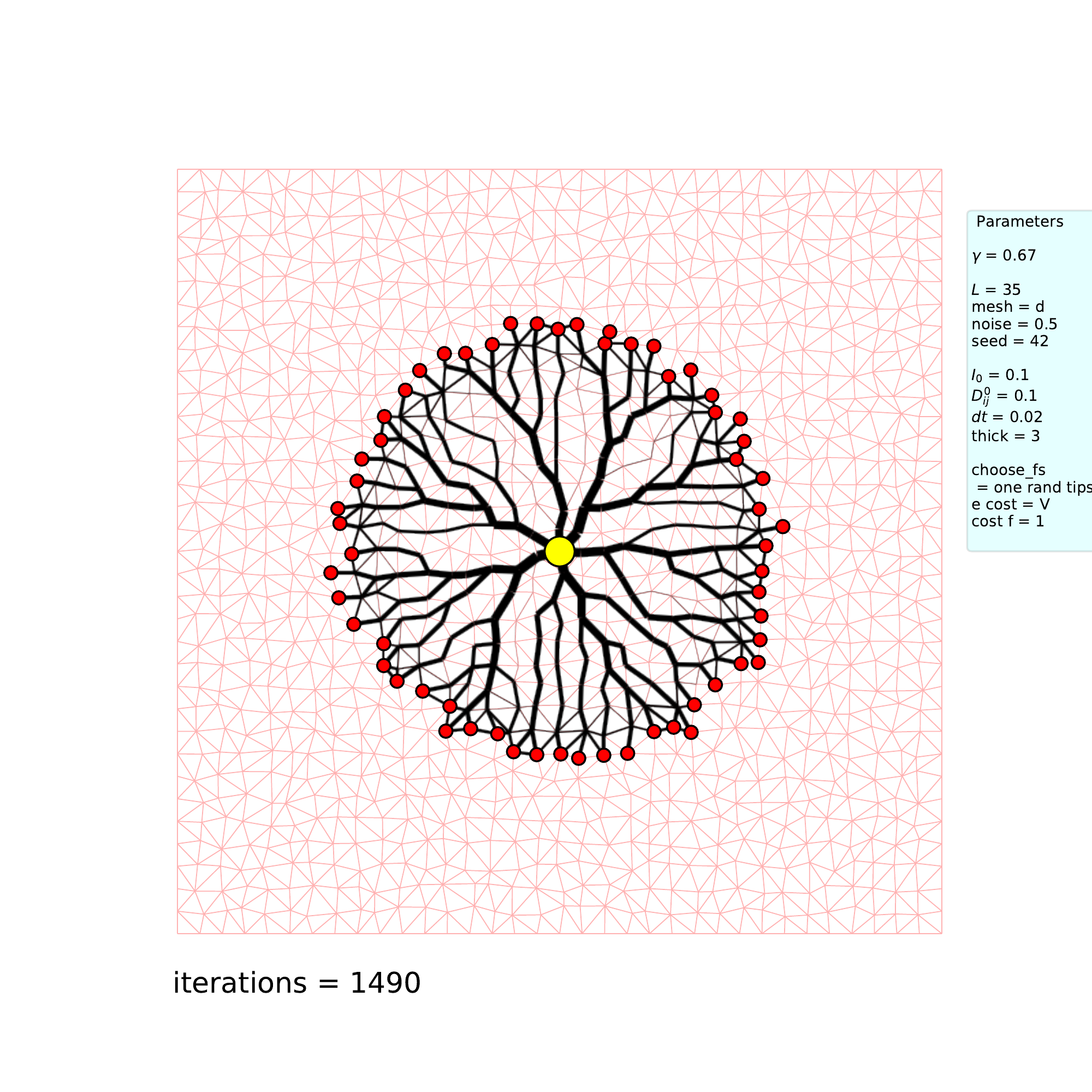} \\

\mysub[$t=2100$]{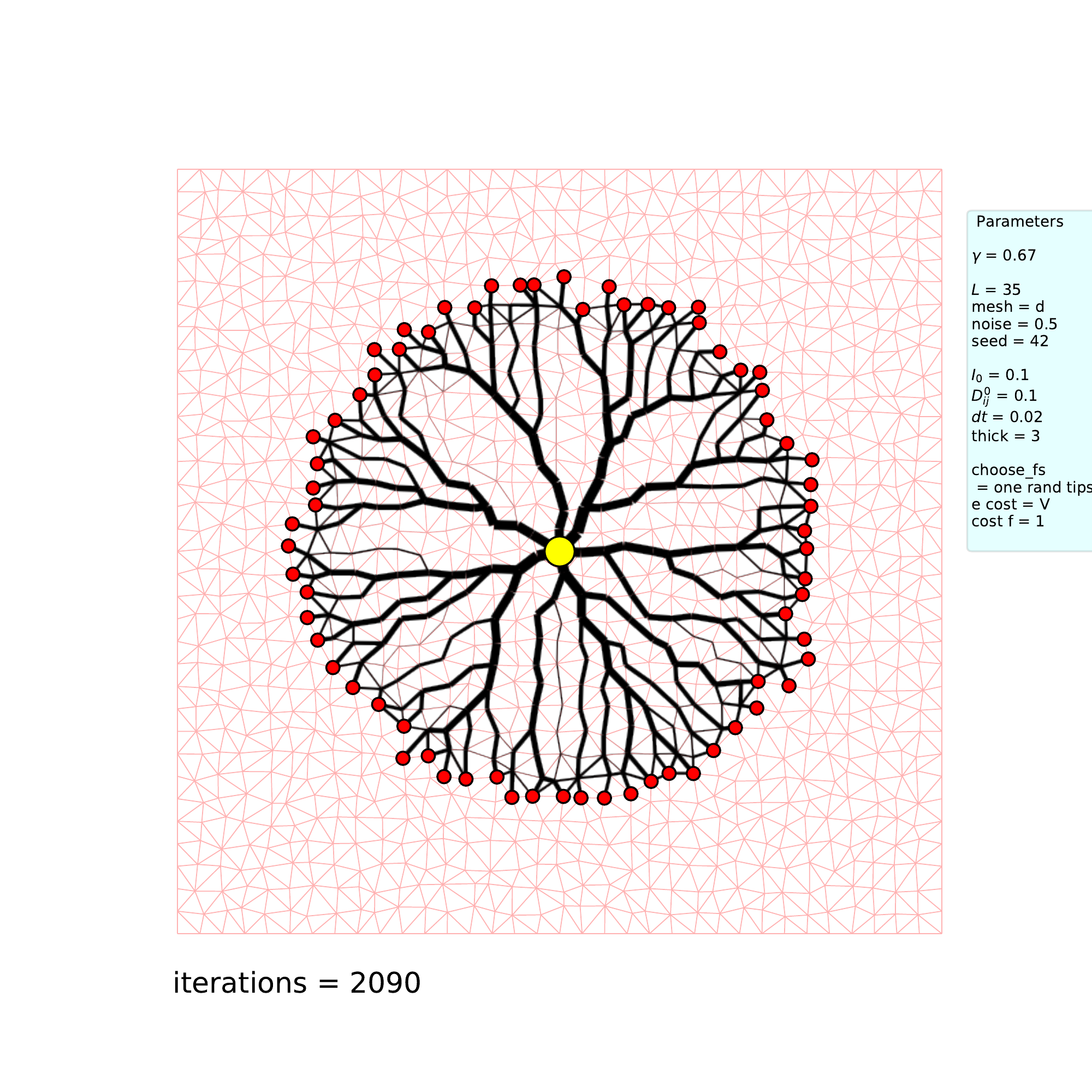} \hfill
\mysub[$t=2700$]{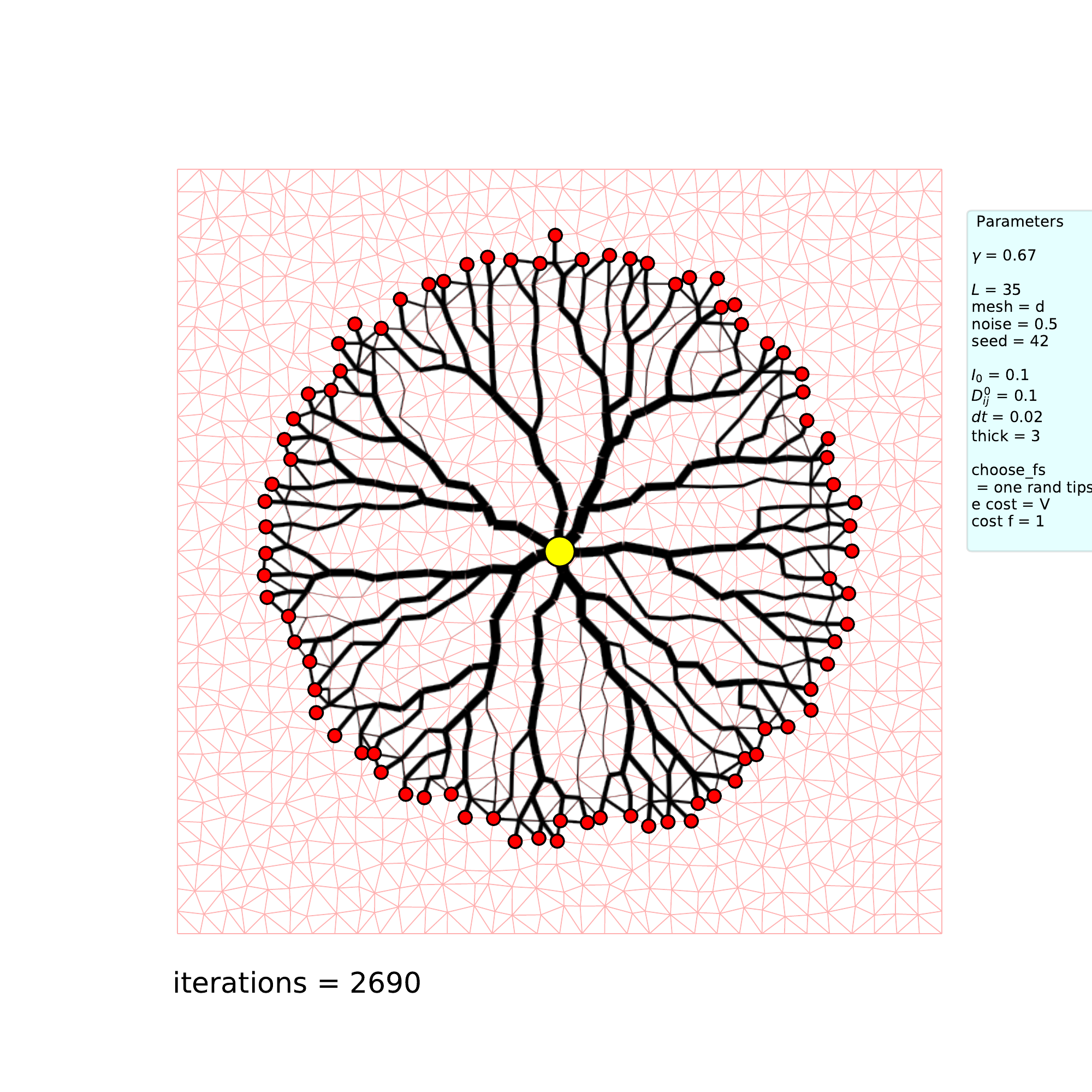} \hfill
\mysub[$t=3900$]{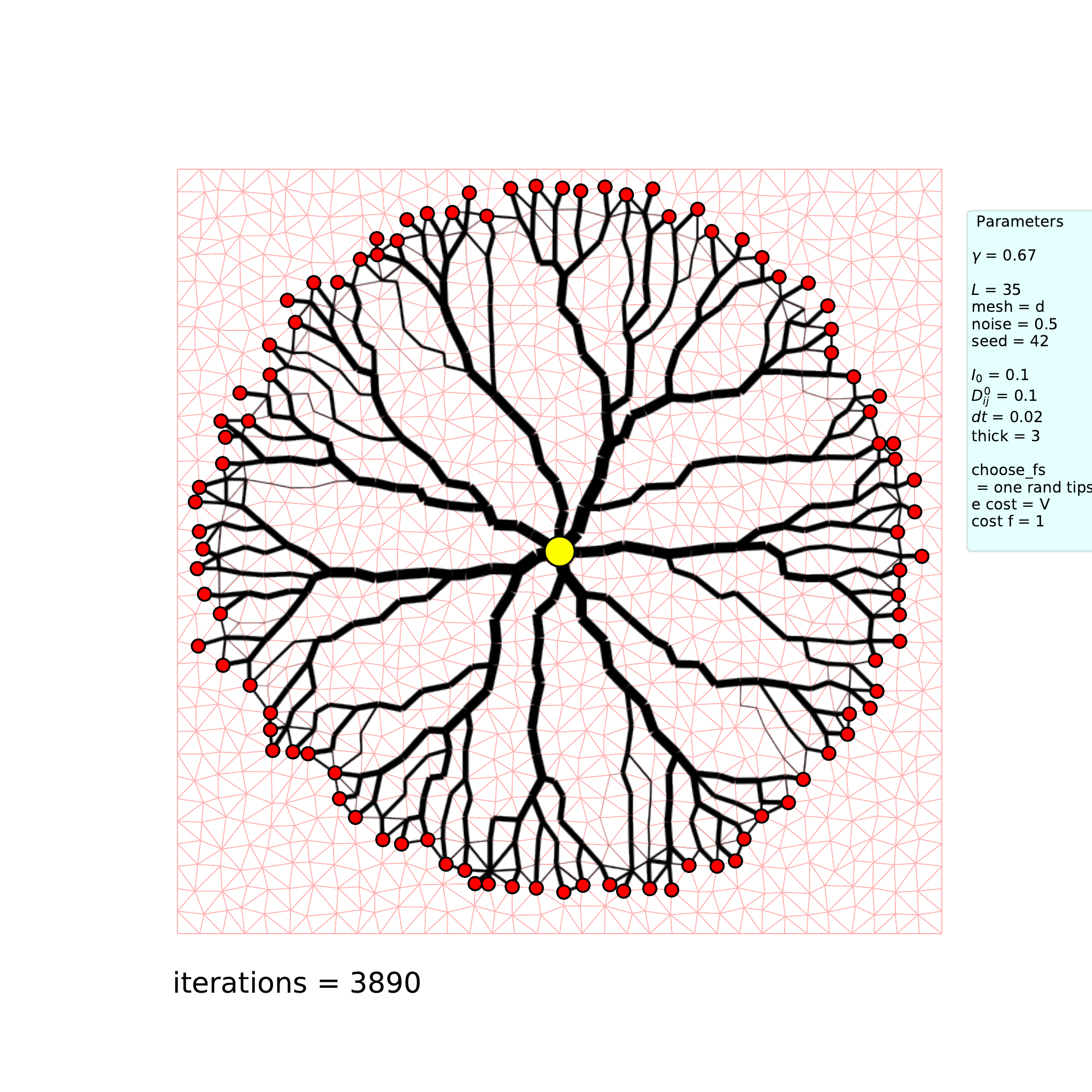} 

\caption{Physarum growing from a food source considering the adaptation dynamics \eqref{eq:new_adapt_rule_gamma} with $\gamma=2/3$ and the sinks receiving a random amount of flux over time. Same settings as in Figure \ref{fig:growth_g0.67}, except that in each time step, the amount of nutrients given by the source is distributed randomly between the boundary sinks. The network evolution is nearly identical to the case of the nodes receiving uniformly, and therefore a random distribution of nutrients can't explain the formation of stable loops observed in \textit{Physarum} networks. 
}
\label{fig:growth_1_random} 
\end{figure}

Previous results of the static optimisation revealed that fluctuations of the nodes fluxes have a great impact on the topology of the final networks, and thus could be at the origin of the stable redundant paths. Until now, we assumed a radial growth driven by a central source and uniform boundary sinks. As discussed previously, the adaptation tends to reinforce the channels along the shortest paths connecting the sources and sinks, which in this case correspond to the connections along the radial direction, as the flux develops preferentially in that direction. Therefore the angular connections end up being the first ones to disappear. 
As an attempt to prevent this behaviour, we considered the hypothesis of each boundary sink receiving a random fraction of the source flux in each time step. In principle, this would create perturbations in the direction of the flow, which could lead to the formation of stable angular connections between the main branches. However, as the Figure \ref{fig:growth_1_random}  shows, the network evolution is very similar to the case of the sinks receiving uniformly (Figure \ref{fig:growth_g0.67}), and redundant paths are still not formed. We conclude that the hypothesis of random sinks is still not sufficient to explain the loops observed in the real \textit{Physarum} networks.

\subsection{Multiple food sources}

Finally, we simulated the case of the \textit{Physarum}
accommodating new food sources as it grows, which is a better representation of its foraging behaviour. However, it's not clear how real organisms manage food consumption when they acquire multiple food sources. We studied  two different methods considering the growth in the presence of two food sources. In both cases, the nutrients supplied by the active food sources are evenly shared among the boundary sinks, and the simulations were performed with $\gamma = 2/3$, $I_0=0.1$, $D_0=0.1$ and $\dt = 0.02$. 

\begin{figure}[hbt!]

\captionsetup[subfloat]{labelformat=empty,position=top,skip=0.5pt}

\newcommand{\mysub}[2][]{%
    \subfloat[#1]{\includegraphics[trim={2cm 2.7cm 2.5cm 3.1cm}, clip, width=0.31\textwidth]{#2}}%
}

\centering

\mysub[$t=300$]{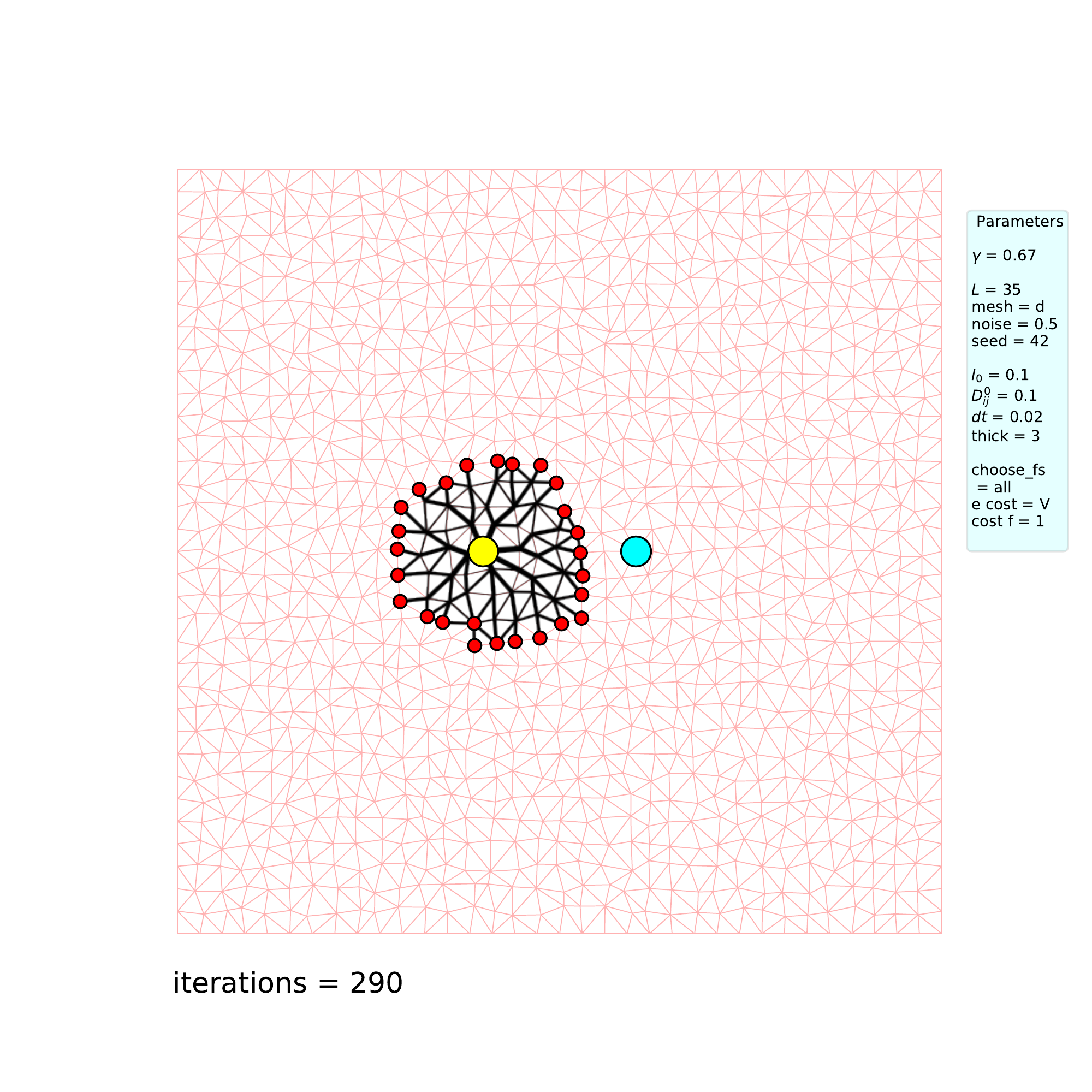} \hfill
\mysub[$t=600$]{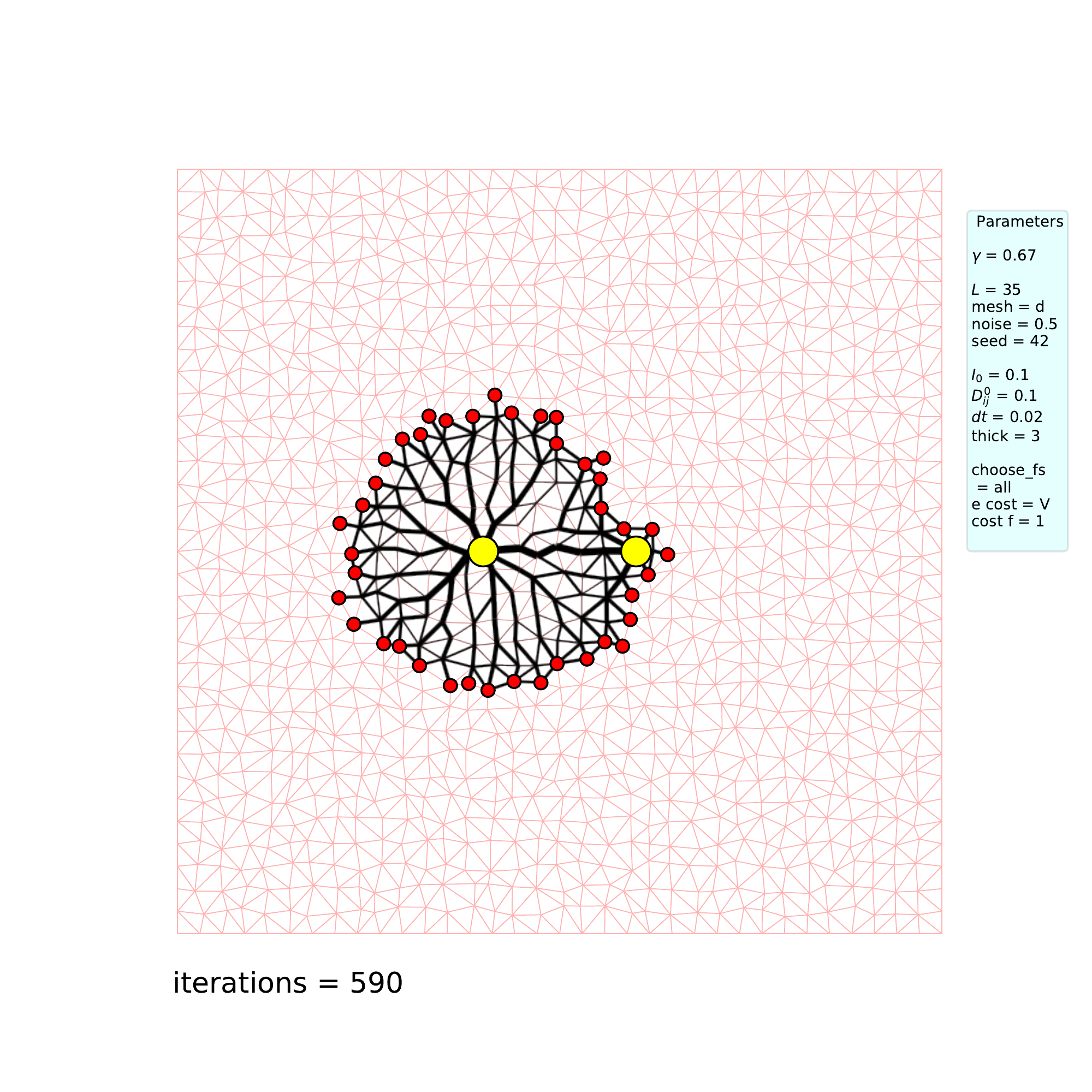} \hfill
\mysub[$t=900$]{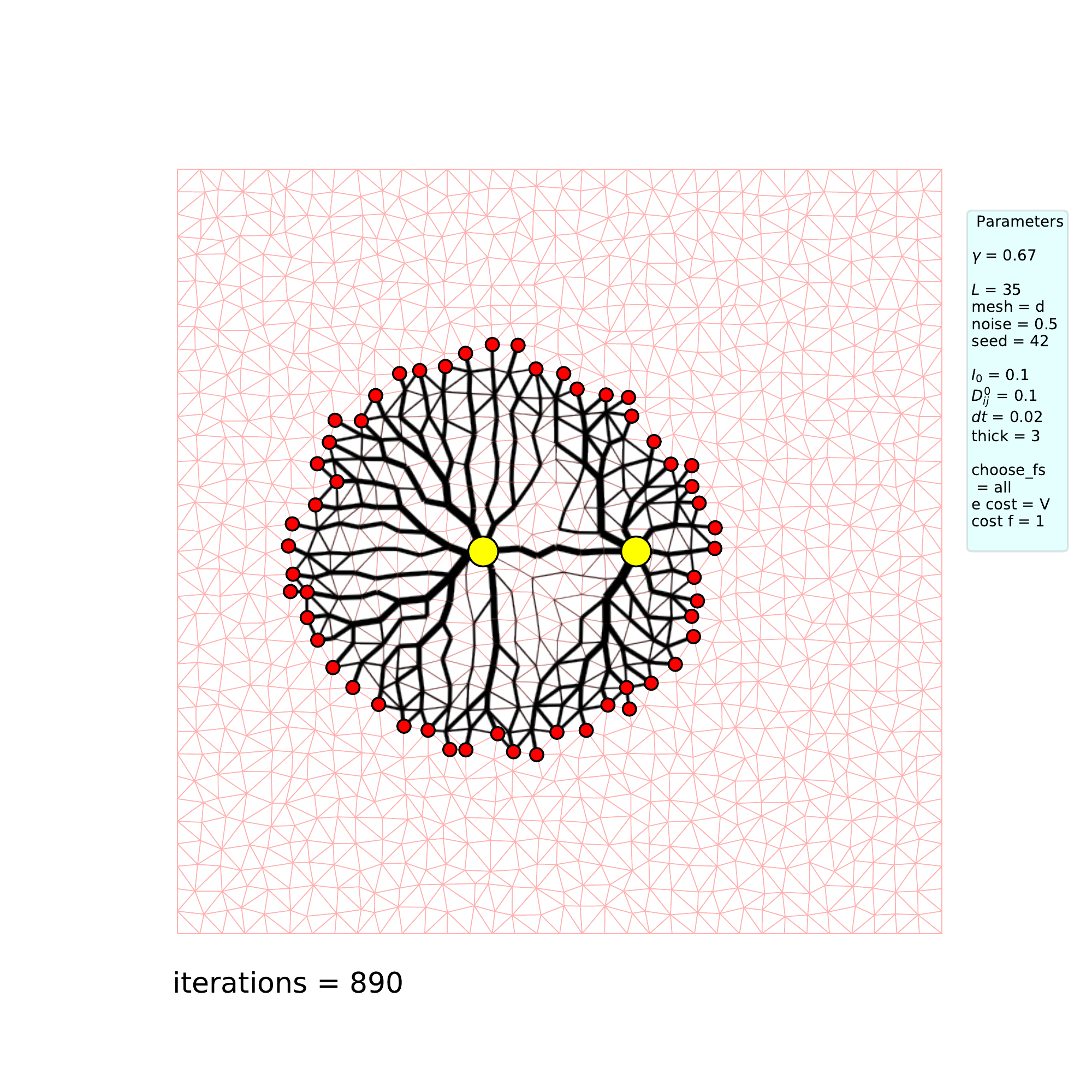} \\

\mysub[$t=1200$]{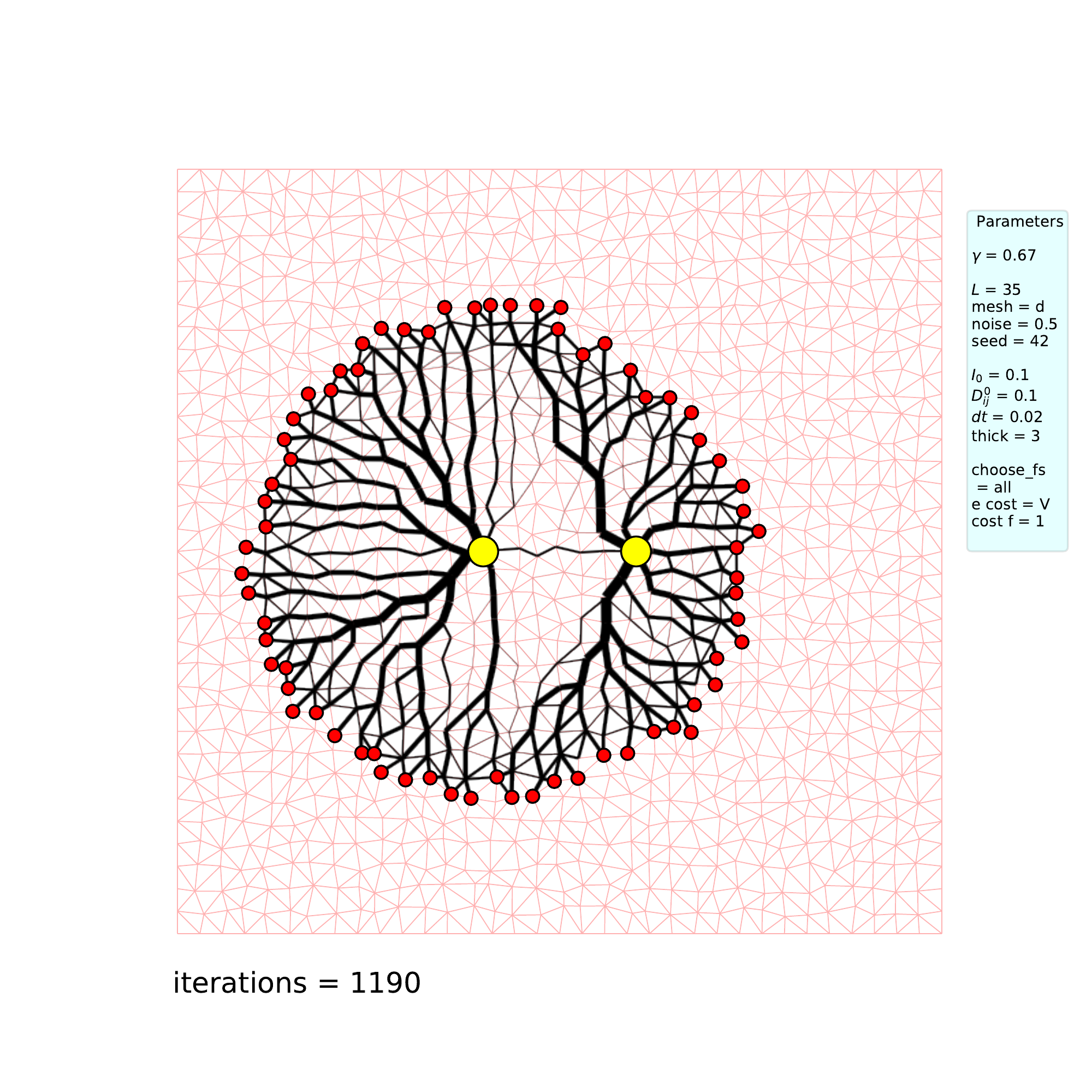} \hfill
\mysub[$t=1500$]{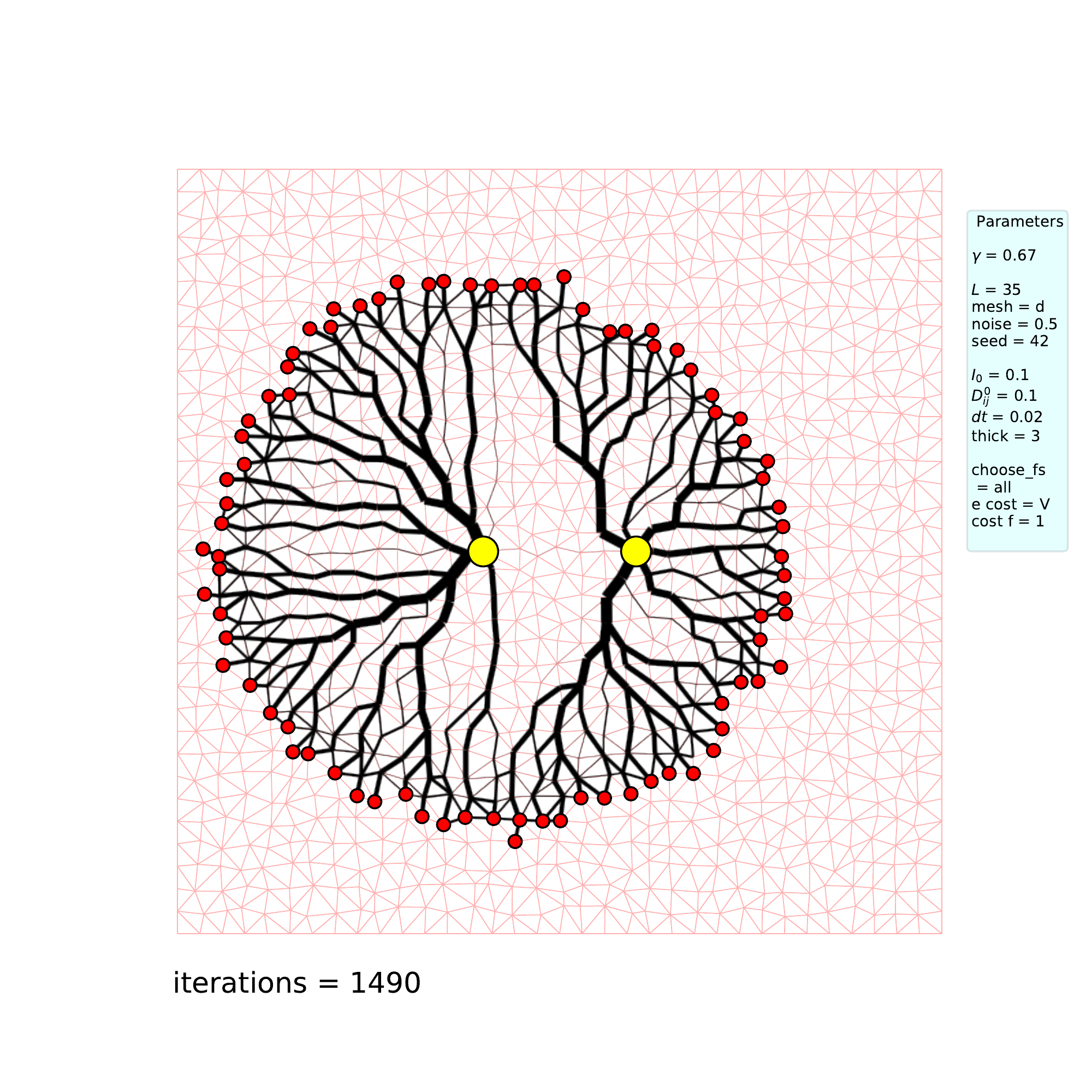} \hfill
\mysub[$t=1800$]{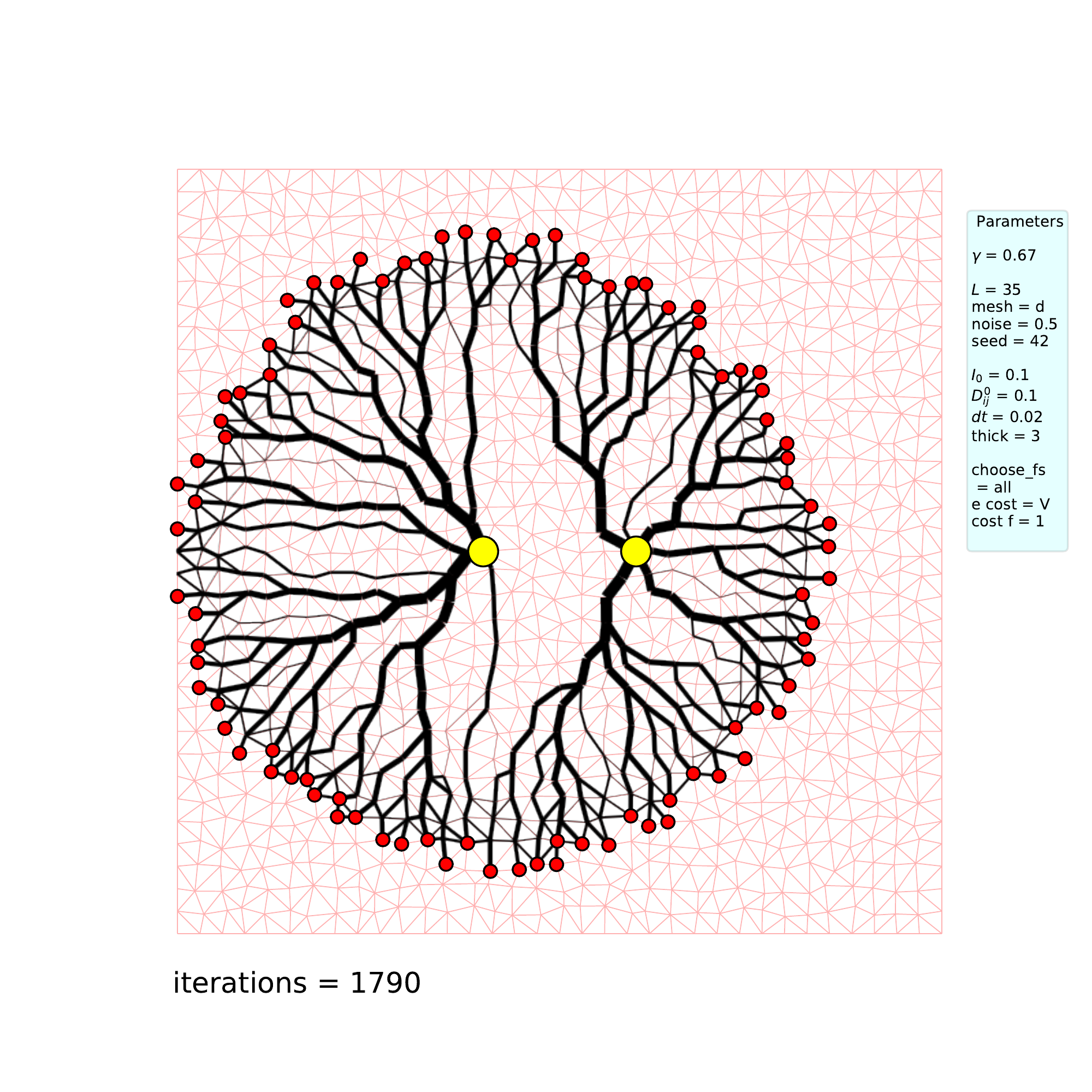} 

\caption{
\textit{Physarum} growing in the presence of two food sources with constant input flux. Same parameters as in Figure \ref{fig:growth_g0.67}. Starting from the left food source, \textit{Physarum} accommodates a second food source (blue circle) as it forages. From that moment, both sources are always active, each injecting a constant $I_0\dt$ amount of nutrients per time step, which is evenly distributed between the boundary sinks (red circles). Yellow circles represent active food sources in a given moment. The synchronous activation of both food sources leads to their ``repulsion'', meaning that no direct connection is formed between them. This is unrealistic in the context of \textit{Physarum}.
}
\label{fig:growth_2_const}
\end{figure}

\begin{figure}
    \centering
    \includegraphics[width=0.51\textwidth]{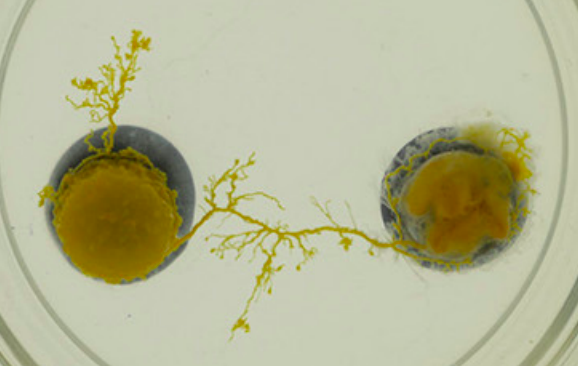}
    \caption{Real \textit{Physarum} growing in a presence of two food sources (agar blocks). A direct connection is established between them. Adapted from \cite{Ray2019}.
    }
    \label{fig:phy_2_FS}
\end{figure}

In the first case,  we considered that as soon as the second food source was reached, both food sources were constantly operational and injected the same amount of nutrients into the network. This means that from that moment, the amount of flux received by each boundary node was doubled. Snapshots of the simulation are depicted in Figure \ref{fig:growth_2_const}. As the images show, after the second food source is accommodated, the short connections between the two are weakened and ultimately collapse. We conclude that 
the synchronous and continuous operation of both food sources leads to their repulsion. This is the opposite of the true \textit{Physarum}'s behaviour, which tends to connect the food sources through short paths to optimise the transport of nutrients and minimise the costs (Figure \ref{fig:phy_2_FS}).

\begin{figure}[hbt!] 
\captionsetup[subfloat]{labelformat=empty,position=top,skip=0.5pt}

\newcommand{\mysub}[2][]{%
    \subfloat[#1]{\includegraphics[trim={2cm 2.7cm 2.5cm 3.1cm}, clip, width=0.31\textwidth]{#2}}%
}

\centering

\mysub[$t=300$]{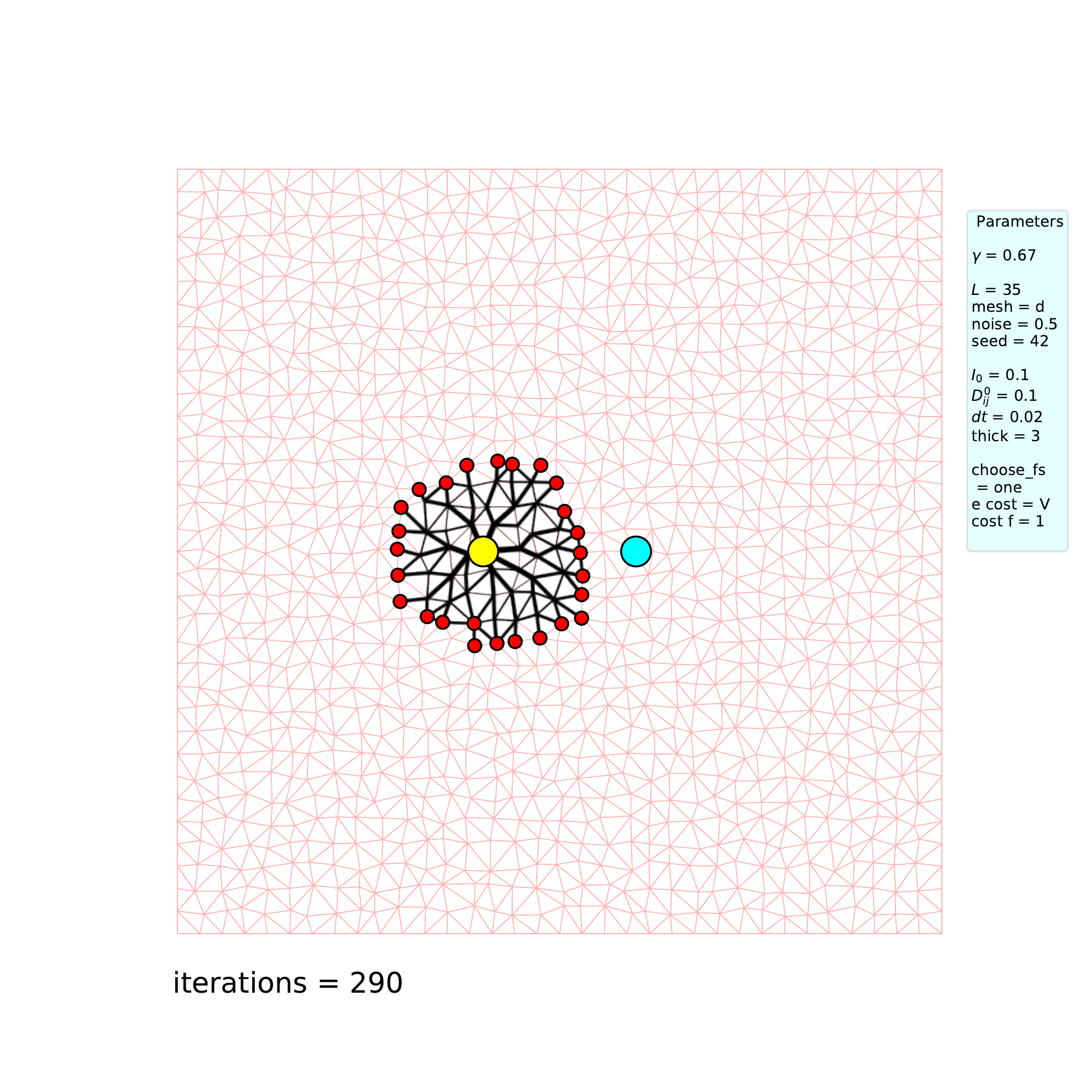} \hfill
\mysub[$t=600$]{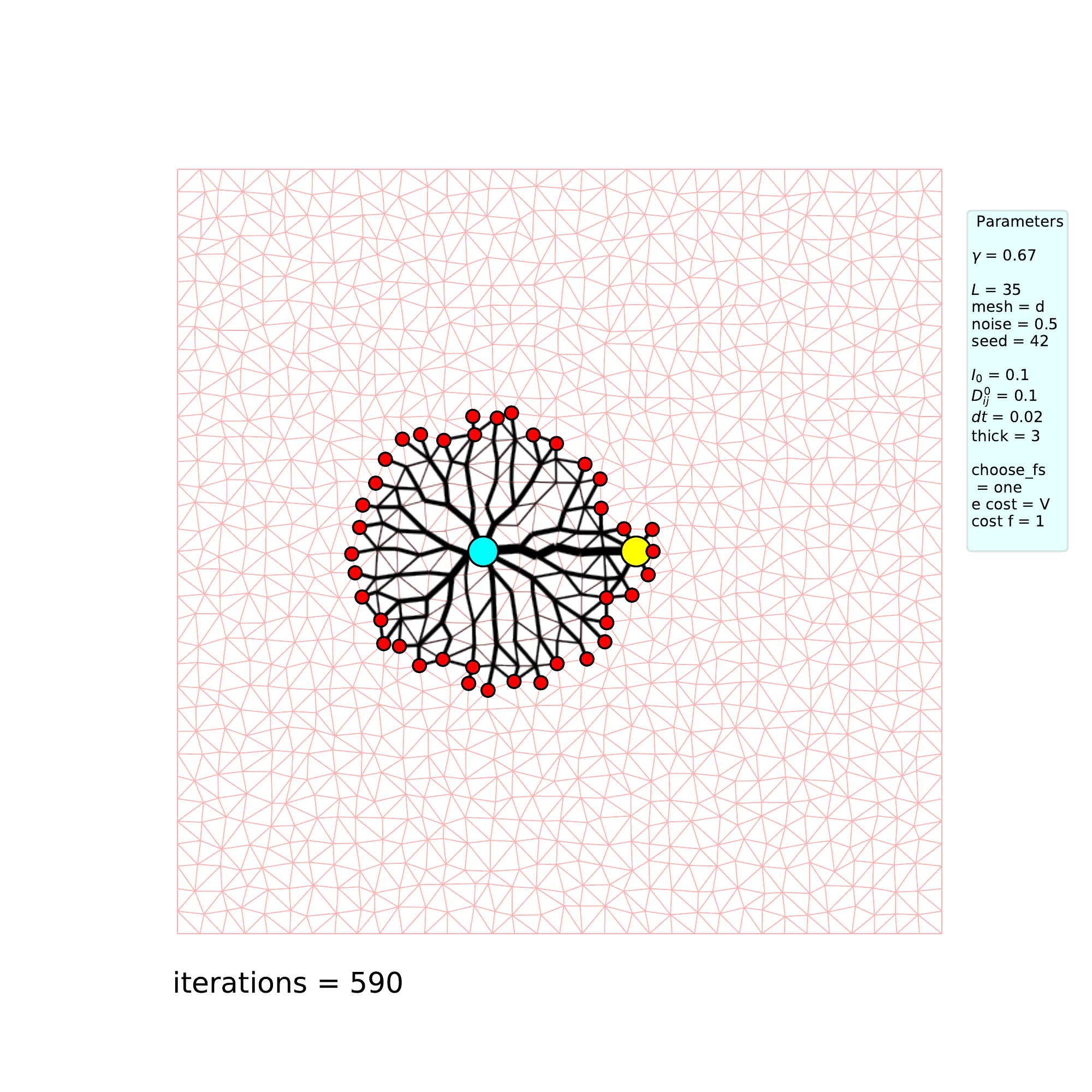} \hfill
\mysub[$t=1200$]{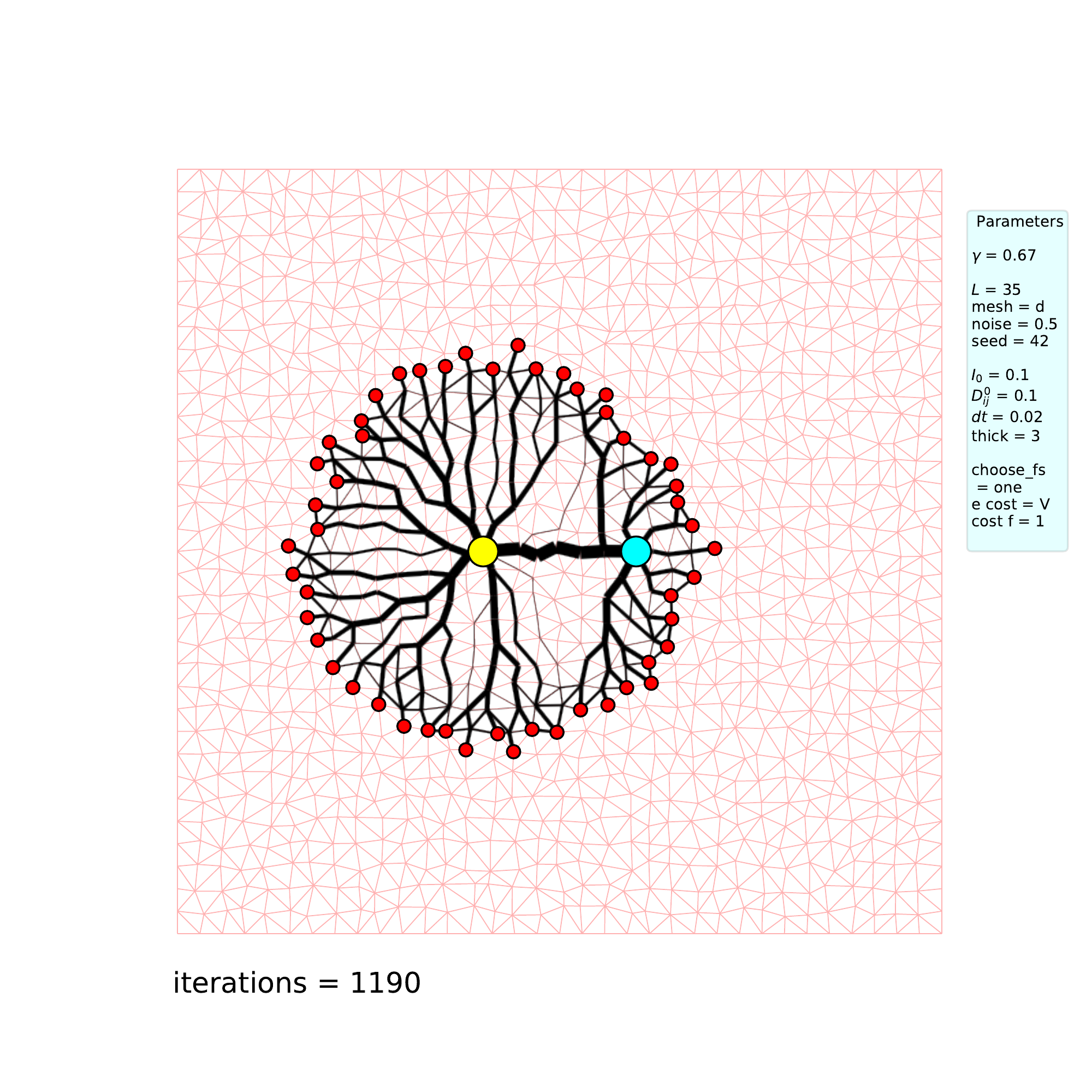} \\

\mysub[$t=1500$]{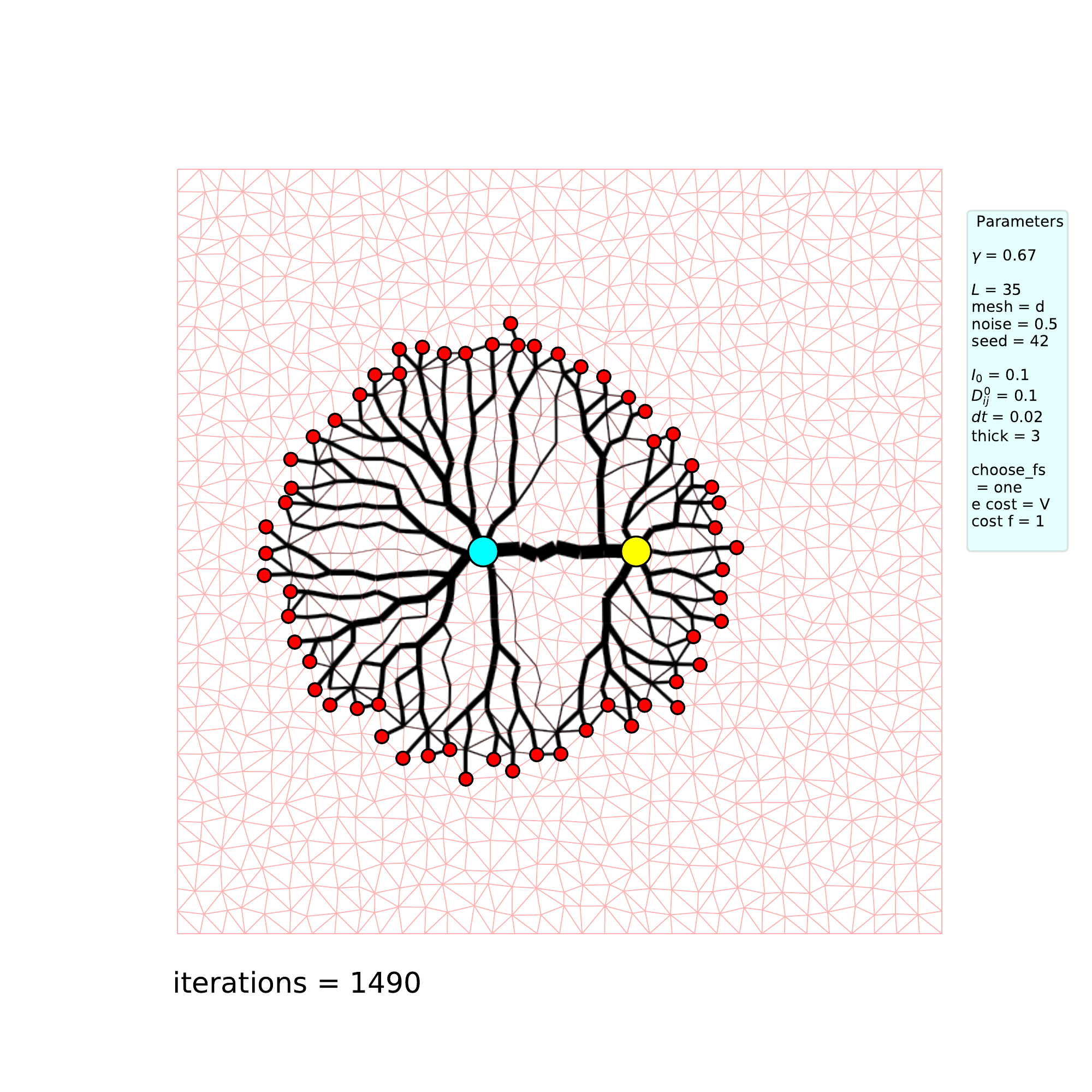} \hfill
\mysub[$t=2100$]{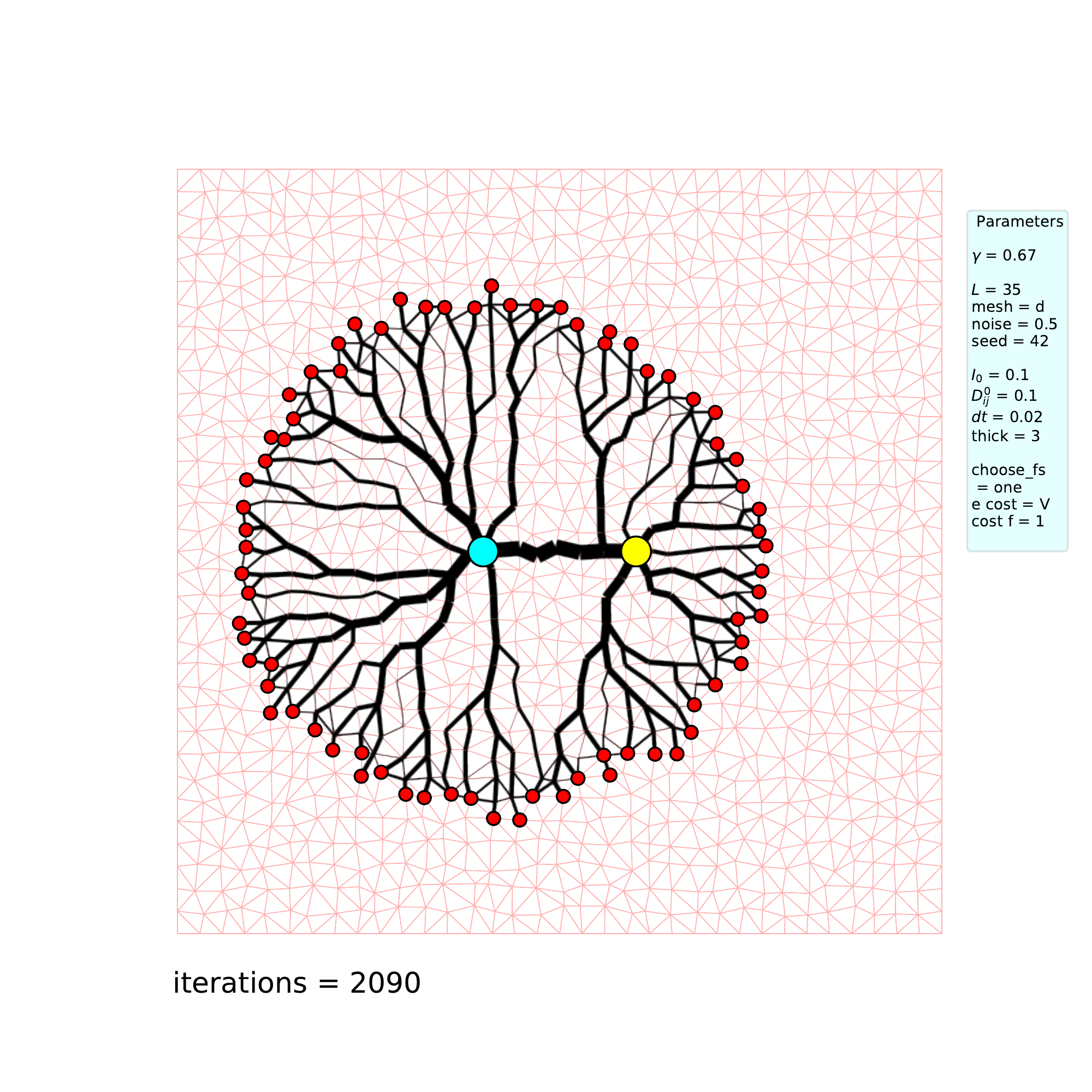} \hfill
\mysub[$t=2700$]{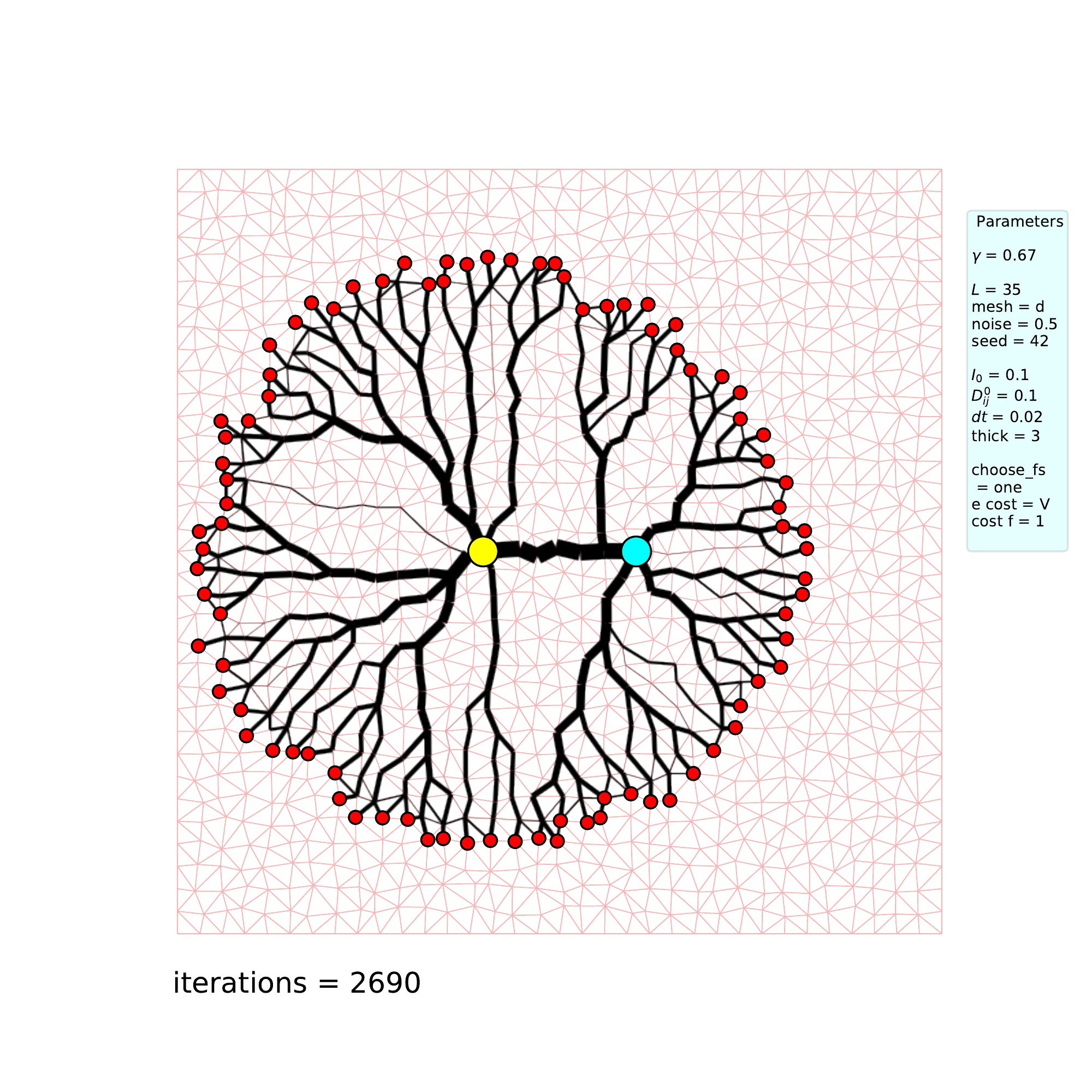} 

\caption{
\textit{Physarum} growing in the presence of two food sources supplying nutrients alternately.
Same settings as in Figure \ref{fig:growth_2_const}, but once the second food source is covered, at a given time step only one of them is randomly selected to supply the nutrients to the boundary sinks (red circles). The yellow circles represent the activated food source in a given instant, while the blue circles represent the inactive one.
This asynchronous operation of the food sources allows the formation of a short direct connection between them, similarly to what is observed in \textit{Physarum} (Figure \ref{fig:phy_2_FS}).
}
\label{fig:growth_2_random}
\end{figure}

In the second case, we considered that after the second food source was activated, only one of them was operational at a time. In each time step,  one of the food sources was randomly selected to supply the nutrients to the boundary nodes. The results can be found in Figure \ref{fig:growth_2_random}. 
The asynchronous operation of the food sources generates flow reversals which are a better approximation of the shuttle streaming behaviour of \textit{Physarum}. Conversely to the previous case and similar to what is observed in the real organism, this mechanism results in the formation of short connections between the food sources (Figure \ref{fig:phy_2_FS}).

\subsection{Discussion}

The results for one food source showed that the simple growth-optimisation model introduced cannot fully reproduce the observed network topology of \textit{Physarum}. Besides the distribution of node fluxes, this could be also related to the relative timescales between the growth and optimisation mechanisms, which are controlled by the channels' production cost, the choice of the update function, $g$, and the time increment $\dt$. The model should be carefully parametrised so that an eventual equilibrium between the processes could be achieved that would allow the persistence  of the redundant connections. 

Regarding the simulations of the growth in the presence of two food sources, the results suggest that the asynchronous operation of the food sources is far more biologically reasonable than the synchronous pumping of both. 

Further improvements to the model can be also made. One can  consider a more realistic distribution of the nodes fluxes, instead of an even distribution or a purely stochastic as considered here. For instance, the sinks can receive a fraction of nutrients proportional to the distance they are from the sources, as the farther away a region is from the source, the longer the nutrient will take to reach it. Alternatively, 
a more realistic approximation to the shuttle streaming might be achieved if the sinks (sources) receive (give) unequal amounts of nutrients which vary periodically in time.
In addition, some interior nodes could be also treated as sinks, as those regions also require nutrients. The cost of producing a channel could also include a second term  describing the metabolic costs inherent to the channels formation. Finally, it would be also interesting to test the dynamics with finite food sources which eventually run out.

\cleardoublepage

\chapter{Concluding remarks}
\label{chapter:conclusions}

\section{Achievements}

In this work, we studied the network development and adaptation of \Pp. It was first analysed the adaptation of static organisms, i.e., neglecting the network growth. We have derived a general class of equations describing the adaptation dynamics of a network comprised of elastic channels filled with an incompressible fluid subject to a \HP{} flow. The adaptation mechanism models the evolution of the channels' radii as a response to the flow flowing through them at a given moment, which is described by an arbitrary function $g$.  This adaptation dynamics was inspired by the previous model proposed by Tero \textit{et al.} \cite{Tero1}, but is more physically realistic, as it provides a correct description of a \HP{} flow by ensuring the conservation of the volume of the fluid over time. 

Following a Lagrangian approach, we proved that the particular choice of $g(|Q_{ij}|)=|Q_{ij}|^{2/3}$ minimises the total power dissipated by the network at the steady-state assuming a fixed set of sources and sinks. Considering these choices, it was shown that the final networks were \textit{trees} connecting the set of terminals, whose geometry is highly dependent on the specific distribution of the nodes' flux. In particular, some configurations led to apparently disconnected solutions which are unrealistic in the context of \textit{Physarum}. It was also observed that the same configuration of terminals could result in different steady-states depending on the initial conductivities. 

The previous choice of the function $g$ led us to study the  adaptation dynamics  of the general class of polynomial functions, $g_\gamma(|Q_{ij}|)=|Q_{ij}|^{\gamma}$, as a function of the parameter $\gamma$. This analysis revealed the existence of a first-order phase transition in the system near $\gamma = 1/2$, marked by the discontinuity of the derivative of dissipated power at the steady state with respect to $\gamma$ and a drastic change in the network topology. 

Some applications of the model were also tested. For graphs describing mazes, with a source and a sink placed at both entrances, it was shown that the adaptation dynamics converged to a single path connecting them, and therefore can reproduce the maze solving abilities of \textit{Physarum} \cite{Nakagaki2000}. However, for a general planar graph, the system doesn't always converge to the shortest path connecting the two terminals. 

Simulations have shown that typical \textit{tree} topologies resulting from the adaptation under fixed sources and sinks are not optimal, as they entail high costs to connect all the terminals and no redundancy, leading to a low transport efficiency. By contrast, the networks displayed by \textit{Physarum} show a good compromise between these metrics, and thus can't be mimicked by constant distribution of nodes fluxes, which naturally can't account for the periodical flow reversals (shuttle streaming). This justified the need of considering the adaptation subject to time-dependent sources and sinks to produce efficient and resilient networks, and as a better approximation to \textit{Physarum}'s shuttle streaming behaviour.

The influence of flux fluctuations on the network fitness was studied using a configuration of terminals that mimicked the geographical locations of major Portuguese cities. Since it's not clear what is the most appropriate choice, different stochastic methods of selecting the driving terminals in each time step were considered. The introduction of flux fluctuations enabled the formation of redundant connections  which provided robustness to the network. As expected, compared to the case of fixed terminals, the stochastic methods achieved a far better performance in every respect. In general, all the stochastic cases also achieved a better overall performance than the MST and the real Portuguese railway. The results also suggest that the choice of a completely random distribution (``All random'') might be a better alternative to the original method proposed by Tero et al. \cite{Tero1} of choosing a source-sink pair (``Random pair''), as it achieved the best overall cost-benefit relationship. Regarding the dependence on $\gamma$, it was  observed that the network resolves towards the MST solution and progressively loses all its redundancy as $\gamma $ is increased. For values of $\gamma \in [0.7, 1[$, the simulations resulted in networks with better overall performance than the MST and the real railway. A more rigorous analysis should be performed to support these results. In particular other values of $\, gamma$ and stochastic distributions can be considered. 

Lastly, we extended the adaptation model to incorporate the network growth, coupling both processes
and providing a better description of \textit{Physarum}'s foraging behaviour. The growth is driven by a nutrient flux from the food sources to the boundary of the organism where the nutrients are accumulated and later used in the formation of new channels. Considering only one source, we observed that this simple mechanism could mimic to some degree the natural growth of \textit{Physarum}, in particular the formation of the fan-shaped fronts. We have also considered the accommodation of new food sources as the organism grows. Since it's not clear how food consumption and distribution is managed in this case, we tested two hypotheses. We concluded that the synchronous operation of the food sources couldn't explain the observed behaviour, as the network optimisation leads to their repulsion, and no short connection  between them remains. Conversely, when the food sources operate alternately, a strong direct connection between the two is established, similarly to what is observed in \textit{Physarum}. This suggests that the second method is a closer approximation to the real mechanism. 

The general form of the adaptation dynamics enables the introduction of different choices for the  function $g$, depending on the specific application, allowing the simulation of other physical and biological  network systems such as fungi networks. It should be noted that the polynomial $g$ adopted here may lead to high conductivity values which are only constrained  by the network's fluid volume.  In this regard, it would be interesting to test other adaptation functions that saturate for large flux values \cite{Rubinow1972, Tero1}. In the context of graph theory, the model can be applied to several network optimisation problems  \cite{Gao2019, PpReview} and as a guide network construction in different domains. Some benchmarks should be carried against the \textit{Physarum Solver} to assess which is the fastest algorithm.

\section{Future work}

The model proposed here couldn't fully mimic the characteristic loopy structure of \textit{Physarum} networks, although different hypotheses of its origin were tested.
In particular, when growth is considered, the dynamics doesn't result in the formation of stable traversal interconnections of the main veins. Nevertheless, the resultant networks share some similarities with the real ones and with other network systems found in nature, such as tumour vascularisation and leaf venation. As an attempt to replicate the loopy patterns, more realistic time- and space-dependent distributions of terminals and other choices of adaptation functions can be considered. The functional form of the latter should be derived from theoretical expectations of the flow dynamics and the model parameters should be calibrated based on available experimental data. 

The contractile activity should also play a prominent role on the channels selection, which isn't explicitly accounted for in the model. The incorporation of the contractions would provide a more realistic description of the shuttle streaming and ultimately explain the self-organisation of the flows \cite{Alim2013}, without relying on an arbitrary distribution of terminals 
as the driving mechanism which the model assumes. This self-organised pulsating behaviour 
may be the key for the observed network structure. Hypothetically, this could be achieved by introducing the transport of signalling molecules released in stimulated regions that would drive the flow and the channels' adaptation based on a similar feedback dynamics \cite{Alim2017}. 

As an improvement to the growth mechanism, one can also consider a reaction-diffusion process at the boundary of the organism to describe the vessel development. The interplay between the nutrients supplied by the food sources and chemical regulators released at the boundary 
would provide a more realistic growth mechanism from where the fan-shaped fronts could naturally emerge.  Finally, the introduction of chemical agents into the model would also open doors to model the migration of the plasmodium which was not here addressed. The mobility should be mediated and directed by chemical stimuli \cite{DirectionalSensing} in order to explain the chemotatic behaviour of \textit{Physarum}. 

Fully modelling \textit{Physarum} remains a challenging and exciting task, due to the complexity of its behaviour and lack of understanding of the underlying mechanism. 
The search for a unified model that can reproduce all the network features and explain the intelligent behaviours of this fascinating brainless organism is still ongoing.

\cleardoublepage

\phantomsection
\addcontentsline{toc}{chapter}{\bibname}
\bibliographystyle{unsrt}
\bibliography{Thesis_0_Bibliography}

\cleardoublepage

\appendix

\end{document}